\documentclass[twocolumn]{aastex631}

\newcommand{\pc}{\textrm{pc}}

\newcommand{\HII}{\textsc{H\,ii} }
\defcitealias{Nebrin2024}{N24}
\defcitealias{Smith2025}{S25}

\usepackage{amsmath}
\usepackage{enumitem}

\makeatletter 
  \patchcmd{\NAT@citex}
    {\@citea\NAT@hyper@{%
      \NAT@nmfmt{\NAT@nm}%
      \hyper@natlinkbreak{\NAT@aysep\NAT@spacechar}{\@citeb\@extra@b@citeb}%
      \NAT@date}}
    {\@citea\NAT@nmfmt{\NAT@nm}%
    \NAT@aysep\NAT@spacechar\NAT@hyper@{\NAT@date}}{}{}

  \patchcmd{\NAT@citex}
    {\@citea\NAT@hyper@{%
      \NAT@nmfmt{\NAT@nm}%
      \hyper@natlinkbreak{\NAT@spacechar\NAT@@open\if*#1*\else#1\NAT@spacechar\fi}%
        {\@citeb\@extra@b@citeb}%
      \NAT@date}}
    {\@citea\NAT@nmfmt{\NAT@nm}%
    \NAT@spacechar\NAT@@open\if*#1*\else#1\NAT@spacechar\fi\NAT@hyper@{\NAT@date}}
    {}{}
\makeatother

\begin{document}

\title{Lyman-$\boldsymbol{\alpha}$ Pressure Strongly Enhances Pre-Supernova Feedback at Cosmic Dawn: \\ The First Multi-Dimensional Lyman-$\boldsymbol{\alpha}$ Radiation Hydrodynamics Simulations}

\correspondingauthor{Olof Nebrin}
\email{olof.nebrin@astro.su.se}

\author[0000-0003-3877-360X]{Olof Nebrin}
\affiliation{Department of Astronomy \& Oskar Klein Centre for Cosmoparticle Physics, \\ AlbaNova, Stockholm University, SE-106 91 Stockholm, Sweden}
\author[0000-0002-2838-9033]{Aaron Smith}
\affiliation{Department of Physics, The University of Texas at Dallas, Richardson, Texas 75080, USA}
\author[0000-0002-2512-6748]{Garrelt Mellema}
\affiliation{Department of Astronomy \& Oskar Klein Centre for Cosmoparticle Physics, \\ AlbaNova, Stockholm University, SE-106 91 Stockholm, Sweden} 
\author[0009-0005-3827-8774]{Kevin Lorinc}
\affiliation{Department of Physics, The University of Texas at Dallas, Richardson, Texas 75080, USA}
\author[0009-0009-1656-7769]{Daniele Manzoni}
\affiliation{Scuola Normale Superiore, Piazza dei Cavalieri 7, I-56126 Pisa, Italy}


\begin{abstract}
The dynamical role of Lyman-$\alpha$ (Ly$\alpha$) radiation pressure feedback has been debated for nearly a century, with recent analytical and 1D numerical studies highlighting its potential dominance over other stellar feedback processes at Cosmic Dawn. Despite this, no multi-dimensional Ly$\alpha$ radiation hydrodynamics (RHD) simulations have been performed to date. In this paper, we present the first 2D Ly$\alpha$ RHD simulations using \textsc{Lydion}, an RHD code with a novel M1 moment method for Ly$\alpha$ transfer, and self-consistent dust dynamics. \textsc{Lydion} yields a $\sim \mathcal{O}(100) \,\times$ speed-up compared to Monte Carlo radiative transfer in simple benchmarks, making 2D Ly$\alpha$ RHD feasible. We perform simulations of star clusters and isolated stars embedded in dense, metal-poor ($Z/Z_\odot \leq 0.01$) clouds, and find that Ly$\alpha$ feedback dramatically boosts outflows and dominates over feedback from direct and infrared radiation pressure. Ly$\alpha$ leakage through lower-column density channels, Doppler shifts, and Ly$\alpha$ photon destruction, while important, cannot prevent the build-up of strong Ly$\alpha$ radiation pressure in \textsc{H\,ii} regions, leading to radiative forces $\sim (2 - 16) \times L_{\rm bol}/c$, and Ly$\alpha$ force multipliers $M_{\rm F} \sim 10-60$. Ly$\alpha$ feedback may not preclude efficient star formation, but raises the threshold gas surface density for this to occur. We conclude that nearly all galaxy and star formation simulations are currently missing the strongest source of radiation pressure feedback in dense and metal-poor environments.
\end{abstract}



\section{Introduction} 
\label{sec:intro}
For nearly a century, great controversy has surrounded the potential dynamical importance of radiation pressure from multiple scatterings of Ly$\alpha$ ($1215.67 \, \text{Å}$) photons in and around \textsc{H\,ii} regions \citep[for early work, see e.g.][]{Ambarzumian1932, Zanstra1934, Chandrasekhar1945, Yada1957, George1973, Cox1985, Haehnelt1995, Henney1998, Odell1998, Oh2002, Dijkstra2008, McKeeTan2008}. In recent years, novel analytical solutions and modelling \citep[][]{Abe2018, Lao2020, Tomaselli2021, Nebrin2022, Kapoor2023, Nebrin2024, Smith2025}, fully coupled 1D Ly$\alpha$ radiation-hydrodynamic (RHD) simulations \citep[][]{George1973, Smith2016, Smith2017}, 1D and 3D simulations with approximate subgrid modelling \citep[][]{Kimm2018, Manzoni2025}, and 3D post-processing results \citep[][]{Smith2019, Menon2026}, suggest that Ly$\alpha$ feedback can dominate the pre-supernova feedback (`early feedback') momentum budget in dust-poor conditions. If true, this would have profound implications for the formation of the first stars, galaxies, and massive black holes, now probed by the \textit{James Webb Space Telescope} (\textit{JWST}):
\begin{itemize}[leftmargin=*]
    \item \textbf{\textit{Population III star formation}:} The very first stars formed at high redshifts in dust-free conditions. It has long been suspected that Ly$\alpha$ feedback could boost radiation pressure from these Population III (Pop III) stars by $\sim 1-3$ orders of magnitude, at least up until Ly$\alpha$ leakage following the breakout of the \textsc{H\,ii} region \citep[e.g.][]{Doroshkevich1976, Oh2002, McKeeTan2008, Stacy2012, Wise2012RadPressure, Jaura2022, Nebrin2022, Klessen2023, Nebrin2024}. This could more easily clear surrounding dense gas, and hence have strong implications for the Pop III star formation efficiency, the Pop III initial mass function, and escape fraction of ionizing photons.
    
    \item \textbf{\textit{Formation of the first Globular Clusters and Ultra-Faint Dwarf galaxies}:} The oldest globular clusters and Ultra-Faint Dwarf (UFD) galaxies are metal-poor, and formed in dust-poor conditions at high redshifts \citep[][]{Brown2014, Ricotti2016, Simon2019, Adamo2024}. \cite{Abe2018} suggested that Ly$\alpha$ feedback could be of critical importance during the formation of metal-poor star clusters, a conclusion which has been reinforced in both analytical modelling \citep[][]{Nebrin2022, Nebrin2024}, and 3D simulations with approximate subgrid prescriptions for Ly$\alpha$ feedback \citep[][]{Kimm2018}. Recent observational evidence of Ly$\alpha$ photon trapping around young metal-poor star clusters further hints at the potential importance of Ly$\alpha$ feedback \citep[][]{Pascale2024, Peng2025}.\footnote{\cite{Salgado2016} also presented observational evidence of significant Ly$\alpha$ heating of dust in the Orion Nebula, although this does not necessarily imply appreciable Ly$\alpha$ radiation pressure. \cite{Baldwin1991}, using \textsc{Cloudy}, estimated that the Ly$\alpha$ radiation pressure is dynamically important in Orion, at values of $\sim 60\%$ of the gas pressure. However, it would take a code with full Ly$\alpha$ RHD like \textsc{Lydion} to better test this. } The same conclusion likely holds for UFDs \citep[][]{Nebrin2022, Nebrin2024}. Unless Ly$\alpha$ feedback is properly accounted for, this could hamper current attempts to constrain $\Lambda$CDM using zoom-in galaxy formation simulations \citep[e.g.][]{Ricotti2016, Wheeler2019, Revaz2023, Andersson2025_EDGE_INFERNO, Brown2025}.

    \item \textbf{\textit{Formation and growth of black holes at Cosmic Dawn}:} Ly$\alpha$ feedback is a crucial missing ingredient in early black hole assembly within dense, primordial gas reservoirs \citep{Inayoshi2020}. In direct-collapse black hole (DCBH) scenarios, the extremely large neutral hydrogen column densities ($N_{\rm HI} \gtrsim 10^{24}\,{\rm cm}^{-2}$) render the environment highly optically thick to Ly$\alpha$ \citep{Dijkstra2016, Ge2017, Johnson2017, Smith2017}. \cite{Mushano2024} have recently shown that, at least for static 1D setups, Ly$\alpha$ radiation pressure can impede super-Eddington accretion onto black holes at Cosmic Dawn. In realistic 3D configurations, anisotropies in density and velocity fields naturally direct where the radiation pressure is most impactful \citep{SmithDCBH2017}. More broadly, the similar nature of Little Red Dots (LRDs) invites explorations of Ly$\alpha$ feedback in these contexts too.
\end{itemize}

\begin{figure*}
\centering
\includegraphics[width=0.96\textwidth]{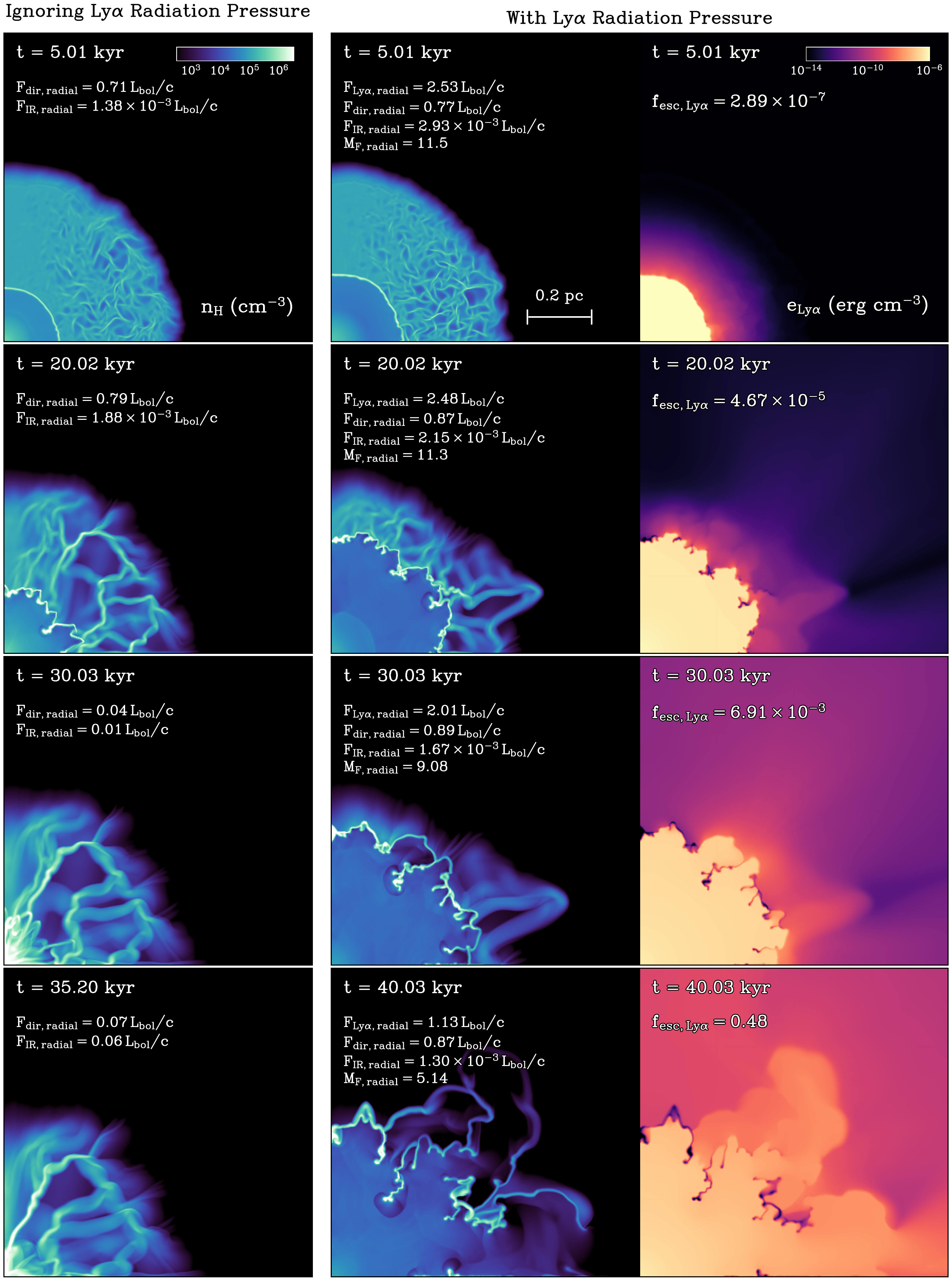}
\caption{ Simulations of a compact ($R_{\rm h} = 0.05 \, \rm pc$), low-mass ($10^4 \, \rm M_\odot$) star cluster in an initially dense ($n_{\rm H} = 10^5 \, \rm cm^{-3}$), dust-poor ($\mathfrak{D}/\mathfrak{D}_\odot = 0.01$) cloud, with (right panels) and without (left panels) Ly$\alpha$ radiation pressure feedback. Plots show the gas density $n_{\rm H}$, and, for the run with Ly$\alpha$ RT, the Ly$\alpha$ energy density $e_{\rm Ly\alpha}$. Also shown are forces from Ly$\alpha$, direct, and IR radiation pressure, as well as the Ly$\alpha$ escape fraction (also see Figs.~\ref{fig: M_F 512 run}--\ref{fig: Feedback comparison 512 sim}). }
\label{Lya vs no Lya figure 512 D0.01}
\end{figure*}

\begin{figure*}
\centering
\includegraphics[width=0.96\textwidth]{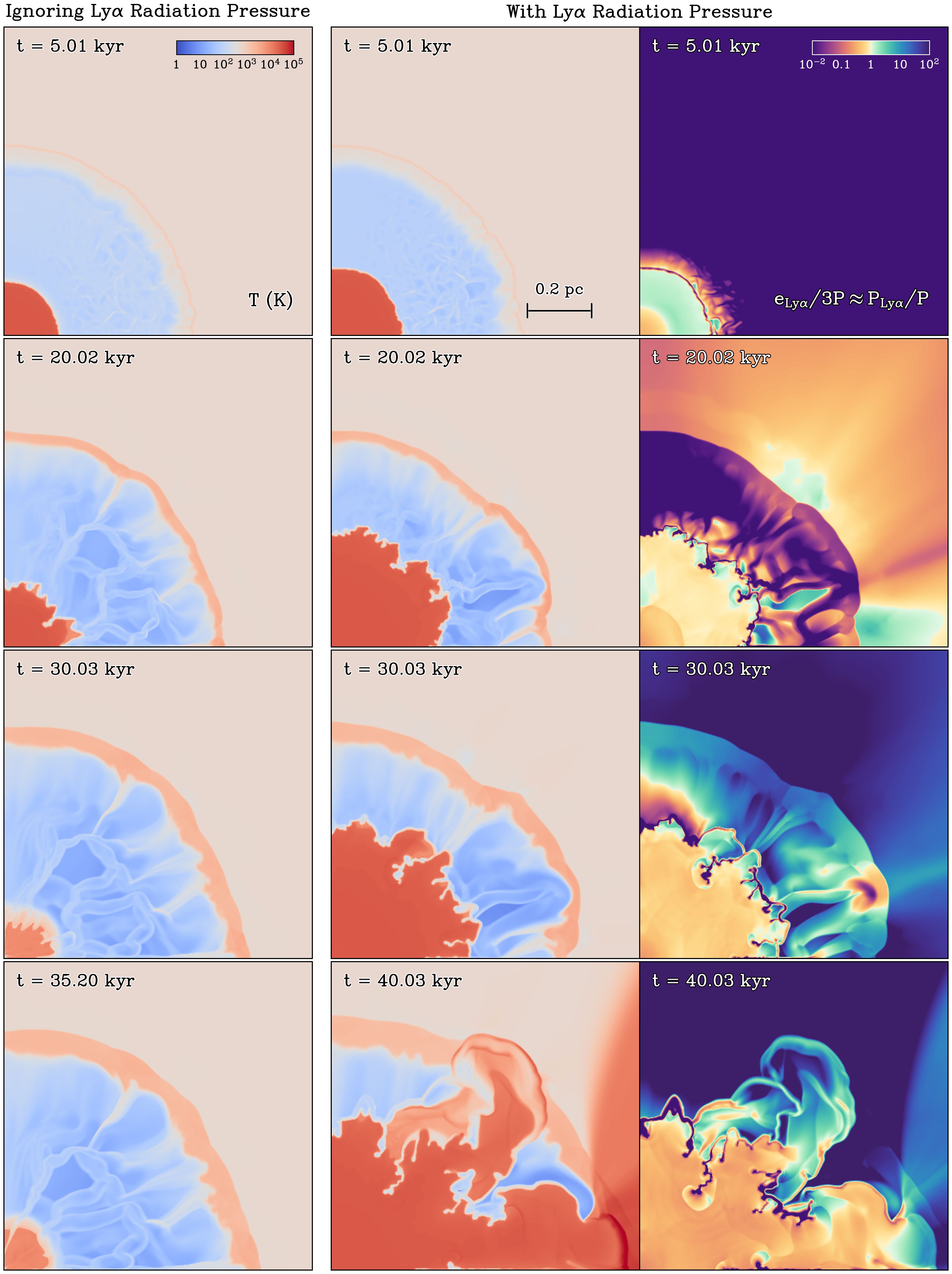}
\caption{ Same as Fig.~\ref{Lya vs no Lya figure 512 D0.01}, but showing the gas temperature and $e_{\rm Ly\alpha}/3P$, which is approximately the ratio between the Ly$\alpha$ pressure and the gas pressure. Near the ionization front, the Ly$\alpha$ pressure is greater than or comparable to the gas pressure. }
\label{Lya vs no Lya figure 512 D0.01 T and PLya}
\end{figure*}

Despite many decades of debate about Ly$\alpha$ feedback, it has proven highly challenging, if not infeasible, to incorporate this process on-the-fly in simulations. This is because conventional Ly$\alpha$ Monte Carlo radiative transfer (MCRT) codes tend to be extremely computationally demanding \citep[see discussion in][]{Kimm2018, Smith2018DDMC, Hopkins2020_Radfeedback, Byrohl2025, Lorinc2025}.\footnote{\cite{Smith2018DDMC} have shown that discrete diffusion MCRT could dramatically speed up MCRT codes in the optically thick regime relevant for Ly$\alpha$ feedback. Recently, \cite{Byrohl2025} have also demonstrated that GPU acceleration could render MCRT Ly$\alpha$ RHD feasible in the near future. } It is mainly for this reason that recent studies have relied on 1D simulations and analytical results. However, it is hard to draw robust conclusions about the importance of Ly$\alpha$ feedback in a realistic clumpy and turbulent interstellar medium (ISM), without the aid of 2D or 3D simulations that can capture Ly$\alpha$ leakage through low-column density channels \citep[][]{Behrens2014, Kimm2019, Kakiichi2021, Yuan2024}. Although recent approximate analytical results predict that strong Ly$\alpha$ feedback can survive in a turbulent ISM \citep[][]{Nebrin2024}, this remains to be tested in self-consistent simulations. Indeed, several authors have highlighted multi-dimensional Ly$\alpha$ RHD as a high-priority goal for understanding star and galaxy formation at Cosmic Dawn \citep[e.g.][]{Smith2018DDMC, Tomaselli2021, Jaura2022, Klessen2023, Thompson2024, Nebrin2024, Peng2025}.

In this paper, we perform the first 2D Ly$\alpha$ RHD simulations, using a new code, \textsc{Lydion} (\textbf{Ly}man-$\alpha$ \textbf{Di}ffus\textbf{ion}). Departing from traditional Ly$\alpha$ MCRT methods, \textsc{Lydion} employs a novel M1 moment method \citep{Levermore1984}, combined with a Fokker--Planck treatment of frequency redistribution \citep{Rybicki2006}. Besides on-the-fly Ly$\alpha$ RHD, \textsc{Lydion} also incorporates comprehensive physical modelling of stellar feedback, dust dynamics, and ISM photo-thermochemistry. In recent years, 2D simulations have elucidated the role of direct UV and indirect infrared (IR) radiation pressure feedback in both idealized wind-tunnel simulations \citep[][]{Davis2014, Zhang2017, Rosdahl2015, Smith2020_Arepo}, and in simulations of young massive (proto-)stars \citep[e.g.][]{Kuiper2018, Fukushima2020, Martini2026}. By including self-consistent Ly$\alpha$ radiation pressure for the first time, along with several additional physical processes often neglected in previous studies (e.g. dust dynamics), \textsc{Lydion} builds on this tradition, and enables more comprehensive exploration of stellar feedback in dense star-forming environments.

As a first exploratory study, we apply \textsc{Lydion} to study the importance of Ly$\alpha$ feedback in metal-poor environments ($0\leq Z/Z_\odot \leq 0.01$), relevant for the formation of old globular clusters, UFD galaxies, and Population III stars. We highlight the results from our highest resolution ($512^2$) simulations in Figs.~\ref{Lya vs no Lya figure 512 D0.01}--\ref{Lya vs no Lya figure 512 D0.01 T and PLya}, where we simulate the feedback from a $10^4 \, \rm M_\odot$ star cluster, embedded in a dense ($n_{\rm H} = 10^5 \, \rm cm^{-3}$), metal/dust-poor gas cloud ($Z/Z_\odot = \mathfrak{D}/\mathfrak{D}_\odot = 0.01$, where $\mathfrak{D}$ is the dust-to-gas ratio). In the simulation ignoring Ly$\alpha$ feedback, photoionization, stellar and IR radiation pressure fail to drive outflows, leading to recollapse (and presumably more star formation). In the simulation incorporating Ly$\alpha$ radiation pressure, the cloud is destroyed. We describe this simulation, and several others, in greater detail in this paper, along with tests of the basic methodology of \textsc{Lydion}.

\begin{table}
\centering
\caption{
A table of contents for this paper, showing where to find more details about a given topic or modelling aspect of \textsc{Lydion}.
}
\begin{tabular}{l l}
\hline
\hline
\noalign{\vskip 2pt}

Topic(s) & Where to find  \\

\noalign{\vskip 3pt}
\hline
\hline
\noalign{\vskip 2pt}

Methodology for Ly$\alpha$ RT$^\dagger$ & Sec.~\ref{M1 Lya RT methodology sec} \& Appendix~\ref{frequency diffusion appendix}  \\

\noalign{\vskip 5pt}

Schematic overview of \textsc{Lydion} & Fig.~\ref{LYDION RHD operator split method} \\

\noalign{\vskip 5pt}

Hydrodynamics \& gravity$^\S$ & Sec.~\ref{sec hydro, dust dynamics, gravity}, Appendix~\ref{Hydro Appendix}, \\

 & Appendix~\ref{Poisson equation Appendix} \\

\noalign{\vskip 5pt}

Dust dynamics \& physics & Sec.~\ref{sec hydro, dust dynamics, gravity} \& Appendix~\ref{Dust physics} \\

\noalign{\vskip 5pt}

Stellar \& infrared RT$^\textrm{\textdaggerdbl}$ & Sec.~\ref{Stellar and IR transfer and dust modelling}, Appendix~\ref{Stellar RT appendix}, \\

 & Appendix~\ref{IR appendix} \\

\noalign{\vskip 5pt}

Photo-thermochemistry & Sec.~\ref{Photothermochemistry section} \& Appendix~\ref{Photochemistry appendix} \\

\noalign{\vskip 5pt}

Tests of Ly$\alpha$ RT methodology & Sec.~\ref{Test section} \\

\noalign{\vskip 5pt}

RHD simulations of embedded & Sec.~\ref{Lya RHD star cluster section} \\

metal-poor star clusters &  \\

\noalign{\vskip 5pt}

RHD simulations of embedded & Sec.~\ref{Lya RHD isolated star section} \\

isolated metal-poor stars &  \\

\noalign{\vskip 5pt}

Discussion & Sec.~\ref{Discussion sec} \\ 

\noalign{\vskip 5pt}

Summary \& conclusions & Sec.~\ref{Summary sec} \\

\noalign{\vskip 2pt}
\hline
\hline
\end{tabular}
\vspace{1 pt}\\
\raggedright
{\footnotesize
$^\dagger$: The overall new scheme is described in Sec.~\ref{M1 Lya RT methodology sec}, with further details about the implicit frequency redistribution method in Appendix~\ref{frequency diffusion appendix}. \\ 
$^\S$: Sec.~\ref{sec hydro, dust dynamics, gravity} gives an overview of the implementation of hydrodynamics and self-gravity. Detailed tests of the hydrodynamics code can be found in Appendix~\ref{Hydro Appendix}. The implementation of self-gravity is described and tested in detail in Appendix~\ref{Poisson equation Appendix}. \\
$^\textrm{\textdaggerdbl}$: Sec.~\ref{Stellar and IR transfer and dust modelling} gives a broad overview. Stellar RT is described and tested in more detail in Appendix~\ref{Stellar RT appendix}. Appendix~\ref{IR appendix} discusses the implementation of infrared RT. }
\label{table:paper_guide}
\end{table}

\begin{figure*}
\centering
\includegraphics[width=0.8\textwidth, trim=0 3cm 0 0, clip]{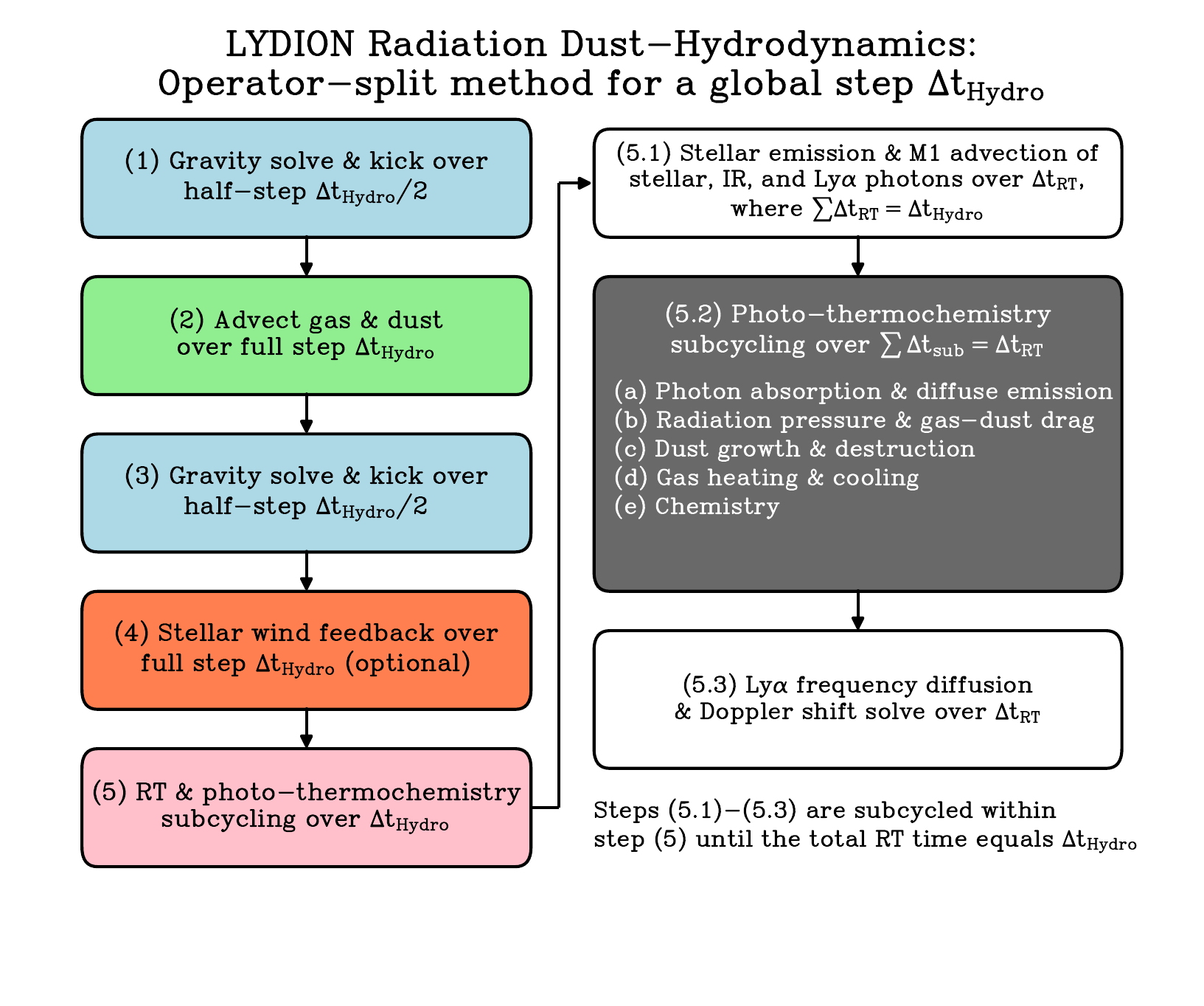}
\caption{A schematic overview of the operator-split method used in \textsc{Lydion} to advance gas, dust, radiation (including Ly$\alpha$), and thermochemistry over each global time-step $\Delta t_{\rm Hydro}$. Each step is described in more detail in the main text and Appendices, although not necessarily in the order they are updated in a simulation time-step (the actual order used is shown in this figure).}
\label{LYDION RHD operator split method}
\end{figure*}

The paper is structured as follows. In Sec.~\ref{Methodology} we outline the methodology underlying \textsc{Lydion}, mostly focused on the new M1 + Fokker--Planck method for Ly$\alpha$ RT. Detailed discussion and tests of the hydrodynamics code can be found in Appendix~\ref{Hydro Appendix}. In Sec.~\ref{Test section} we verify the M1 Ly$\alpha$ RT method by comparing predictions against analytical Ly$\alpha$ RT solutions and MCRT results. In Sec.~\ref{Sec: RHD simulations} we apply \textsc{Lydion} to study feedback-driven outflows from metal-poor star clusters and isolated stars, embedded in dense, metal/dust-poor gas clouds. We discuss the robustness and implications of our results in Sec.~\ref{Discussion sec}. Finally, we conclude and summarize our results in Sec.~\ref{Summary sec}. To aid the reader in navigating the paper, a more detailed table of contents can also be found in Table~\ref{table:paper_guide}. A schematic overview of the \textsc{Lydion} methodology for RHD simulations can be found in Fig.~\ref{LYDION RHD operator split method}.

\section{Methodology}
\label{Methodology}

In this section we discuss the methodology underlying the \textsc{Lydion} code. \textsc{Lydion} is a CPU-based code, written in the \textsc{Julia} programming language \citep{Julia2017},\footnote{We also intend to explore implementations of \textsc{Lydion} in other codebases or languages, for example C$^{++}$, as well as to implement GPU acceleration. } which has facilitated a combination of fast code development and good performance. 

\subsection{A new M1 moment method for Lyman-$\alpha$\\ radiative transfer}
\label{M1 Lya RT methodology sec}

In \textsc{Lydion}, we solve the Ly$\alpha$ RT equation using a new two-moment method, working in the comoving frame. To proceed, we define the frequency-dependent zeroth moment of the intensity (the mean intensity) $J(\boldsymbol{r},\nu) \equiv \int \text{d}\Omega \, I(\boldsymbol{r}, \nu,\boldsymbol{n}) / 4\pi$, first moment $\boldsymbol{H}(\boldsymbol{r},\nu) \equiv \int \text{d}\Omega \, I(\boldsymbol{r}, \nu,\boldsymbol{n}) \, \boldsymbol{n} / 4\pi$ (related to the flux, $\boldsymbol{F} = 4 \pi \boldsymbol{H}$), and second moment $\boldsymbol{\mathsf{K}}(\boldsymbol{r},\nu) \equiv \int \text{d}\Omega \, I(\boldsymbol{r}, \nu,\boldsymbol{n}) \, \boldsymbol{n} \otimes \boldsymbol{n} / 4\pi$ (related to the radiation pressure tensor, $\boldsymbol{\mathsf{P}} = 4 \pi \boldsymbol{\mathsf{K}} / c$).

To order $\mathcal{O}(\boldsymbol{u}/c)$ in the gas velocity, the mean intensity in the comoving frame is then governed by \citep[][]{Castor2004, Nebrin2024, Smith2025}:\footnote{Here we are using the short-hand notation \citep[e.g.][]{Castor2004}: \newline $\boldsymbol{\mathsf{K}}\boldsymbol{:}\boldsymbol{\nabla u} = \sum_{i,j} \boldsymbol{\mathsf{K}}_{i,j}\partial_i u_j$.} 
\begin{align}
    \dfrac{1}{\Tilde{c}} \dfrac{\partial J}{\partial t}  ~&=~ \underbrace{j_{\rm s}}_{\text{Emission}} -  \underbrace{\boldsymbol{\nabla}\boldsymbol{\cdot}\left(\boldsymbol{H} + \dfrac{\boldsymbol{u}}{c} J \right)}_{\text{Spatial flux/advection}} \label{Full J equation} \\ ~&+~ \underbrace{ \dfrac{\partial}{\partial \nu} \left\{ \dfrac{1}{2} \Delta \nu_{\rm D}^2 \alpha \mathcal{H} \left( \dfrac{\partial J}{\partial \nu} + \mathfrak{R}J \right) \right\} }_{\text{Frequency diffusion \& atomic recoil}}  \nonumber \\ ~&+~ \underbrace{\left[\dfrac{\partial (\nu \boldsymbol{\mathsf{K}})}{\partial \nu} -  \boldsymbol{\mathsf{K}} \right] \boldsymbol{:} \dfrac{\boldsymbol{\nabla}\boldsymbol{u}}{c} }_{\text{Doppler shift \& radiation pressure work}} \nonumber \\ ~&-~ \underbrace{ [\alpha_{\rm c} (1-\omega) + p_{\rm d} \alpha \mathcal{H} ] J }_{\text{Ly}\alpha~\text{absorption/destruction}} \nonumber
\end{align}
where $\Tilde{c} \leq c$ is the possibly reduced speed of light in numerical simulations, $\alpha \equiv n_{\rm HI} \sigma_{0}$ is the Ly$\alpha$ absorption opacity at line center, $\mathcal{H}$ the dimensionless Hjerting-Voigt profile, $\alpha_{\rm c}$ and $\omega$ the continuum extinction opacity and albedo, respectively, and $p_{\rm d} \ll 1$ the Ly$\alpha$ destruction probability per scattering event. A Fokker--Planck approximation to frequency redistribution has been employed in Eq.~(\ref{Full J equation}), following \cite{Rybicki2006}. With the Fokker--Planck approximation we avoid having to solve an integro-partial differential equation, which is significantly more expensive to do.

The first moment, in the comoving frame, is governed by \citep[][]{Castor2004, Nebrin2024}:\footnote{Following discussion and arguments in \cite{Buchler1983}, \cite{Castor2004}, and \citet{Nebrin2024}, we have neglected other velocity terms in the equation for $\boldsymbol{H}$ that are sub-dominant in all regimes of interest. This is further tested in Test 2 of Sec.~\ref{Test section}. }
\begin{align}
    \dfrac{1}{\Tilde{c}} \dfrac{\partial \boldsymbol{H}}{\partial t} ~&=~ - \underbrace{\boldsymbol{\nabla}\boldsymbol{\cdot}\left( \boldsymbol{\mathsf{K}} + \dfrac{\boldsymbol{u}}{c} \boldsymbol{H} \right)}_{\text{Spatial flux/advection}}  -\underbrace{\alpha_{\rm F} \boldsymbol{H}}_{\text{Ly$\alpha$ flux absorption}} \, , \label{Full H equation}  \\ \alpha_{\rm F} &=~ \alpha \mathcal{H} + \alpha_{\rm c}(1 - \omega g) \, .
\end{align}
where $g$ is the continuum scattering asymmetry (e.g. by dust). To close the system and get a two-moment method, we adopt the M1 moment closure approximation of \cite{Levermore1984}:
\begin{equation}
    \boldsymbol{\mathsf{K}} = \left[ \dfrac{1-\chi}{2} \boldsymbol{\mathsf{I}} + \dfrac{3\chi-1}{2} \boldsymbol{h}\otimes\boldsymbol{h} \right] J = \boldsymbol{\mathsf{D}} J \, , \label{M1 closure}
\end{equation}
where $\boldsymbol{\mathsf{I}}$ is the identity matrix, $\boldsymbol{h} \equiv \boldsymbol{H} / \lvert \boldsymbol{H} \rvert$, and $\chi$ is a function of the reduced flux $f \equiv \lvert \boldsymbol{H} \rvert /J \leq 1$:\footnote{Other M1 closures exist, which have slightly different functional forms for $\chi (f)$. However, these other closures have the same qualitative behaviour, and are therefore not expected to change any conclusions here.}
\begin{equation}
    \chi = \dfrac{3+ 4 f^2}{5 + 2 \sqrt{4 - 3 f^2}} \, .
\end{equation}
The M1 moment closure, although not perfect, can model both the diffusive optically thick regime for Ly$\alpha$ transfer ($\boldsymbol{\mathsf{D}} \simeq \boldsymbol{\mathsf{I}}/3$), and beam-like Ly$\alpha$ escape through low-column density channels ($\boldsymbol{\mathsf{D}} \simeq \boldsymbol{h} \otimes\boldsymbol{h}$). The M1 closure in Eq.~(\ref{M1 closure}) has also been used in similar contexts to study infrared radiation pressure feedback \citep[][]{Rosdahl2015, Rosdahl_Agertz2015, Skinner2015, Kannan2019}. 

The use of an M1 moment method also has the advantage over flux-limited diffusion models that the spatial transfer of Ly$\alpha$ photons can be computed with an explicit rather than an implicit scheme, which will offset some of the numerical costs of 2D Ly$\alpha$ RT. However, unlike the spatial transfer, the frequency diffusion and Ly$\alpha$ absorption/destruction must still be treated implicitly to avoid severe time-step constraints.\footnote{If we were to solve the frequency diffusion step (Eq.~\ref{Frequency step}) explicitly, the time-step constraint would be $\Delta t_{\rm RT} < \min[(\Delta \nu / \Delta \nu_{\rm D})^2 / (\Tilde{c} \alpha \mathcal{H})]$. Since $\mathcal{H} = 1$ and $(\Delta \nu / \Delta \nu_{\rm D})^2 = (30 \, \textrm{K} / T)$ at line center (for our discretization), this would give a tiny time-step, $\Delta t_{\rm RT} < (\Delta r / \Tilde{c}) (30 \, \textrm{K} /T) / \tau_{\rm cell}$, where $\tau_{\rm cell} = \alpha \Delta r$ is the cell optical depth at Ly$\alpha$ line center. Since cells can have, e.g., $\tau_{\rm cell} \sim 10^8$, this is tiny compared to the CFL time-step, $\sim \Delta r / \Tilde{c}$. This rules out an explicit treatment. } As a result, we solve Eqs.~(\ref{Full J equation})--(\ref{Full H equation}) using an operator split, implicit--explicit (IMEX) method. We discretize $J$ and $\boldsymbol{H}$, as $J \rightarrow J_{i,j,k}$ and $\boldsymbol{H} \rightarrow \boldsymbol{H}_{i,j,k}$, where $(i,j)$ denote the spatial cell in cylindrical $(R,Z)$ coordinates (assuming axisymmetry), and $k$ denotes the frequency bin. For the frequency bins, we follow \cite{Smith2018DDMC} and use piecewise-constant bin-averaged values of $\mathcal{H}$ (see Appendix~\ref{frequency diffusion appendix} for details).

\begin{figure*}
\centering
\includegraphics[width=0.8\textwidth]{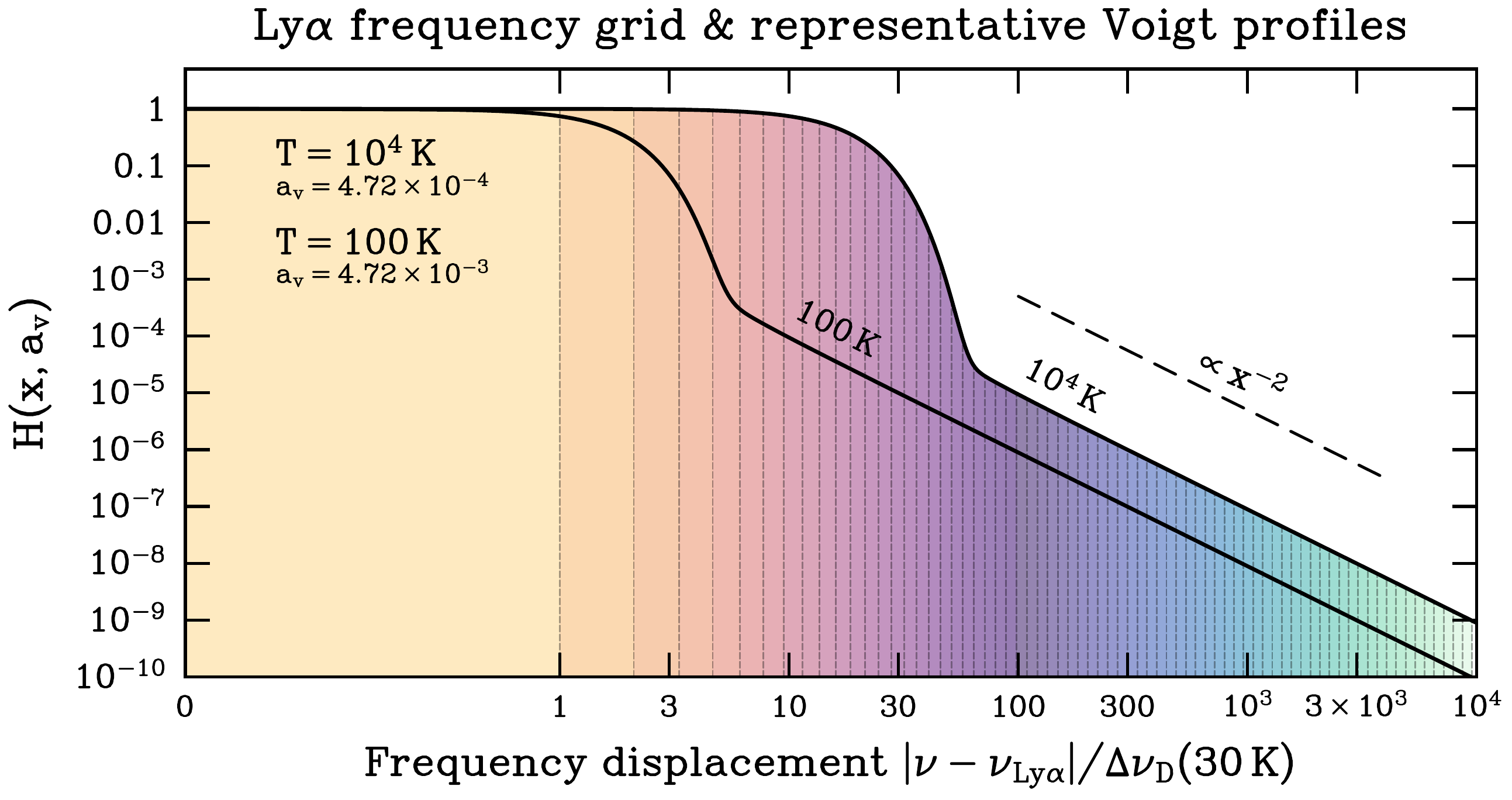}
\caption{ The default Ly$\alpha$ frequency grid adopted in \textsc{Lydion}, along with the Voigt profile for two representative temperatures, $T = 100 \, \rm K$ and $T = 10^4 \, \rm K$. Dashed lines show the frequency bin interfaces, highlighting that the core of the line is always resolved. Note that the full frequency grid is not shown here, which extends to $\lvert \nu - \nu_{\rm Ly\alpha} \rvert = 10^5 \, \Delta \nu_{\rm D} (30 \, \rm K)$.}
\label{Frequency grid plot}
\end{figure*}

The required frequency range for Ly$\alpha$ transfer is difficult to determine
\textit{a priori} in an RHD simulation. Analytically, it is expected to be $\lvert \nu - \nu_{\rm Ly\alpha} \rvert \lesssim \textrm{few} \times (a_{\rm v}\tau_0)^{1/3} \, \Delta \nu_{\rm D}$, where $a_{\rm v} \simeq 4.7 \times 10^{-4} \, T_4^{-1/2}$ is the Voigt parameter, and $\tau_0 \simeq 5.9 \times 10^{-14} \, T_4^{-1/2} N_{\rm HI}$ is the Ly$\alpha$ optical depth at line center \citep[e.g.][]{Neufeld1990, Lao2020, Nebrin2024, Smith2025}. Ensuing velocity gradients, variations in temperature and the \textsc{H\,i} column density $N_{\rm HI}$ can therefore greatly affect the required range of the frequency grid. Furthermore, the core of the line must be well-resolved to capture most scatterings, as well as Ly$\alpha$ destruction processes (e.g. $2p \rightarrow 2s$ transitions). We therefore adopt a non-uniform fixed frequency grid, similar to \cite{Mushano2024}. In particular, we use a geometrically stretched grid in frequency-displacement space $\delta \nu = \nu - \nu_{\rm Ly\alpha}$, symmetric around line center. The two central bins have width $\Delta \nu_{\rm D}(T_{\rm min})$, and the bin widths increase according to $\Delta \nu_{k+1} = q \,\Delta \nu_k$ when moving outwards into the wings, where $q=1.1$ by default. The full frequency grid covers $\lvert \delta \nu\rvert \leq \min[ 10^5 \, \Delta \nu_{\rm D}(T_{\rm min}), \,  0.99 \, \nu_{\rm Ly\alpha}, \, \nu_{\rm LyC} - \nu_{\rm Ly\alpha}]$. We choose $T_{\rm min} = 30 \, \rm K$ by default,\footnote{$T_{\rm min}$ only determines the target frequency resolution for Ly$\alpha$ RT, not the actual minimum gas temperature in RHD simulations. For Ly$\alpha$ feedback, most scatterings take place in gas of temperature $T \sim 50 - \textrm{few} \times 10^4 \, \rm K$, which informs our choice for $T_{\rm min}$. However, the gas in our RHD simulations can, in principle, cool down to $0.01 \, \rm K$, the actual numerically imposed temperature floor in \textsc{Lydion}. In practice however, photoelectric heating, inefficient cooling (below $\sim 10 \, \rm K$), and the CMB background, prevents cooling to such low temperatures. } which yields $N_\nu = 194$ frequency bins. In Fig.~\ref{Frequency grid plot} we show the adopted frequency grid, along with the Voigt profile for two representative temperatures, $T = (100, 10^4 ) \, \rm K$.

Our method is formally first-order accurate in time, and second-order accurate in space and frequency. Thus, our overall update scheme for a radiative transfer time-step $\Delta t_{\rm RT}$ is as follows, in order:   
\begin{enumerate}[leftmargin=*]
    \item \textbf{Advection step:} We first treat advection of Ly$\alpha$ photons by solving: 
    \begin{align}
         \dfrac{1}{\Tilde{c}} \dfrac{\partial J}{\partial t}  ~&=~ - \boldsymbol{\nabla}\boldsymbol{\cdot}\left(\boldsymbol{H} + \dfrac{\boldsymbol{u}}{c} J \right) - \dfrac{\boldsymbol{\nabla}\boldsymbol{u}}{c} \boldsymbol{:}\boldsymbol{\mathsf{K}}    \, , \\ \dfrac{1}{\Tilde{c}} \dfrac{\partial \boldsymbol{H}}{\partial t} ~&=~ - \boldsymbol{\nabla}\boldsymbol{\cdot}\left( \boldsymbol{\mathsf{K}} + \dfrac{\boldsymbol{u}}{c} \boldsymbol{H} \right) \, .
    \end{align}
    We solve these equations explicitly for every active frequency bin (see below), using a finite-volume method.\footnote{The radiation pressure work term, $-\boldsymbol{\nabla}\boldsymbol{u}\boldsymbol{:}\boldsymbol{\mathsf{K}}/c$, has a magnitude small enough to permit an explicit treatment. } We use a dimensionally unsplit approach, with first-order time integration. While higher-order time integration methods exist (e.g. RK2) for M1 advection \citep[e.g.][]{Melon2019, He2024_AsymptoticPreserving}, they also reduce to first-order accuracy in the optically thick regime of interest to us, while being at least twice as numerically expensive for fixed $\Delta t_{\rm RT}$, and harder to combine with subcycling. For these reasons, we have chosen to use a first-order time integration. In contrast, second-order spatial reconstruction has been shown to be more important for accuracy in M1 transport \citep[][]{Kannan2019}. To ensure second-order accuracy in space, we linearly reconstruct $J$ and $\boldsymbol{f} =  \boldsymbol{H}/J$ at cell interfaces \citep[see e.g.][]{Wikbing2022}, and use a MinMod slope limiter.\footnote{Early in testing we also tried the related approach of \cite{Kannan2019}, wherein the magnitude $f$ is extrapolated to the interface, but not the flux direction. However, with this approach we found greater deviations from spherical symmetry in tests of expanding \textsc{H\,ii} regions in uniform media. We have found better symmetry preservation when the flux directions are reconstructed, which is the approach adopted in \textsc{Lydion} for M1 transport of both Ly$\alpha$ and other photons. In case $f > 1$ at an interface, we scale down the components accordingly \citep[][]{Wikbing2022}. } We adopt the Global Lax-Friedrichs (GLF) approximate Riemann solver,\footnote{We have tried the HLL flux too, but find it to be slightly inferior at maintaining spherical symmetry, consistent with e.g. \cite{Rosdahl2013}. Furthermore, with the adopted second-order spatial reconstruction, there is little advantage in using HLL over GLF to capture shadows \citep{Kannan2019}. Finally, with the GLF flux we avoid expensive calculations of wave speeds \citep{Skinner2013}, which can impact the performance for Ly$\alpha$ RT with many frequency bins.} with corrections for the optically thick regime, described below. Thus, at a given interface $i+1/2$ we write the $R$-component of $\boldsymbol{H}$ as:
    \begin{align}
        (H_R)_{i+1/2} &=~ \epsilon \, \dfrac{1}{2}[(H_R)_{i+1/2}^{\rm L}  + (H_R)_{i+1/2}^{\rm R}] \label{Corrected GLF flux J}\\ &-~ \epsilon \, \dfrac{1}{2}(J_{i+1/2}^{\rm R} - J_{i+1/2}^{\rm L}) \nonumber \\ &-~ (1-\epsilon) \, \dfrac{J_{i+1} - J_i}{3 \tau_{i+1/2}} \nonumber \, ,
    \end{align}
    where superscripts L and R denote the reconstructed values on the left and right sides of the interface $i+1/2$, respectively. The top two lines are simply the GLF flux multiplied by the factor $\epsilon$, while the final line is $(1-\epsilon)$ times the discretized diffusion flux $-\boldsymbol{\nabla}J/3 \alpha_{\rm F}$, valid in the highly optically thick regime. As many authors have pointed out, for $\epsilon = 1$, there will be too much numerical diffusion in case the cell optical depth is $\gtrsim 1$, leading to a failure to obtain the correct flux (and hence radiation pressure) in the highly optically thick regime \citep[e.g.][]{OConnor2013, Just2015, Rosdahl2015, Skinner2019, Mezzacappa2020, Bloch2021}. We tackle this problem by adopting an interpolating factor $\epsilon$ given by:
    \begin{equation}
        \epsilon = \dfrac{0.2 + \tau_{i+1/2}}{0.2 + \tau_{i+1/2} + \tau_{i+1/2}^2} \, .
    \end{equation}
    Here the flux optical depth between cells $i$ and $i+1$ is:
    \begin{align}
        \tau_{i+1/2} &=~ (\alpha_{\rm F})_{i} \, (R_{i+1/2} - \Bar{R}_i) \\ &+~ (\alpha_{\rm F})_{i+1} \, (\Bar{R}_{i+1} - R_{i+1/2}) \nonumber \, ,
    \end{align}
    where $\Bar{R}_i$ denotes the cell centroid (in the $R$-direction) for cell $(i,j)$, and $R_{i+1/2}$ the cell interface. Thus, when cells are optically thin, $\epsilon \simeq 1$, and we recover the uncorrected GLF flux. In the highly optically thick regime, $\epsilon \simeq 1/\tau_{i+1/2} \rightarrow 0$, and we recover the correct diffusion limit. The specific scheme in Eq.~(\ref{Corrected GLF flux J}) was adopted for M1 transport of neutrinos by \cite{OConnor2015} and \cite{OConnor2018}, although we adopt the slightly different interpolation $\epsilon$ from \cite{Hopkins2017_Anisotropic} which is cheaper to evaluate, and was found to give slightly better results for e.g. the emergent spectra.\footnote{Several forms for $\epsilon$ have been suggested in the literature. \cite{OConnor2015} uses $\epsilon = \tanh(1/\tau_{i+1/2})$, \cite{Skinner2019} adopt $\epsilon = \min(1/\tau_{i+1/2},1)$, and \cite{Rosdahl2015} and \cite{Bloch2021} suggest $\epsilon = 1/ (1 + \eta \tau_{i+1/2})$, with $\eta$ dependent on the wave speeds. All these proposals give $\epsilon = 1$ for $\tau_{i+1/2} \rightarrow 0$, and $\epsilon \sim 1/\tau_{i+1/2}$ for $\tau_{i+1/2} \gg 1$, with a transition around $\tau_{i+1/2} \sim 1$.} For the components of $\boldsymbol{\mathsf{K}}$ at cell interfaces, we take the following modification of the GLF flux \citep{OConnor2015, Mezzacappa2020}:
    \begin{align}
        \boldsymbol{\mathsf{K}}_{i+1/2} &=~ \dfrac{1}{2}(\boldsymbol{\mathsf{K}}_{i+1/2}^{\rm L} + \boldsymbol{\mathsf{K}}_{i+1/2}^{\rm R}) \\ &-~ \epsilon \, \dfrac{1}{2}[(H_R)_{i+1/2}^{\rm R} - (H_R)_{i+1/2}^{\rm L}] \, . \nonumber
    \end{align}
    Finally, for the advection terms $\boldsymbol{\boldsymbol{\nabla}} \boldsymbol{\cdot}(\boldsymbol{u} J/c)$ and $\boldsymbol{\boldsymbol{\nabla}} \boldsymbol{\cdot}(\boldsymbol{u} \boldsymbol{H}/c)$ we use an upwind scheme, using the star gas velocities at interfaces, taken from the HLLC Riemann solver during the hydrodynamics time-step update \citep[see e.g.][]{Skinner2019}.

    \item \textbf{Ly$\boldsymbol{\alpha}$ emission and absorption step:} Next we treat Ly$\alpha$ emission and absorption. Because Ly$\alpha$ couples to chemistry and dust, we subcycle Ly$\alpha$ emission and absorption together with the rest of the photochemistry, gas heating/cooling, and dust growth/destruction (see Fig.~\ref{LYDION RHD operator split method} for an overview). Thus, we split the RT advection time-step $\Delta t_{\rm RT}$ into smaller substeps $\Delta t_{\rm sub}$. In each substep $\bullet \rightarrow \bullet \bullet$ we update $J$ and $\boldsymbol{H}$ implicitly:
    \begin{align}
        \dfrac{J^{\bullet\bullet} - J^\bullet}{\Tilde{c} \Delta t_{\rm sub}} &=~ j_{\rm s} - [\alpha_{\rm c} (1-\omega) + p_{\rm d} \alpha \mathcal{H} ] J^{\bullet\bullet} \, , \\
        \dfrac{\boldsymbol{H}^{\bullet\bullet} - \boldsymbol{H}^\bullet}{\Tilde{c} \Delta t_{\rm sub}} &=~ - [\alpha_{\rm c} (1-\omega g) + \alpha \mathcal{H} ] \boldsymbol{H}^{\bullet\bullet} \, ,
        \label{Absorption step}
    \end{align}
    which can be easily re-arranged to obtain the updated values $J^{\bullet\bullet}$ and $\boldsymbol{H}^{\bullet\bullet}$. In RHD simulations, we accept the substep update if the Ly$\alpha$ energy density $e_{\rm Ly\alpha} = (4\pi/c) \int \textrm{d}\nu \, J$ changed by $< 10 \%$ --- if not, we reduce the substep in half and redo the update.\footnote{More accurately stated, we accept the update if $\lvert e_{\rm Ly\alpha}^{\bullet\bullet} - e_{\rm Ly\alpha}^{\bullet}\rvert / (e_{\rm Ly\alpha}^{\bullet} + e_{\rm floor}) < 0.1$, where $e_{\rm floor} = 4 \pi J_{\rm floor}/c$, with $J_{\rm floor} = 10^{-15} \rm \, erg \, cm^{-2} \, s^{-1} \, sr^{-1}$. } When the sum of all substeps equals $\Delta t_{\rm RT}$, the subcycling ends, and we have our final updated values for $J$ and $\boldsymbol{H}$ from this stage. For the Ly$\alpha$ emissivity $j_{\rm s}$ (in erg cm$^{-3}$ s$^{-1}$ Hz$^{-1}$ sr$^{-1}$) we include emission from hydrogen recombinations and collisional excitations, so that:
    \begin{align}
       j_{\rm s}(\boldsymbol{r},\nu) &=~   \dfrac{E_{\rm Ly\alpha}}{4\pi} \, \mathcal{P}_{\rm Ly\alpha} k_{\rm rec} n_{\rm HII} n_{\rm e}  \, \phi_\nu \\ &+~ \dfrac{E_{\rm Ly\alpha}}{4\pi} \, k_{\rm coll} n_{\rm HI}n_{\rm e}  \, \phi_\nu \, . \nonumber
    \end{align}
    Here $E_{\rm Ly\alpha} \simeq 10.2 \, \rm eV$ is the Ly$\alpha$ photon energy, $\mathcal{P}_{\rm Ly\alpha}$ is the probability that a recombination event will produce a Ly$\alpha$ photon, $k_{\rm rec}(T)$ is the recombination coefficient, $k_{\rm coll}(T)$ the collisional excitation coefficient \citep[see][]{Smith2022_disc}, and $\phi_\nu = \mathcal{H}(\nu)/\int \textrm{d}\nu' \, \mathcal{H}(\nu') = \mathcal{H}(\nu)/\sqrt{\pi}$ the normalized Voigt emission line profile. By default we simulate diffuse ionizing recombination radiation on-the-fly with M1 transport (Sec.~\ref{Stellar and IR transfer and dust modelling}), in which case we adopt the Case A recombination coefficient from \cite{Hui1997}. We adopt the Ly$\alpha$ emission probability $\mathcal{P}_{\rm Ly\alpha}$ from \cite{Storey1995}, which is a weak function of $T$ and $n_{\rm e}$, as shown in Fig.~\ref{fig: Emission probability}. For Case A recombinations and typical conditions studied in this paper, $\mathcal{P}_{\rm Ly\alpha} \sim 0.4$. 

    \begin{figure}
    \centering
    \includegraphics[width=1.0\columnwidth]{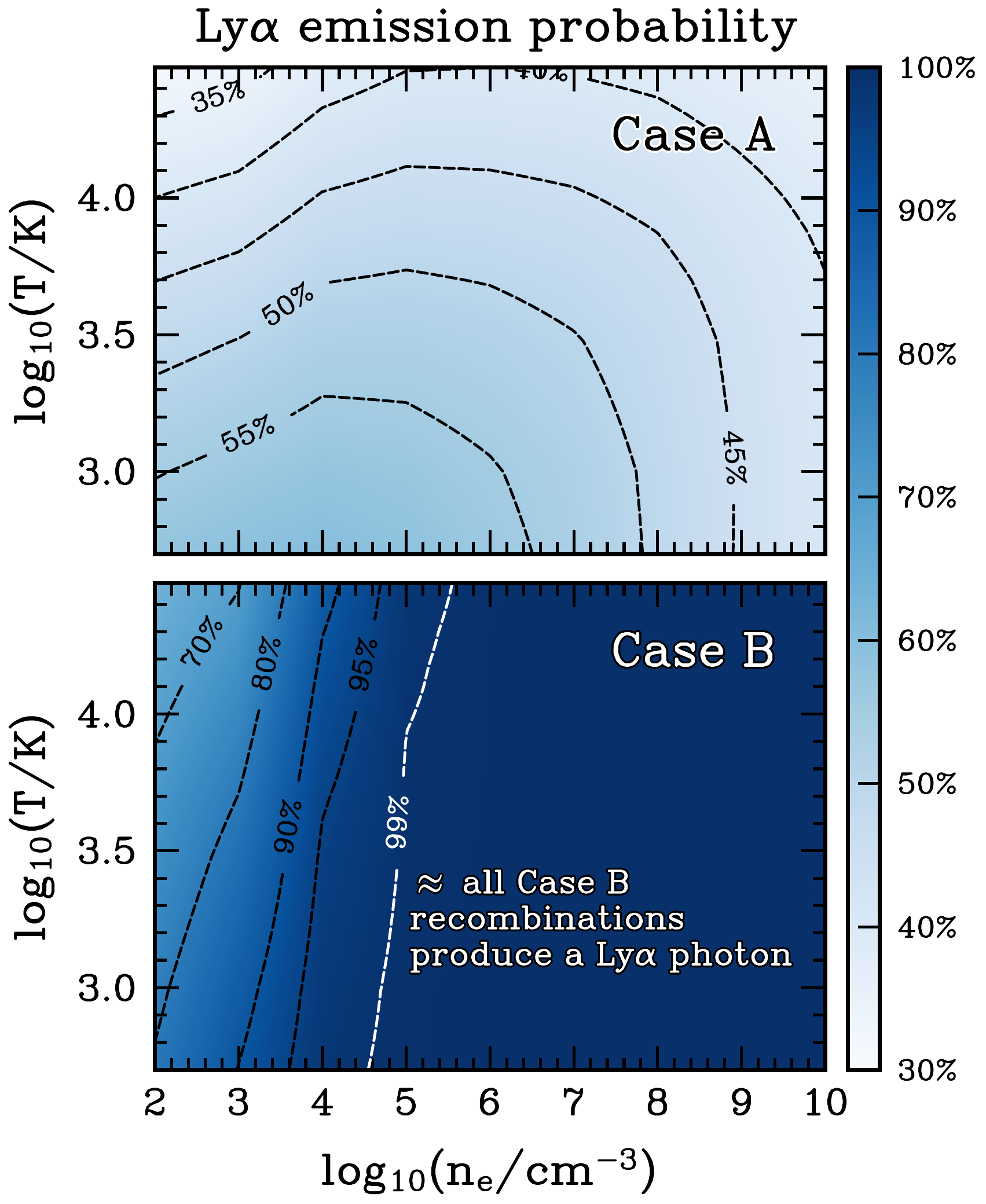}
    \caption{The Ly$\alpha$ emission probability $\mathcal{P}_{\rm Ly\alpha}$ per recombination event as a function of temperature and electron density for Case A (top panel) and Case B recombinations (bottom panel), as computed by \citet{Storey1995}. }
    \label{fig: Emission probability}
    \end{figure}

    For continuum absorption and scattering, we consider dust absorption and scattering (Sec.~\ref{Stellar and IR transfer and dust modelling}, Table~\ref{Dust optical properties non IR}), Thomson scattering by electrons, inelastic Raman scattering by H$_2$ \citep[][]{Nebrin2024}, and absorption by the pseudo-molecules OH$_{\rm x}$ and CH$_{\rm x}$ (representing e.g. OH and CH), which leads to their photodissociation (for details, see Appendix~\ref{Carbon chemistry appendix}). Besides continuum absorption, other destruction mechanisms for Ly$\alpha$ photons can become important in dense gas \citep{Nebrin2024}. The Ly$\alpha$ destruction probability $p_{\rm d}$ is computed following \cite{Nebrin2024}. In short, we take into account $2p \rightleftharpoons 2s$ transitions in collisions with $e^-$, \textsc{H\,i}, \textsc{H\,ii}, H$_2$, and He\textsc{\,i}, as well as transitions induced by the Cosmic Microwave Background. Also incorporated is Ly$\alpha$ absorption by neighboring lines in H$_2$ \citep{Neufeld1990, Nebrin2024}. 

    \item \textbf{Frequency diffusion and Doppler shift:} Finally, after subcycling Ly$\alpha$ emission/absorption and photo-thermochemistry, we update $J$ using the terms describing frequency diffusion, atomic recoil, and Doppler shifts: 
    \begin{align}
        \dfrac{1}{\Tilde{c}} \dfrac{\partial J}{\partial t}  ~&=~ \dfrac{\partial}{\partial \nu} \left\{ \dfrac{1}{2} \Delta \nu_{\rm D}^2 \alpha \mathcal{H} \left( \dfrac{\partial J}{\partial \nu} + \mathfrak{R}J \right) \right\} \label{Frequency step} \\ ~&+~ \dfrac{1}{c} \dfrac{\partial }{\partial \nu}  [\nu \, (\boldsymbol{\mathsf{D}} \boldsymbol{:} \boldsymbol{\nabla}\boldsymbol{u})J]  \nonumber \, .
    \end{align}
    We solve this equation separately for each cell $(i,j)$ over the time-step $\Delta t_{\rm RT}$, using an implicit backward Euler method.\footnote{ While there are second-order implicit methods, we have settled on backward Euler (BDF1). The Crank--Nicholson method, while formally second-order accurate, is only A-stable, which tends to produce numerical oscillations. In contrast, the BDF1 method is L-stable, and can quickly damp numerical oscillations. There is also the second-order L-stable TR-BDF2 method \citep[e.g.][]{Edwards2011}, but it requires two implicit stages, hence making the overall step twice as expensive. Furthermore, it is slightly less damping of oscillations than BDF1, and not unconditionally monotonic \citep[][]{TRBDF2_Paper}. For reasons of performance, stability, simplicity, and consistency with the convergence order of the spatial advection method, we have opted for BDF1. } We use an exponential reconstruction scheme to handle, in a stable manner, the presence of both diffusion and frequency advection terms. A detailed discussion of the solution method is provided in Appendix~\ref{frequency diffusion appendix}. We use the updated values for $J$ from step (2) as initial conditions in Eq.~(\ref{Frequency step}), along with the latest values for $\boldsymbol{\mathsf{D}}$ and $\boldsymbol{\nabla}\boldsymbol{u}$. Causality demands that $\lvert \boldsymbol{H} \rvert/J \leq 1$, and we have assumed a constant $\boldsymbol{\mathsf{D}}$ in the solution of Eq.~(\ref{Frequency step}).\footnote{We note that in the optically thick limit, $\boldsymbol{\mathsf{D}} \rightarrow \boldsymbol{\mathsf{I}}/3$, and this approximation becomes exact.} We therefore update $\boldsymbol{H}$ according to $\boldsymbol{H}^{\rm new} = (J^{\rm new}/J^{\rm old}) \boldsymbol{H}^{\rm old}$, consistent with a fixed $\boldsymbol{\mathsf{D}}$.
    
\end{enumerate}
This completes the time-step update of $J$ and $\boldsymbol{H}$. To ensure numerical stability, we choose the time-step for Ly$\alpha$ RT to satisfy the Courant–Friedrichs–Lewy (CFL) condition \citep[e.g.][]{Rosdahl2013, Wikbing2022}:
\begin{equation}
    \Delta t_{\rm RT} = \textrm{CFL}_{\rm RT}  \times \dfrac{1}{D} \min_{i,j} \left[ \dfrac{\min(\Delta R_i, \Delta Z_j)}{\Tilde{c} + \lvert (\boldsymbol{u})_{i,j} \rvert} \right] \, , \nonumber
\end{equation}
with a CFL factor $\textrm{CFL}_{\rm RT} < 1$ ($0.4$ by default), and where the number of dimensions in our case is $D = 2$. To ensure accuracy, we choose the Ly$\alpha$ RSL $\Tilde{c}$ to be large enough that it always significantly exceeds the maximum gas velocity. Furthermore, to ensure accuracy we must choose $\Tilde{c}$ such that the photon diffusion or free-streaming time does not become unphysically long (which would otherwise yield a too slow build-up and response time for $J$) \citep[][]{Skinner2013}. Taken together, we choose $\Tilde{c}$ on-the-fly in RHD simulations according to:\footnote{A similar method for a dynamic RSL was recently employed by \cite{Rosdahl2025} in their moment method for cosmic rays.}
\begin{equation}
    \Tilde{c} = \min\left[\max\left( \Tilde{c}_{\rm min}, \, \Tilde{c}_{\rm dyn} \right), \,c \right] \, ,
\end{equation}
where:
\begin{align}
    \Tilde{c}_{\rm dyn} &=~ \epsilon_{\rm RSL} \max_{i,j} \Bigg\{ \max \left[1,\, (a_{\rm v}\tau_{\rm cell,0})^{1/3} f_{\rm abs} \right] \label{Dynamic RSL} \\ &\times~ \max\big[\, \lvert(\boldsymbol{u})_{i,j}\rvert, \, (c_{\rm s})_{i,j} \,\big] \Bigg\} \nonumber \, , \\
    f_{\rm abs} &=~ \dfrac{1}{1 + (a_{\rm v} \tau_{\rm cell,0})^{1/3}\tau_{\rm cell,c} + 0.6 \, p_{\rm d} \tau_{\rm cell,0}} \, ,
\end{align}
with $\epsilon_{\rm RSL} = 4$ by default. The expression for $\Tilde{c}_{\rm dyn}$ is motivated as follows. The time-scale for Ly$\alpha$ photons to cross a cell is expected to be at most $\sim (a_{\rm v}\tau_{\rm cell,0})^{1/3} \max(\Delta R_i, \,\Delta Z_j)/\Tilde{c}$ in the optically thick limit \citep[e.g.][]{Lao2020, McClellan2022}, where $\tau_{\rm cell,0} \sim \alpha \max(\Delta R_i, \,\Delta Z_j)$ is the maximum cell optical depth at line center. In contrast, in the free-streaming limit we instead have a photon crossing time-scale $\sim \max(\Delta R_i, \,\Delta Z_j)/\Tilde{c}$. Equation~(\ref{Dynamic RSL}) is then derived from the requirement that the free-streaming or effective diffusion velocity be at least $\sim \epsilon_{\rm RSL}$ times faster than the gas velocity and sound speed $c_{\rm s}$, for all cells. Continuum absorption and a finite destruction probability $p_{\rm d}$ can destroy Ly$\alpha$ photons, and hence reduce the effective trapping time and force. In cells where this is the case, Ly$\alpha$ photons are unlikely to be dynamically important, and should not count much in the estimated $\Tilde{c}$. This is captured by the factor $f_{\rm abs}$, which scales with the cell continuum absorption optical depth $\tau_{\rm cell,c}$, and $p_{\rm d}$, in a manner predicted by recent analytical solutions \citep[][]{Nebrin2024}.\footnote{We have used scalings motivated by the force multiplier $M_{\rm F} \sim t_{\rm trap}/t_{\rm light}$ for the dependence on $p_{\rm d}$ and the continuum absorption optical depth $\tau_{\rm c}$. More specifically, eqs.~(68) and (77) in \cite{Nebrin2024}, for extended sources. Physically speaking, Ly$\alpha$ photons are effectively destroyed if $p_{\rm d} \tau_{0} \gtrsim 1$ or $(a\tau_0)^{1/3}\tau_{\rm c} \gtrsim 1$. Velocity gradients will also reduce the force and trapping time, but the effect is more modest, and analytical expressions only exist for simplified setups \citep{Nebrin2024, Smith2025}. For this reason, and to be conservative, we neglect velocity gradients in the estimate of the required RSL.} 

The quantity $\Tilde{c}$ is set to be at most $c$, and always larger than $\Tilde{c}_{\rm min} = 100 \, \rm km \, s^{-1}$, so that ionization front propagation is always well-captured \citep[e.g.][]{Geen2015, Grudic2021}. Finally, we allow $\Tilde{c}$ to decrease (increase) by at most $25\%$ (a factor $4$) from its previous value, to avoid any rapid changes following thermochemical evolution.

\subsection{Parallelization, and code optimization strategies}

Despite improvements in performance over MCRT, brute-force Ly$\alpha$ M1 RT with $\mathcal{O}(200)$ frequency bins is, by far, the most numerically expensive ingredient of \textsc{Lydion}. To boost performance of \textsc{Lydion}, we parallelize the M1 step over frequency bins, and the frequency diffusion step over spatial cells. We use two additional optimization techniques to speed up both the M1 advection and frequency diffusion steps:
\begin{enumerate}[leftmargin=*]
    \item \textbf{\textit{Only advect active bins}:} Since we use a very wide frequency grid, it turns out that many frequency bins often contain negligible $J$, and hence have negligible impact on both the radiation pressure, and the emergent spectra. Thus, we only advect `active' bins containing $>10^{-10}$ of the total Ly$\alpha$ energy (over the entire simulation volume). We have found that this alone can reduce the number of bins advected by up to a factor $\sim 2$.

    \item \textbf{\textit{Optimize Voigt profile calculation}:} The (bin-averaged) Voigt profile $\mathcal{H}$ is needed for every cell and frequency bin, and at every time-step. This quickly becomes time-consuming as a result of several numerically expensive functions involved \citep[e.g. Dawson function, and $e^{-x^2}$, see][]{Smith2015, Smith2018DDMC}, which, if not done with care, can take up a significant portion of the overall CPU time for the Ly$\alpha$ RT update. As further discussed in Appendix~\ref{frequency diffusion appendix}, we therefore create look-up tables for these functions of the dimensionless frequency $x \equiv (\nu-\nu_{\rm Ly\alpha})/\Delta \nu_{\rm D}$, augmented by asymptotic expansions at large arguments. With these look-up tables, and other related optimizations, we have found up to $\sim 8 \times$ speed-up of this step compared to earlier non-optimized versions. 
    
\end{enumerate}
Future development of \textsc{Lydion} will target GPU architectures, domain distributed workloads, and adaptive mesh refinement. These developments are expected to significantly speed up Ly$\alpha$ RHD simulations at a given target resolution.

\subsection{Lyman-$\alpha$ radiation pressure feedback}
\label{Lya radiation pressure numerical method}

The force per unit volume on the gas and dust from Ly$\alpha$ radiation pressure is $\boldsymbol{f}_{\rm Ly\alpha} = 4 \pi \int \textrm{d}\nu \, [\alpha \mathcal{H} + \alpha_{\rm c}(1-\omega g)] \boldsymbol{H}/c$. This in turn, for an emission/absorption substep $\bullet \rightarrow \bullet \bullet$ of size $\Delta t_{\rm sub}$, can be related to the change in $\boldsymbol{H}$, to obtain \citep[e.g.][]{Skinner2019}:
\begin{equation}
    (\boldsymbol{f}_{\rm Ly\alpha})_{i,j} = \dfrac{4 \pi}{c \Tilde{c}}  \sum_{k=1}^{N_{\nu}} \Delta \nu_k \, \dfrac{(\boldsymbol{H}^{\bullet}_{i,j,k} - \boldsymbol{H}^{\bullet \bullet}_{i,j,k})}{\Delta t_{\rm sub}} \, , \label{Lya rad force general}
\end{equation}
where we have written the result in discretized form, for the force per unit volume in cell $(i,j)$. We update the gas momentum density $(\rho \boldsymbol{u})_{i,j}$ by simply adding $\epsilon_{\rm gas} (\boldsymbol{f}_{\rm Ly\alpha})_{i,j} \, \Delta t_{\rm sub}$, where $\epsilon_{\rm gas}$ is the ratio between the gas flux opacity and the total opacity from both gas and dust. The dust bin momentum densities are updated in a similar fashion.

It is straightforward to check that Eq.~(\ref{Lya rad force general}) has the correct behavior. In the optically thin limit, the change in $\boldsymbol{H}$ goes to zero, and there is negligible force exerted on the gas. In the optically thick limit, the change in $\boldsymbol{H}$ is (using Eq.~\ref{Absorption step}): 
\begin{align}
  \dfrac{(\boldsymbol{H}^{\bullet}_{i,j,k} - \boldsymbol{H}^{\bullet\bullet}_{i,j,k})}{\Delta t_{\rm sub}} ~&=~ \dfrac{(\alpha_{\rm F})_{i,j,k} \Tilde{c} \, \boldsymbol{H}^{\bullet}_{i,j,k}}{1 + (\alpha_{\rm F})_{i,j,k} \Tilde{c} \Delta t_{\rm sub}} \\ \underbrace{\longrightarrow}_{\tau_{\rm cell} \gg 1}~&\simeq~  - \, \Tilde{c} \, (\boldsymbol{\boldsymbol{\nabla}}\boldsymbol{\cdot}\boldsymbol{\mathsf{K}})_{i,j,k}^n \, , \nonumber
\end{align}
where $\alpha_{\rm F} = [\alpha \mathcal{H} + \alpha_{\rm c}(1-\omega g)]$ is the flux opacity. The second line follows because $(\alpha_{\rm F})_{i,j,k} \Tilde{c} \Delta t \sim \textrm{CFL}_{\rm RT} \times \tau_{\rm cell} \gg 1$, and $\boldsymbol{H}^{\rm adv} \simeq \boldsymbol{H}^n - \Tilde{c} \Delta t \, (\boldsymbol{\boldsymbol{\nabla}}\boldsymbol{\cdot}\boldsymbol{\mathsf{K}})^n$ from the explicit advection update (ignoring velocity terms here). Thus, if $\tau_{\rm cell} \gg 1$, the Ly$\alpha$ radiation pressure force relaxes to the divergence of the radiation pressure tensor:
\begin{equation}
    (\boldsymbol{f}_{\rm Ly\alpha})_{i,j} \simeq - \sum_{\textrm{optically thick }k} \Delta \nu_k \, (\boldsymbol{\nabla}\boldsymbol{\cdot}\boldsymbol{\mathsf{P}})_{i,j,k} \, .
\end{equation}
Our IMEX method is therefore expected to predict the correct behavior for the radiation pressure force in all regimes. We confirm this in Sec.~\ref{Test section}, where the predicted radiation pressure force is compared against analytical solutions and MCRT simulations.

\subsection{Hydrodynamics, dust dynamics, and gravity}
\label{sec hydro, dust dynamics, gravity}

We couple Ly$\alpha$ and stellar feedback to a 2D hydrodynamics and dust dynamics code. Dust dynamics is potentially important because radiation pressure on grains can, in some cases, partially drive out the dust from \textsc{H\,ii} regions \citep[][]{Akimkin2015, Akimkin2017, Ishiki2018}, and presumably lead to a stronger build-up of Ly$\alpha$ radiation pressure. The gas density $\rho$, momentum density $\rho \boldsymbol{u}$, and energy density $E$ are obtained by solving: 
\begin{align} 
    \dfrac{\partial \rho }{\partial t} + \boldsymbol{\nabla}\boldsymbol{\cdot}(\rho \boldsymbol{u}) &= 0 \, , \label{gas continuity eq} \\
    \dfrac{\partial (\rho \boldsymbol{u})}{\partial t} + \boldsymbol{\nabla}\boldsymbol{\cdot}(\rho \boldsymbol{u}\otimes\boldsymbol{u} + P \boldsymbol{\mathsf{I}}) &= \boldsymbol{f}_{\rm g} - \rho \boldsymbol{\nabla}\Phi \\ &+ \sum_\beta \mathcal{K}_\beta(\boldsymbol{u}_{\rm d,\beta} - \boldsymbol{u}) \nonumber \, , \\  \dfrac{\partial E }{\partial t} + \boldsymbol{\nabla}\boldsymbol{\cdot}[(E+P) \boldsymbol{u}] &= \boldsymbol{f}_{\rm g}\boldsymbol{\cdot}\boldsymbol{u}- \rho \boldsymbol{\nabla}\Phi \boldsymbol{\cdot} \boldsymbol{u} \\ &- \Lambda_{\rm net}   \, , \nonumber
\end{align}
where $\boldsymbol{f}_{\rm g}$ is the total radiation pressure force per unit volume on the gas (by Ly$\alpha$ and photons in other bands), $P$ the gas pressure, $\Phi$ is the gravitational potential, $\Lambda_{\rm net}$ the net gas cooling rate, and $\boldsymbol{u}$ ($\boldsymbol{u}_{\rm d,\beta}$) the gas (dust bin) velocity, with $\mathcal{K}_\beta$ being the drag coefficient for dust bin $\beta$. To ensure accurate thermal evolution in the high-Mach regime, we use the dual energy formalism \citep[see e.g.][]{Bryan1995, Byran2014, Teyssier2015}, which is implemented closely following \cite{Byran2014}. We consider three dust bins/types: silicate grains (Sil), graphite grains (C), and polycyclic aromatic hydrocarbons (PAH). Each dust bin/type $\beta \in \rm (Sil,C,PAH)$ is treated as a pressureless fluid, governed by:
\begin{align} 
    \dfrac{\partial \rho_{\rm d,\beta} }{\partial t} + \boldsymbol{\nabla}\boldsymbol{\cdot}(\rho_{\rm d} \boldsymbol{u}_{\rm d})_\beta &= \textrm{Growth}_{\beta} \label{dust continuity eq} \\ &- \textrm{Destruction}_\beta \, , \nonumber \\
    \dfrac{\partial (\rho_{\rm d} \boldsymbol{u}_{\rm d})_\beta}{\partial t} + \boldsymbol{\nabla}\boldsymbol{\cdot}(\rho_{\rm d} \boldsymbol{u}_{\rm d}\otimes\boldsymbol{u}_{\rm d})_\beta &= \boldsymbol{f}_{\rm d,\beta} - \rho_{\rm d,\beta} \boldsymbol{\nabla}\Phi  \label{dust momentum eq} \\ &+ \mathcal{K}_{\beta}(\boldsymbol{u} - \boldsymbol{u}_{\rm d,\beta}) \nonumber \, ,
\end{align}
where the right-hand side of Eq.~(\ref{dust continuity eq}) encapsulates dust growth and destruction terms (discussed below in detail),\footnote{Formally, the same terms with opposite signs should enter the continuity equation for $\rho$, as destroyed dust grains return their mass to the gas, and dust growth by accretion drains the gas-phase metal reservoir. } and $\boldsymbol{f}_{\rm d,\beta}$ is the total radiation pressure force per unit volume on dust bin $\beta$. 

To solve Eqs.~(\ref{gas continuity eq})--(\ref{dust momentum eq}), we operator split the source terms (radiation pressure, gravity, drag, dust growth/destruction, and gas cooling/heating). We start the time update $t^n \rightarrow t^{n+1} =t^n+\Delta t_{\rm hydro}$ by performing a gravity solve + kick over a half-step $\Delta t_{\rm hydro}/2$. Thus, we first solve the Poisson equation for $\Phi$:
\begin{equation}
    \nabla^2 \Phi^{n} = 4\pi G(\rho^{n} + \rho_{\rm d}^{n} +\rho_{\star}) \, , \label{Poisson equation}
\end{equation}
where $\rho^{n}$ and $\rho_{\rm d}^n$ are the gas and (total) dust densities at the old time-step $n$, and $\rho_\star$ the fixed stellar mass density. We solve Eq.~(\ref{Poisson equation}) using an iterative ADI method  \citep[e.g.][]{Black1975, Stone1992}, described and tested in Appendix~\ref{Poisson equation Appendix}. The gas and dust momenta, and the gas energy, are then updated according to \citep{Truelove1998}:
\begin{align}
    \rho \boldsymbol{u}^{\triangle} &=~ \rho\boldsymbol{u}^{n} - \rho^{n} \boldsymbol{\nabla}\Phi^{n} \, \dfrac{\Delta t_{\rm hydro}}{2} \label{Gravity kick 1} \, ,\\ (\rho_{\rm d} \boldsymbol{u}_{\rm d}^{\triangle})_\beta &=~ (\rho_{\rm d}\boldsymbol{u}_{\rm d}^{n})_\beta - \rho_{\rm d,\beta}^{n} \boldsymbol{\nabla}\Phi^{n} \, \dfrac{\Delta t_{\rm hydro}}{2} \, , \\ E^{\triangle} &=~ \underbrace{\left[ E^{n} -  \dfrac{(\rho \boldsymbol{u}^{n})^2}{2 \rho^{n}}\right]}_{\text{Fixed gas internal energy }e} + \,\dfrac{(\rho \boldsymbol{u}^{\triangle})^2}{2 \rho^{\triangle}} \, , \label{Gravity kick 3}
\end{align}
where the densities remain the same over the gravity half-kick (e.g. $\rho^\triangle = \rho^n$).

After the gravity solve + kick over $\Delta t_{\rm hydro}/2$, we update the hydro- and dust-dynamical conservative variables over a global step $\Delta t_{\rm hydro}$,\footnote{We also advect passive scalars, e.g. abundance fractions, during this step.} considering only the advection terms, and using the partially updated ($\triangle$)--state from Eqs.~(\ref{Gravity kick 1})--(\ref{Gravity kick 3}) as initial conditions. The resulting equations are solved using a dimensionally unsplit finite-volume method, with WENO3 spatial reconstruction \citep[closely following][]{Mignone2014}, and third-order RK3 time-integration \citep{Mignone2007}. While WENO3 offers third-order accurate interface reconstruction in smooth regions, we do not distinguish between fluxes at the center of an interface and the interface area-averaged fluxes. As a result of this, the code is formally second-order spatially accurate. By default we use the HLLC flux function for hydrodynamics, and the HLLgd flux function of \cite{Verrier2025} for dust dynamics. For stability reasons, near strong shocks, and regions with steep gradients, we revert to second-order linear reconstruction (with a MinMod slope-limiter), and the HLL flux function for the gas. More details of the WENO3 method, and tests of the hydrodynamics code, can be found in Appendix~\ref{Hydro Appendix}. 

After the advection of gas and dust, we perform another gravity solve + kick over a half-step $\Delta t_{\rm hydro}/2$, to complete the full update $n \rightarrow n+1$. As described in the next sections, other source terms (e.g. radiation pressure, drag, cooling/heating) are subcycled over (typically many smaller) substeps, the sum of which add up to $\Delta t_{\rm hydro}$. An overview of the operator-split ordering in \textsc{Lydion} is shown in Fig.~\ref{LYDION RHD operator split method}.

\subsection{Stellar and infrared radiative transfer, feedback \& dust modelling}
\label{Stellar and IR transfer and dust modelling}

\begin{table*}
\centering
\setlength{\tabcolsep}{12pt}
\caption{Dust optical properties in the non-IR bands adopted in \textsc{Lydion}, computed assuming the \cite{Weingartner2001} $R_{\rm V} = 5.5$ ($b_{\rm C} = 3 \times 10^{-5}$) dust grain size distribution for each dust bin/type. The table lists the band-averaged mass absorption coefficient $\kappa_{\rm abs} = \kappa_{\rm ext}(1-\omega)$ and flux-mean absorption coefficient $\kappa_{\rm F} = \kappa_{\rm ext}(1-\omega g)$ for silicate grains, graphite grains, and PAHs. The (non-Ly$\alpha$) band-averaging assumes a $4.8\times10^4 \, \rm K$ black-body spectrum within each band, to approximate the spectrum of a $\sim 30 \, \rm M_{\odot}$ star \citep[estimated using fits from][with $Z_{\star}/Z_{\odot} = 0.01$]{Tanikawa2020}. For the IR band we refer the reader to Appendix~\ref{IR appendix}. The initial fractions of the total dust mass in each bin is $f_{\rm Sil} = 0.7264$, $f_{\rm C} = 0.2517$, and $f_{\rm PAH} = 0.0218$, for silicate, graphite, and PAH dust, respectively. }
\begin{tabular}{c cc cc cc}
\hline
\hline
\noalign{\vskip 2pt}

 & \multicolumn{2}{c}{Silicate grains} & \multicolumn{2}{c}{Graphite grains} & \multicolumn{2}{c}{PAHs} \\

Band 
& $\kappa_{\rm abs}$ & $\kappa_{\rm F}$ 
& $\kappa_{\rm abs}$ & $\kappa_{\rm F}$ 
& $\kappa_{\rm abs}$ & $\kappa_{\rm F}$ \\

 & \multicolumn{6}{c}{(cm$^2$ g$^{-1}$ of dust in bin)} \\

\noalign{\vskip 3pt}
\hline
\hline
\noalign{\vskip 2pt}

IR ($< 1\, \rm eV$)  & \multicolumn{6}{c}{Appendix~\ref{IR appendix} (see Fig.~\ref{Planck Rosseland opacities fig})} \\

Optical ($1-5.8 \, \rm eV$)  
& $5.98 \times 10^3$ & $1.69 \times 10^4$ & $2.09 \times 10^4$ & $2.61 \times 10^4$ & $2.27 \times 10^5$ & $2.27 \times 10^5$ \\

FUV ($5.8 - 11.2 \, \rm eV$) 
& $1.71 \times 10^4$ & $2.21\times 10^4$ & $2.29 \times 10^4$ & $2.57 \times 10^4$ & $2.86 \times 10^5$ & $2.86 \times 10^5$ \\

LW ($11.2 - 13.6 \, \rm eV$) 
& $1.80 \times 10^4$ & $2.18 \times 10^4$ & $3.88 \times 10^4$ & $4.48 \times 10^4$ & $5.92 \times 10^5$ & $5.92 \times 10^5$ \\

EUV1 ($13.6-24.6 \, \rm eV$) 
& $1.53 \times 10^4$ & $1.85 \times 10^4$ & $5.60 \times 10^4$ & $6.22 \times 10^4$ & $1.08 \times 10^6$ & $1.08 \times 10^6$ \\

EUV2 ($24.6-54.4 \, \rm eV$) 
& $1.37 \times 10^4$ & $1.46 \times 10^4$ & $1.90 \times 10^4$ & $2.12 \times 10^4$ & $2.83 \times 10^5$ & $2.84 \times 10^5$ \\

EUV3 ($54.4-\infty \, \rm eV$) 
& $1.30 \times 10^4$ & $1.32 \times 10^4$ & $1.04 \times 10^4$ & $1.09 \times 10^4$ & $8.81 \times 10^4$ & $8.82 \times 10^4$ \\

Ly$\alpha$ ($10.2 \, \rm eV$)  
& $1.84 \times 10^4$ & $2.22 \times 10^4$ & $2.48 \times 10^4$ & $2.84 \times 10^4$ & $3.31 \times 10^5$ & $3.31 \times 10^5$ \\

\noalign{\vskip 2pt}
\hline
\hline

\end{tabular}
\vspace{1 pt}\\
\raggedright
\label{Dust optical properties non IR}
\end{table*}

To model \textsc{H\,ii} regions and stellar feedback, we implement RT in seven broad frequency bands following \cite{Kannan2020} and \cite{Deng2024}, ranging from infrared (IR) and optical (opt), to far-UV (FUV), Lyman-Werner (LW), \textsc{H\,i}-ionizing (EUV1), He\textsc{\,i}-ionizing (EUV2), and He\textsc{\,ii}-ionizing (EUV3) photons (see Table~\ref{Dust optical properties non IR} for photon energy intervals):
\begin{equation}
    \mathcal{B} \in \rm(IR,opt,FUV,LW,EUV1,EUV2,EUV3) \, .
\end{equation}
As for Ly$\alpha$, we employ the M1 moment method to solve for $J_{\mathcal{B}}$ and $\boldsymbol{H}_{\mathcal{B}}$ in each band $\mathcal{B}$, but ignoring frequency derivative terms. Stellar emission is restricted to the non-IR bands. \textsc{Lydion} allows for stellar emission from a single star, or a star cluster. These are modelled as follows:
\begin{itemize}[leftmargin=*]
    \item \textit{\textbf{Single star:}} When modelling emission from a single star, the stellar emissivity is taken to have a Gaussian spatial distribution, with $\sigma_{\star} = 3 \, \rm cells$ by default. Our experiments have shown that $\sigma_\star \gtrsim 2-3$ is required to obtain approximately spherical \textsc{H\,ii} regions in uniform media \citep[also see][]{Skinner2015, Chan2021, Menon2023}. We estimate stellar luminosities in each band following \cite{Deng2024}, who provide SED-derived fits to photon emission rates in all bands, as a function of the zero-age main sequence stellar mass $m_{\star}$ and stellar metallicity $10^{-8} \lesssim Z_{\star}/Z_{\odot} \lesssim 1$. This covers the full range from primordial Pop III stars to present-day metal-enriched stars.

    \item \textit{\textbf{Star clusters:}} When modelling emission from star clusters, we also consider a Gaussian density profile for simplicity, but determine $\sigma_\star$ from the desired half-mass radius of the star cluster, $R_{\rm h} \simeq 1.538 \, \sigma_\star$. Here too we adopt photon emission rates for individual stars from \cite{Deng2024}, but IMF-average these rates and multiply by the mass of the star cluster. We assume a fully sampled Kroupa IMF \citep[][]{Kroupa2001}, extending from $0.08 - 100 \, \rm M_\odot$, and only consider emission/feedback from stars more massive than $8 \, \rm M_\odot$ \citep[see][]{Deng2024}.
\end{itemize}
By default, we also model diffuse emission of ionizing photons from recombinations of \textsc{H\,ii}, He\textsc{\,ii}, and He\textsc{\,iii}, following \cite{Rosdahl2013} and \cite{Kannan2019}, in which case we use Case A recombination coefficients for these species.\footnote{The Case B recombination coefficients are appropriate if we did \textit{not} model recombination radiation, and the gas was optically thick ($\tau \gtrsim \textrm{few}$) to ionizing photons \citep{Nebrin2023_rec}. An advantage of the M1 moment method over ray-tracing methods is that we can model the diffuse recombination radiation self-consistently at no extra performance cost.} In Appendix~\ref{Stellar RT appendix} we test the stellar radiative transfer implementation in \textsc{Lydion}, and its coupling to thermochemistry, by running a D-type photoionization test \citep[test 5 of][]{Iliev2009}.

\begin{figure}
    \centering
    \includegraphics[width=1.0\columnwidth]{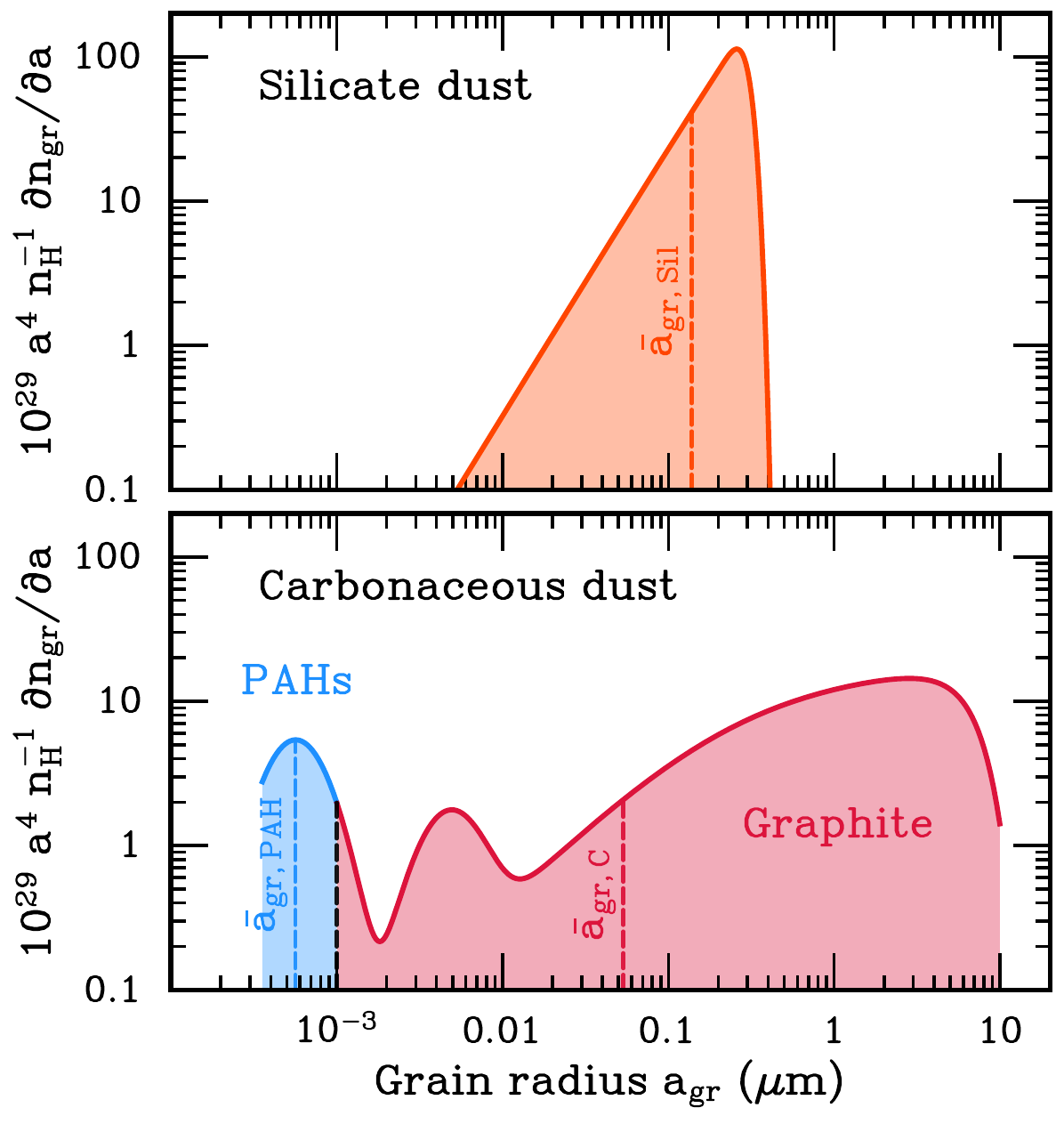}
    \caption{The dust grain size distribution for silicate grains (top panel) and carbonaceous grains (bottom panel) for the $R_{\rm V} = 5.5$, $b_{\rm C} = 3 \times 10^{-5}$, case B dust model of \cite{Weingartner2001}, for their fiducial Milky Way dust abundance (cf. their fig.~6). We split up the latter into PAHs ($< 10^{-3} \, \rm \mu m$) and graphite dust ($> 10^{-3} \, \rm \mu m$). We have also marked the area-weighted mean grain sizes, $\Bar{a}_{\rm gr} \equiv \langle a_{\rm gr}^3\rangle / \langle a_{\rm gr}^2\rangle$, for each dust bin/type by dashed lines (see Appendix~\ref{IR appendix} for values).  }
    \label{fig: Dust distribution}
\end{figure}

We include absorption of photons by all tracked gas species and dust. Scattering by dust and electrons (Thomson scattering) is also implemented. \textsc{Lydion} improves on earlier M1 modelling in simulations by taking into account anisotropic dust scattering, rather than assuming that dust scattering is isotropic. Absorption and scattering of photons exert radiation pressure, just as for Ly$\alpha$ (Sec.~\ref{Lya radiation pressure numerical method}). We compute band-averaged dust opacities for all dust bins/types (silicate, graphite, and PAH dust), assuming the $R_{\rm V} = 5.5$ (case B, $b_{\rm C} = 3 \times 10^{-5}$) dust model of \cite{Weingartner2001} as initial conditions, appropriate for denser, star-forming regions (see Fig.~\ref{fig: Dust distribution}).\footnote{We leave a more complete model of dust physics (e.g. with dust coagulation and additional dust bins) to future work.} More specifically, we use the data made available by B. T. Draine for smoothed astrosilicate dust, graphite dust, and PAHs as a function of grain size and wavelength.\footnote{See the files \texttt{PAHion\_30.gz}, \texttt{Gra\_81.gz}, and \texttt{suvSil\_81.gz}, available at \url{https://www.astro.princeton.edu/~draine/dust/dust.diel.html}. We use opacities for ionized PAHs, since they best represent the PAH charges in \textsc{H\,ii} regions (see Fig.~\ref{Dust properties figure 512 Lya simulation}), where Ly$\alpha$ absorption is most consequential. We use the data from \texttt{PAHion\_30.gz} up to $a_{\rm gr} = 0.01 \, \rm \mu m$, since it smoothly interpolates between PAH and graphite optical properties. } The relevant dust opacities are summarized in Table~\ref{Dust optical properties non IR}. We stress that, because \textsc{Lydion} incorporates dust growth and destruction (see below) as well as dust dynamics, the total dust opacity in a given region will evolve dynamically, beyond what is expected from mere gas dynamics. 

Absorption of stellar and Ly$\alpha$ photons by dust will lead to re-emission in the IR, and, in case $\tau_{\rm IR} \gtrsim 1$, subsequent IR absorption and re-emission. The resulting IR radiation pressure can, potentially, be important in very dense, dusty clouds \citep[e.g.][]{Skinner2015, Rosdahl2015, Menon2022, Menon2023}. For completeness, we therefore model IR transport and feedback. We assume that IR emission is solely by dust, such that $J_{\rm IR}$ evolve according to:
\begin{align}
    \dfrac{1}{\Tilde{c}}\dfrac{\partial J_{\rm IR}}{\partial t} &=~ \sum_\beta \rho_{\rm d,\beta} \left( \kappa_{\rm P,\beta}  \dfrac{\sigma_{\rm SB}T_{\rm d,\beta}^4}{\pi} - \kappa_{\rm P,\beta} J_{\rm IR} \right ) \\ &-~ \boldsymbol{\nabla}\boldsymbol{\cdot}\left(\boldsymbol{H}_{\rm IR} + \dfrac{\boldsymbol{u}}{c}J_{\rm IR}\right)  - \boldsymbol{\mathsf{K}}_{\rm IR} \boldsymbol{:}\dfrac{\boldsymbol{\nabla}\boldsymbol{u}}{c} \, , \nonumber
\end{align}
where $T_{\rm d,\beta}$ is the dust temperature for dust bin $\beta$, and $\rho_{\rm d,\beta}\kappa_{\rm P,\beta}$ is the Planck-mean dust absorption opacity for this bin, evaluated at the dust bin temperature $T_{\rm d,\beta}$ (for the emission term), or IR radiation temperature $T_{\rm rad}$ (for the absorption term $\propto J_{\rm IR}$). The dust temperatures are computed self-consistently by balancing dust emission, and dust absorption of photons in all bands (including IR and Ly$\alpha$), and collisional heating/cooling.\footnote{\cite{Menon2022} have highlighted that the use of temperature-dependent IR dust opacities is crucial for accurately modelling IR radiation pressure feedback. This is naturally done in \textsc{Lydion}, with fully self-consistent dust physics and heating.} The IR radiation temperature is estimated using a prescription introduced by \cite{Grudic2021}. We refer the reader to Appendix~\ref{IR appendix} for a more complete discussion of the IR transport implementation in \textsc{Lydion}. 

To model growth and destruction of dust, and its coupling to gas dynamics, we implement the following processes:
\begin{itemize}[leftmargin=*]
    \item \textbf{\textit{Dust growth by accretion}:} We allow the dust density in each bin to grow by accretion of metals in the ISM. The rate for a dust bin/type $\beta \in \rm (Sil,C,PAH)$ is proportional to the availability of the limiting gas-phase metal:
    \begin{equation}
        \dfrac{\partial \rho_{\rm d,\beta}}{\partial t}\bigg\lvert_{\rm growth} = \dfrac{\rho_{\rm d,\beta}}{t_{\rm gr,\beta,0}}\left(1 - \dfrac{\rho_{\rm d,\beta}}{\rho_{\rm d,\beta} + \rho_{\rm X}} \right) \, . 
    \end{equation}
    The accretion time-scale $t_{\rm gr,\beta,0}$ is estimated following \cite{Hirashita2019}, adopting the recent results for the sticking coefficient on amorphous carbon grains by \cite{Bossion2024}. For accretion onto graphite and PAH dust, $\rho_{\rm X} = \rho_{\rm C}$ (the gas-phase carbon density), and for silicate dust we take $\rho_{\rm X} = \rho_{\rm Si}$ (the gas-phase Si density). The accretion onto silicate dust depletes the gas-phase Si abundance, and similarly for the carbon abundance with respect to growth of graphite and PAH dust.

    \item \textbf{\textit{Dust destruction}:} Dust in our simulations can be destroyed by thermal sputtering \citep{Micelotta2010, Choban2026}, non-thermal sputtering \citep[][]{Hu2019}, sublimation \citep[][]{Guhathakurta1989, Waxman2000}, chemical sputtering of graphite grains \citep[][]{Lenzuni1995, Borderies2025}, and photodissociation of PAHs \citep[][]{Murga2019, Murga2020}. \textit{Thermal sputtering} of silicate and graphite grains is only important for gas temperatures $T \gtrsim 10^5 \, \rm K$, which is typically only encountered in case stellar wind feedback is simulated. However, thermal sputtering of PAHs can be important down to $T \sim 10^4 \, \rm K$ \citep[][]{Micelotta2010}. \textit{Sublimation} of dust can be important in hyper-compact \textsc{H\,ii} regions \citep{Arthur2004}, where dust temperatures can reach the threshold $\gtrsim 1200 - 1800 \, \rm K$ for rapid sublimation. \textit{Non-thermal sputtering} can destroy dust moving at relative velocities $\gtrsim 50 \, \rm km \, s^{-1}$ with respect to the gas. Such high drift velocities are common near massive stars as a result of radiation pressure on dust \citep[e.g.][]{Draine2011_HII, Akimkin2015, Akimkin2017, Ishiki2018, Soliman2024}. 
    
    \textit{Chemical sputtering} of graphite grains can occur when they accrete hydrogen atoms and produce hydrocarbons that leave the grain \citep[][]{Draine1979_ChemicalSputtering, BarNun1980, Sorrell1992, Lenzuni1995, Borderies2025}. This erosion mechanism for graphite grains can be important in dense \textsc{H\,ii} regions \citep[][]{Draine1979_ChemicalSputtering}. \textit{Photodissociation} of small PAHs can occur following absorption by UV photons, leading to loss of carbon atoms in case cooling by infrared emission is inefficient \citep[][]{Allain1996, Murga2019, Murga2020, Murga2022, CALIMA2026}. Observations indeed show depletion of PAHs in \textsc{H\,ii} regions, connected to the strong UV background in these regions \citep[][]{Pavlyuchenkov2013, Egorov2023}.\footnote{During testing of the PAH photodissociation implementation, we found that the simulated PAHs, with $N_{\rm C} = 82$ carbon atoms (for the assumed characteristic PAH size), were efficiently destroyed in \textsc{H\,ii} regions (see $q_{\rm PAH}$ in Fig.~\ref{Dust properties figure 512 Lya simulation} for an example). This is consistent with observations, although has been hard to explain theoretically for $N_{\rm C} \gtrsim 60-80$ \citep[e.g.][]{Pavlyuchenkov2013, Murga2019, Kirsanova2023}. We are greatly indebted to Maria S. Murga for carefully checking our modelling assumptions and calculations. In short, we find efficient PAH destruction primarily because (\textit{i}) energetic He-ionizing photons give large estimated dissociation probabilities, and (\textit{ii}) because the photoelectric emission probability estimated with the model of \cite{Kimura2016} is not unity at these energies, which allows energetic photons to partake in photodissociation of PAHs. \cite{Kirsanova2023} suggested that extreme UV photons could potentially explain the destruction/depletion of large PAHs in \textsc{H\,ii} regions. Our modelling, albeit simplified, supports their hypothesis. } 
    
    As described in detail in Appendix~\ref{Dust growth destruction appendix}, we estimate the dust destruction time-scale $t_{\rm dest,\beta}$ taking the aforementioned processes into account, and add the corresponding dust destruction term to the evolution of the dust density for each dust bin/type $\beta$:
    \begin{equation}
        \dfrac{\partial \rho_{\rm d,\beta}}{\partial t}\bigg\lvert_{\rm destruction} = -\dfrac{\rho_{\rm d,\beta}}{t_{\rm dest,\beta}} \, . 
    \end{equation}

    \item \textbf{\textit{Dust drag}:} Dust is coupled to the gas via drag:
    \begin{equation}
        \dfrac{\partial(\rho_{\rm d}\boldsymbol{u}_{\rm d})_{\beta}}{\partial t} \bigg\lvert_{\rm drag} = \mathcal{K}_{\beta} (\boldsymbol{u} - \boldsymbol{u}_{\rm d,\beta}) \, .
    \end{equation}
    We compute the drag coefficients $\mathcal{K}_\beta$ taking into account normal gas-dust collisions, as well as Coulomb drag between charged dust grains, ions, and electrons \citep[][]{DraineSalpeter1979_dustdrag, Akimkin2015, Akimkin2017, Ishiki2018}. The grain charges are computed self-consistently taking into account photoelectric charging by UV photons (including Ly$\alpha$) \citep[][]{Weingartner2001_Photoelectric, Kimura2016}, and collisional charging by ions and electrons \citep{DraineSutin1987}. Typically, the Coulomb drag dominates in \textsc{H\,ii} regions, but can still be overwhelmed by radiation pressure on larger grains \citep[][]{Draine2011_HII, Akimkin2015,Akimkin2017, Ishiki2018}. 
\end{itemize}
We subcycle dust growth/destruction and drag (together with radiation pressure) in the overall photo-thermochemistry substeps (see Fig.~\ref{LYDION RHD operator split method}). More detailed discussions of gas-dust drag and dust growth/destruction can be found in Appendix~\ref{Dust dynamics appendix} and \ref{Dust growth destruction appendix}, respectively. 

Finally, for completeness, we have implemented stellar wind feedback in \textsc{Lydion}, following \cite{Deng2024}, although by default it is turned off to avoid severe CFL time-step constraints on $\Delta t_{\rm Hydro}$, caused by hot and fast-moving shock-heated gas. Furthermore, we note that stellar wind feedback is only important at high metallicities. If turned on, stellar wind momentum and mass is injected each global step $\Delta t_{\rm Hydro}$ in a finite volume centered on the star, weighted by the local smoothed stellar density. This injection takes place after the final gravity update, but before subcycling of radiative transfer and photo-thermochemistry (see Fig.~\ref{LYDION RHD operator split method}).

\subsection{Photo-thermochemistry}
\label{Photothermochemistry section}

To model the thermo-chemical state of both primordial and metal-enriched gas in detail, we explicitly track the non-equilibrium photochemistry of $e^-$, \textsc{H\,i}, \textsc{H\,ii}, H$_2$, \textsc{D\,i}, \textsc{D\,ii}, HD, He\textsc{\,i}, He\textsc{\,ii}, He\textsc{\,iii}, \textsc{C\,i}, \textsc{C\,ii}, CO, \textsc{O\,i}, \textsc{O\,ii}, CH$_{\rm x}$ (representing CH, CH$_2$, CH$^+$, CH$_2^+$, CH$_3^+$), and OH$_{\rm x}$ (representing OH, H$_2$O, OH$^+$, H$_2$O$^+$, H$_3$O$^+$).\footnote{We also model the equilibrium chemistry of H$^-$ (Appendix~\ref{Mol hydrogen chemistry appendix}) and H$^+_3$ (Appendix~\ref{Carbon chemistry appendix}). These species are important in the formation of H$_2$ and CO, respectively.} For electrons, atomic hydrogen, and helium, we use recombination and collisional ionization rates summarized in \cite{Rosdahl2013}. Heating from photoionization of \textsc{H\,i}, H$_2$, He\textsc{\,i}, and He\textsc{\,ii} follows \cite{Kannan2020}. To better model the ISM outside \textsc{H\,ii} regions, we also implement photoelectric heating \citep{Weingartner2001_Photoelectric}, heating from photodissociation and vibrational excitation of H$_2$ \citep{Hollenbach1979, Baczynski2015}, chemical heating (cooling) from H$_2$ formation (collisional dissociation) \citep[][]{Hollenbach1979, Yoshida2006}, and heat exchange from dust-gas collisions (see Appendix~\ref{IR appendix}). For photoelectric heating, because we evolve the dust (though in a coarse manner), we cannot adopt fits that assume a fixed grain size distribution. Thus, for self-consistency we instead compute the heating directly, as outlined in Appendix~\ref{PhotoelectricHeatingAppendix}.

Heating and cooling is then subcycled in a manner following \cite{Rosdahl2013}, by solving $\Dot{e} = - \Lambda_{\rm net}$ over (typically small) subcycle steps for the gas internal energy density $e = nk_{\rm B}T/(\gamma -1)$,\footnote{We recover $e$ from the dual energy variable $e_{\rm dual}$, which is equal to $E - (\rho \boldsymbol{u})^2/2$ in case truncation errors in this subtraction are small. As for the adiabatic index $\gamma$, we estimate it using a simplified method wherein $\gamma_{\rm H_2} = 7/5$, and $\gamma_i = 5/3$ for $i \in (\textrm{\textsc{H\,i}}, \textrm{\textsc{H\,ii}}, \textrm{He\textsc{\,i}}, \textrm{He\textsc{\,ii}}, \textrm{He\textsc{\,iii}}, e^-)$ \citep[and using this in eq.~5 of][]{Omukai1Nishi998}. In reality, $\gamma_{\rm H_2}$ depends on temperature, but the constant--$\gamma_{\rm H_2}$ approximation has been shown to be adequate in most circumstances \citep{Sharda2019_GammaH2}. } where $\Lambda_{\rm net}$ is the \textit{net} cooling rate. As in \cite{Rosdahl2013}, we use a linearized BDF1 method for stability, such that over a subcycle step $\bullet \rightarrow \bullet \bullet$ of size $\Delta t_{\rm sub}$ we solve:
\begin{equation}
    \dfrac{e^{\bullet\bullet} - e^{\bullet}}{\Delta t_{\rm sub}} = - \left[\Lambda_{\rm net}(e^\bullet) + \dfrac{\partial \Lambda_{\rm net}}{\partial e}\Big\lvert_{e^{\bullet}} (e^{\bullet\bullet} - e^{\bullet}) \right] \, ,
\end{equation}
where $\partial_e \Lambda_{\rm net}$ is computed numerically. The update $e^{\bullet\bullet}$ is accepted if $e$ changed by $< 10\%$, and if $\Delta t_{\rm sub} < 0.1 \, e^{\bullet}/\lvert \Lambda_{\rm net}(e^{\bullet})\rvert$.\footnote{We also check that $e^{\bullet\bullet}$ indeed was a good approximation to a fully non-linear BDF1 update, and only accept $e^{\bullet \bullet}$ if that is the case.} If not, we retry with a smaller substep $\Delta t_{\rm sub} \rightarrow \Delta t_{\rm sub} /2$ \citep{Rosdahl2013}.

We model H$_2$ chemistry mostly following \cite{Hopkins2023} \citep[for similar modelling, see][]{Katz2017, Nickerson2018, Kannan2020, Park2021}, including H$_2$ formation in the gas-phase via H$^-$, on dust grains, in three-body reactions, and H$_2$ destruction by collisional dissociation (via $e^-$, \textsc{H\,i}, \textsc{H\,ii}, H$_2$), Lyman-Werner photodissociation \citep[including self-shielding, following][]{Draine1996}, and photoionization by EUV1, EUV2, and EUV3 photons \citep[][]{Baczynski2015, Kannan2020}. The adopted rates for the chemical reactions are mostly the same as in \cite{Hopkins2023}, with a few exceptions where we have opted for more accurate and/or recent fits.\footnote{For H$_2$ formation in the gas-phase via H$^-$, \cite{Hopkins2023} adopts rates compiled in \cite{GloverJappsen2007}. Where there are differences, we instead adopt the rates compiled in \cite{Glover2010}, except for $\textrm{H}+e^- \rightarrow \textrm{H}^- + h\nu$ and  $\textrm{H}^- + \textrm{H} \rightarrow \textrm{H}_2 + e^-$. For the former rate we use the fit from \cite{Galli1998}, following discussion of different fits in \cite{Glover2015}. For the latter we use the accurate fit from \cite{Kreckel2010}. Finally, to model collisional dissociation of H$_2$, we adopt the rates and fits from \cite{Glover2010} rather than the fits from \cite{Glover2008} used by \cite{Hopkins2023}.} A detailed discussion of H$_2$ photochemistry and cooling in \textsc{Lydion} can be found in Appendix~\ref{Mol hydrogen chemistry appendix}.

Cooling by HD can be important in primordial gas \citep[e.g.][]{McGreer2008, Hirano2015, Lenoble2024}.\footnote{\textsc{D\,i} can also scatter Ly$\alpha$ photons and leave an imprint in the Ly$\alpha$ spectrum \citep[][]{Dijkstra2006, RASCAS, Stace2025}. However, \cite{Stace2025} found that \textsc{D\,i} has only a minor impact on Ly$\alpha$ feedback. We therefore leave the implementation of \textsc{D\,i} scattering to future work. } We therefore include a simple model of deuterium chemistry, tracking the non-equilibrium abundances of \textsc{D\,i}, \textsc{D\,ii}, and HD, in a similar manner to \cite{McGreer2008}. We use updated rate coefficients where possible \citep[e.g.][]{Faure2024}, and also add a few additional processes to those considered by \cite{McGreer2008}, including photodissociation of HD, and photo- and collisional ionization of \textsc{D\,i} (see Table~\ref{Deuterium chemistry rates}). We account for self-shielding by HD \citep[][]{Draine1996}, shielding of HD by H$_2$ \citep[][]{Wolcott2011_HD}, and shielding of HD by \textsc{H\,i} and dust in the calculation of $J_{\rm LW}$. We refer the reader to Appendix~\ref{Deuterium chemistry appendix} for more details about the implementation of deuterium photochemistry and cooling in \textsc{Lydion}. 

Carbon and oxygen chemistry (\textsc{C\,i}, \textsc{C\,ii}, \textsc{O\,i}, \textsc{O\,ii}, CH$_{\rm x}$, OH$_{\rm x}$, and CO) is implemented following the simplified networks of \cite{Gong2017} and \cite{Khatri2024}. We model photoionization of \textsc{O\,i} and \textsc{C\,i}, and photodissociation of CH$_{\rm x}$, OH$_{\rm x}$, and CO. Cooling by \textsc{O\,i} ($63.18\, \rm \mu m$, $145.5\, \rm \mu m$), \textsc{C\,i} ($609.13 \, \rm \mu m$, $370.41 \, \rm \mu m$), and \textsc{C\,ii} ($157.75 \, \rm \mu m$) is properly implemented by solving for the relevant level populations, following \cite{Kim2023} and \cite{Deng2024}. CO-cooling becomes important in dense metal-enriched clouds, and is therefore also included, based on modelling of \cite{Hollenbach1979} and \cite{Whitworth2018}, which includes suppression at high column densities. Equilibrium cooling by other metals at high temperatures is modelled following \cite{Kim2023}.\footnote{\cite{Kim2023} provides a fit for cooling by nebular lines, valid at $< 3.5 \times 10^4 \, \rm K$. At higher temperatures they smoothly interpolate to cooling by metals in collisional ionization equilibrium (CIE). We follow the same method, using data from \cite{Orly2012} for CIE cooling, for the species C, N, O, Ne, Mg, Si, S, and Fe. We assume the abundance ratios of \cite{Asplund2009}, and scale both the nebular and CIE cooling rates by $Z/Z_{\odot}$. } More details about the carbon and oxygen photochemistry and cooling in \textsc{Lydion} can be found in Appendix~\ref{Carbon chemistry appendix}. To estimate the column densities for self-shielding of H$_2$ and HD, and for CO-cooling, we use a Sobolev-like estimate, e.g. $(N_{\rm H_2})_{i,j} = (n_{\rm H_2})_{i,j} (\Delta \ell_{\rm Sob})_{i,j}$, where the path length is estimated according to:\footnote{Since the Lyman-Werner band in our model does not resolve individual lines, an approximate self-shielding treatment is necessary. The Sobolev-like estimate of the column density is a common, numerically cheap method \citep[e.g.][]{Gnedin2009, WolcottGreen2011_H2, Chiaki2023, Hopkins2023}, and has been shown to give satisfactory agreement with ray tracing results \citep[][]{WolcottGreen2011_H2, Chiaki2023}, albeit with some caution \citep[][]{Chiaki2023, Nguyen2026}.}
\begin{align}
(\Delta \ell_{\rm Sob})_{i,j} &= \min\left[ (\Delta \ell_{\rm cell})_{i,j} + \dfrac{\rho_{i,j}}{\lvert (\boldsymbol{\nabla}\rho) _{i,j}\rvert}, \, \Delta \ell_{\rm max} \right] \, , \label{Sobolev path length}
\end{align}
where $(\Delta \ell_{\rm cell})_{i,j} = \max(\Delta R_i,\Delta Z_j)$, and we choose $\Delta \ell_{\rm max} = \min(\Delta \ell _{\rm box},  1 \, \rm kpc)$, where $\Delta \ell_{\rm box} = \max(R_{\rm max}, Z_{\rm max})$ is the box size, to avoid unphysical column densities if $\lvert \boldsymbol{\nabla}\rho \rvert \rightarrow 0$.\footnote{The upper limit of $1 \, \rm kpc$ is consistent with our target applications, ranging from individual star-forming clouds, to the virial radii of low-mass halos hosting early star formation \citep[][]{Nebrin2023_starbursts}.} 

Taken together, the detailed photo-thermochemistry network in \textsc{Lydion} allows for a realistic treatment of atomic and molecular cooling over a wide range of gas densities, metallicities, and dust-to-gas ratios, while also tracking key species like $e^-$, H$_2$, \textsc{H\,i} that can destroy Ly$\alpha$ photons in dense environments, by boosting $p_{\rm d}$ \citep[][]{Nebrin2024}.

\subsection{Summary of operator split RHD method}

We have now described the physics implemented in \textsc{Lydion}, and the corresponding methods used. To advance by a global time-step $\Delta t_{\rm Hydro}$, we use the operator split order shown in Fig.~\ref{LYDION RHD operator split method}. To summarize, hydro and gas-dynamical advection and gravity (and stellar winds, if on) are first considered, after which we sub-cycle radiative transport, emission/absorption, and various other stiff processes (e.g. gas cooling) within the global hydro advection step. This ensures stability, as well as good performance, since we avoid a hydrodynamical solve every time we advect photons (the CFL time-step for the latter is generally much smaller than the former). To further ensure accuracy and stability, we limit $\Delta t_{\rm Hydro}$ as follows:
\begin{equation}
    \Delta t_{\rm Hydro} = \min( \Delta t_{\rm CFL}, \Delta t_{\rm CFL,d}, \Delta t_{\rm grav}, \Delta t_{\rm RP} ) \, ,
\end{equation}
where the CFL (for gas and dust) and gravity-limited time-steps are:
\begin{align}
    \Delta t_{\rm CFL} &=~ \textrm{CFL}_{\rm Hydro}  \times \dfrac{1}{D} \min_{i,j} \left[ \dfrac{\Delta X_{i,j}}{(c_{\rm s})_{i,j} + \lvert (\boldsymbol{u})_{i,j} \rvert} \right] \, , \nonumber \\ \Delta t_{\rm CFL,d} &=~ \textrm{CFL}_{\rm Hydro}  \times \dfrac{1}{D} \min_{i,j, \beta} \left[ \dfrac{\Delta X_{i,j}}{\lvert (\boldsymbol{u}_{\rm d,\beta})_{i,j} \rvert} \right] \, , \nonumber \\
    \Delta t_{\rm grav} &=~ C_{\rm grav} \times \min_{i,j}\left[ \sqrt{\dfrac{2 \Delta X_{i,j}}{\lvert (\boldsymbol{\nabla}\Phi)_{i,j} \rvert}} \,, (t_{\rm ff})_{i,j} \right] \, , \nonumber
\end{align}
where $t_{\rm ff} = \sqrt{3 \pi /32G(\rho + \rho_{\rm d})}$ is the free-fall time-scale, $\Delta X_{i,j} =\min(\Delta R_i, \Delta Z_j)$, and $D = 2$ is the number of dimensions. By default we choose $\textrm{CFL}_{\rm Hydro} = 0.4$, and $C_{\rm grav} = 0.1$. Finally, we take into account the radiation pressure acceleration time-scale \citep[see e.g.][]{Byran2014}: 
\begin{equation}
    \Delta t_{\rm RP} = \min( \Delta t_{\rm RP,g}, \Delta t_{\rm RP,g+d}, \Delta t_{\rm RP,d}) \, ,
\end{equation}
where:\footnote{We note that because we inject radiation pressure momentum \textit{after} hydro advection, our scheme is always stable, since the next CFL-limited time-step will automatically self-adjust. The reason we use $C_{\rm RP} < 1$ is rather to ensure accuracy. }
\begin{align}
    \Delta t_{\rm RP,g} &=~ C_{\rm RP}^{\rm g} \min_{i,j}\left[ \sqrt{\dfrac{2 \Delta X_{i,j}}{\lvert (\boldsymbol{a}_{\rm rad, g})_{i,j} \rvert}} \,\right] \,, \\ \Delta t_{\rm RP,g+d} &=~ C_{\rm RP}^{\rm g+d} \min_{i,j}\left[ \sqrt{\dfrac{2 \Delta X_{i,j}}{\lvert (\boldsymbol{a}_{\rm rad, g+d})_{i,j} \rvert}} \,\right] \,, \\
    \Delta t_{\rm RP,d} &=~ C_{\rm RP}^{\rm d} \min_{i,j,\beta}\left[ \sqrt{\dfrac{2 \Delta X_{i,j}}{\lvert (\boldsymbol{a}_{\rm rad, d})_{i,j}^\beta \rvert}} \,\right] \,.
\end{align}
Here $\boldsymbol{a}_{\rm rad, g}$, $\boldsymbol{a}_{\rm rad, g+d}$, and $\boldsymbol{a}_{\rm rad, d}^{\beta}$ are the radiation pressure accelerations experienced by gas, gas + dust (assuming perfect coupling), and dust bin $\beta$, respectively. By default, we choose $C_{\rm RP}^{\rm g} = C_{\rm RP}^{\rm g+d} = 0.1$, and $C_{\rm RP}^{\rm d} = 0.4$. Finally, we allow for at most a $25\%$ increase in $\Delta t_{\rm Hydro}$ from the last time-step \citep{Stone1992}.

\section{Testing the new Lyman-$\alpha$ Radiative transfer method in Lydion}
\label{Test section}

\begin{figure*}
    \centering
    \includegraphics[width=0.85\textwidth]{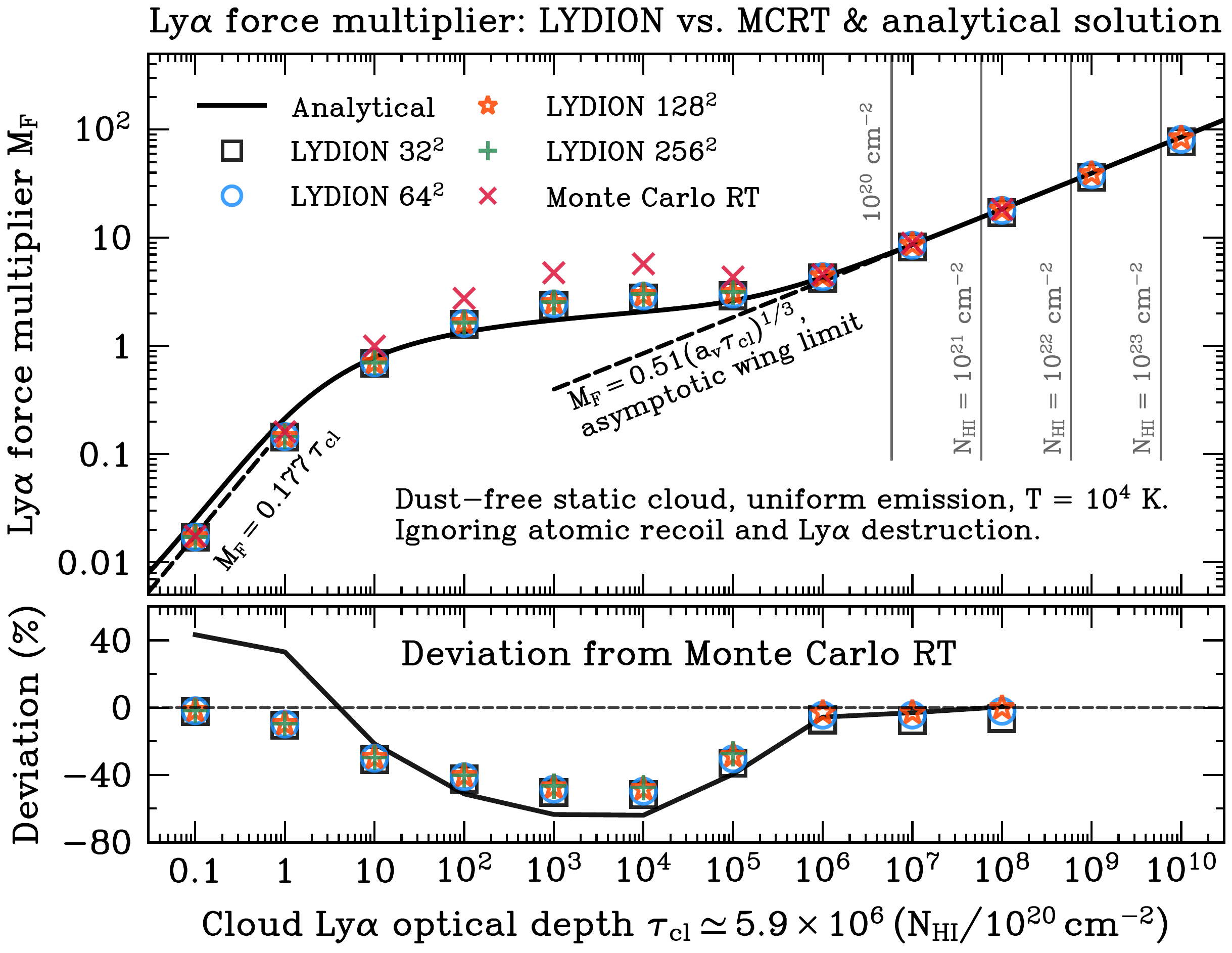}
    \caption{ The predicted force multiplier $M_{\rm F}$ for clouds of varying Ly$\alpha$ optical depth $\tau_{\rm cl}$ (Test 1). \textsc{Lydion} predictions for different spatial resolutions $32^2 - 256^2$ are shown (symbols), together with analytical predictions \citep[solid line, using the closed-form solution of][]{Tomaselli2021}, and MCRT results from \textsc{colt} (diamonds) for $0.1\leq \tau_{\rm cl} \leq 10^8$. In the top panel, dashed black lines show the asymptotic relations for optically thin clouds (Eq.~\ref{M_F tau goes to zero limit}), and highly optically thick clouds, $M_{\rm F} = 0.51 \, (a_{\rm v}\tau_{\rm cl})^{1/3}$ \citep[][]{Lao2020}. A few reference values of the \textsc{H\,i} column density $N_{\rm HI}$ are also shown (vertical gray lines). We note that the clouds in our RHD simulations have $N_{\rm HI} \gtrsim 10^{23} \, \rm cm^{-2}$ before breakout of the ionization front (see Table~\ref{RHD scenarios table}).  The bottom panel shows the deviations of the \textsc{Lydion} and analytical predictions from the MCRT results. The significant deviations for $10 \lesssim \tau_{\rm cl} \lesssim 10^5$ arise because of the breakdown in the Fokker--Planck approximation. }
    \label{fig: MF test, static dust-free}
\end{figure*}

In this section we test the accuracy of the new M1 + Fokker--Planck Ly$\alpha$ RT solver in \textsc{Lydion}, by comparing predictions to accurate MCRT results, and analytical solutions that are valid in the optically thick regime. We obtain the MCRT results using the \textsc{colt} code \citep[][]{Smith2015}. For verification tests of other code modules in \textsc{Lydion} --- e.g. hydrodynamics, gravity, and stellar and IR RT --- we refer the reader to the Appendix. 

In the tests below, as well as in our RHD simulations, we will often study various force multipliers, which measures the boost of the Ly$\alpha$ radiation pressure force with respect to $L_{\rm Ly\alpha}/c$ \citep[e.g.][]{Dijkstra2008, Smith2019, Lao2020, Tomaselli2021, Nebrin2024}. Thus, we define:
\begin{align}
    M_{\rm F} &\equiv~ \dfrac{1}{L_{\rm Ly\alpha}/c} \int \textrm{d}V \, \lvert \boldsymbol{f}_{\rm Ly\alpha} \rvert \,, \label{Naive M_F def} \\
    M_{\rm F,radial} &\equiv~ \dfrac{1}{L_{\rm Ly\alpha}/c} \int \textrm{d}V \,  \boldsymbol{f}_{\rm Ly\alpha} \boldsymbol{\cdot} \boldsymbol{\hat{r}} \,, \label{Radial M_F def} 
\end{align}
Among these definitions, $M_{\rm F}$ is the original definition of the force multiplier found in the literature, but does not take the directionality of the force into account. For this reason we also consider $M_{\rm F,radial}$, measuring the Ly$\alpha$ force directed radially outwards from the center of the cloud (or the stellar source in RHD simulations). If $M_{\rm F} \gg 1$ or $M_{\rm F,radial} \gg 1$, Ly$\alpha$ feedback is expected to be strong. In spherically symmetric settings, $M_{\rm F} = M_{\rm F, radial}$.

\subsection{Lyman-$\alpha$ source in a uniform spherical cloud}

As a first set of tests, we consider spherically symmetric and uniform \textsc{H\,i} clouds of temperature $T$, and line center optical depth $\tau_{\rm cl} = \alpha R_{\rm cl}$, surrounded by a vacuum. For all these tests, the simulation box extends to $1.1 \times R_{\rm cl}$ in each direction, with the cloud center placed at the origin (i.e. we assume midplane symmetry). Outflow boundary conditions are assumed at the outer edges of the simulation box, and reflective (i.e. zero flux) boundary conditions at the cylindrical axis $R = 0$. We set $\Tilde{c} = c$, and evolve up until $t_{\rm max} = 4 \max[\, (a_{\rm v} \tau_{\rm cl})^{1/3}, 3] \, R_{\rm cl} /c$, corresponding to at least a few trapping times in the dust-free, static cloud limit \citep{Smith2018DDMC, McClellan2022}.  We perform the following tests:\footnote{All the \textsc{Lydion} tests in this section were run on a Macbook Pro laptop, with 8 CPU threads. With this setup, a handful of $128^2$ simulations could be run in one day.}

\begin{figure}
    \centering
    \includegraphics[width=1.0\columnwidth]{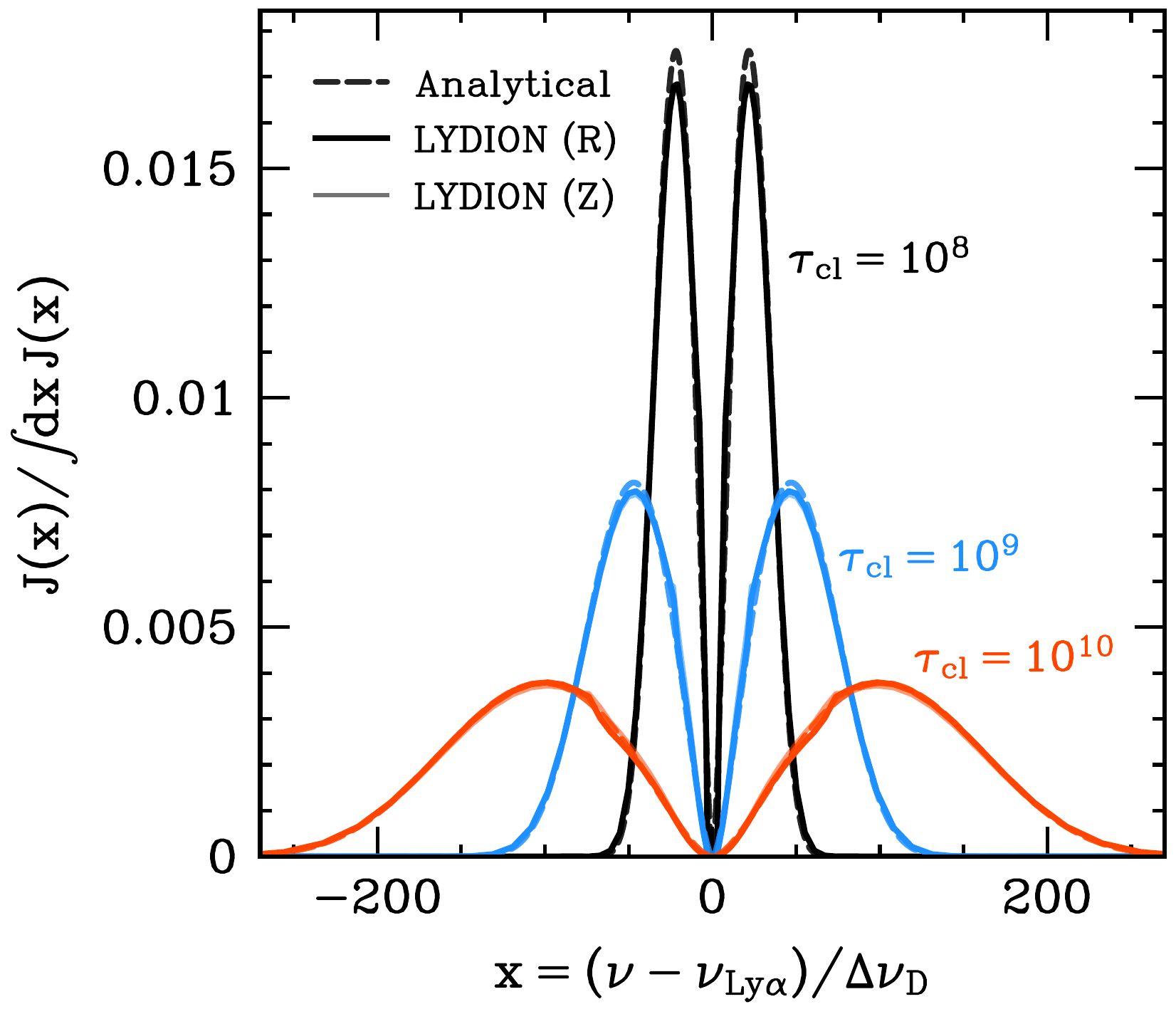}
    \caption{ Predicted normalized emergent spectra from static dust-free clouds, of temperature $T = 10^4 \, \rm K$, with uniform Ly$\alpha$ emission, and ignoring atomic recoil (Test 1). Dashed lines show the analytical solution of \citet{Lao2020}. Solid lines show the spectra obtained from \textsc{Lydion} (for spatial resolution $128^2$), at the outer $R$ and $Z$-boundaries. }
    \label{fig: Spectra test, static dust-free}
\end{figure}

\begin{figure}
    \centering
    \includegraphics[width=0.9\columnwidth]{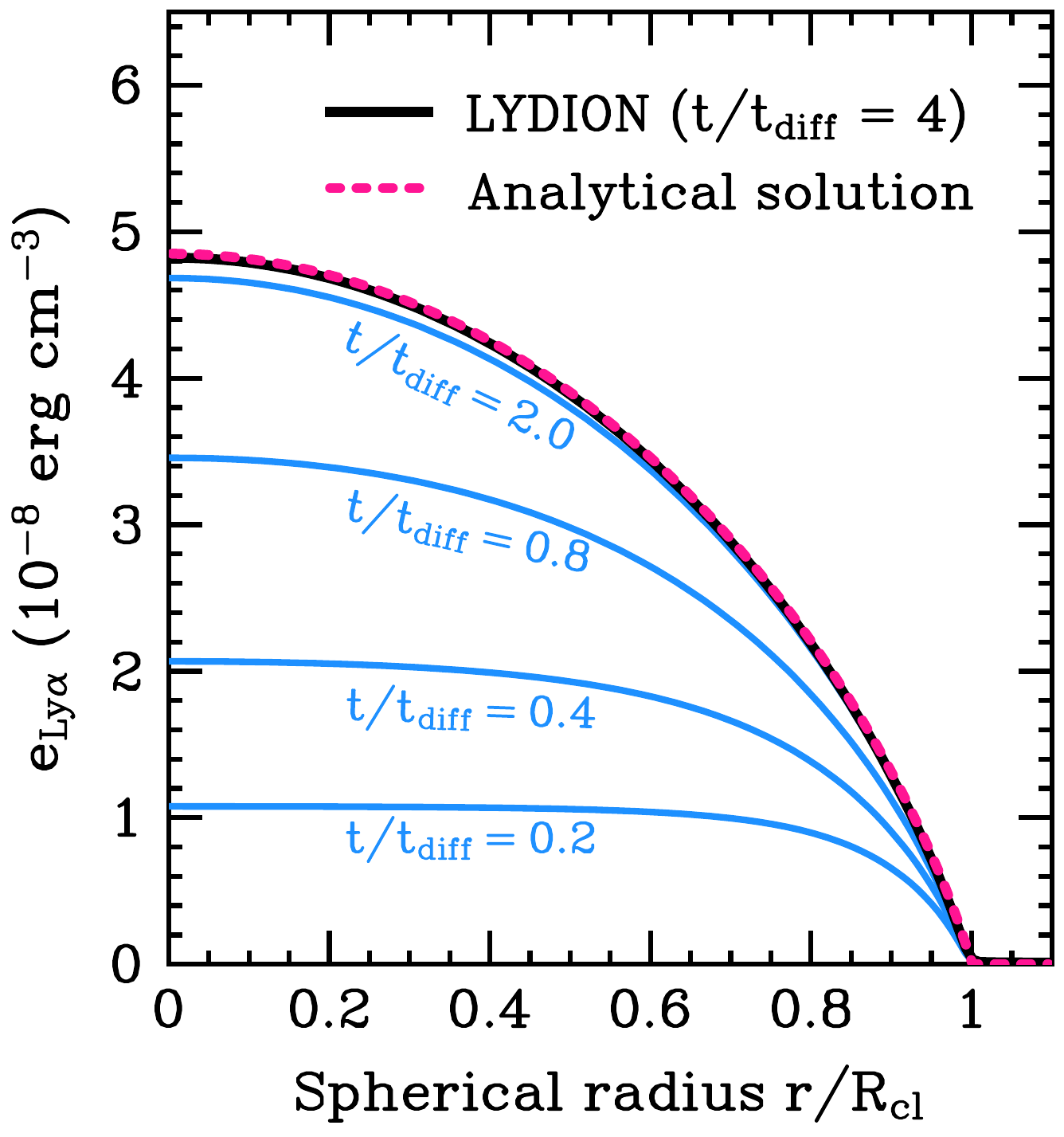}
    \caption{ The predicted spherically averaged Ly$\alpha$ energy density profile $e_{\rm Ly\alpha}(r)$, for the dust-free static cloud setup (Test 1: $\tau_{\rm cl} = 10^{10}$, $T = 10^4 \, \rm K$, uniform Ly$\alpha$ emission, ignoring atomic recoil). Solid lines show the \textsc{Lydion} results (for spatial resolution $128^2$) at different times $t$, in units of $t_{\rm diff} \equiv (a_{\rm v} \tau_{\rm cl})^{1/3} R_{\rm cl}/c$. The dashed pink line shows the analytical solution of \citet{Lao2020} for the same setup. The numerical results converge to the analytical solution as time progresses. }
    \label{fig: e_Lya, static dust-free}
\end{figure}

\begin{enumerate}[leftmargin=*]
    \item \textbf{Static dust-free clouds (Test 1):} First we consider dust-free static clouds (no velocity gradients), and further ignore Ly$\alpha$ destruction, and atomic recoil. We first consider clouds of temperature $T = 10^4 \, \rm K$, with uniform Ly$\alpha$ emission, and Ly$\alpha$ optical depth spanning 11 orders of magnitude, $0.1 \leq \tau_{\rm cl} \leq 10^{10}$. This range covers both the optically thin and highly optically thick regimes. In Fig.~\ref{fig: MF test, static dust-free} we plot the Ly$\alpha$ force multiplier $M_{\rm F}$ predicted by \textsc{Lydion}, for spatial resolutions $32^2 - 256^2$. Also shown are the results from MCRT tests for the same setup, which we obtained with \textsc{colt} using $10^6$ photons, and without any core-skipping approximation. Since the MCRT simulations become expensive to run at high $\tau_{\rm cl}$ without core-skipping, we only run these simulations up to $\tau_{\rm cl} = 10^8$.
    
    Finally, to test \textsc{Lydion} predictions at higher $\tau_{\rm cl}$, we also plot the analytical solution \citep{Lao2020, Tomaselli2021}. Analytical solutions for $M_{\rm F}$ are now known to be highly accurate for $\tau_{\rm cl} \gg 10^3/ a_{\rm v} \simeq 2 \times 10^6 \, T_4^{1/2}$, where the intensity is nearly isotropic (justifying the Eddington approximation), and the Fokker--Planck approximation becomes increasingly accurate \citep[for tests, see][]{Lorinc2025, Nebrin2024, Smith2025}. For very optically thin clouds ($\tau_{\rm cl} \lesssim 1$), we find that \textsc{Lydion} is in good agreement with MCRT, with both approaching the exact limit:
    \begin{equation}
        M_{\rm F}(\tau_{\rm cl} \ll 1) = \dfrac{\tau_{\rm cl}}{4\sqrt{2}} \, . \label{M_F tau goes to zero limit}
    \end{equation}
    This is the result for pure scattering and negligible frequency redistribution.\footnote{The result can be derived by solving $\boldsymbol{\nabla}\boldsymbol{\cdot}\boldsymbol{H} = j_{\rm s}$ for a uniform source, with an emission profile $\phi_\nu$. This yields $\boldsymbol{H} = L_{\rm Ly\alpha} \phi_\nu r \,\boldsymbol{\hat{r}}/ R_{\rm cl}^3$, and $M_{\rm F} = \tau_{\rm cl} \eta /4$, where $\eta \equiv \int \textrm{d}\nu \, \mathcal{H} \phi_\nu$. If all photons are emitted at line center, $\eta = 1$. In contrast, for a Voigt emission profile, $\eta \simeq 1/\sqrt{2}$ (for small $a_{\rm v}$).} The analytical solution of \cite{Tomaselli2021} is larger than Eq.~(\ref{M_F tau goes to zero limit}) by a factor $\sqrt{2}$ because their derivation assumed a Dirac delta emission profile, rather than the full Voigt profile $\phi_\nu \propto \mathcal{H}$ \citep[this is to simplify analytical calculations, see][]{Harrington1973, Nebrin2024}. In the opposite limit of high $\tau_{\rm cl}$ ($\gtrsim 10^6$), we find excellent agreement between \textsc{Lydion}, MCRT, and the analytical solution, as expected. We note that the clouds in our RHD simulations have $N_{\rm HI} \gtrsim 10^{23} \, \rm cm^{-2}$, which falls into this regime (see Table~\ref{RHD scenarios table}). 
    
    In the intermediate regime $10 \lesssim \tau_{\rm cl} \lesssim 10^5$, we find that \textsc{Lydion} and the analytical solution underestimate $M_{\rm F}$, albeit slightly less so for \textsc{Lydion}. The error in the \textsc{Lydion} predictions for this regime is $\lesssim 50\%$, and likely stems from the Fokker--Planck (FP) approximation. The FP approximation is expected to break down for scattering near the core-wing transition \citep{Lorinc2025}, which becomes relevant at these optical depths. Ditching the FP approximation would likely reduce this error, but at significant numerical cost. We therefore stick with the FP approximation for this paper, and note that the deviation from MCRT is only expected to be relevant when the ionization front breaks out in RHD simulations, at which point Ly$\alpha$ feedback becomes less important regardless. Furthermore, because \textsc{Lydion} \textit{underestimates} $M_{\rm F}$ in this regime, our conclusions regarding the importance of Ly$\alpha$ feedback in this paper will tend to be more conservative.  

    Finally, in Fig.~\ref{fig: Spectra test, static dust-free} we plot the emergent spectra from the clouds with the highest optical depth ($\tau_{\rm cl} = 10^8, 10^9, 10^{10}$), and in Fig.~\ref{fig: e_Lya, static dust-free} we plot the spherically averaged Ly$\alpha$ energy density profile $e_{\rm Ly\alpha}(r)$ for $\tau_{\rm cl} = 10^{10}$. We compare both predictions to the analytical solution of \cite{Lao2020}, and find excellent agreement. 

    \item \textbf{Impact of velocity gradients (Test 2):} Next we test the ability of \textsc{Lydion} to capture velocity gradients (Doppler shifts). In Fig.~\ref{fig: Spectra expanding clouds test} we plot emergent comoving spectra, from clouds expanding radially in a Hubble-like manner $\boldsymbol{u} = (\Dot{R}_{\rm cl}/R_{\rm cl}) r\, \boldsymbol{\hat{r}}$, with $\Dot{R}_{\rm cl} = (10, 200) \, \rm km \, s^{-1}$. We compare to the MCRT results from \textsc{colt}, obtained for the same setup in \cite{Smith2025}. The results are in good overall agreement. Minor deviations are visible for $200 \, \rm km \, s^{-1}$, likely caused by several factors, including the $\tau_{\rm cell} \sim 1$ transition between the GLF and diffusion fluxes in the M1 treatment, differences in where and how the spectra are taken (e.g. $J$ vs. $\boldsymbol{H}$, or at the simulation box edges vs. the cloud edge), chosen frames (comoving vs. lab frame), the Fokker--Planck approximation, and frequency binning. To test the latter, we run the same \textsc{Lydion} simulations with finer frequency bins ($q = 1.03$, giving $N_\nu = 594$ bins), and find better agreement with MCRT for $\Dot{R}_{\rm cl} = 200 \, \rm km \, s^{-1}$.

    \begin{figure}
    \centering
    \includegraphics[width=1.0\columnwidth]{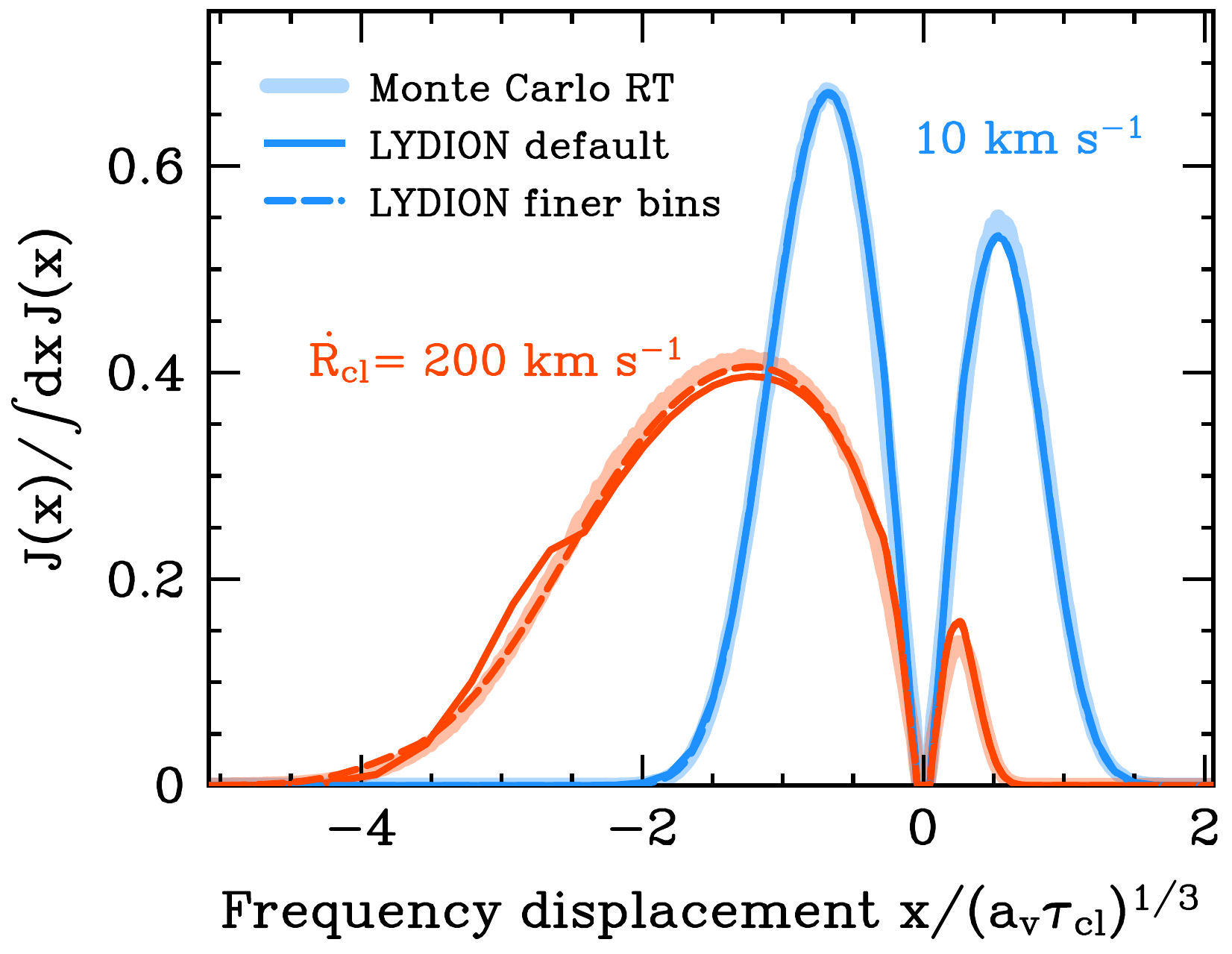}
    \caption{ Emergent spectra (averaged over all boundary cells), for expanding dust-free uniform clouds of optical depth $\tau_{\rm cl} = 5 \times 10^8$, temperature $T = 9084 \, \rm K$, and uniform Ly$\alpha$ emission (Test 2). Solid and dashed lines show spectra from \textsc{Lydion} with the default and higher number of frequency bins, respectively. Bands show MCRT results from \citet{Smith2025}. }
    \label{fig: Spectra expanding clouds test}
    \end{figure}

    \begin{figure}
    \centering
    \includegraphics[width=1.0\columnwidth]{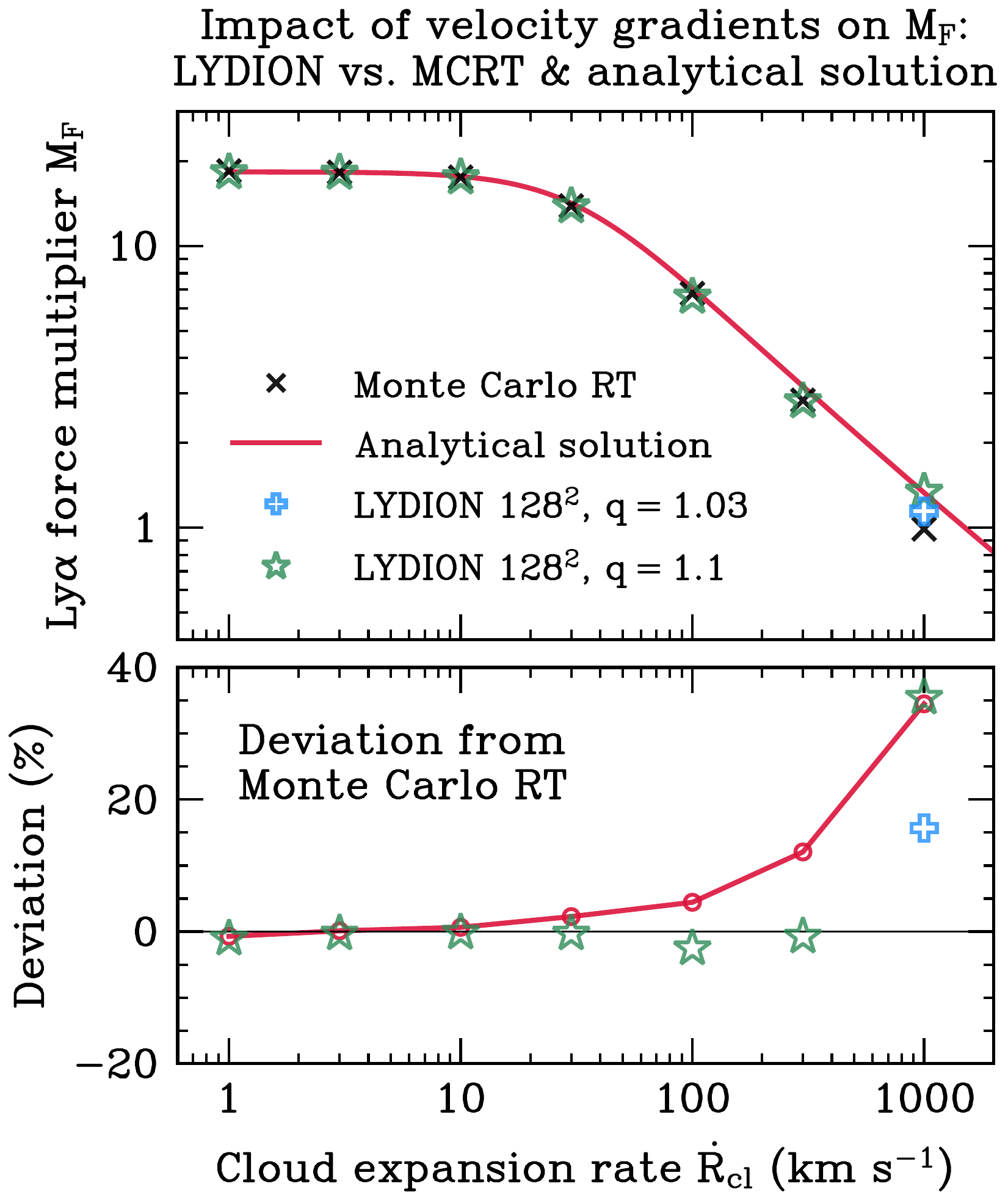}
    \caption{Suppression of the Ly$\alpha$ force multiplier $M_{\rm F}$ with increasing cloud expansion rates $\Dot{R}_{\rm cl}$, for a uniform dust-free cloud with temperature $T = 10^4 \, \rm K$, Ly$\alpha$ optical depth $\tau_{\rm cl} = 10^8$ at line center, and uniform Ly$\alpha$ emission (Test 2). Atomic recoil is ignored. \textbf{Top panel:} \textsc{Lydion} predictions are shown as symbols for a fixed spatial resolution of $128^2$. We run one simulation with more frequency bins ($q = 1.03$) for the highest considered expansion velocity. MCRT results from \textsc{colt} are shown as crosses. The analytical solution of \citet{Nebrin2024} and \citet{Smith2025}, valid in the diffusion limit, is shown as the solid red line. \textbf{Bottom panel:} The deviation (in $\%$) of the \textsc{Lydion} and analytical predictions from the MCRT results.  }
    \label{fig: Velocity suppression MF test}
    \end{figure}

    More important for \textsc{Lydion}, in Fig.~\ref{fig: Velocity suppression MF test} we plot the suppression of the Ly$\alpha$ force multiplier with increasing cloud expansion rate, and compare against the analytical solution of \cite{Nebrin2024} and \cite{Smith2025}, and MCRT results without core-skipping from \textsc{colt}. The predictions of \textsc{Lydion} agree with MCRT to within $< 5\%$ for $\Dot{R}_{\rm cl} \leq 300 \, \rm km \, s^{-1}$. At the highest considered cloud expansion rate, $\Dot{R}_{\rm cl} \leq 1000 \, \rm km \, s^{-1}$, we find significantly larger deviation, of $\sim 35\%$. This discrepancy is reduced to $\sim 15\%$ if we increase the number of frequency bins, with $q = 1.03$. The remaining discrepancy may be due to the neglect of higher-order moment terms in the equation for $\boldsymbol{H}$ (Eq.~\ref{Full H equation}), which \cite{Buchler1983} and \cite{Castor2004} neglect. \cite{Nebrin2024} argue that these extra terms are small in Ly$\alpha$ RT for expansion velocities $\lesssim \textrm{few} \times 100 \, \rm km \, s^{-1}$, consistent with our numerical results. 
    
    These terms could be implemented in future versions of \textsc{Lydion}, following M1 treatments of neutrino transfer \citep[e.g.][]{Just2015}. For now, we just note that discrepancies with respect to MCRT only appear for velocities much greater than those relevant to the applications in this paper (closer to $\sim 10 \, \rm km \, s^{-1}$). We therefore conclude that \textsc{Lydion} correctly captures the impact of velocity gradients on Ly$\alpha$ radiation pressure in relevant regimes. 

    \begin{figure}
    \centering
    \includegraphics[width=0.9\columnwidth]{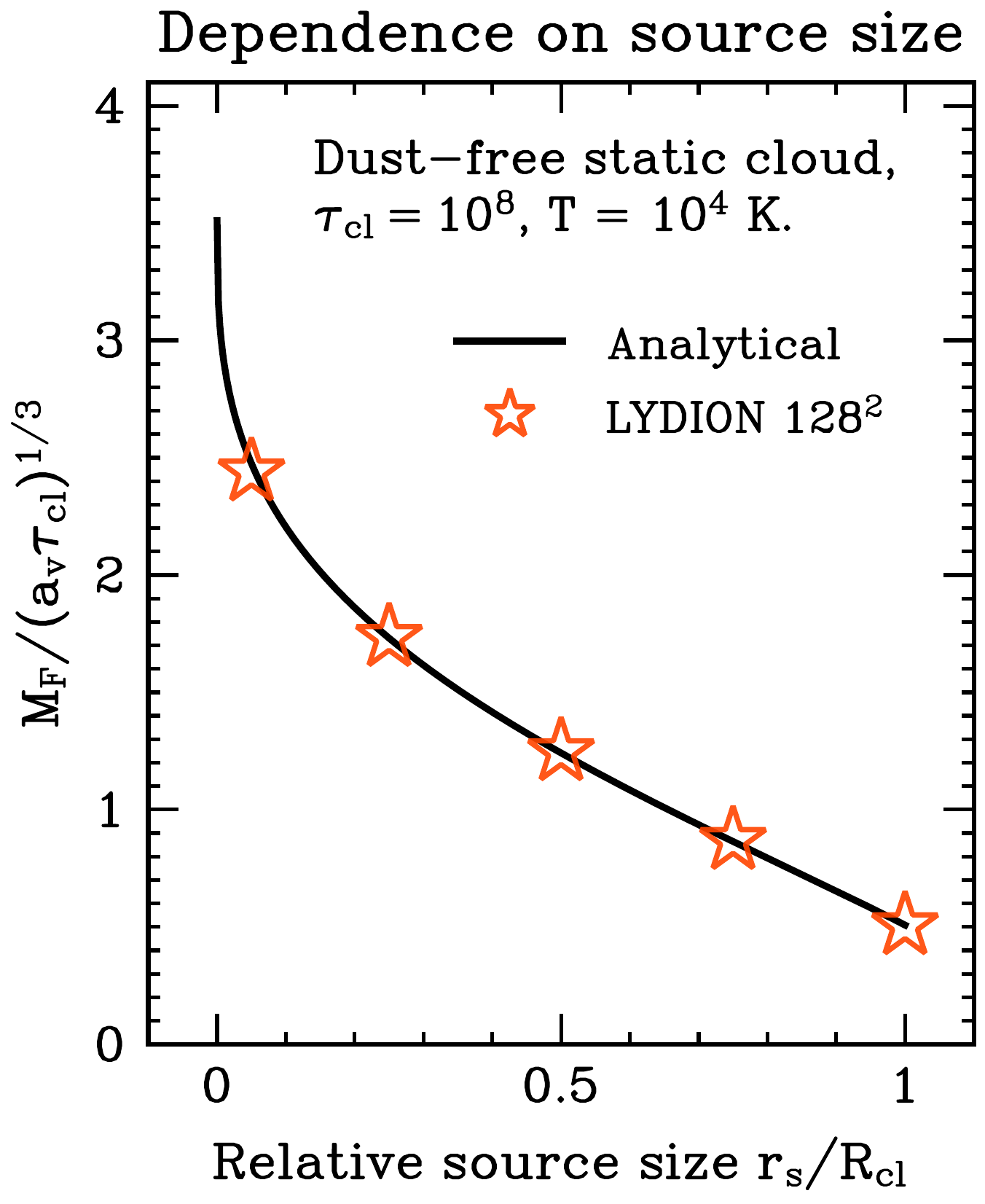}
    \caption{ Predicted dependence of the force multiplier on the relative size of the emission region (Test 3). Solid line shows the analytical solution of \citet{Nebrin2024}, valid in the diffusion limit. Symbols show the predictions of \textsc{Lydion}. }
    \label{fig: Source size test}
    \end{figure}

    \item \textbf{Impact of source distribution (Test 3):} So far we have focused on clouds with uniform Ly$\alpha$ emission, to avoid resolution effects for spatial point-like sources. \cite{Nebrin2024} generalized earlier analytical Ly$\alpha$ RT solutions for uniform and point sources \citep{Dijkstra2006, Lao2020, Tomaselli2021}, to study sources of finite extent $0 < r_{\rm s} \leq R_{\rm cl}$. In Fig.~\ref{fig: Source size test} we compare the predictions for the Ly$\alpha$ force multiplier $M_{\rm F}$ for dust-free static clouds with $\tau_{\rm cl} = 10^8$ and $T = 10^4 \, \rm K$, but varying source radius $0.05 \leq r_{\rm s}/R_{\rm cl} \leq 1$. All simulations are run at a resolution of $128^2$, and we ignore atomic recoil for an ``apples-to-apples'' comparison. We find good agreement between \textsc{Lydion} and the analytical solution.

\begin{figure}
    \centering
    \includegraphics[width=1.0\columnwidth]{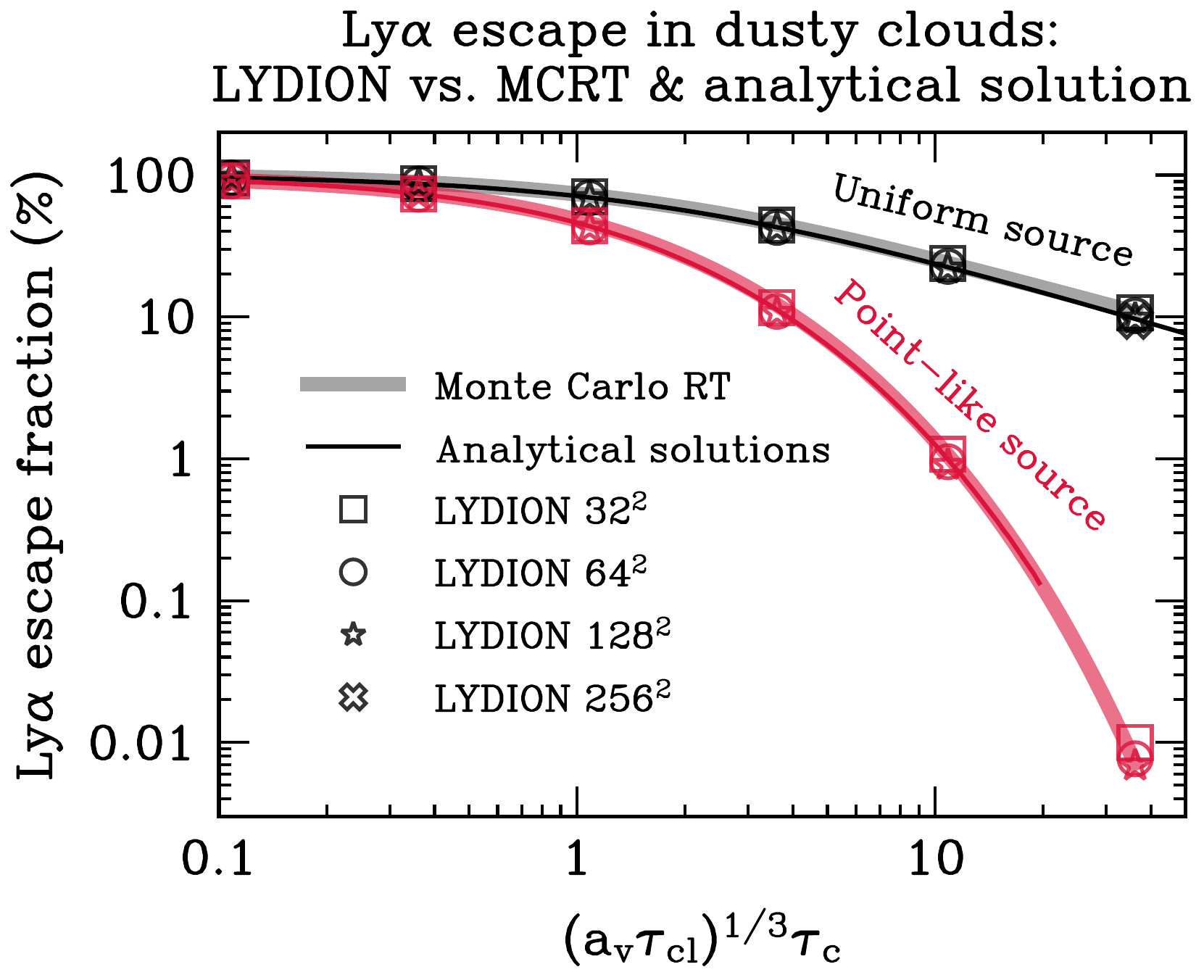}
    \caption{The Ly$\alpha$ escape fraction for different cloud continuum absorption optical depths $\tau_{\rm c}$, and for different source distributions (Test 4): uniform Ly$\alpha$ emission (gray), and  central point-like emission (red). Symbols show results from \textsc{Lydion} for a few values of $\tau_{\rm c}$, and for different spatial resolutions ($32^2 - 128^2$, plus one simulation at $256^2$). Bands show MCRT results from \textsc{colt} assuming a modest static core-skipping ($x_{\rm crit} = 2$), originally presented in fig.~3 of \cite{Nebrin2024}. Solid lines show the predictions of the analytical solution in \cite{Nebrin2024}. For the point-source case, the analytical solution for $f_{\rm esc,Ly\alpha}$ converges extremely slowly for $(a_{\rm v} \tau_{\rm cl})^{1/3} \tau_{\rm c} \gtrsim 20$, so we only plot the solution up to that point. }
    \label{fig: Escape fraction test}
    \end{figure}

    \begin{figure}
    \centering
    \includegraphics[width=1.0\columnwidth]{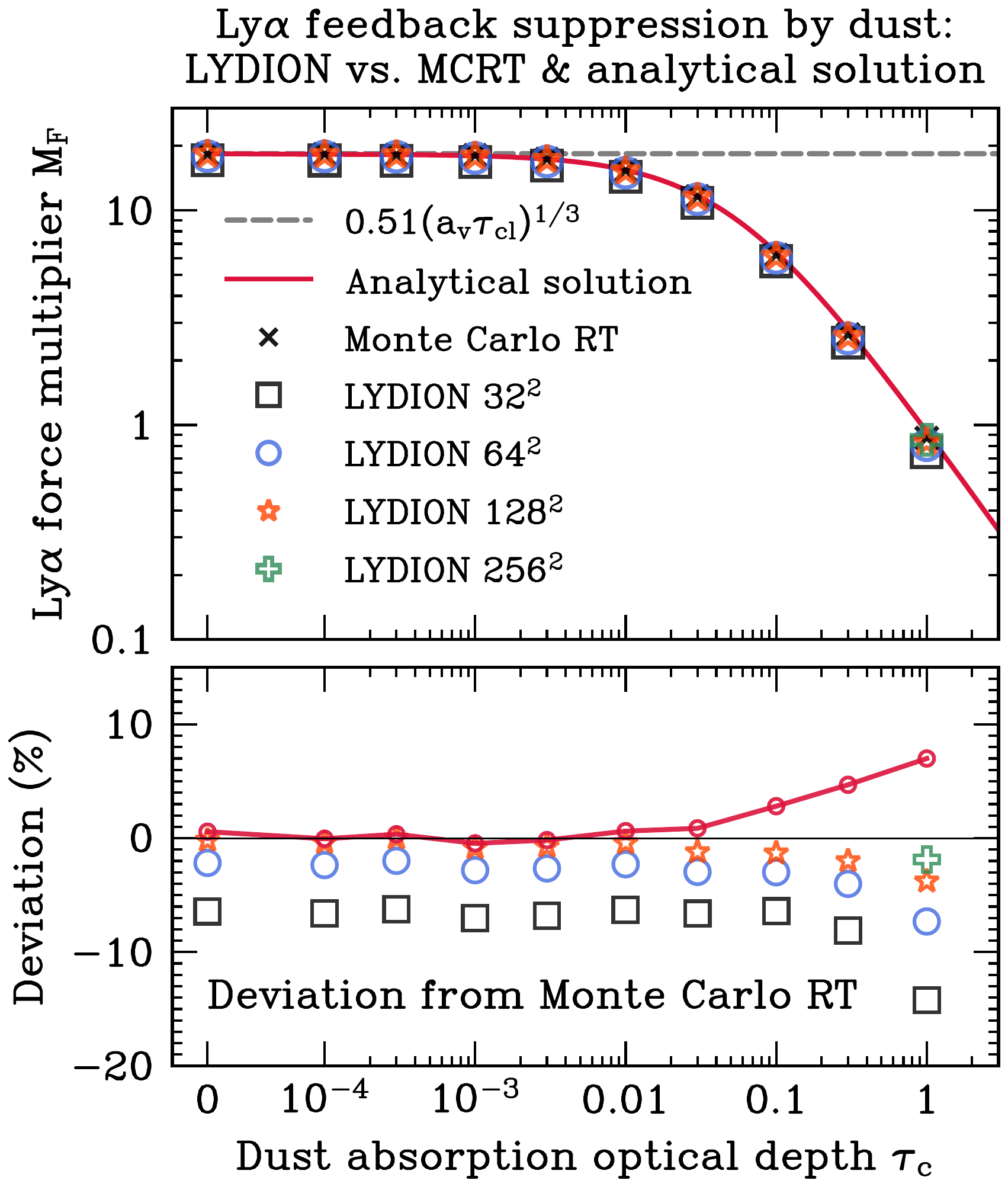}
    \caption{Suppression of the Ly$\alpha$ force multiplier $M_{\rm F}$ with increasing dust absorption optical depth $\tau_{\rm c}$, for a uniform cloud with temperature $T = 10^4 \, \rm K$, Ly$\alpha$ optical depth $\tau_{\rm cl} = 10^8$ at line center, and uniform Ly$\alpha$ emission (Test 5). \textbf{Top panel:} \textsc{Lydion} predictions are shown as symbols for varying spatial resolution, $32^2 - 256^2$. MCRT results from \textsc{colt} are shown as crosses. The analytical solution of \citet{Nebrin2024}, valid in the diffusion limit, is shown as the solid red line. \textbf{Bottom panel:} The deviation (in $\%$) of the \textsc{Lydion} and analytical predictions from the MCRT results.  }
    \label{fig: Dust suppression MF test}
    \end{figure}
    
    \item \textbf{Lyman-$\boldsymbol{\alpha}$ escape from dusty clouds (Test 4):} To test the accuracy of \textsc{Lydion} in capturing dust absorption, we compute the Ly$\alpha$ escape fraction from optically thick, static, uniform clouds with $\tau_{\rm cl} = 10^8$, and $T = 10^4 \, \rm K$. We ignore atomic recoil and Ly$\alpha$ destruction ($p_{\rm d} = 0$). The dust absorption optical depth ranges from $\tau_{\rm c} \sim 0$ to $\tau_{\rm c} \sim 1$, and we consider both uniform Ly$\alpha$ emission, and a point-like source (emission from the innermost cell). This setup is identical to the test in \cite{Nebrin2024}, where analytical predictions were compared to MCRT results from \textsc{colt} \citep{Smith2015}. We estimate the Ly$\alpha$ escape fraction as:
    \begin{equation}
        f_{\rm esc,Ly\alpha} = \dfrac{1}{L_{\rm Ly\alpha}} \int_{0}^{\infty} \textrm{d}\nu  \int_{\rm boundary} \textrm{d}\boldsymbol{A} \boldsymbol{\cdot} 4\pi \boldsymbol{H} \, . \label{fesc Lya definition}
    \end{equation}
    We compare the results from \textsc{Lydion} to the \textsc{colt} MCRT results and analytical solution from \cite{Nebrin2024} in Fig.~\ref{fig: Escape fraction test}. We find that the predictions of \textsc{Lydion} are in excellent agreement with the MCRT and analytical predictions, even at a low spatial resolution of $32^2$. We conclude that \textsc{Lydion} can correctly capture Ly$\alpha$ escape in dusty clouds.

    \item \textbf{Suppression of Lyman-$\boldsymbol{\alpha}$ radiation pressure in dusty clouds (Test 5):} Next, we check whether \textsc{Lydion} can capture the suppression of the Ly$\alpha$ force multiplier by dust. We use the simulation results for the uniform emission scenario in the escape fraction test, and compare to the analytical solution of \cite{Nebrin2024}. In the dust-free high optical depth case, the Ly$\alpha$ force multiplier is expected to be $\simeq 0.51 \, (a_{\rm v} \tau_{\rm cl})^{1/3}$ \citep[see Fig.~\ref{fig: MF test, static dust-free} and][]{Lao2020, Tomaselli2021}. \cite{Nebrin2024} extended the uniform-source solution of \cite{Lao2020} to consider dust absorption, and found that $M_{\rm F}$ drops approximately as $\propto 1 / \tau_{\rm c}$ for dust (or continuum) absorption optical depths $\tau_{\rm c}$ satisfying $(a_{\rm v} \tau_{\rm cl})^{1/3}\tau_{\rm c} \gtrsim 1$. In Fig.~\ref{fig: Dust suppression MF test} we compare \textsc{Lydion} predictions to the analytical solution of \cite{Nebrin2024}, and MCRT results from \textsc{colt} obtained without  core-skipping. Overall we find that all methods broadly agree, with \textsc{Lydion} showing better agreement with MCRT at higher dust absorption optical depths. As dust absorption becomes important, there are fewer Ly$\alpha$ scatterings and hence a less isotropic intensity. Since the analytical solution assumes the Eddington approximation, this will render it less accurate. In contrast, the M1 method of \textsc{Lydion} can naturally capture this regime, yielding better agreement with MCRT. We conclude that \textsc{Lydion} correctly captures the impact of dust absorption on Ly$\alpha$ radiation pressure.
\end{enumerate}

\begin{figure*}
    \centering
    \includegraphics[width=0.95\textwidth]{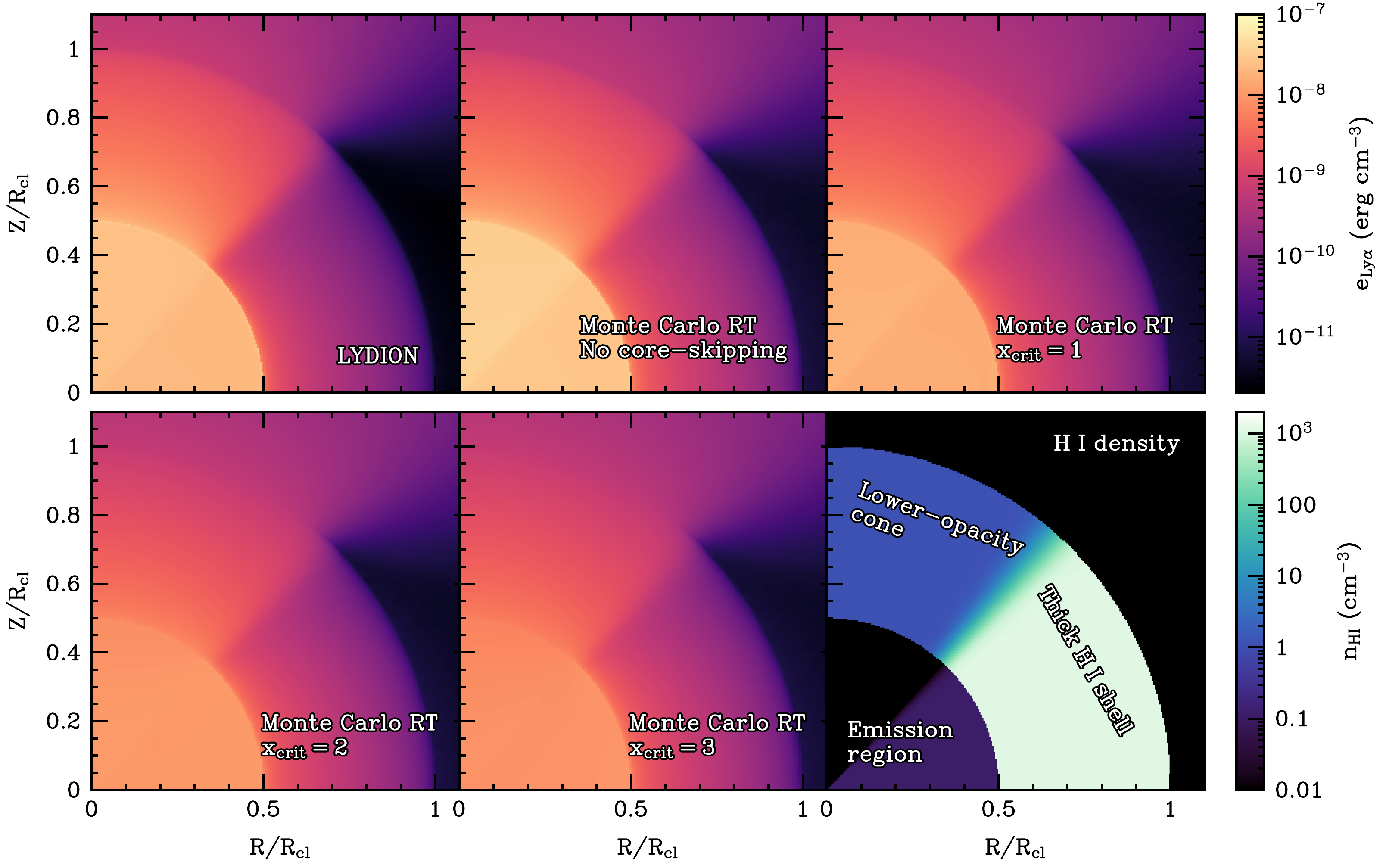}
    \caption{ The Ly$\alpha$ energy density $e_{\rm Ly\alpha}(R,Z)$ and \textsc{H\,i} density for the anisotropic cone test (Test 6). \textbf{Top left panel:} The predicted $e_{\rm Ly\alpha}$ from \textsc{Lydion}. \textbf{Top middle and right panels:} The predicted $e_{\rm Ly\alpha}$ from \textsc{colt}, without core-skipping (middle panel), and with modest core-skipping (right panel, $x_{\rm crit} = 1$). \textbf{Bottom left and middle panels:} The predicted $e_{\rm Ly\alpha}$ from \textsc{colt}, with increasing levels of core-skipping. \textbf{Bottom right panel:} The assumed \textsc{H\,i} density in the cone test, for reference. }
    \label{fig: cone test e_Lya}
\end{figure*}
    
\begin{figure*}
    \centering
    \includegraphics[width=0.95\textwidth]{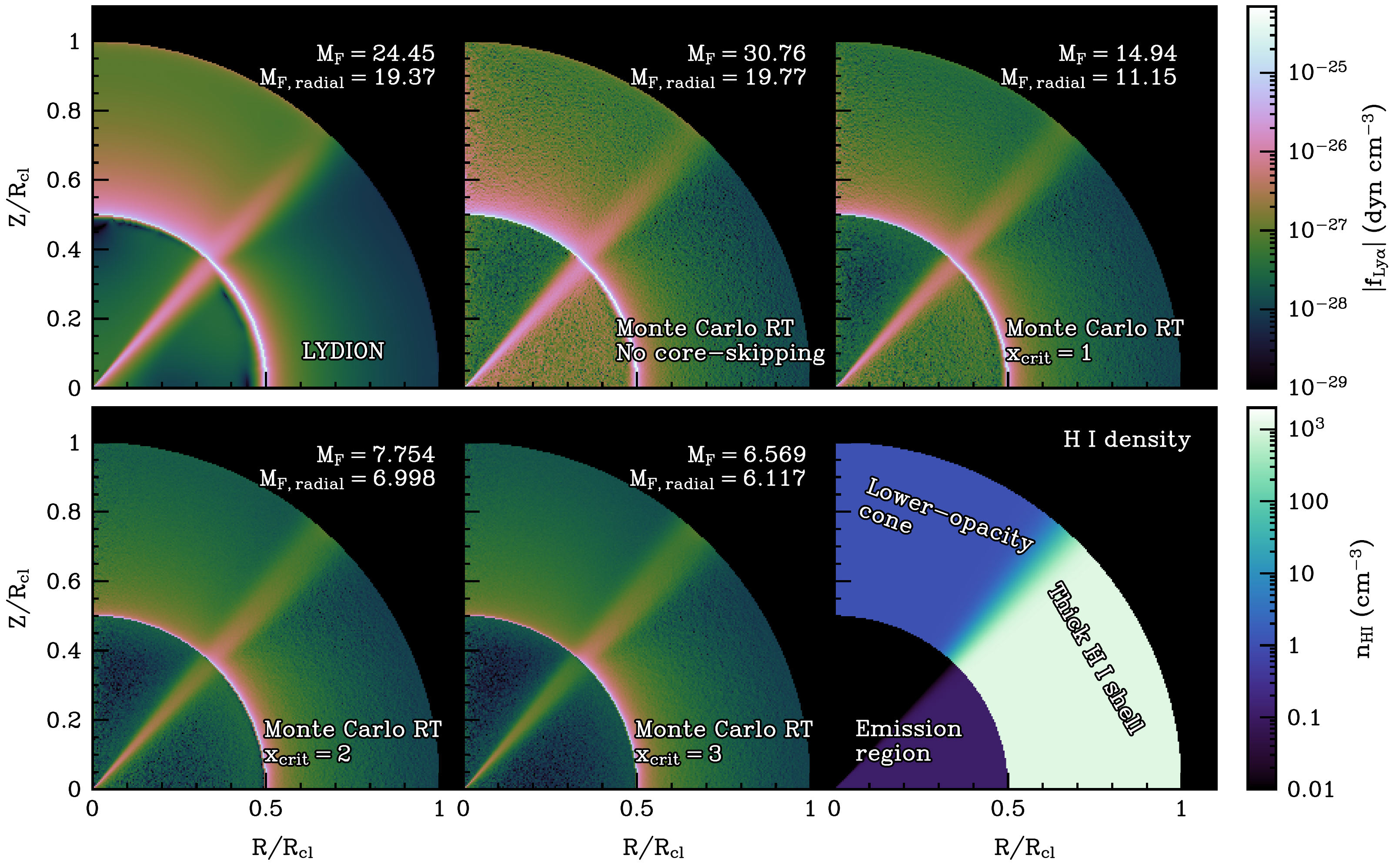}
    \caption{ Same as Fig.~\ref{fig: cone test e_Lya}, but showing the magnitude of the Ly$\alpha$ radiative force per unit volume, $\lvert \boldsymbol{f}_{\rm Ly\alpha} \rvert$, for the anisotropic cone test (Test 6). Also shown are the predicted force multipliers, $M_{\rm F}$ and $M_{\rm F,radial}$. }
    \label{fig: cone test f_Lya}
\end{figure*}

\begin{figure}
    \centering
    \includegraphics[width=0.95\columnwidth]{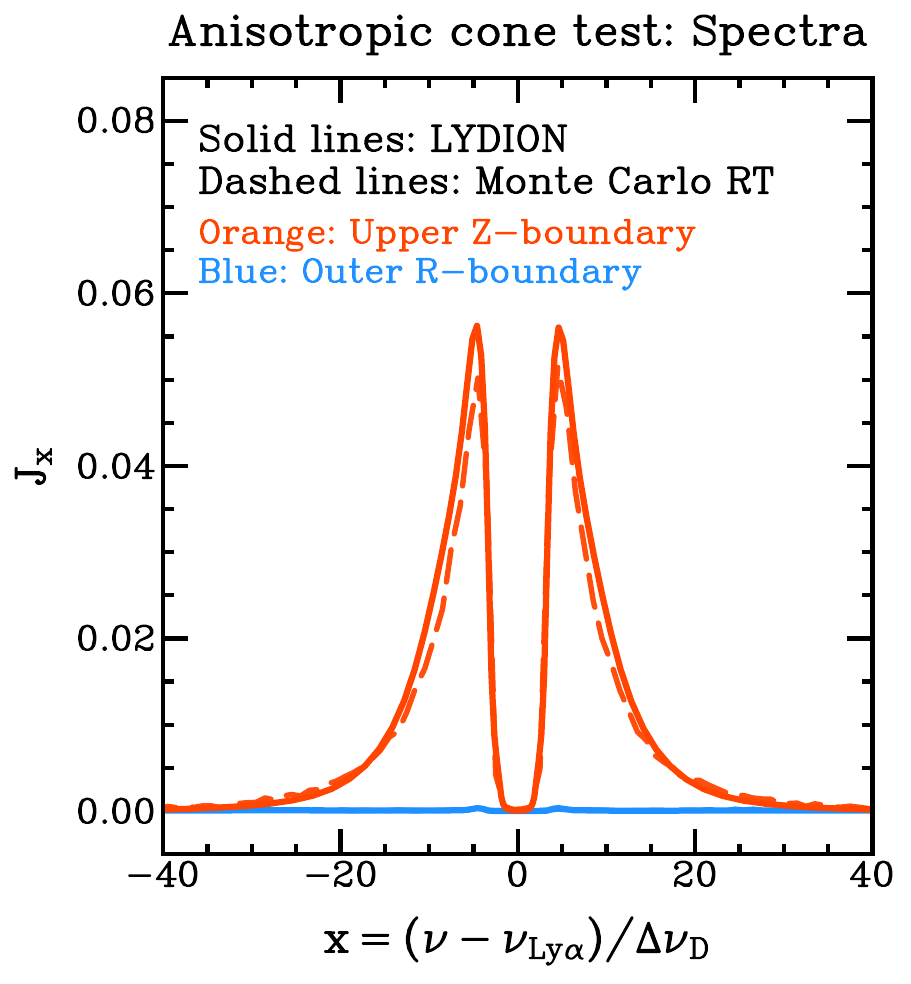}
    \caption{ Raw emergent spectra, $J_x \equiv \Delta \nu_{\rm D} \, J(\nu)$, from the anisotropic cone test (Test 6), obtained near the middle of the upper $Z$-boundary (orange lines), and outer $R$-boundary (blue lines). Solid and dashed lines show the results from \textsc{Lydion} and MCRT, respectively. }
    \label{fig: cone test spectra}
\end{figure}

\subsection{Anisotropic Lyman-$\alpha$ escape (Test 6)}


Next, we test the ability of \textsc{Lydion} to handle anisotropic Ly$\alpha$ escape. To do so, we consider an otherwise spherically symmetric uniform cloud as before, except that we introduce a bipolar lower opacity region aligned with the $Z$-axis. This geometry provides a controlled comparison with MCRT in a configuration where photons can either diffuse through the optically thick cloud or preferentially escape through lower opacity polar channels ($\tau_{\rm cone} \ll \tau_{\rm cl}$).

The cloud is axisymmetric about the $Z$-axis and is specified in cylindrical coordinates $(R,Z)$, with spherical radius $r = (R^2 + Z^2)^{1/2}$ and angular coordinate $\mu = |Z| / r$. The low-opacity region is symmetric above and below the midplane. We take the fiducial half-opening angle of the polar channel to be $\theta_{\rm cone} = \pi/4$ ($45^{\circ})$, corresponding to $\mu_{\rm cone} = \cos\theta_{\rm cone} = 1/\sqrt{2}$. We take the \textsc{H\,i} density within the cloud (spherical radius $r < R_{\rm cl}$) to be $n_{\rm HI} = \epsilon n_{\rm H}$, where:
\begin{align}
    n_{\rm H}(R,Z) &=~ n_{\rm cone} \\ &+~ \frac{n_{\rm cl} - n_{\rm cone}}{2} \left[ 1 - \tanh\left( \frac{\mu - \mu_{\rm cone}}{\Delta\mu} \right) \right] \, , \nonumber 
\end{align}
with $\Delta \mu = 0.025$ controlling the angular width of the transition between the lower opacity polar region and the high-opacity equatorial cloud. We further take $\epsilon = 1$ for $0.5 \,R_{\rm cl} < r \leq R_{\rm cl}$ (the \textsc{H\,i} shell), and $\epsilon = 10^{-4}$ for $r \leq 0.5 \,R_{\rm cl}$ (the emission region). We take the Ly$\alpha$ emission to be uniform in the emission region. For simplicity, the gas has a fixed temperature $T = 10^{4} \, \rm K$, and we choose $n_{\rm cl}$ and $n_{\rm cone}$ such that $\tau_{\rm cl} = 10^8$, and $\tau_{\rm cone} = 10^{5}$. Atomic recoil is turned on, but we ignore Ly$\alpha$ continuum absorption and destruction. The \textsc{Lydion} test is performed at a spatial resolution of $256^2$ cells. For comparison, we also run the same test with \textsc{colt}, with different levels of constant core-skipping, $x_{\rm crit} = (0,1,2,3)$, where $x_{\rm crit} = 0$ corresponds to no core-skipping approximation. To check MCRT convergence, we also vary the photon count from $4\times 10^5$ to $10^7$. To make the cases as similar as possible, we directly feed the \textsc{Lydion} density and emissivity grids into \textsc{colt} to isolate the anisotropic transport effects.

In Fig.~\ref{fig: cone test e_Lya} we plot the predicted Ly$\alpha$ energy density from \textsc{Lydion} and \textsc{colt} (for $10^7$ photons), as well as the \textsc{H\,i} density for reference. Despite the strong anisotropy in the setup, we find that \textsc{Lydion} is in broad agreement with \textsc{colt}, showing very similar anisotropic escape through the cone. The energy density in the emission region is slightly higher in the MCRT run without core-skipping than in \textsc{Lydion}, but the factor is modest, $\sim \textrm{few} \times 10\%$. In contrast, with core-skipping we find $e_{\rm Ly\alpha}$ to be more strongly suppressed in the emission region compared to \textsc{Lydion} and MCRT without core-skipping.

In Fig.~\ref{fig: cone test f_Lya} we plot the magnitude of the Ly$\alpha$ radiation pressure force per unit volume, $\boldsymbol{f}_{\rm Ly\alpha}$. Also shown are the computed force multipliers $M_{\rm F}$ and $M_{\rm F,radial}$. Both \textsc{Lydion} and MCRT predict that the force is concentrated at the inner edge of the \textsc{H\,i} shell. As a result, \textsc{Lydion} finds a remarkably accurate radial force multiplier, $M_{\rm F,radial} = 19.37$, differing by only $2\%$ from the MCRT results with no core-skipping ($M_{\rm F,radial} = 19.77$). Larger errors arise for the force multiplier $M_{\rm F}$, owing to differences in the non-radial forces that are most apparent in the emission region. Here, the results of \textsc{Lydion} underestimate $M_{\rm F}$ by $\sim 20\%$ ($M_{\rm F} = 24.45$ vs. $30.76$). In contrast, the MCRT calculations with core-skipping substantially suppress the force multiplier. Relative to the no-core-skipping MCRT result, $M_{\rm F}$ drops to $14.94$, $7.754$, and $6.569$ for $x_{\rm crit}=1$, $2$, and $3$, respectively. Thus, core-skipping reduces $M_{\rm F}$ by factors of $\simeq 2.1$, $4.0$, and $4.7$. We regard errors of this magnitude as unacceptable in Ly$\alpha$ RHD simulations.

Finally, in Fig.~\ref{fig: cone test spectra} we compare the emergent spectra from the upper $Z$-boundary (near the cone), and near the midplane. These spectra are not normalized to $1$ to highlight anisotropic photon escape. We find that \textsc{Lydion} and MCRT both predict that the vast majority of photons escape through the cone, with similar emergent spectra. In summary, we conclude that \textsc{Lydion} can correctly capture anisotropic Ly$\alpha$ escape.

\subsection{Performance scaling of \textsc{Lydion} and comparison to MCRT}

\begin{figure}
    \centering
    \includegraphics[width=1.0\columnwidth]{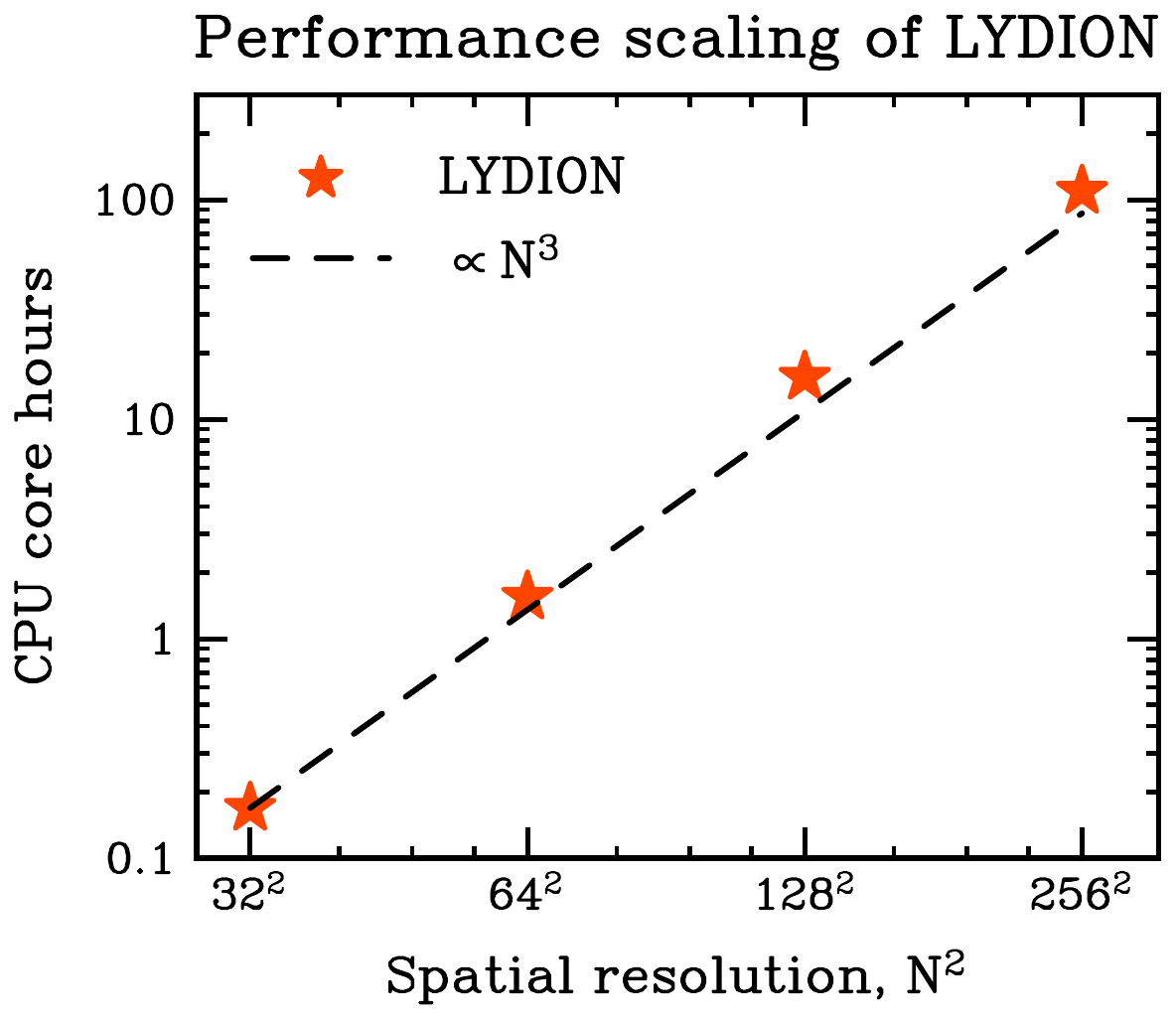}
    \caption{CPU core hours for the anisotropic cone test run on a MacBook Pro laptop (Apple M2 chip with 16 GB RAM) using $8$ CPU threads, for varying spatial resolutions $32^2$--$256^2$. The final time, $t_{\rm max} = 4\,t_{\rm diff}$, and the number of frequency bins are fixed. The dashed line shows the expected $N^3$ scaling for a spatial resolution of $N^2$.}
    \label{fig:cpu_core_time_scaling}
\end{figure}

\begin{figure}
    \centering
    \includegraphics[width=1.0\columnwidth]{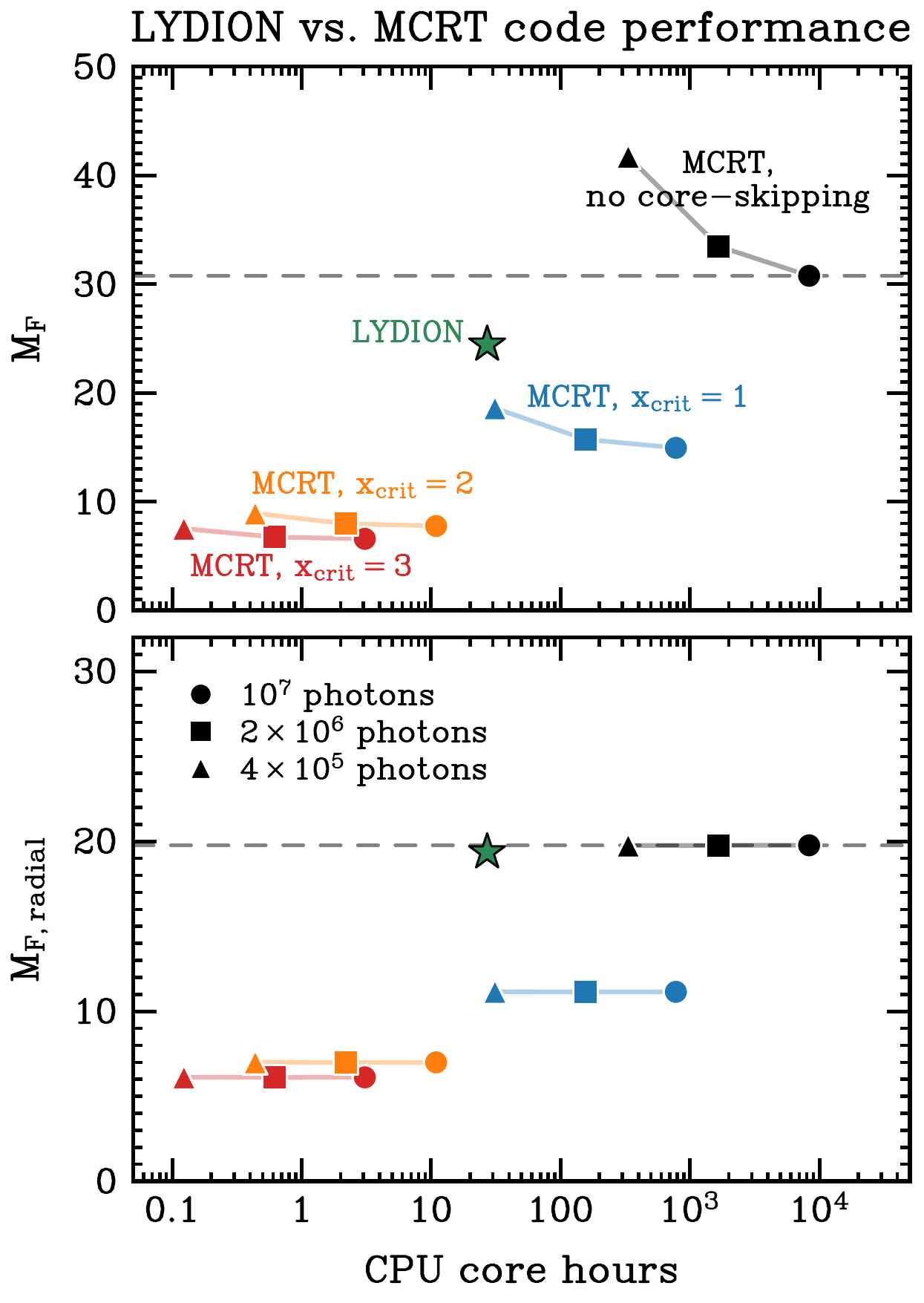}
    \caption{ The naive and radial force multipliers $M_{\rm F}$ (top panel) and $M_{\rm F,radial}$ (bottom panel), respectively, for the anisotropic cone test. The results are plotted against the CPU core hours, as reported in \textsc{Lydion} and all the MCRT calculations. For the latter, we show results with and without core-skipping, and varying number of photons ($4\times10^5 -10^7$). The dashed gray lines highlight the force multipliers for the most accurate MCRT run without core-skipping. }
    \label{fig: Cone test performance LYDION vs MCRT}
\end{figure}

Finally, we quantify the computational performance of \textsc{Lydion} using the anisotropic cone test introduced above. We first examine the scaling with spatial resolution by re-running the same calculation at resolutions $32^2$--$256^2$, while keeping the final time and frequency resolution fixed. For a spatial grid with resolution $N^2$, the work per timestep scales as $\propto N^2$, while the number of radiation timesteps scales as $t_{\rm max} / \Delta t_{\rm RT} \propto N$ for a CFL-limited timestep. We therefore expect the total CPU cost to scale approximately as $N^3$. As shown in Fig.~\ref{fig:cpu_core_time_scaling}, the measured timings closely follow this expectation, indicating the cost per cell update remains approximately constant over this resolution range.

In Fig.~\ref{fig: Cone test performance LYDION vs MCRT} we compare the accuracy of \textsc{Lydion} against MCRT, measuring the total ($M_{\rm F}$) and radial ($M_{\rm F,radial}$) force multipliers against the CPU core hours. The reference solution is taken to be the highest-photon-count MCRT calculation without core-skipping, using $10^7$ photon packets.  This provides the cleanest available MCRT baseline because core-skipping can bias internal radiation-field quantities and derived forces. We also include lower-photon-number MCRT runs and core-skipping accelerated calculations with different thresholds to illustrate the trade-off between Monte Carlo noise, systematic bias, and computational cost. For \textsc{Lydion}, we plot the CPU core hours when both $M_{\rm F}$ and $M_{\rm F,radial}$ have reached within $99\%$ of their final time values, since convergence to the steady state occurs well before the maximum time evolved. 

We see in Fig.~\ref{fig: Cone test performance LYDION vs MCRT} that \textsc{Lydion} is comparable in speed to MCRT with modest core-skipping ($x_{\rm crit} = 1$), and relatively few photons ($4\times10^5$). Despite this, \textsc{Lydion} is significantly more accurate than all the core-skipping runs, which severely underestimate both $M_{\rm F}$ and $M_{\rm F,radial}$. To achieve competitive accuracy with MCRT, one has to discard core-skipping and use $\gtrsim 10^6$ photons, in which case the simulation would be  $\gtrsim 30 \,\times$ more costly to run than \textsc{Lydion}, and still not converged in $M_{\rm F}$ when compared to the baseline $10^7$ photon MCRT calculation. Compared to the latter, \textsc{Lydion} is $\sim 300 \, \times$ faster to run. This benchmark shows that \textsc{Lydion} reaches accurate force estimates at substantially lower computational cost than direct MCRT for this optically thick, dust-free problem.\footnote{MCRT without core-skipping required $\simeq 3\,{\rm sec}$ per photon packet,  while core-skipping accelerates the calculation by factors of $10.7$, $762$, and $2710$ for $x_{\rm crit} = 1$, $2$, and $3$, respectively. Timings were measured on nodes with AMD EPYC 9334 2.7 GHz processors.}

This comparison should be interpreted as a representative example that may miss other aspects of these methods. In MCRT, the cost depends on the number of scatterings (set by the optical depth, dust absorption, destruction processes, and velocity gradients), the core-skipping prescription, the estimator used for momentum deposition, and the target observable. In the present test we use the scattering-based force estimator, which converges more rapidly than path-based momentum estimators, and exploit the midplane symmetry of the setup to boost statistics. For static clouds, MCRT scales as $N_{\rm scat}^{\rm core} \sim \tau_0$ without core-skipping and reduces to $N_{\rm scat}^{\rm wing} \sim (a\tau_0)^{2/3}$ with core-skipping, although additional optimizations are possible with discrete diffusion \citep[][Kimura et al., in prep.]{Smith2018DDMC}. In \textsc{Lydion}, the cost is set primarily by the grid size, timestep constraint (e.g., diffusion $\propto (a\tau_0)^{1/3}$), and number of frequency bins, and does not suffer from Monte Carlo sampling noise. Thus, \textsc{Lydion} excels in optically thick, dust-poor systems where direct MCRT becomes extremely expensive, but where accurate time-dependent radiation forces are still required.

\section{The first 2D Lyman-$\alpha$ RHD simulations:  metal-poor star clusters and isolated stars}
\label{Sec: RHD simulations}

\begin{figure*}
\centering
\includegraphics[width=0.9\textwidth]{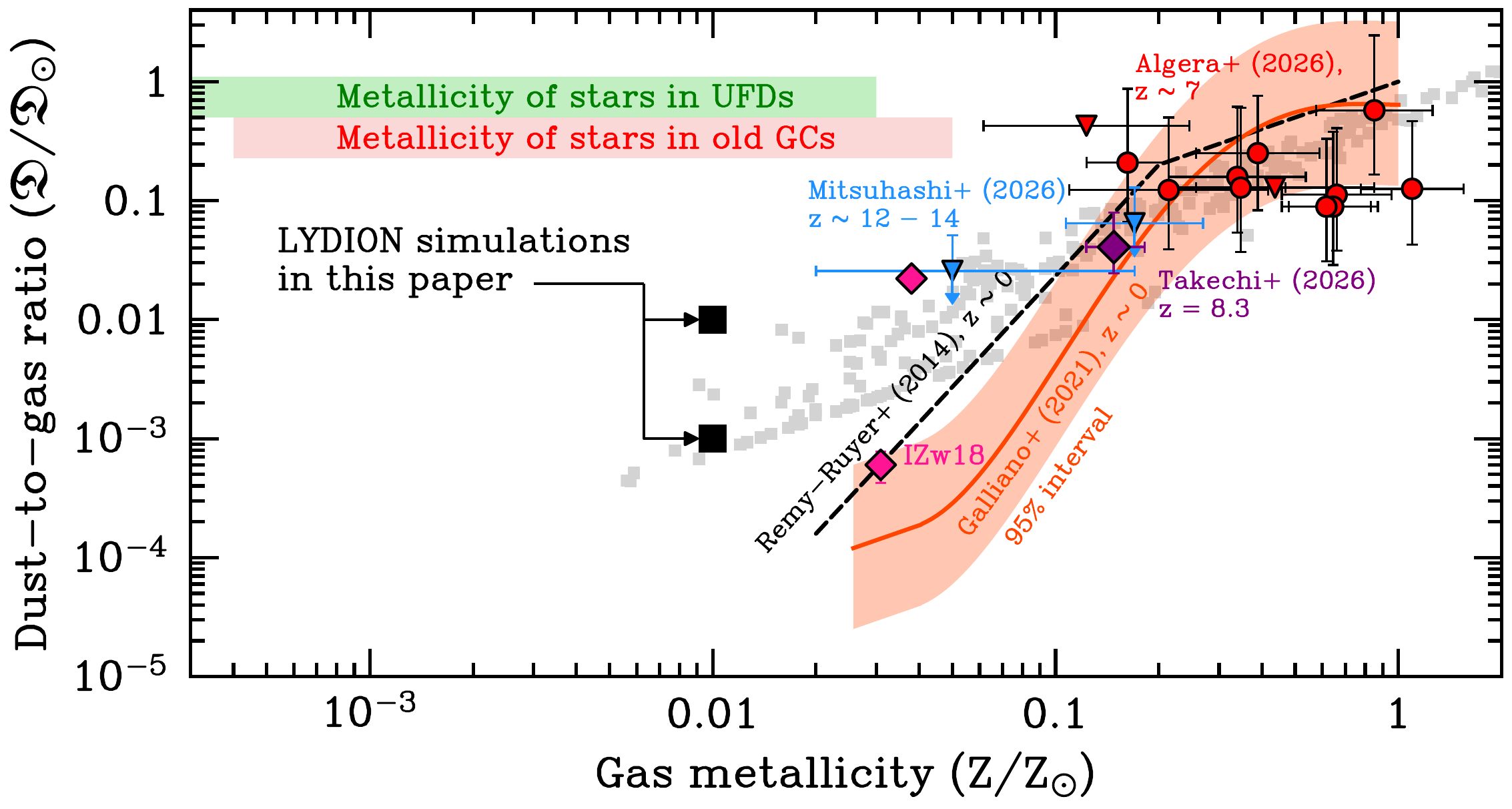}
\caption{ In this plot we show the assumed initial DtG ratios and gas metallicities for the subset of \textsc{Lydion} runs with dust, and how they compare to various low and high-redshift objects. We show low-redshift trends for galaxies from \citet{Remy2014}, and \citet{Galliano2021} (black dashed, and orange solid line, respectively). Gray squares show quasar absorbers at redshift $z \sim 0-5$ from \citet{Peroux2020}. Data for the local low-metallicity galaxies I Zw 18 and SBS 0335-052 are shown as pink diamonds \citep[from][]{Schneider2016}. Recent data for high-redshift galaxies is also shown, from \citet{Algera2026}, \citet{Mitsuhashi2026}, and \citet{Takechi2026}. We note that, for some of these galaxies, there are only upper limits on the DtG ratio (inverted triangle symbols in the plot). Furthermore, \citet{Algera2026} constrains $M_{\rm dust}/M_{\rm H_2}$ rather than $M_{\rm dust}/M_{\rm gas}$, which the authors point out may lead to a factor $\sim 3$ overestimate of $\mathfrak{D}/\mathfrak{D}_\odot$. Finally, we show the relevant range of metallicities for Ultra-Faint Dwarf galaxies (UFDs) and old globular clusters (GCs), as inferred from their stellar metallicities \citep[e.g.][]{Harris2010_GC, Simon2019, Larsen2020_GC, Martin2020_GC, Fu2023_UFD, Weisz2023_GCEridanus, Adamo2024, Nakajima2025}. Meaningful constraints on $\mathfrak{D}/\mathfrak{D}_\odot$ in the birth environments of these objects are not available at the present. }
\label{Dust-to-gas vs observations}
\end{figure*}

\begin{table*}
\centering
\caption{This table contains a summary of the different RHD simulation setups we consider. Our simulations consider both low-mass ($10^4 \, \rm M_\odot$) star clusters (\texttt{SC} in the name), and isolated main-sequence stars of mass $35 \, \rm M_\odot$ (\texttt{Star} in the name). The table lists, among other things, the initial cloud \textsc{H\,i} column density $N_{\rm HI}$,$^\dagger$ initial dust absorption optical depth $\tau_{\rm c,abs}$ of the cloud to Ly$\alpha$,$^\dagger$ the initial turbulent RMS velocity $u_{\rm RMS}$, the initial maximum gas density $n_{\rm H,max}$, the initial dust-to-gas ratio in Solar units, $\mathfrak{D}/\mathfrak{D}_\odot$, and the maximum time $t_{\rm max}$ the simulation was evolved. Each simulation assumes a redshift $z = 10$. The gas clouds are initialized with an H$_2$ fraction $x_{\rm H_2} = 0.1$, and the temperature is initialized at $500 \, \rm K$ everywhere. For the simulations with dust, the gas and stellar metallicity are both set to $0.01 \, Z_\odot$. In the dust-free simulations, both the gas and the star are metal-free. Stellar wind feedback is ignored in all simulations for performance reasons, and because it is negligible at the assumed low metallicities.}
\begin{tabular}{c c c c c c c c c c c}
\hline
\hline
\noalign{\vskip 2pt}
Name & Ly$\alpha$ RT & $N_{\rm HI}$ & $\tau_{\rm c,abs}$ & $u_{\rm RMS}$ & $n_{\rm H,max}$ & $\mathfrak{D}/\mathfrak{D}_\odot$  & $t_{\rm max}$ & Box size & \# of cells & Resolution  \\

 & & (cm$^{-2}$) &  &   (km s$^{-1}$) & (cm$^{-3}$) & & (kyr) & (pc) & ($N_R \times N_Z$) & (AU) \\
\noalign{\vskip 3pt}
\hline
\hline
\noalign{\vskip 2pt}

\texttt{SCLyaR128}  & Yes  & $1.2 \times 10^{23}$ & $0.57$  & $4$ & $10^5$ & $0.01$ & $141.8$ & $1.2$ & $128^2$ & $1930$  \\

\texttt{SCLyaR181}  & Yes  & $1.2 \times 10^{23}$ & $0.57$ & $4$ & $10^5$ & $0.01$ & $147.6$ & $1.2$ & $181^2$ & $1370$  \\

\texttt{SCLyaR256}  & Yes  & $1.2 \times 10^{23}$ & $0.57$ & $4$ & $10^5$ & $0.01$ & $81.6$ & $1.2$ & $256^2$ & $967$  \\

\texttt{SCLyaR362}  & Yes  & $1.2 \times 10^{23}$ & $0.57$  & $4$ & $10^5$ & $0.01$ & $91.6$ & $1.2$ & $362^2$ & $684$  \\

\texttt{SCLyaR512}  & Yes & $1.2 \times 10^{23}$ & $0.57$  & $4$ & $10^5$ & $0.01$ & $75.4$ & $1.2$ & $512^2$ & $483$  \\

\texttt{SCNoLyaR128}  & No & $1.2 \times 10^{23}$ & $0.57$ & $4$ & $10^5$ & $0.01$ & $137.4$ & $1.2$ & $128^2$ & $1930$  \\

\texttt{SCNoLyaR181}  & No & $1.2 \times 10^{23}$ & $0.57$ & $4$ & $10^5$ & $0.01$ & $72.9$ & $1.2$ & $181^2$ & $1370$  \\

\texttt{SCNoLyaR256}  & No & $1.2 \times 10^{23}$ & $0.57$ & $4$ & $10^5$ & $0.01$ & $54.2$ & $1.2$ & $256^2$ & $967$  \\

\texttt{SCNoLyaR362}  & No & $1.2 \times 10^{23}$ & $0.57$ & $4$ & $10^5$ & $0.01$ & $45.2$ & $1.2$ & $362^2$ & $684$  \\

\texttt{SCNoLyaR512}  & No & $1.2 \times 10^{23}$ & $0.57$  & $4$ & $10^5$ & $0.01$ & $35.2$ & $1.2$ & $512^2$ & $483$  \\

\\

\texttt{StarLyaRMS0}  & Yes & $2.5 \times 10^{23}$ & $0.11$  & $0$ & $10^6$ & $10^{-3}$ & $5$ & $0.1$ & $256^2$ & $80.6$  \\

\texttt{StarLyaRMS4}  & Yes & $2.5 \times 10^{23}$ & $0.11$  & $4$ & $10^6$ & $10^{-3}$ & $5$ & $0.1$ & $256^2$ & $80.6$ \\

\texttt{StarLyaRMS8}  & Yes & $2.5 \times 10^{23}$ & $0.11$ & $8$ & $10^6$ & $10^{-3}$ & $5$ & $0.1$ & $256^2$ & $80.6$ \\

\texttt{StarNoLyaRMS0}  & No  & $2.5 \times 10^{23}$ & $0.11$ & $0$ & $10^6$ & $10^{-3}$ & $5$ & $0.1$ & $256^2$ & $80.6$  \\

\texttt{StarNoLyaRMS4}  & No  & $2.5 \times 10^{23}$ & $0.11$ & $4$ & $10^6$ & $10^{-3}$ & $5$ & $0.1$ & $256^2$ & $80.6$  \\

\texttt{StarNoLyaRMS8}  & No  & $2.5 \times 10^{23}$ & $0.11$ & $8$ & $10^6$ & $10^{-3}$ & $5$ & $0.1$ & $256^2$ & $80.6$  \\

\\

\texttt{StarLyaNoDust}  & Yes & $2.5 \times 10^{23}$ &  0 & $4$ & $10^6$ & $0$ & $5$ & $0.1$ & $256^2$ & $80.6$ \\

\texttt{StarNoLyaNoDust}  & No & $2.5 \times 10^{23}$ & 0  & $4$ & $10^6$ & $0$ & $5$ & $0.1$ & $256^2$ & $80.6$ \\

\texttt{StarLyaNoDest}$^\S$  & Yes & $2.5 \times 10^{23}$ & 0  & $4$ & $10^6$ & $0$ & $5$ & $0.1$ & $256^2$ & $80.6$ \\

\noalign{\vskip 2pt}
\hline
\hline
\end{tabular}
\vspace{1 pt}\\
\raggedright
{\footnotesize
$^\dagger$: The initial \textsc{H\,i} column density and initial dust absorption optical depth are estimated using $N_{\rm HI} \simeq (1 - 2x_{\rm H_2}) n_{\rm H,max} \,\Delta L$, and $\tau_{\rm c,abs} = (\kappa_{\rm abs,Sil}f_{\rm Sil} + \kappa_{\rm abs,C} f_{\rm C} + \kappa_{\rm abs,PAH} f_{\rm PAH}) \mathfrak{D} \rho \,\Delta L$, respectively. The dust bin opacities (in the Ly$\alpha$ band) and initial bin mass fractions are given in Table~\ref{Dust optical properties non IR}. For the star cluster simulations, $\Delta L = R_{\rm cl} = 0.5 \, \rm pc$. For the simulations of isolated stars, we have set $\Delta L$ equal to the box size in our estimate. \\ 
$^\S$: Besides no dust, this simulation also ignores Ly$\alpha$ destruction, i.e. $p_{\rm d} = 0$. }
\label{RHD scenarios table}
\end{table*}

In this section we present the results of the first Ly$\alpha$ RHD simulations in multiple dimensions, and the first ones with dust dynamics \citep[for earlier dust-free 1D simulations, see][]{George1973, Smith2017}. As a first application, we consider metal-poor star clusters and isolated stars embedded in metal and dust-poor, dense gas clouds. Since our simulations do not incorporate star formation itself, but rather a fixed stellar source, we did not choose a realistic stellar source. Instead, the more basic questions we aim to answer are the following:
\begin{enumerate}[leftmargin=*]
    \item What is the relative importance of Ly$\alpha$ feedback compared to other early (pre-supernova) feedback processes in a \textit{typical} dust-poor environment?
    \item Can Ly$\alpha$ feedback drive outflows in such dust-poor environments?
\end{enumerate}
To address these questions, we therefore consider the following setups:
\begin{enumerate}
    \item \textbf{Embedded metal-poor star clusters:} For the main runs with the highest resolution (in terms of cell count), we consider low-mass ($10^4 \, \rm M_\odot$), metal-poor ($Z_\star / Z_\odot = 0.01$) star clusters, embedded in a dense ($n_{\rm H} = 10^5 \, \rm cm^{-3}$), metal and dust-poor cloud ($Z/Z_\odot = \mathfrak{D}/\mathfrak{D}_\odot = 0.01$) of radius $R_{\rm cl} \simeq 0.5 \, \rm pc$. Outside this radius (distance $r > R_{\rm cl}$ from cloud center), the initial gas density decays as $n_{\rm H} \propto \exp[- (r / 0.05 \, R_{\rm cl})^2]$, until reaching the initial external density $n_{\rm H} = 1 \, \rm cm^{-3}$. The simulation volume extends to $(R_{\rm max},Z_{\rm max}) = (1.2, \,1.2) \, \rm pc$, and mirror symmetry is imposed at the midplane $Z = 0$.
    
    All the gas is initialized with a temperature $T = 500 \, \rm K$, although in practice the dense cloud gas quickly cools down to $T \lesssim 100 \, \rm K$. The gas cloud is initialized with some turbulence, $u_{\rm RMS} = 4 \, \rm km \, s^{-1}$, corresponding to a Mach number of $\mathcal{M} \sim 5$ for gas of temperature $T \sim 100 \, \rm K$. The star cluster has a fixed half-mass radius $R_{\rm h} = 0.05 \, \rm pc$, giving a stellar surface density $\Sigma_{\star} = M_\star / 2 \pi R_{\rm h}^2 \simeq 6 \times 10^5 \, \rm M_{\odot} \, pc^{-2}$. This is consistent with the large stellar surface densities recently observed in high-$z$ star clusters \citep[][]{Adamo2024}, and theoretically predicted (pre-\textit{JWST}) in simulations \citep[][]{Kimm2016, Ricotti2016} and analytical modelling \citep[][]{Nebrin2022}. Our compact-cluster setup here is also numerically convenient, because it will more easily create a single, morphed \textsc{H\,ii} region around the star cluster, as well as maximize direct and infrared radiation pressure feedback by reducing flux cancellation effects \citep[][]{Menon2023, Menon2026}. The latter point will render our conclusions about the \textit{relative} importance of Ly$\alpha$ feedback more conservative.  

    \item \textbf{Embedded isolated metal-poor stars:} We also perform a series of simulations of isolated metal-poor massive stars ($35 \, \rm M_\odot$), embedded in even denser gas ($n_{\rm H} = 10^6 \, \rm cm^{-3}$), with lower dust-to-gas ratios of $\mathfrak{D}/\mathfrak{D}_\odot = (0, 10^{-3})$. For the dust-enriched scenarios ($\mathfrak{D}/\mathfrak{D}_\odot = 10^{-3}$) we vary the level of initial turbulence $u_{\rm RMS}$ between $0$ and $8 \, \rm km \, s^{-1}$ to check the impact on the Ly$\alpha$ radiative force. We also run a few metal/dust-free simulations, as a preliminary assessment of Ly$\alpha$ feedback during Pop III star formation. The luminosity of the star is $L_{\rm bol} = (1.6, \,2.0)\times 10^5 \, \rm L_\odot$ for $Z_\star/Z_\odot = (0.01,\, 0)$, with most of the starlight in the ionizing and Lyman-Werner bands ($\sim 70\%$ and $\sim 90\%$ for the Pop II star and Pop III star, respectively).
\end{enumerate}
The assumed dust-to-gas ratios in our runs deserve some comments and context. Our setups are partly motivated by the low metallicities of old and metal-poor globular clusters (GCs), and of Ultra-Faint Dwarf galaxies. Typical stellar metallicities of these objects are $10^{-4} \lesssim Z/Z_\odot \lesssim \textrm{few} \times 0.01$, with old GCs being biased to the higher end of this range \citep[e.g.][]{Harris2010_GC, Simon2019, Larsen2020_GC, Martin2020_GC, Fu2023_UFD, Weisz2023_GCEridanus, Adamo2024, Nakajima2025}. Based on a naive, but commonly assumed, linear scaling between the dust-to-gas (DtG) ratio $\mathfrak{D}$ and gas metallicity, this would suggest that these objects were born in dust-poor clouds with $10^{-4} \lesssim \mathfrak{D}/\mathfrak{D}_\odot \lesssim \textrm{few} \times 0.01$.  

In reality, observations of local metal-poor dwarf galaxies, and damped Ly$\alpha$ systems, find that they are significantly more dust-poor than expected from a linear extrapolation \citep[e.g.][]{Fisher2014, Remy2014, Schneider2016, DeVis2019, Peroux2020, Galliano2021, RomanDuval2022}. For example, based on the observational results of \cite{Remy2014}, a galaxy of metallicity $Z/Z_\odot = 0.05$ would be expected to have a DtG ratio of $\mathfrak{D}/\mathfrak{D}_\odot \sim 3 \times 10^{-3}$. Indeed, I Zw 18, a local metal-poor and star-forming dwarf galaxy, has an estimated gas metallicity $Z/Z_\odot \sim 0.03$, yet an extremely low DtG ratio $\mathfrak{D}/\mathfrak{D}_\odot \sim 8 \times 10^{-4}$ \citep[][]{Schneider2016}. \cite{Pascale2024} also report evidence of significant Ly$\alpha$ trapping in the dense ionized gas of Godzilla, a possible young star cluster at redshift $z = 2.37$,\footnote{Whether Godzilla is a star cluster or not is debated, see \citet{Vanzella2020_Sunburst} and \citet{Choe2025} for alternative hypotheses.} despite a gas metallicity $Z/Z_\odot \simeq 0.25$. As a result of this inference, \cite{Pascale2024} argue for a near-absence of dust in the \textsc{H\,ii} region where Ly$\alpha$ is produced.\footnote{To explain this, \cite{Pascale2024} show that small dust grains can be sublimated in the compact \textsc{H\,ii} regions. They also argue that radiation pressure can evacuate larger grains. However, these authors neglected Coulomb drag on charged grains, which is the dominant source of drag in \textsc{H\,ii} regions. It therefore remains to be seen if dust drift is a viable explanation. \textsc{Lydion} could be used to explore this in future work.}

Recent constraints from higher redshift ($z \gtrsim 6$) galaxies are still scarce, and mostly concentrated at higher metallicities ($Z /Z_\odot \gtrsim 0.1$) \citep[][]{Algera2025_HZ10, Algera2026, Mitsuhashi2026, Takechi2026}. Still, these galaxies have inferred dust-to-metal ratios that tend to be lower than local galaxies of the same metallicity. Lower metallicity ($Z_\star/Z_\odot \sim 0.01$) high-$z$ galaxies have also been detected \citep[e.g.][]{Hsiao2025, Nakajima2025, Morishita2025}, and although they provide no constraint on the DtG ratio, they have very low inferred dust attenuation ($A_{\rm V} \sim 0 - 0.2$). 

Simulations of high-redshift galaxy formation, incorporating dust growth and destruction, predict similar results, with $\mathfrak{D}/\mathfrak{D}_\odot \sim (\textrm{few}\times 0.01 - 0.1) \times Z/Z_\odot$ at high redshifts \citep[][]{Choban2025, Narayanan2026}. \cite{Narayanan2026} also finds that the grain size distribution is biased to larger grains at high redshifts, which would further reduce dust absorption of Ly$\alpha$, as well as make dust more susceptible to evacuation by radiation pressure \citep[][]{Akimkin2015,Akimkin2017, Ishiki2018}. All of these factors would tend to boost Ly$\alpha$ feedback. In Fig.~\ref{Dust-to-gas vs observations} we plot the adopted initial DtG ratios in the \textsc{Lydion} simulations of this paper, and how they compare to observed low and high-redshift objects.  

\subsection{Lyman-$\alpha$ feedback from embedded star clusters}
\label{Lya RHD star cluster section}

We start by presenting the results for the simulations of embedded star clusters. We will mainly focus on our highest resolution run, \texttt{SCLyaR512}, but add results from the lower-resolution runs to test convergence, and better check long-term trends. 

\subsubsection{Qualitative results: Lyman-$\alpha$ driven winds and suppressed star formation?}

\begin{figure}
    \centering
    \includegraphics[width=0.95\columnwidth]{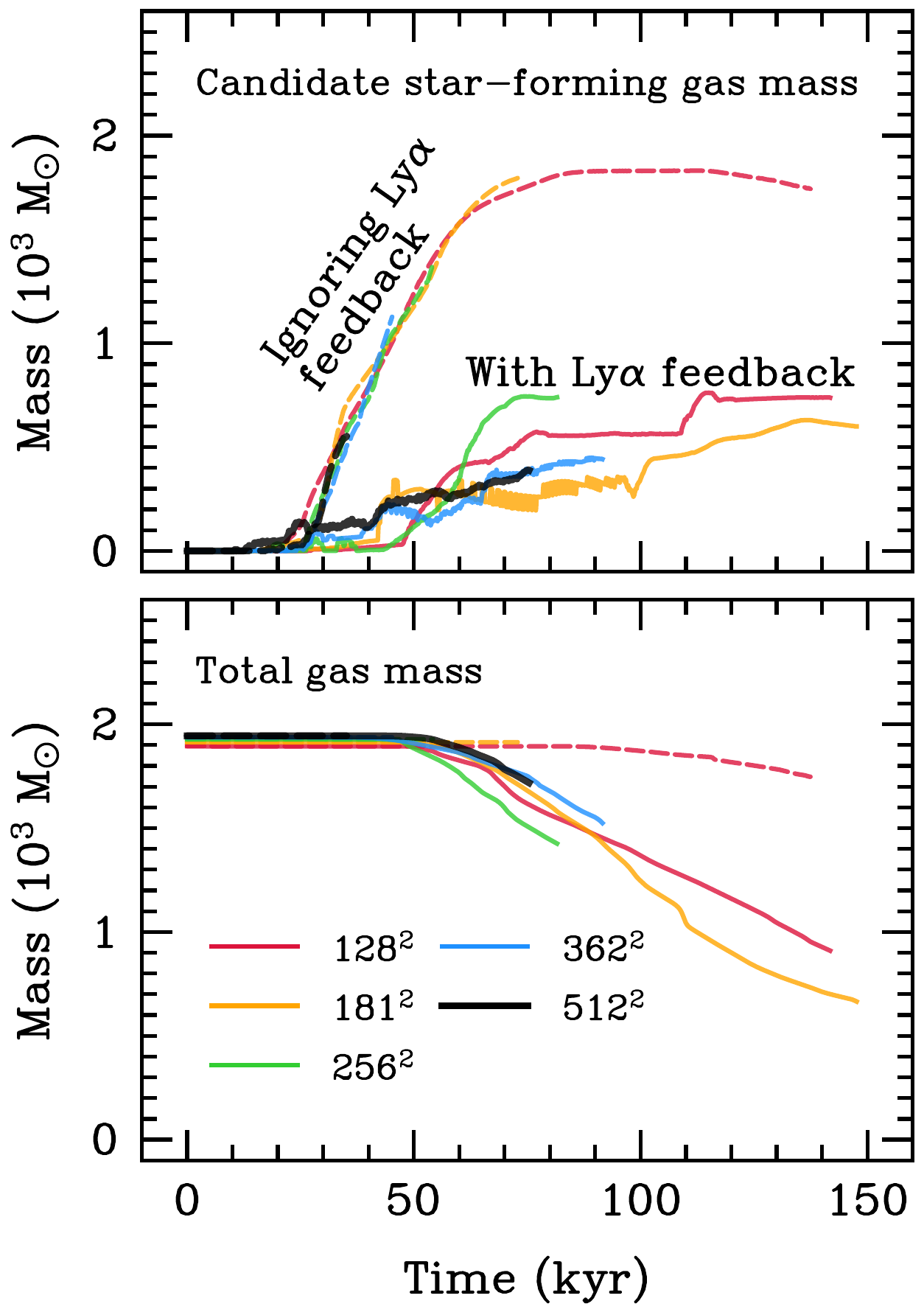}
    \caption{ The evolution of the potentially star-forming gas mass (top panel), and total gas mass remaining in the simulation volume (bottom panel), for the star cluster simulations. Solid (dashed) lines show the results when incorporating (ignoring) Ly$\alpha$ pressure. Ly$\alpha$ feedback is expected to suppress star formation for this particular setup.  }
    \label{fig: starforming gas}
\end{figure}

The time evolution of the gas density, temperature, Ly$\alpha$ energy density, and ratio between Ly$\alpha$ and gas pressures is highlighted in Figs.~\ref{Lya vs no Lya figure 512 D0.01}--\ref{Lya vs no Lya figure 512 D0.01 T and PLya}. In the simulation without Ly$\alpha$ feedback, the gas recollapses rapidly, which would presumably trigger additional star formation in a full, more self-consistent 3D simulation. In contrast, the introduction of Ly$\alpha$ pressure launches outflows that ultimately destroy the cloud and enhance LyC leakage. While star formation itself is not currently modelled in \textsc{Lydion}, we can estimate the mass of dense gas that would be susceptible to star formation. To do so, in Fig.~\ref{fig: starforming gas} we plot the time evolution of the mass of gas that satisfies two criteria: (\textit{i}) density $n_{\rm H} > 10^7 \, \rm cm^{-3}$, and (\textit{ii}) gravity dominates over Ly$\alpha$ and direct radiation pressure. The density threshold is high enough to isolate collapsing regions with densities in excess of what is expected from isothermal shocks, $n_{\rm H,shock} \simeq \mathcal{M}^2 \, n_{\rm H} \sim 3 \times 10^6 \, \rm cm^{-3}$, for $\mathcal{M} \sim 5$, $n_{\rm H} \sim 10^5 \, \rm cm^{-3}$. To quantify criterion (\textit{ii}), we define the Eddington ratio for Ly$\alpha$ + direct radiation pressure, opposite to the local gravitational force:
\begin{equation}
    f_{\rm Edd}^{\rm Ly\alpha + dir}(\textrm{antigrav}) \equiv \dfrac{ (\boldsymbol{f}_{\rm Ly\alpha} + \boldsymbol{f}_{\rm dir})\boldsymbol{\cdot} (-\boldsymbol{\hat{f}}_{\rm grav})}{\lvert \boldsymbol{f}_{\rm grav} \rvert} \, , \label{dir + Lya fEdd}
\end{equation}
where $\boldsymbol{f}_{\rm grav} = - \rho \boldsymbol{\nabla}\Phi$ is the local gravitational force per unit volume on the gas, and $\boldsymbol{f}_{\rm dir}$ is the force per unit volume from direct radiation pressure. Only if this ratio is $<1$ do we consider a cell a candidate site of star formation. 

Overall, we see that the simulations ignoring Ly$\alpha$ feedback predict that nearly all remaining gas should be converted into stars. This is expected in the absence of Ly$\alpha$ feedback for the high total (star + gas) surface densities encountered in our simulations \citep[e.g.][]{Kimm2016, Grudic2018, Fukushima2021, Nebrin2022, Dekel2023, Nebrin2024, Manzoni2025}. In contrast, the simulations with Ly$\alpha$ feedback predict much slower star formation, only converting a fraction of the remaining gas into stars. These trends are consistent among all simulation resolutions, although they are most evident for the lower-resolution runs that can be evolved for longer at acceptable numerical cost.\footnote{Our higher-resolution runs without Ly$\alpha$ feedback effectively stall as the gas collapses to high densities, shortening the time-steps in the simulations. With Ly$\alpha$ feedback, the gas is ejected, but the simulations are costly to run at the highest resolutions.}

\subsubsection{Which feedback process(es) dominate?}

\begin{figure*}
    \centering
    \includegraphics[width=0.85\textwidth]{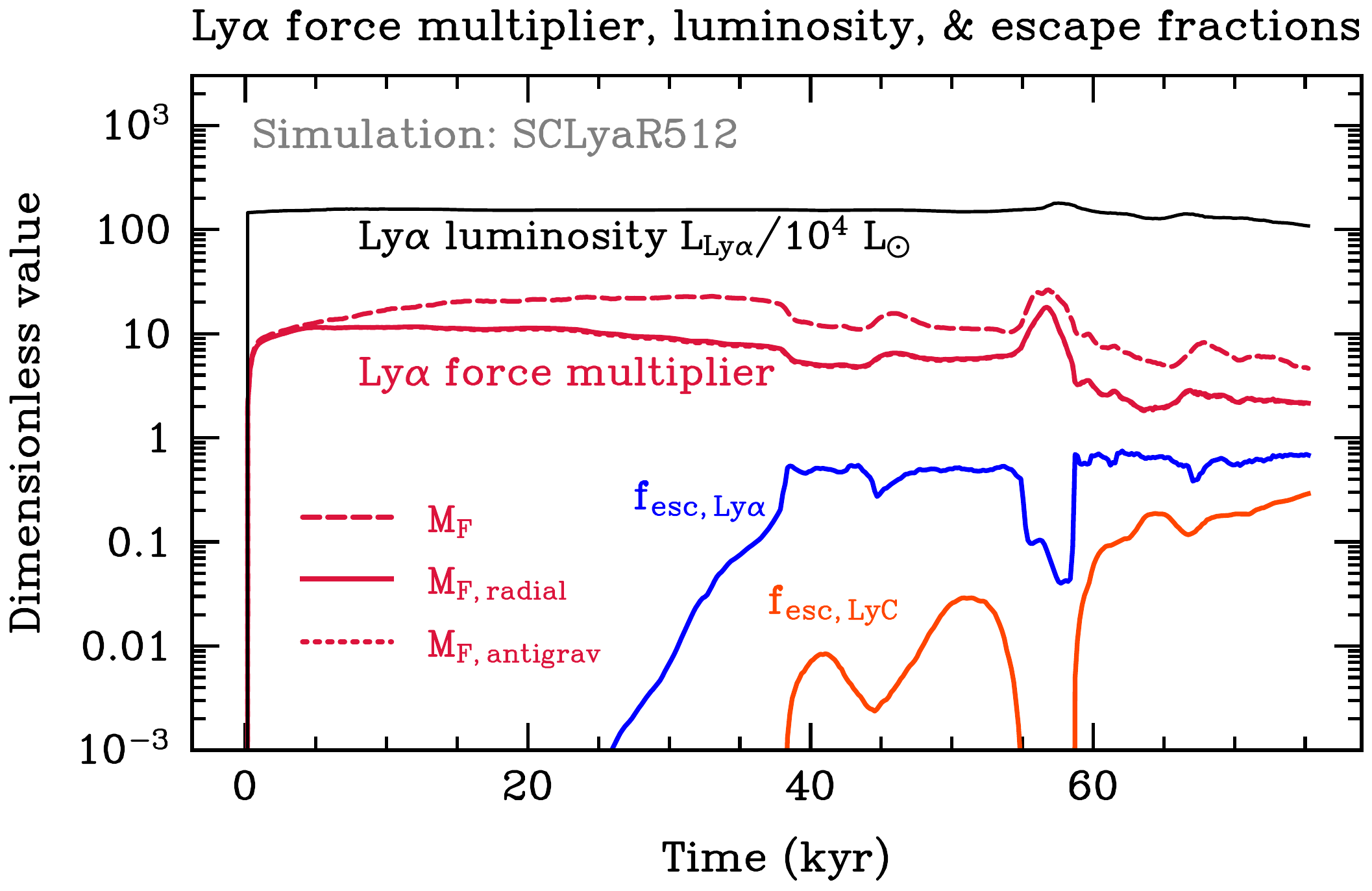}
    \caption{ Plot of the evolution of the Ly$\alpha$ force multiplier, Ly$\alpha$ luminosity, Ly$\alpha$ and LyC escape fractions, for the star cluster simulation \texttt{SCLyaR512}. The radial and anti-gravity Ly$\alpha$ force multipliers ($M_{\rm F,radial}$ and $M_{\rm F,antigrav}$) remain nearly identical, and around $\sim 10$ before breakout of the ionization front. The naive force multiplier, $M_{\rm F}$ (dashed red line), is somewhat higher, reaching $\sim 20$, owing to significant non-radial Ly$\alpha$ forces.  }
    \label{fig: M_F 512 run}
\end{figure*}

\begin{figure*}
    \centering
    \includegraphics[width=0.85\textwidth]{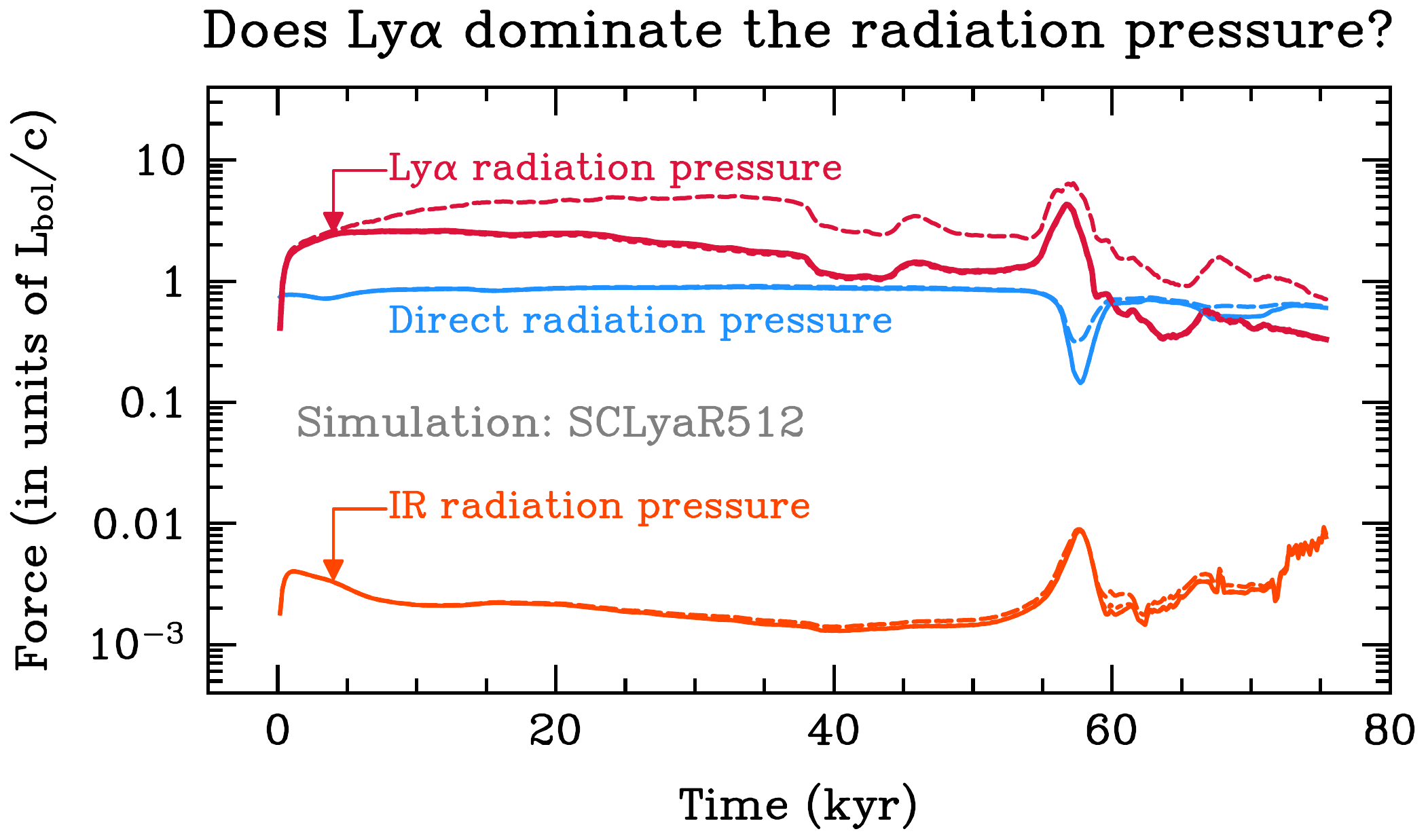}
    \caption{ Comparison of forces from Ly$\alpha$, direct, and IR radiation pressure, for the star cluster simulation \texttt{SCLyaR512}. Naive, radially outward, and antigravity forces are shown as dashed, solid, and dotted lines, respectively. Radial and antigravity forces largely overlap, whereas naive and radial forces can differ significantly for Ly$\alpha$ pressure.   }
    \label{fig: Feedback comparison 512 sim}
\end{figure*}

\begin{figure}
    \centering
    \includegraphics[width=0.95\columnwidth]{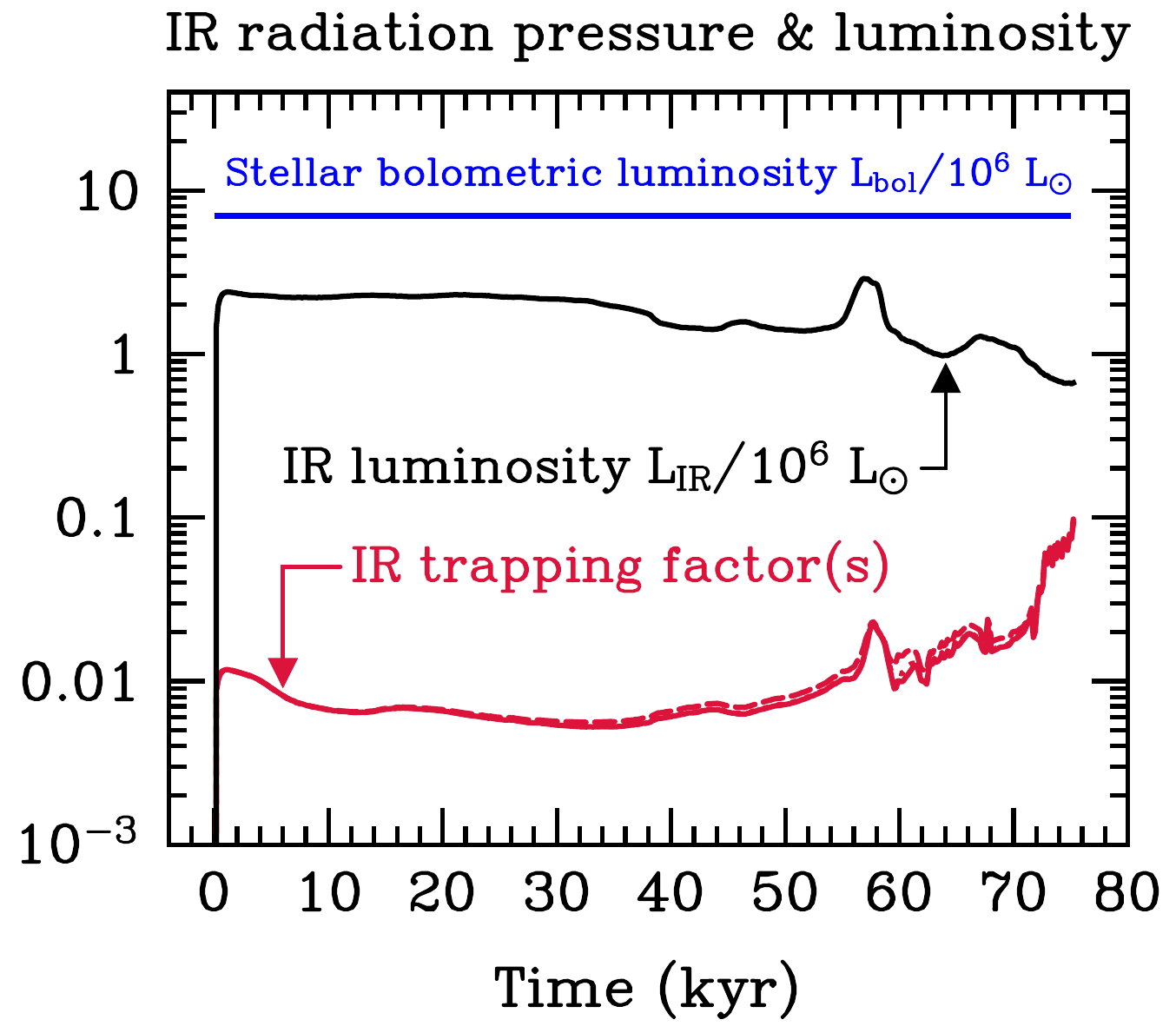}
    \caption{ Plot of the IR luminosity (black line), stellar bolometric luminosity (blue line), and IR radiation force trapping factors (red lines). The latter are shown for the naive, radial, and antigravity forces (c.f. Fig.~\ref{fig: Feedback comparison 512 sim}). Luminosities are normalized by $10^6 \, \rm L_\odot$.  }
    \label{fig: IR feedback force}
\end{figure}

Next we quantify the strength of Ly$\alpha$ feedback, and how it compares to other feedback forces and gravity. In Fig.~\ref{fig: M_F 512 run} we plot the evolution of the Ly$\alpha$ luminosity, escape fractions of both Ly$\alpha$ and LyC ($h\nu > 13.6 \, \rm eV$) photons, the Ly$\alpha$ force multipliers $M_{\rm F}$ and $M_{\rm F,radial}$ (Eqs.~\ref{Naive M_F def}--\ref{Radial M_F def}), and:
\begin{equation}
    M_{\rm F,antigrav} \equiv \dfrac{1}{L_{\rm Ly\alpha}/c} \int \textrm{d}V \,  \boldsymbol{f}_{\rm Ly\alpha} \boldsymbol{\cdot} (-\boldsymbol{\hat{f}}_{\rm grav}) \, . \label{Antigrav M_F def}
\end{equation}
Here $M_{\rm F,antigrav}$ measures the Ly$\alpha$ force in the opposite direction of the local gravitational force. Initially, when the cloud is spherically symmetric, we expect that $M_{\rm F} \simeq M_{\rm F,radial} \simeq M_{\rm F,antigrav}$. As non-sphericity and overdensities develop, we expect $M_{\rm F,radial}$ and $M_{\rm F,antigrav}$ to drop below $M_{\rm F}$. Figure~\ref{fig: M_F 512 run} shows the results for the simulation \texttt{SCLyaR512}, where we see that at early times the force multiplier measures in Eqs.~(\ref{Naive M_F def})--(\ref{Radial M_F def}) and (\ref{Antigrav M_F def}) give nearly identical values, as expected. The values are around $\sim 10$, indicative of significant Ly$\alpha$ trapping. As time progresses, we find that $M_{\rm F}$ slowly increases to $\sim 20$, whereas $M_{\rm F,radial}$ and $M_{\rm F,antigrav}$ are somewhat lower ($\sim 10$), and nearly identical. Around $t \sim 39 \, \rm kyr$, the ionization front breaks out, which causes a sharp increase in the Ly$\alpha$ escape fraction to $f_{\rm esc,Ly\alpha} \gtrsim 50 \%$. However, the force multipliers only drop by a factor $< 2$, showing a remarkable resilience to Ly$\alpha$ leakage.

In Fig.~\ref{fig: Feedback comparison 512 sim} we compare the forces from Ly$\alpha$, direct, and infrared radiation pressure (also see Fig.~\ref{Lya vs no Lya figure 512 D0.01}). Direct radiation pressure exerts a radially outward force of magnitude $\sim 0.8 \, L_{\rm bol}/c$ for most of the simulation time, which is close to the maximum value for single-scattering radiation pressure around a point-like source ($L_{\rm bol}/c$). A temporary dip in the force occurs when a fraction of the gas recollapses at $t \sim 57 \, \rm kyr$, causing the force to be imparted on scales comparable to the star cluster, giving rise to significant flux (and hence force) cancellation \citep[][]{Kim2018, Menon2023}. Feedback from IR radiation pressure is negligible, $\sim (10^{-3} - 0.01) \times \, L_{\rm bol}/c$, since the dust-poor cloud is optically thin to IR photons. In Fig.~\ref{fig: IR feedback force} we further plot the IR luminosity and radiation force trapping factors, $f_{\rm IR,trap} \equiv F_{\rm IR} /(L_{\rm IR}/c)$. We see that only a fraction of the stellar and Ly$\alpha$ luminosities are converted into IR photons. This is because \textsc{H\,i}, H$_2$, He\textsc{\,ii}, and He\textsc{\,iii} compete for absorption of ionizing and Lyman-Werner photons, and the relatively low dust attenuation at lower photon energies facilitate escape (except for resonantly scattered Ly$\alpha$). The IR force trapping factors are of order $\sim 0.007$ at early times, close to what is analytically expected ($f_{\rm IR,trap} \sim \tau_{\rm IR}$).\footnote{Analytically, for uniform clouds, we expect that $f_{\rm IR,trap} = \eta \, \tau_{\rm IR}$, where $\eta =1$ for a central source, and $1/4\sqrt{2}$ for a uniform source (see Eq.~\ref{M_F tau goes to zero limit}). For a typical dust temperature of $\sim 300 \, \rm K$ (Fig.~\ref{Dust properties figure 512 Lya simulation}), and the initial cloud configuration, we estimate $\tau_{\rm IR} \sim 0.01$, which is close to what we see in Fig.~\ref{fig: IR feedback force}, if most of the IR photons are produced in the dense shock front (so that $\eta \simeq 1$).} 

Ly$\alpha$ pressure is seen to dominate over both direct and IR radiation pressure in Fig.~\ref{fig: Feedback comparison 512 sim}, exerting radial forces $\sim 2.5 \, L_{\rm bol}/c$ before breakout of the \textsc{H\,ii} region. The naive force from Ly$\alpha$ (dashed line), i.e. including non-radial forces, is even greater, reaching $\sim 5 \, L_{\rm bol}/c$. This highlights the dynamical importance of Ly$\alpha$ pressure. In summary, including Ly$\alpha$ feedback in our star cluster simulation boosts the total radially outward radiation pressure force by a factor $\sim 4$, and the naive force by a factor $\sim 7$. 

The next major feedback process is photoionization (PI), which overpressurizes the gas. As seen in Fig.~\ref{Lya vs no Lya figure 512 D0.01 T and PLya}, Ly$\alpha$ pressure is greater than the gas pressure throughout most of the \textsc{H\,ii} region at early times. This is crucial for launching the outflows during the phase where the gas is most gravitationally bound. At later times, the gas pressure becomes increasingly important, but we see that Ly$\alpha$ pressure can still exceed the gas pressure near the ionization front, and be comparable to the gas pressure in a significant portion of the \textsc{H\,ii} region. As we will discuss later (Sec. \ref{Discussion: How robust are our conclusions}), the fact that we are studying low-mass star clusters ($10^4 \, \rm M_\odot$) here will tend to overemphasize the relative importance of PI feedback during star cluster formation. For the formation of massive star clusters ($\sim 10^5 - 10^6 \, \rm M_\odot$), we expect PI feedback to become increasingly unimportant in comparison to both Ly$\alpha$ and direct radiation pressure \citep[][]{Krumholz2009_HII}.  

\begin{figure*}
\centering
\includegraphics[width=1.0\textwidth]{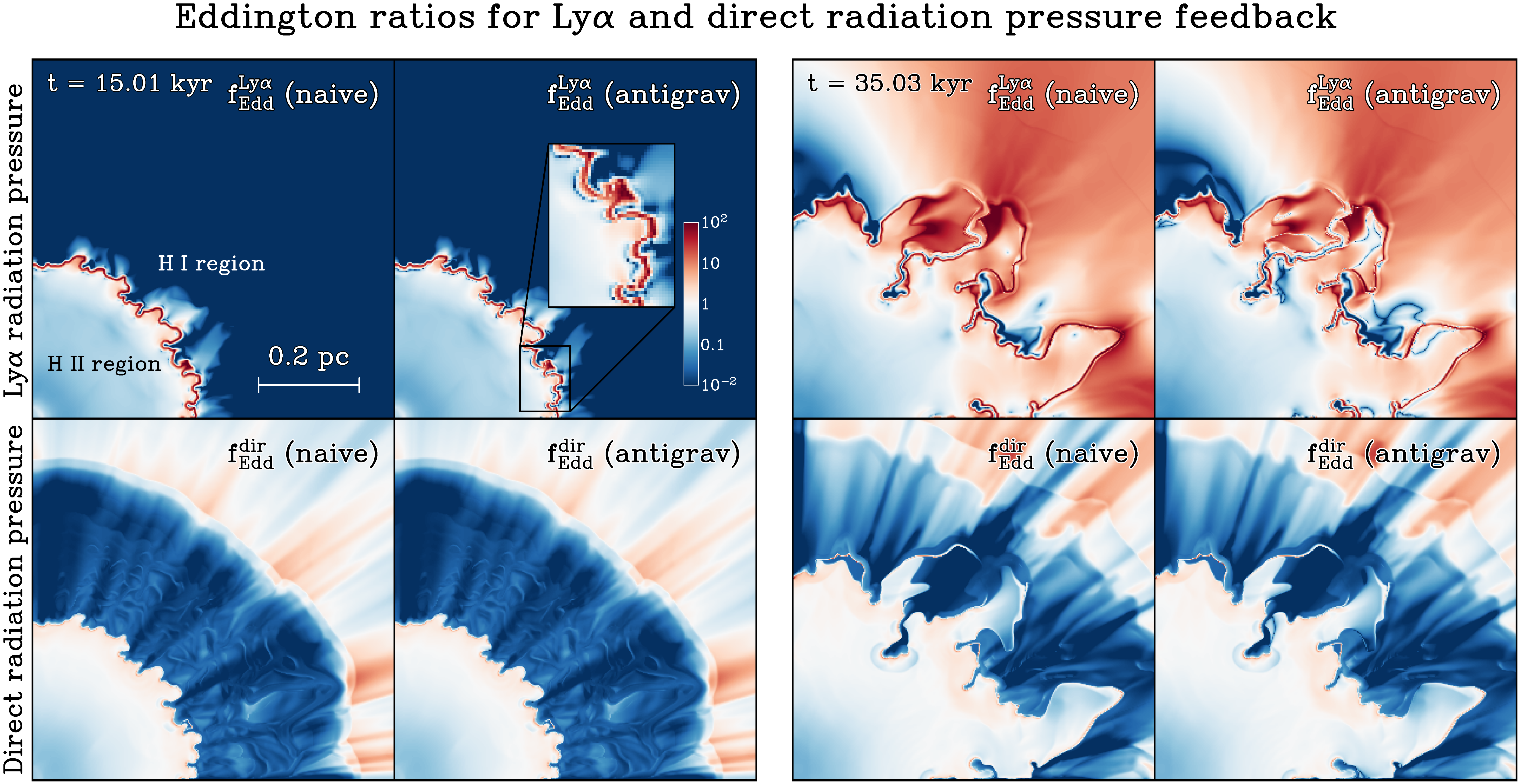}
\caption{Maps from \texttt{SCLyaR512}, at $t = 15 \, \rm kyr$ (left panels) and $t = 35 \, \rm kyr$ (right panels), of the Eddington ratio $f_{\rm Edd} \sim (\textrm{Radiation pressure force})/(\textrm{Gravitational force})$ for both Ly$\alpha$ radiation pressure (top row), and direct (non-IR) radiation pressure (bottom row). The Ly$\alpha$ radiation pressure force is highly concentrated near the shock/ionization front, with extreme Eddington ratios reaching $\textrm{few} \times 10 \lesssim f_{\rm Edd}^{\rm Ly\alpha} \lesssim 100$. In contrast, for direct radiation pressure, $f_{\rm Edd}^{\rm dir}$ is more modest and smoothly distributed throughout the \textsc{H\,ii} region. }
\label{Eddington ratios}
\end{figure*}

Next, in Fig.~\ref{Eddington ratios} we plot maps of Eddington ratios at $t = (15,35) \, \rm kyr$, measuring the relative strength of Ly$\alpha$ and direct radiation pressure compared to gravity. Specifically, we plot the naive and antigravity Eddington ratios, for both direct (dir) and Ly$\alpha$ radiation pressure (c.f. Eq.~\ref{dir + Lya fEdd}):
\begin{align}
    f_{\rm Edd}^{\rm Ly\alpha/dir}(\textrm{naive})  &\equiv~ \dfrac{\lvert \boldsymbol{f}_{\rm Ly\alpha/dir} \rvert}{\lvert \boldsymbol{f}_{\rm grav} \rvert} \, , \\ f_{\rm Edd}^{\rm Ly\alpha/dir}(\textrm{antigrav}) &\equiv~ \dfrac{\boldsymbol{f}_{\rm Ly\alpha/dir} \boldsymbol{\cdot} (-\boldsymbol{\hat{f}}_{\rm grav})}{\lvert \boldsymbol{f}_{\rm grav} \rvert} \, .
\end{align}
We see in Fig.~\ref{Eddington ratios} that, for Ly$\alpha$ radiation pressure, extreme Eddington ratios of $\sim 10-100$ are highly concentrated near the ionization front. At $t = 35 \, \rm kyr$, shortly before breakout of the ionization front, there is greater cloud anisotropy, which causes a more uniform distribution of $f_{\rm Edd}^{\rm Ly\alpha}$. In contrast, for direct radiation pressure we find more modest maximum values of $f_{\rm Edd}^{\rm dir} \lesssim \textrm{few}$, and more evenly distributed forces overall. 

\subsubsection{Emergent Lyman-$\alpha$ spectra}

\begin{figure*}
\centering
\includegraphics[width=0.48\textwidth]{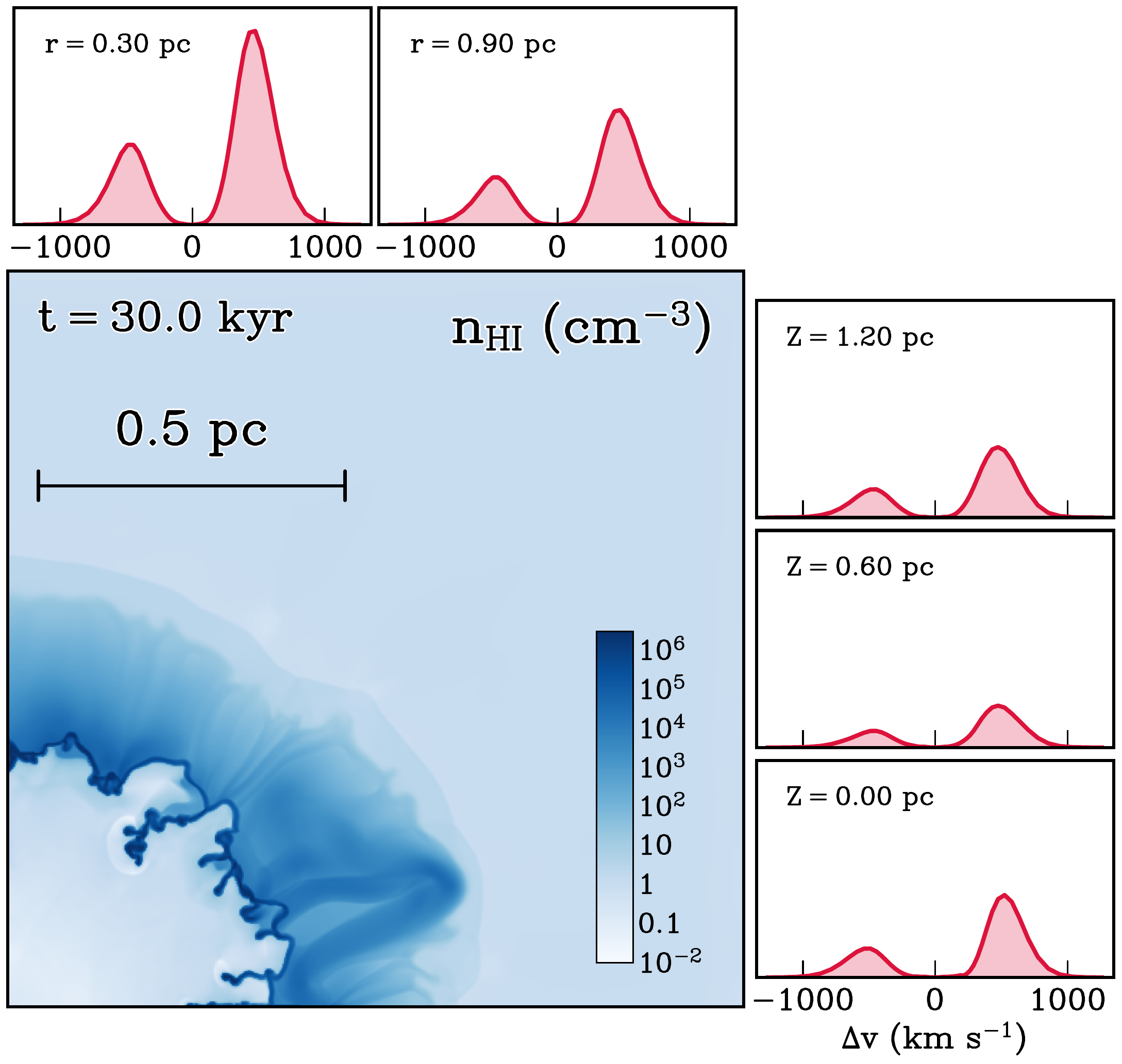}
\hspace{0.02\textwidth}
\includegraphics[width=0.48\textwidth] {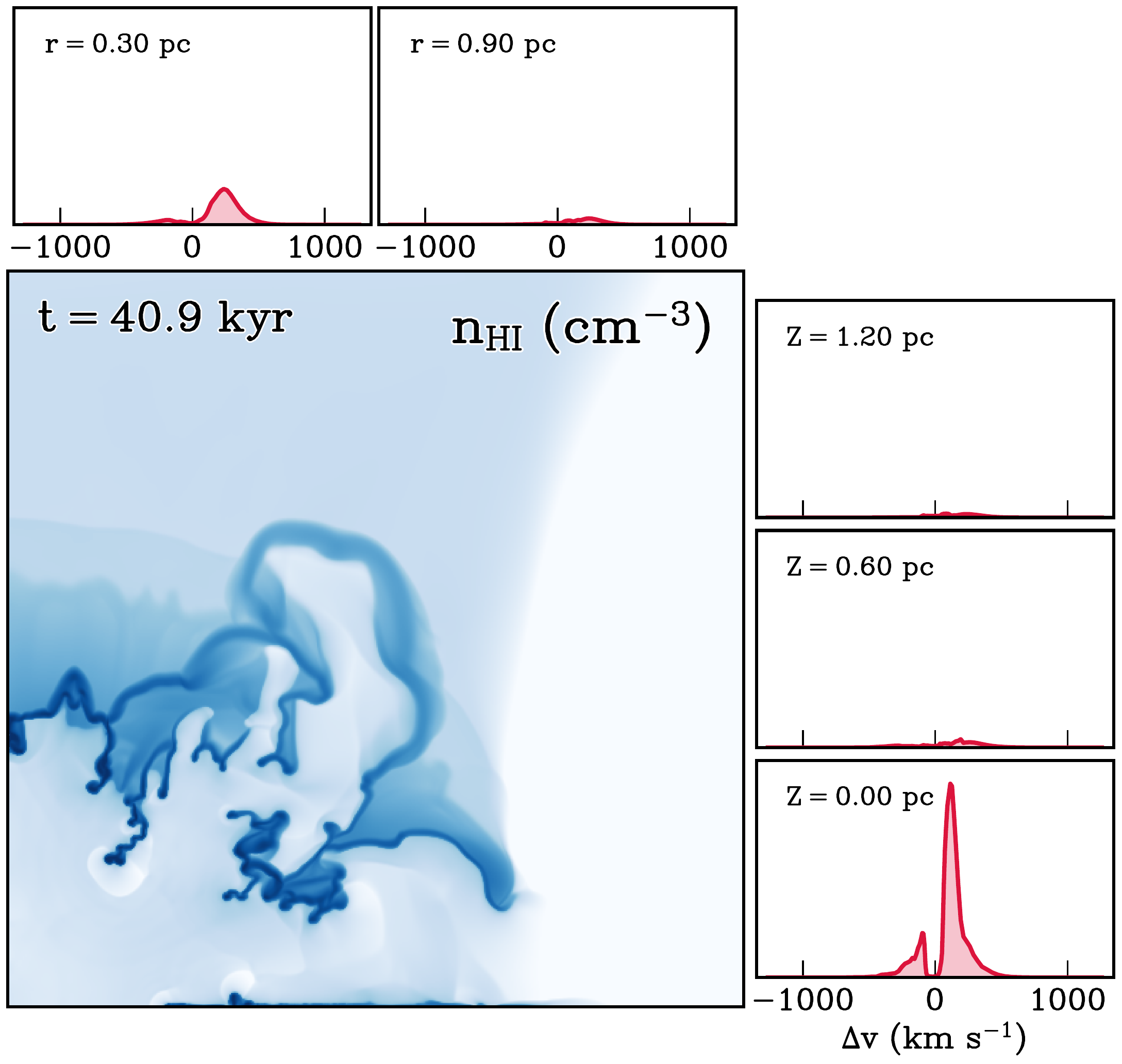}
\vspace{0.03\textwidth}

\includegraphics[width=0.48\textwidth]{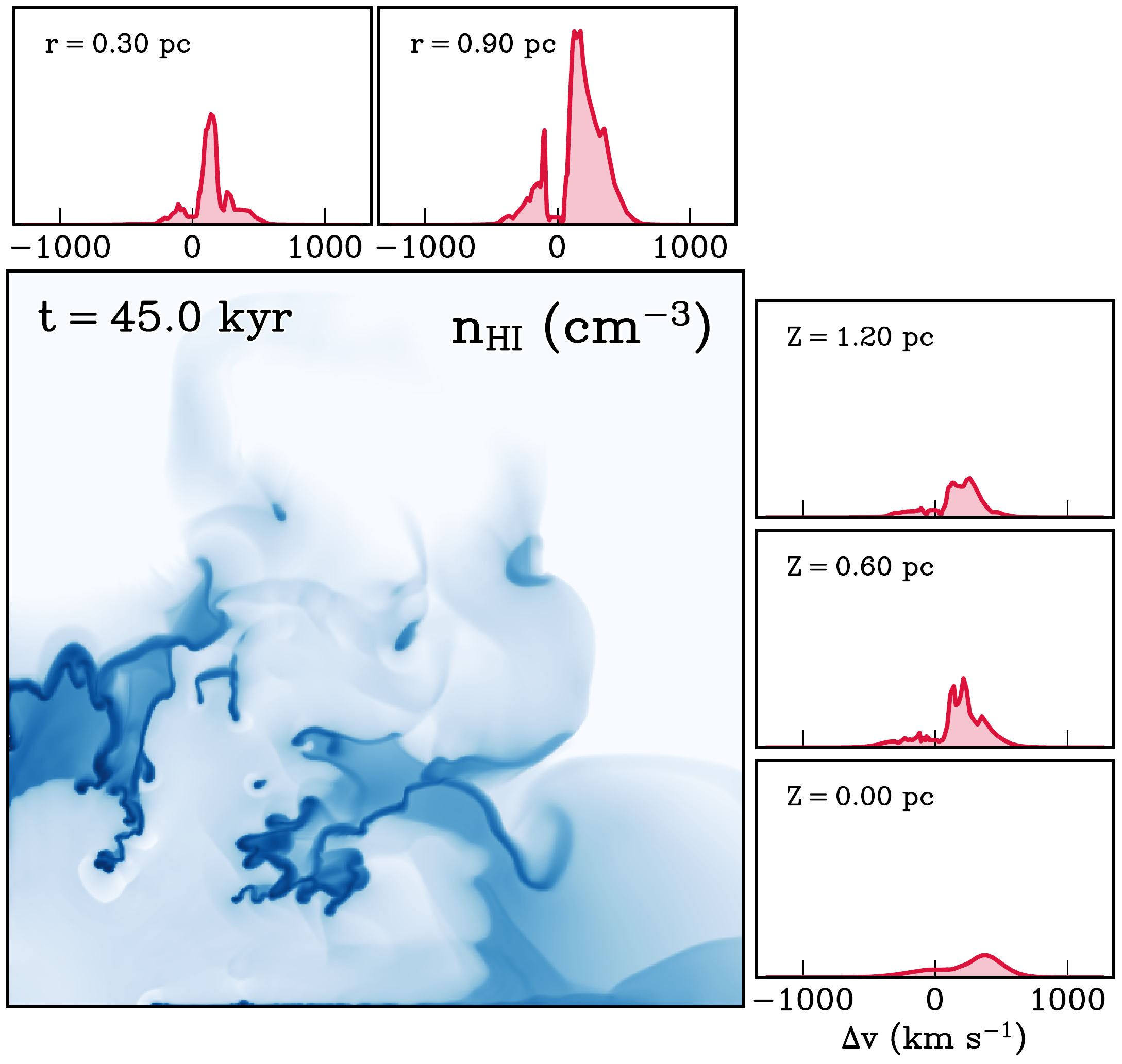}
\hspace{0.02\textwidth}
\includegraphics[width=0.48\textwidth] {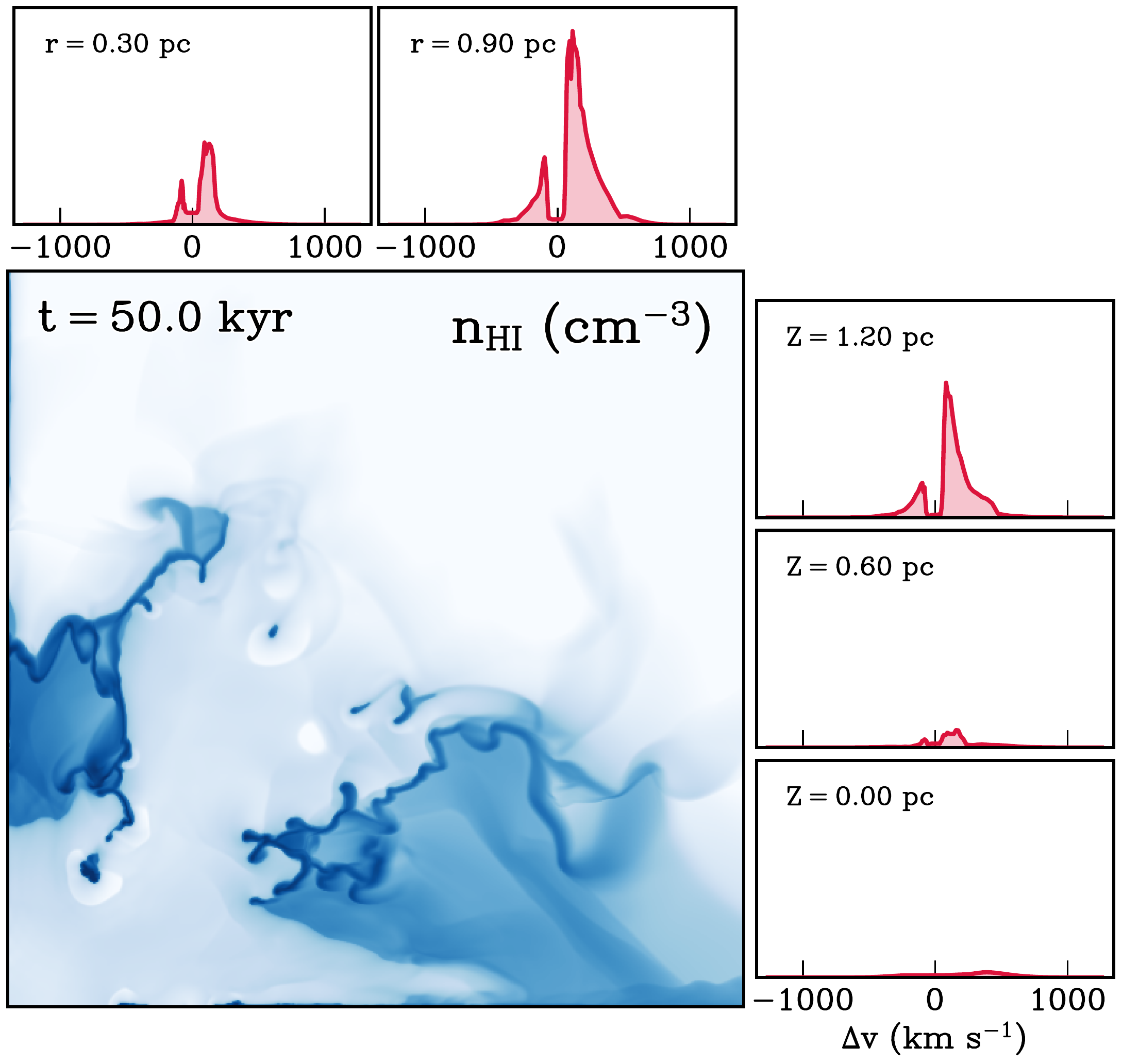}
\caption{ Ly$\alpha$ emergent spectra before ($t = 30 \, \rm kyr$) and after ($t \gtrsim 41 \, \rm kyr$) breakout of the ionization front in the simulation \texttt{SCLyaR512}. Color maps show the \textsc{H\,i} number density for the full $1.2 \, \rm \pc \times 1.2 \, \rm pc$ box, with dark regions highlighting dense \textsc{H\,i} gas. Emergent Ly$\alpha$ spectra, taken near simulation box boundaries, are shown above (for the upper $Z$-boundary) and to the right (for the outer $R$-boundary) of the respective density panel. For each time $t$, the amplitude of the spectra are scaled relative to the maximum among them. Wavelengths in the spectra panels are in velocity units, i.e. $\Delta v \equiv c\,(\lambda - \lambda_{\rm Ly\alpha})/\lambda_{\rm Ly\alpha}$. }
\label{RHD simulation spectra}
\end{figure*}

In Fig.~\ref{RHD simulation spectra} we plot emergent Ly$\alpha$ spectra, before and after breakout of the ionization front. Before breakout, we find characteristic double-peaked spectra, biased towards the red (positive $\Delta v \equiv c\,\Delta \lambda/\lambda_{\rm Ly\alpha}$). This is in line with expectations for an expanding \textsc{H\,i} gas cloud \citep[e.g.][]{Zheng2002, Tasitsiomi2006_z8, Verhamme2006, Laursen2009, Nebrin2024, Smith2025}. For a dust-free, static cloud, we expect the spectra to peak at $\Delta v = \pm \eta \, (2 k_{\rm B} T /m_{\rm H})^{1/2} \, (a_{\rm v} \tau_{\rm cl})^{1/3}$ with $\eta \simeq 0.60 - 0.93$ \citep{Lao2020}, which evaluates to:
\begin{equation}
    \lvert \Delta v \rvert = 837 \, \, \eta \,T_{100}^{1/6} N_{\rm HI,23}^{1/3} ~ \textrm{km s}^{-1} \, ,
\end{equation}
where $T_{100} \equiv T/100 \, \rm K$, and $N_{\rm HI,23} \equiv N_{\rm HI} / 10^{23} \, \rm cm^{-2}$, when normalized to a typical \textsc{H\,i} temperature and column density. This is within a factor $\sim 2$ of the wavelength peak offsets we observe before breakout, $\sim 400-500 \, \rm km \, s^{-1}$ at $t = 30 \, \rm kyr$. The observed peak separations are a factor $\sim 1.5-2$ smaller than analytically predicted for a central point source ($\eta = 0.93$), but dust absorption and turbulence is expected to reduce the frequency offset by a similar factor \citep{Kimm2019, Kakiichi2021, Nebrin2024}. Shortly after breakout of the ionization front in the $R$-direction, we find that most Ly$\alpha$ photons escape via the newly created low-opacity channel. The emergent spectra are narrower (peaking at $\lvert \Delta v \rvert \sim 100-200 \, \rm km \, s^{-1}$), indicative of fewer scatterings by \textsc{H\,i}.

\subsubsection{Dust and chemical state}

\begin{figure*}
\centering
\includegraphics[width=0.9\textwidth]{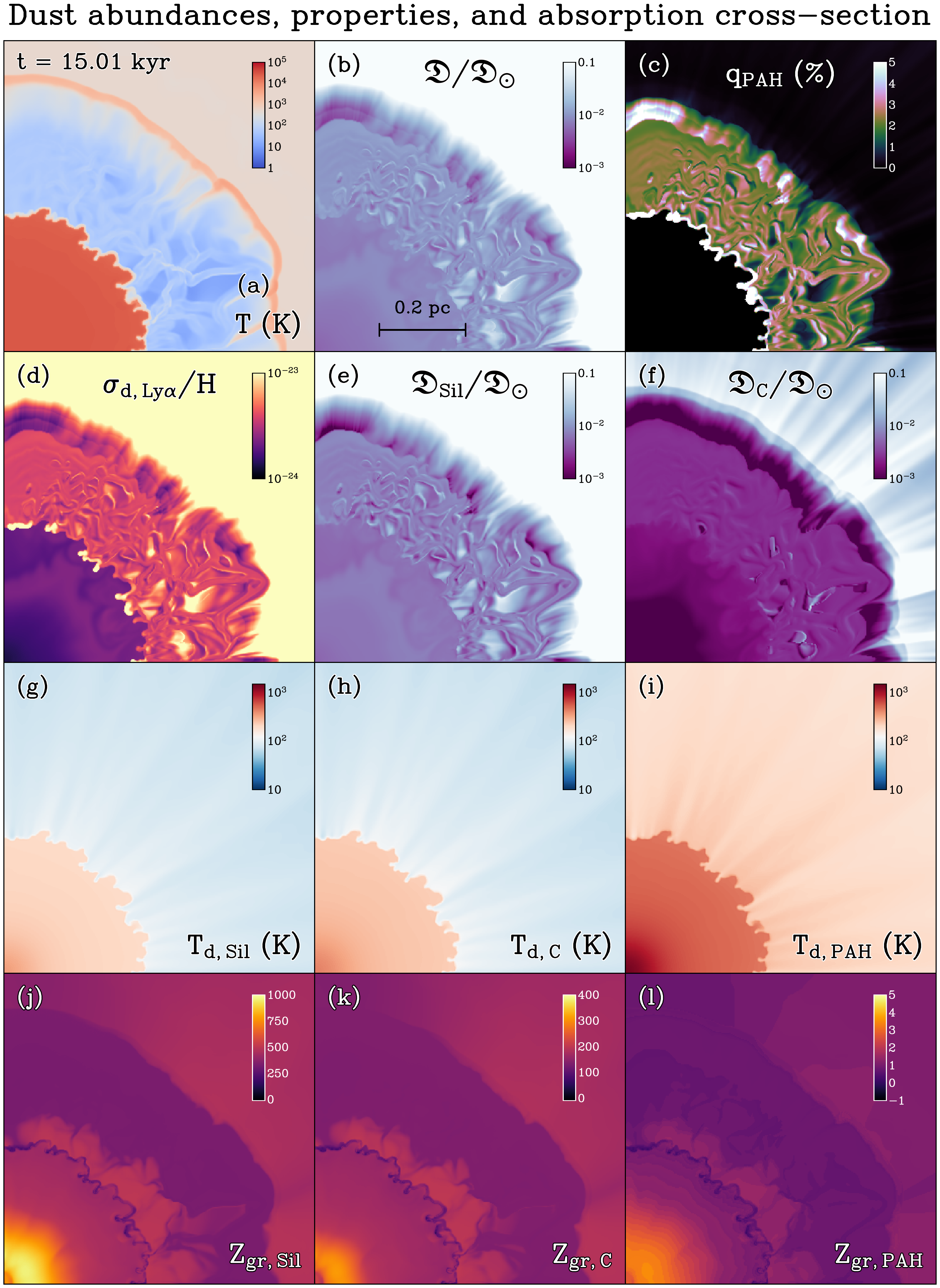}
\caption{ Dust properties (and the gas temperature for comparison) roughly at $t = 15 \, \rm kyr$ in the high-resolution Ly$\alpha$ RHD simulation (\texttt{\texttt{SCLyaR512}}). \textbf{Panel (a):} The gas temperature.  \textbf{Panel (b):} The total dust-to-gas ratio in Solar units. \textbf{Panel (c):} The fraction of dust mass in PAHs. \textbf{Panel (d):} The Ly$\alpha$ dust absorption cross-section per H nucleus (in $\rm cm^2 \, H^{-1}$). \textbf{Panels (e)--(f):} The dust-to-gas ratios of silicate (Sil) and graphite (C) dust, in units of $\mathfrak{D}_\odot = 1/162$. \textbf{Panels (g)--(i):} The dust temperature for each dust type. \textbf{Panels (j)--(l):} The dust grain charges for each dust type. We note that the grain charges are positive even outside the \textsc{H\,ii} region. This reflects the low dust attenuation, which causes efficient photoelectric charging. }
\label{Dust properties figure 512 Lya simulation}
\end{figure*}

\begin{figure*}
\centering
\includegraphics[width=0.9\textwidth]{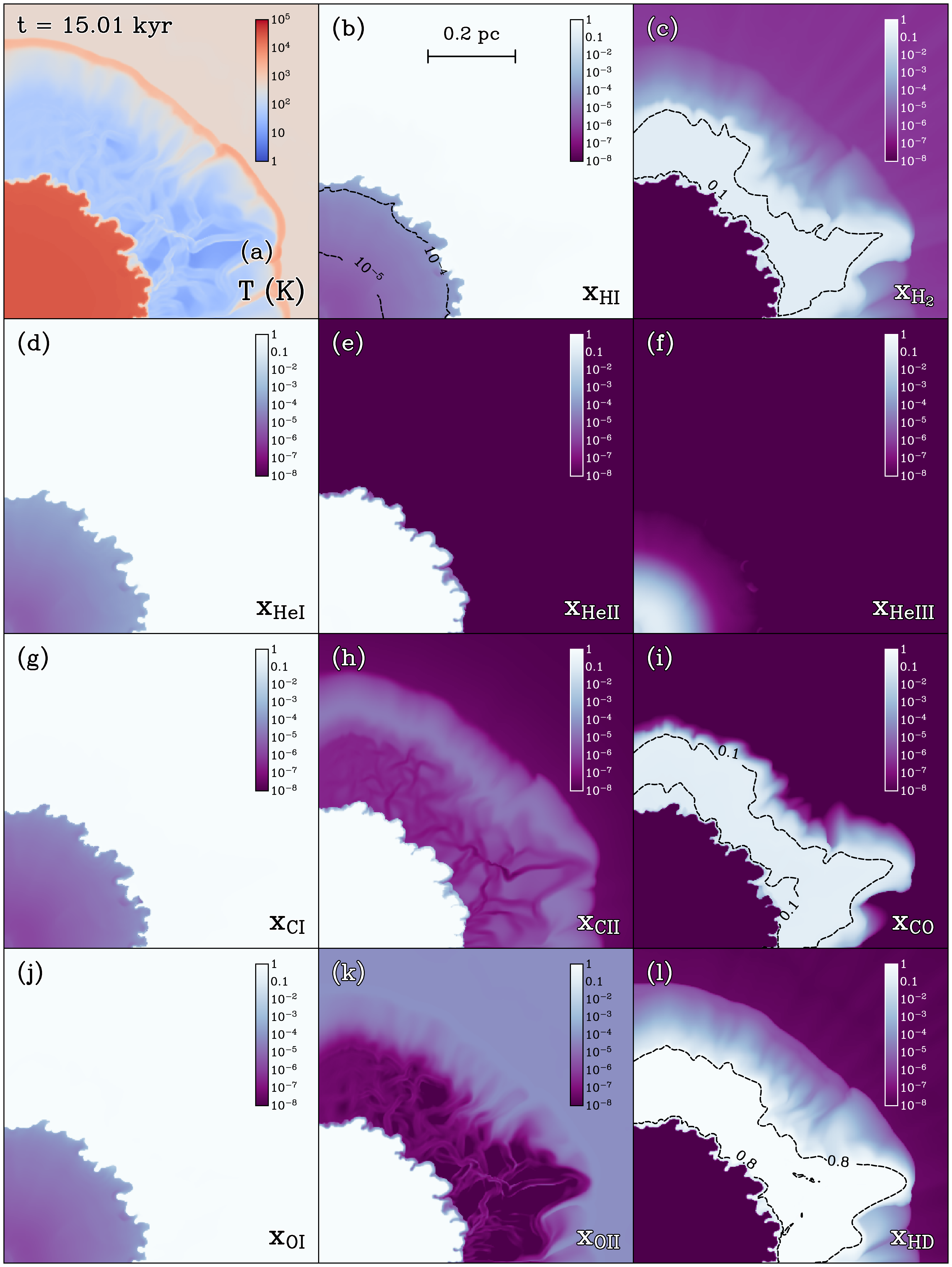}
\caption{ Gas abundances (and the gas temperature for comparison) at $t = 15 \, \rm kyr$ in the high-resolution Ly$\alpha$ RHD simulation (\texttt{\texttt{SCLyaR512}}). \textbf{Panel (a):} The gas temperature.  \textbf{Panel (b):} The \textsc{H\,i} abundance $x_{\rm HI} \equiv n_{\rm HI} /n_{\rm H}$. \textbf{Panel (c):} The H$_2$ abundance $x_{\rm H_2} \equiv n_{\rm H_2} /n_{\rm H}$. \textbf{Panel (d):} The He\textsc{\,i} abundance $x_{\rm HeI} \equiv n_{\rm HeI} /n_{\rm He}$. \textbf{Panels (e):} The He\textsc{\,ii} abundance $x_{\rm HeII} \equiv n_{\rm HeII} /n_{\rm He}$. \textbf{Panel (f):} The He\textsc{\,iii} abundance $x_{\rm HeIII} \equiv n_{\rm HeIII} /n_{\rm He}$. \textbf{Panel (g):} The \textsc{C\,i} abundance $x_{\rm CI} \equiv n_{\rm CI} /n_{\rm C}$. 
\textbf{Panel (h):} The \textsc{C\,ii} abundance $x_{\rm CII} \equiv n_{\rm CII} /n_{\rm C}$. \textbf{Panel (i):} The CO abundance $x_{\rm CO} \equiv n_{\rm CO} /n_{\rm C}$. \textbf{Panel (j):} The \textsc{O\,i} abundance $x_{\rm OI} \equiv n_{\rm OI} /n_{\rm O}$. \textbf{Panel (k):} The \textsc{O\,ii} abundance $x_{\rm OII} \equiv n_{\rm OII} /n_{\rm O}$. \textbf{Panel (l):} The HD abundance $x_{\rm HD} \equiv n_{\rm HD} /n_{\rm D}$.}
\label{Gas abundances figure 512 Lya simulation}
\end{figure*}

Lyman-$\alpha$ feedback is strongly coupled to the dust and gas chemical state, primarily via dust absorption, and Ly$\alpha$ destruction by H$_2$ line absorption and $2p \rightarrow 2s$ transitions \citep[][]{Neufeld1990, Nebrin2024}. In Fig.~\ref{Dust properties figure 512 Lya simulation}, we plot the dust abundances and properties (e.g. dust temperatures and charges), at $t = 15 \, \rm kyr$ in our highest resolution Ly$\alpha$ RHD simulation (\texttt{SCLyaR512}). We see that the dust-to-gas ratio is somewhat lower in the \textsc{H\,ii} region, mostly due to depletion of graphite dust (PAHs, although important for the dust opacity, make up only a tiny fraction of the dust mass). PAHs are also practically absent in the \textsc{H\,ii} region. The depletion of graphite dust and PAHs causes a noticeable reduction in the Ly$\alpha$ dust absorption opacity, by a factor $\sim 2-3$, which further facilitates a build-up of Ly$\alpha$ radiation pressure. The depletion of graphite dust and PAHs is mainly due to destruction processes. In particular, the PAHs are easily photodissociated by ionizing photons in the \textsc{H\,ii} region, with the higher gas temperatures also promoting thermal sputtering of PAHs. Graphite dust is mainly depleted by chemical sputtering, caused by the relatively high dust temperatures ($T_{\rm d, C} \sim \textrm{few} \times 100 \, \rm K$) and gas densities (see Fig.~\ref{Lya vs no Lya figure 512 D0.01}) in the \textsc{H\,ii} region \citep[][]{Draine1979_ChemicalSputtering, Lenzuni1995}.

Dust dynamics, which primarily affects the larger silicate grains, has only a moderate impact on the reduction in the dust-to-gas ratio and the dust opacity. This is because of the high grain charges in the \textsc{H\,ii} region of the star cluster, which causes strong Coulomb drag on the dust. Instead, dust dynamics is mainly important in compact, isolated \textsc{H\,ii} regions around individual stars, where grain charges are more modest. 

In Fig.~\ref{Gas abundances figure 512 Lya simulation}, we plot the chemical abundance ratios of various species followed in \textsc{Lydion}. Outside the \textsc{H\,ii} region, the H$_2$ abundance is near its initial assumed value ($x_{\rm H_2} \sim 0.1$), owing to the inefficient rate of H$_2$ formation in the dust-poor gas, as well as the efficient self-shielding against the Lyman-Werner radiation from the star cluster, by H$_2$ column densities $N_{\rm H_2} \sim 10^{22} \, \rm cm^{-2}$.\footnote{This may change somewhat had we assumed that the hydrogen gas was fully atomic initially. The reason for assuming a partial initial molecular abundance $x_{\rm H_2} = 0.1$ is to be conservative with respect to Ly$\alpha$ feedback. In particular, the H$_2$ promotes Ly$\alpha$ destruction, promotes further H$_2$ formation by self-shielding against Lyman-Werner radiation, and lowers the \textsc{H\,i} column densities responsible for Ly$\alpha$ scattering. } The large surviving \textsc{H\,i} reservoir further traps Ly$\alpha$ photons in the cloud. Inside the \textsc{H\,ii} region itself, there is also residual \textsc{H\,i} ($x_{\rm HI} \sim 10^{-5} - 10^{-4}$) which can scatter Ly$\alpha$. We also find that most deuterium ($\gtrsim 80\%$), and a significant fraction of carbon ($\sim 20\%$), is in the form of HD and CO, respectively. These molecules are efficient coolants, and together with \textsc{C\,i},\footnote{We find negligible \textsc{C\,ii} in the neutral gas, owing again to the self-shielding (mainly by H$_2$) against the Lyman-Werner radiation that is responsible for the photoionization of \textsc{C\,i} in neutral gas (see Appendix~\ref{Carbon chemistry appendix}).} \textsc{O\,i}, and H$_2$, cool the neutral gas down to $T \sim 30 - 100 \, \rm K$, with the lower limit being set by the CMB temperature at the assumed redshift $z = 10$.

\begin{figure}
    \centering
    \includegraphics[width=0.9\columnwidth]{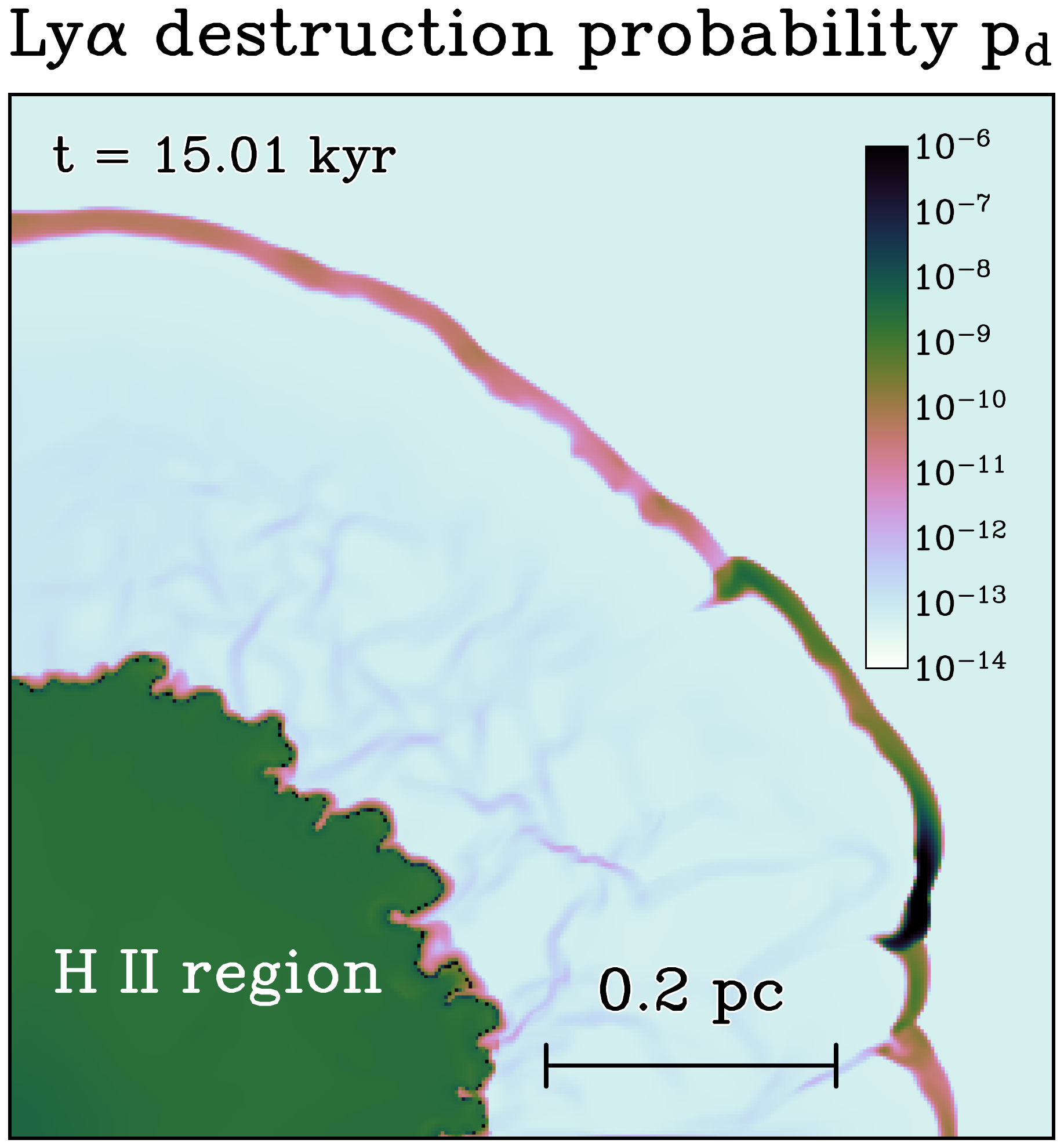}
    \caption{The Ly$\alpha$ destruction probability $p_{\rm d}$ at $t = 15 \, \rm kyr$ in the Ly$\alpha$ RHD simulation \texttt{\texttt{SCLyaR512}}. The destruction probability is $\sim 10^{-9}$ in the \textsc{H\,ii} region, and drops to $\sim 10^{-13} - 10^{-11}$ in the neutral gas. }
    \label{fig: Destruction probability}
\end{figure}

Finally, we plot the Ly$\alpha$ destruction probability $p_{\rm d}$ in Fig.~\ref{fig: Destruction probability}. The destruction probability is set by $2p \rightarrow 2s$ transitions, primarily in collisions with \textsc{H\,i}, He\textsc{\,i}, \textsc{H\,ii}, and electrons \citep[][]{Nebrin2024}, and H$_2$ line absorption.\footnote{Since the relevant H$_2$ lines lie close to Ly$\alpha$ line center, they can be treated as an effective destruction probability \citep[][]{Neufeld1990, ChiuDraine1998, Nebrin2024}.} We find that $p_{\rm d} \sim \textrm{few} \times 10^{-9}$ in the \textsc{H\,ii} region, which is around the maximum value that can be attained from $2p \rightarrow 2s$ transitions \citep[see eq. 99 and footnote 21 in][]{Nebrin2024}:
\begin{equation}
    p_{\rm d,HI} \lesssim \dfrac{A_{2\gamma}}{3A_{\rm Ly\alpha}} = 4.37 \times 10^{-9} \, .
\end{equation}
This high value is driven by collisions with \textsc{H\,ii} and $e^-$, which have relatively high collisional rate coefficients. However, destruction is only efficient if $p_{\rm d} \tau_0 \gtrsim 1$ in the region of interest. Since the Ly$\alpha$ optical depth in the \textsc{H\,ii} region is $\tau_0 \sim 10^4 - 10^5$, Ly$\alpha$ destruction is inefficient here, despite the significant destruction probability.   

\subsubsection{Are the results converged?}
\begin{figure*}
    \centering
    \includegraphics[width=0.8\textwidth]{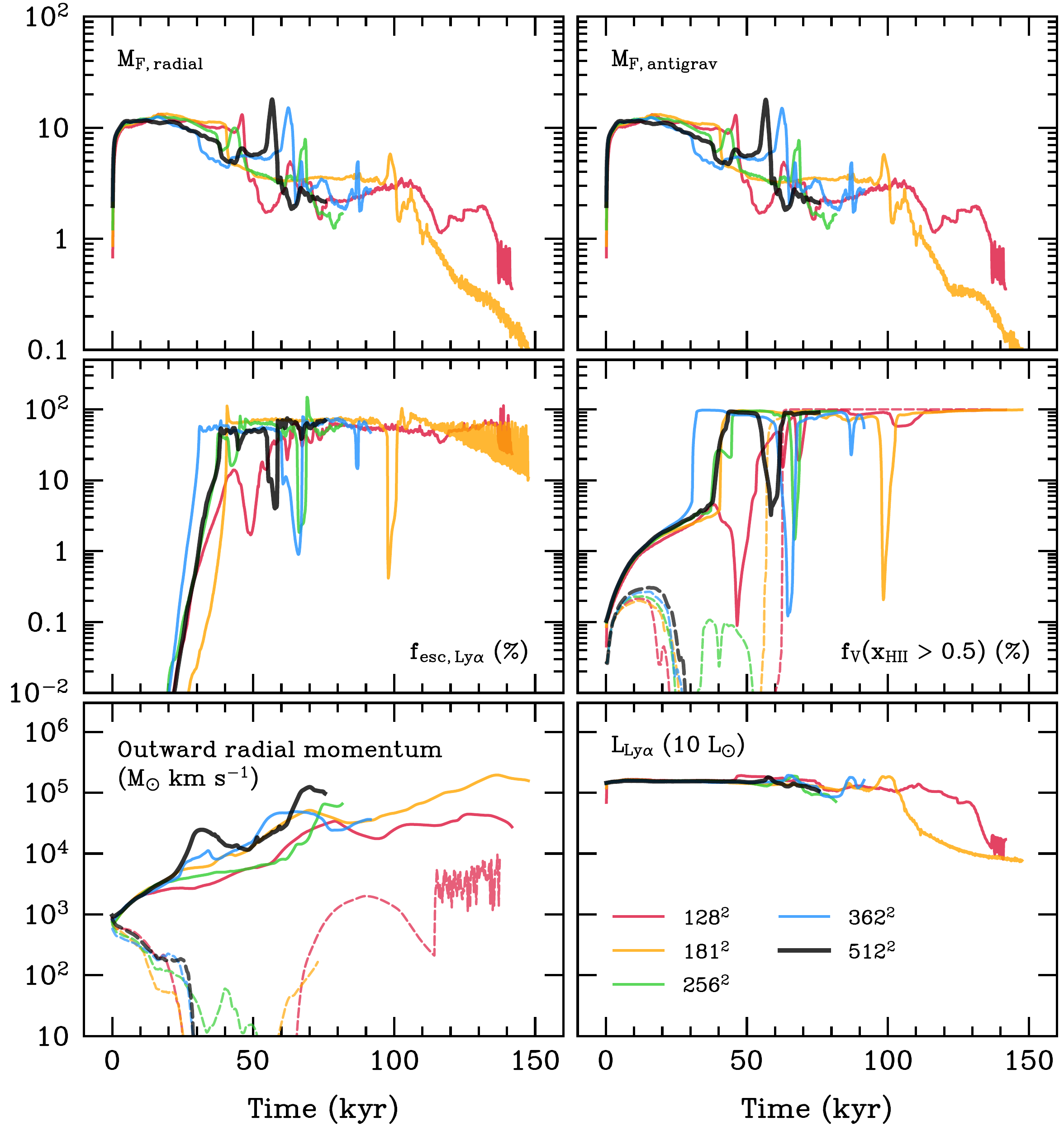}
    \caption{ The evolution of various quantities (e.g. force multipliers), for all spatial resolutions ($128^2 - 512^2$) in the star cluster simulations. \textbf{Top left panel:} The net radial Ly$\alpha$ force multiplier $M_{\rm F,radial}$ for the Ly$\alpha$ RHD simulations. \textbf{Top right panel:} The net Ly$\alpha$ force multiplier $M_{\rm F,antigrav}$ in the opposite direction of gravity, for the Ly$\alpha$ RHD simulations. \textbf{Middle left panel:} The Ly$\alpha$ escape fraction. We note that $f_{\rm esc,Ly\alpha}$ can formally reach $> 100\%$ at some instances. This is primarily because the reduced speed of light leads to a delay in reaching the simulation box boundaries, $t_{\rm esc} = 3.9 \, (\Delta r/1.2 \, \textrm{pc}) (\Tilde{c}/300 \, \rm km \, s^{-1})^{-1} \, \rm kyr$ for free-streaming photons (longer for trapped Ly$\alpha$). This numerical effect is mainly important following a rapid (even if small) decrease in $L_{\rm Ly\alpha}$, which causes an increase in $f_{\rm esc,Ly\alpha} \propto 1/L_{\rm Ly\alpha}$ (see Eq.~\ref{fesc Lya definition}). \textbf{Middle right panel:} The fraction of the simulation volume that is ionized (defined as $x_{\rm HII}>0.5$), with Ly$\alpha$ feedback (solid lines), and without (dashed lines). \textbf{Bottom left panel:} The outward radial momentum of the gas in the simulations with Ly$\alpha$ feedback (solid lines), and without (dashed lines). \textbf{Bottom right panel:} The Ly$\alpha$ luminosity, for the Ly$\alpha$ RHD simulations. }
    \label{fig: Convergence plot}
\end{figure*}

In our discussions above, we have focused on our highest resolution run (\texttt{SCLyaR512}), and shown that Ly$\alpha$ feedback both dominates among feedback processes, and drives outflows for our particular setup. In contrast, the simulation ignoring Ly$\alpha$ feedback predicts collapse. Because higher resolution simulations resolve more low-opacity channels and gas instabilities, one might wonder whether our results are converged.\footnote{We thank Mike Grudi{\'c} for stressing the importance of convergence checks here.} To check this, in Fig.~\ref{fig: Convergence plot} we plot the results for the Ly$\alpha$ force multipliers $M_{\rm F,radial}$ and $M_{\rm F,antigrav}$, for resolutions ranging from $128^2$ to $512^2$ (the main run). Also plotted are the Ly$\alpha$ luminosities and escape fractions, the fraction of the simulation volume that is ionized, and the outward radial momentum of the gas:
\begin{equation}
    p_{\rm radial,out} \equiv \int \textrm{d}V \, \max(\rho \boldsymbol{u} \boldsymbol{\cdot} \boldsymbol{\hat{r}},\, 0) \, .
\end{equation}
Prior to breakout of the ionization front, we find barely any difference in the force multiplier between runs, i.e. all have $M_{\rm F,radial/antigrav}  \simeq 11$. The precise timing of breakout varies between the runs, but there is no obvious trend with resolution. For example, our $512^2$ simulation predicts breakout before the $181^2$ simulation, but \textit{after} the $362^2$ simulation. Following breakout there is more scatter in the predicted force multipliers, with precise evolution likely tied to both the particular realization of turbulent initial conditions, and the spatial resolution. However, our results are nearly identical before breakout, and qualitatively the same after breakout. The basic prediction for our assumed setup, of strong Ly$\alpha$ pressure with $M_{\rm F,radial/antigrav} \sim 10$, is therefore likely converged. 

We also see that simulations largely agree on trends in $p_{\rm radial,out}$ and the volume fraction of ionized gas, $f_{\rm V}(x_{\rm HII}>0.5)$. The simulations with Ly$\alpha$ feedback experience growing $p_{\rm radial,out}$, with magnitudes consistent with expectations from Ly$\alpha$-driven feedback, i.e. $p_{\rm radial,out} \simeq M_{\rm F,radial} \, L_{\rm Ly\alpha} t /c$, or (using relevant values from Fig.~\ref{fig: M_F 512 run} \& \ref{fig: Convergence plot}):
\begin{align}
    p_{\rm radial,out} &\simeq~ 5800 \, \left( \dfrac{M_{\rm F,radial}}{11} \right) \left( \dfrac{L_{\rm Ly\alpha}}{1.3\times10^6 \, \textrm{L}_\odot} \right) \\ &\times~ \left( \dfrac{t}{20 \, \rm kyr} \right) \, \rm M_{\odot} \, km \, s^{-1} \, . \nonumber
\end{align}
The simulations without Ly$\alpha$ feedback have significantly lower residual outward radial momenta, owing to large-scale collapse. Finally, the simulations with Ly$\alpha$ feedback experience large-scale ionization significantly earlier. Only a subset of lower-resolution simulations could be evolved for long, as a result of time-step constraints during the collapse. However, we see that there can be delayed LyC leakage in the simulations without Ly$\alpha$ feedback. We note that because we treat diffuse ionizing radiation from recombinations, this can give values of $f_{\rm V}(x_{\rm HII}>0.5) \sim 100\%$ even if the LyC escape fraction is relatively low, since the low-density external gas is easily ionized by diffuse photons. 

\subsection{Lyman-$\alpha$ feedback from isolated stars, the impact of turbulence, dust, and Ly$\alpha$ destruction}
\label{Lya RHD isolated star section}

Next we discuss the results for the isolated star setup, which probe the impact of Ly$\alpha$ feedback during the earliest (compact) stages of the \textsc{H\,ii} region. We vary the level of turbulence ($u_{\rm RMS} = 0-8 \, \rm km \ s^{-1}$), and the dust abundance ($\mathfrak{D}/\mathfrak{D}_\odot = 0, 10^{-3}$). To check the importance of Ly$\alpha$ destruction by $2p \rightarrow 2s $ transitions and H$_2$ line absorption at these high densities ($n_{\rm H} = 10^6 \, \rm cm^{-3}$ initially), we also run reference simulations ignoring these processes.

In Figs.~\ref{Isolated stars, nH plot} and \ref{Isolated stars, T, PLya/P plot}, we show the final snapshots of the gas density, temperature, Ly$\alpha$ energy density, and ratio between Ly$\alpha$ and gas pressures. We see that in all these simulations, Ly$\alpha$ feedback is dynamically important, causing a more rapid growth of the \textsc{H\,ii} region. At the final snapshot, the \textsc{H\,ii} regions are a factor $\sim 1.5 - 2 \, \times$ larger in the simulations with Ly$\alpha$ feedback than in those ignoring the process. We further see that the Ly$\alpha$ pressure is $\sim 0.5 - \textrm{few} \,\times$ the gas pressure near the outer edges of the \textsc{H\,ii} region, with the most dramatic effect in the dust-free simulation. 

These results run counter to generally held expectations that radiation pressure feedback is negligible in the \textsc{H\,ii} region dynamics of isolated stars, at least for the densities probed here \citep[e.g.][]{Krumholz2009_HII, Sales2014, Hollenbach2026_HII}. While this is true for direct stellar radiation pressure feedback (and in more dust-enriched clouds), we find that Ly$\alpha$ feedback is $\sim 4 - 16\, \times$ stronger in these dust-poor conditions, as we show in Fig.~\ref{fig: Lya vs direct RP isolated stars}. As seen in Fig.~\ref{fig: MF isolated star runs}, we find $M_{\rm F,radial} \sim 15 - 40$ for the simulations with $\mathfrak{D}/\mathfrak{D}_\odot = 10^{-3}$, and $M_{\rm F,radial} \sim 35-60$ for the dust-free simulation (with Ly$\alpha$ destruction). Our simulations also inform us about the impact of turbulence, dust, and Ly$\alpha$ destruction:
\begin{itemize}[leftmargin=*]
    \item \textbf{Impact of turbulence on Ly$\boldsymbol{\alpha}$ feedback, and comparison to analytical predictions:} As we vary the initial turbulence from $u_{\rm RMS} = 0 \, \rm km \, s^{-1}$ to $u_{\rm RMS} = 8 \, \rm km \, s^{-1}$, we find that Ly$\alpha$ feedback becomes weaker, but only by a modest factor of $\sim 2$. This relatively weak suppression is approximately in line with the analytical predictions of \cite{Nebrin2024}, who generalized uniform-cloud solutions using methods of asymptotic homogenization to treat Ly$\alpha$ diffusion in turbulent inhomogeneous media.
    Ignoring dust, large-scale expansion, and Ly$\alpha$ destruction and dust absorption, \cite{Nebrin2024} find that the radial force multiplier should drop as:
    \begin{equation}
        M_{\rm F,radial} \propto [1 + (b_{\rm s}\mathcal{M})^2]^{-4/9} \, , \label{M_F analytical drop with Mach}
    \end{equation}
    where $b_{\rm s} = 1/3$ for solenoidal turbulence, and $\mathcal{M}$ is the Mach number. This would predict a suppression by a factor $\sim 3$. Adding large-scale cloud expansion, dust absorption, and Ly$\alpha$ destruction yields a weaker dependence on $\mathcal{M}$ than predicted in Eq.~(\ref{M_F analytical drop with Mach}), as a result of easier Ly$\alpha$ escape, reducing the efficiency of destructive processes \citep[][]{Nebrin2024}. The net effect can therefore be small (factor $< 2$), consistent with our numerical results. We leave a more detailed comparison against analytical predictions to future work. 

    \item \textbf{Impact of dust absorption:} We find that going from $\mathfrak{D}/\mathfrak{D}_\odot = 0.01$ (our star cluster simulations) to $\mathfrak{D}/\mathfrak{D}_\odot = 10^{-3}$ boosts $M_{\rm F,radial}$ by a factor $\sim 2-4$, for fixed $u_{\rm RMS} = 4 \, \rm km \, s^{-1}$. Thus, Ly$\alpha$ feedback is sensitive to even small abundances of dust, although the scaling appears to be sublinear with respect to $\mathfrak{D}/\mathfrak{D}_\odot$. Finally, removing dust increases $M_{\rm F, radial}$ further by a factor of $\sim 2$ compared to $\mathfrak{D}/\mathfrak{D}_\odot = 10^{-3}$.

    \item \textbf{Impact of Ly$\boldsymbol{\alpha}$ destruction:} We find Ly$\alpha$ destruction ($p_{\rm d} > 0$) to be important for the densities probed in the isolated star simulations ($n_{\rm H} \sim 10^6 - 10^7 \, \rm cm^{-3}$). This is very evident in Fig.~\ref{fig: MF isolated star runs}, where we see that for dust-free conditions, taking Ly$\alpha$ destruction into account reduces $M_{\rm F,radial}$ from $\sim 100-300$ to $\sim 35-60$. It is therefore crucial to account for Ly$\alpha$ destruction by $2p \rightarrow 2s$ transitions and H$_2$ line absorption in future simulations, as also shown analytically in \cite{Nebrin2024}. We stress that this not only includes previously considered electron and proton-mediated $2p \rightarrow 2s$ transitions \citep[e.g.][]{McKeeTan2008}, but also transitions induced by collisions with neutral particles like \textsc{H\,i}, which can easily dominate in the cold gas \citep[][]{Nebrin2024}.
    
\end{itemize}

\begin{figure*}
\centering
\includegraphics[width=0.9\textwidth]{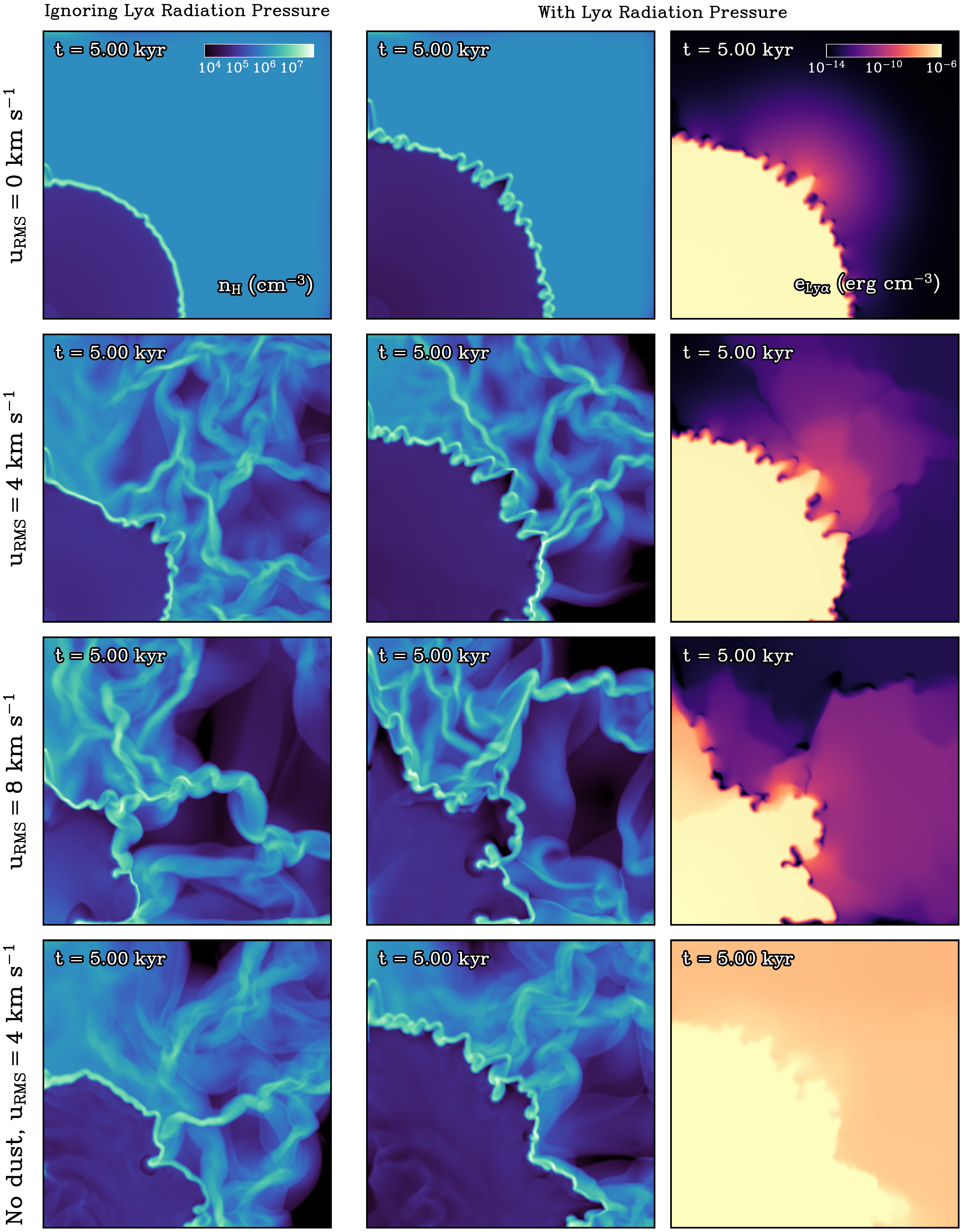}
\caption{ The gas density and Ly$\alpha$ energy density $e_{\rm Ly\alpha}$ for the simulations of the isolated stars of mass $35 \, \rm M_\odot$, at $t = 5 \, \rm kyr$, for different levels of initial turbulent velocities $u_{\rm RMS} = 0 - 8 \, \rm km \, s^{-1}$. Each panel is $0.1 \, \textrm{pc} \times 0.1 \, \rm pc$. The bottom row shows the results for the dust-free simulations (\texttt{StarNoLyaNoDust}, \texttt{StarLyaNoDust}), whereas the rest of the rows assume an initial DtG ratio of $\mathfrak{D}/\mathfrak{D}_\odot = 10^{-3}$. Ly$\alpha$ feedback is $\sim 4 - 16 \times$ stronger than direct radiation pressure over the duration of these runs (see Fig.~\ref{fig: Lya vs direct RP isolated stars}).  }
\label{Isolated stars, nH plot}
\end{figure*}

\begin{figure*}
\centering
\includegraphics[width=0.9\textwidth]{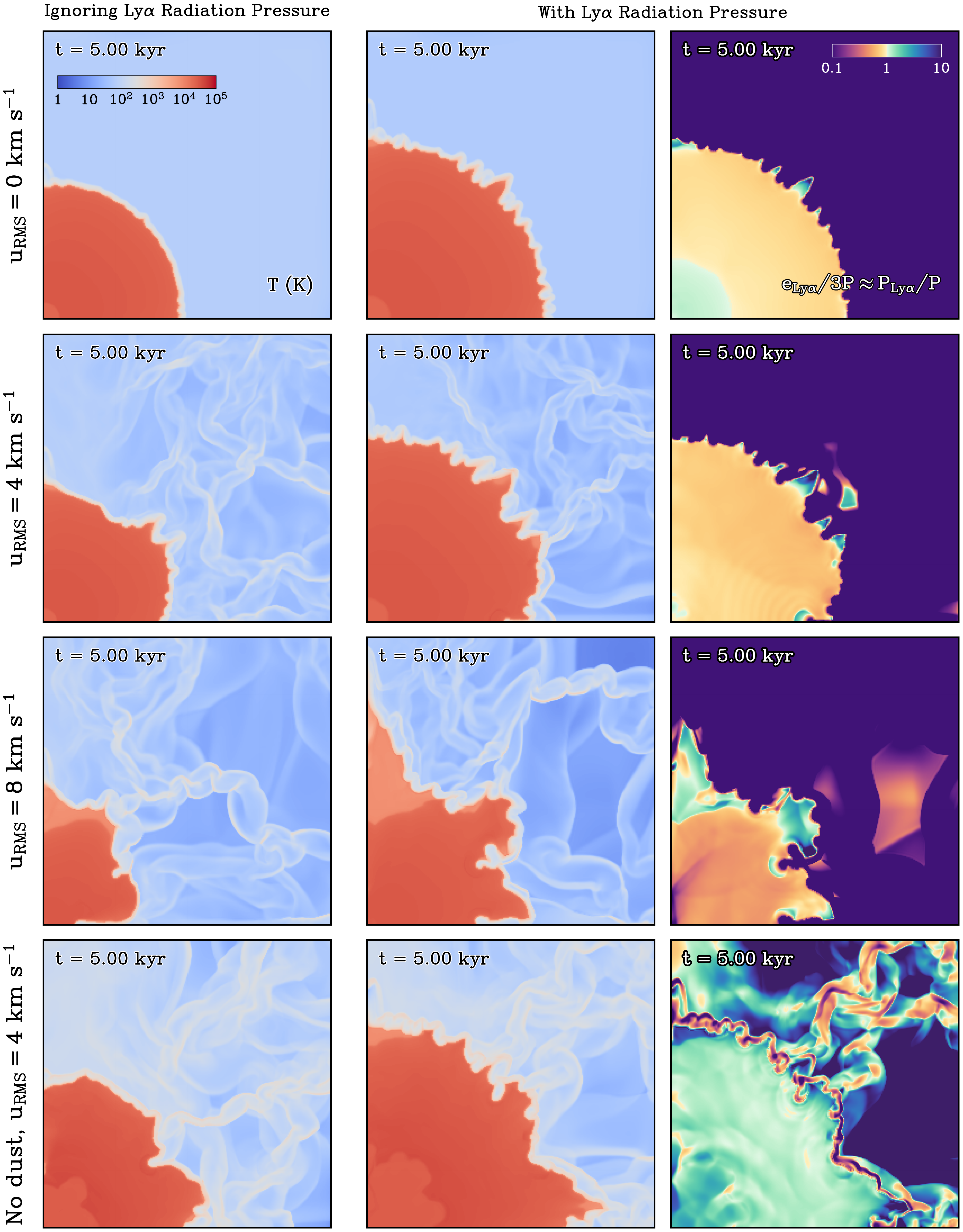}
\caption{ Same as Fig.~\ref{Isolated stars, nH plot} for the isolated star simulations, but showing the gas temperature and the approximate ratio between the Ly$\alpha$ pressure and gas pressure, at the final time $t = 5 \, \rm kyr$.  }
\label{Isolated stars, T, PLya/P plot}
\end{figure*}

\begin{figure}
    \centering
    \includegraphics[width=1.0\columnwidth]{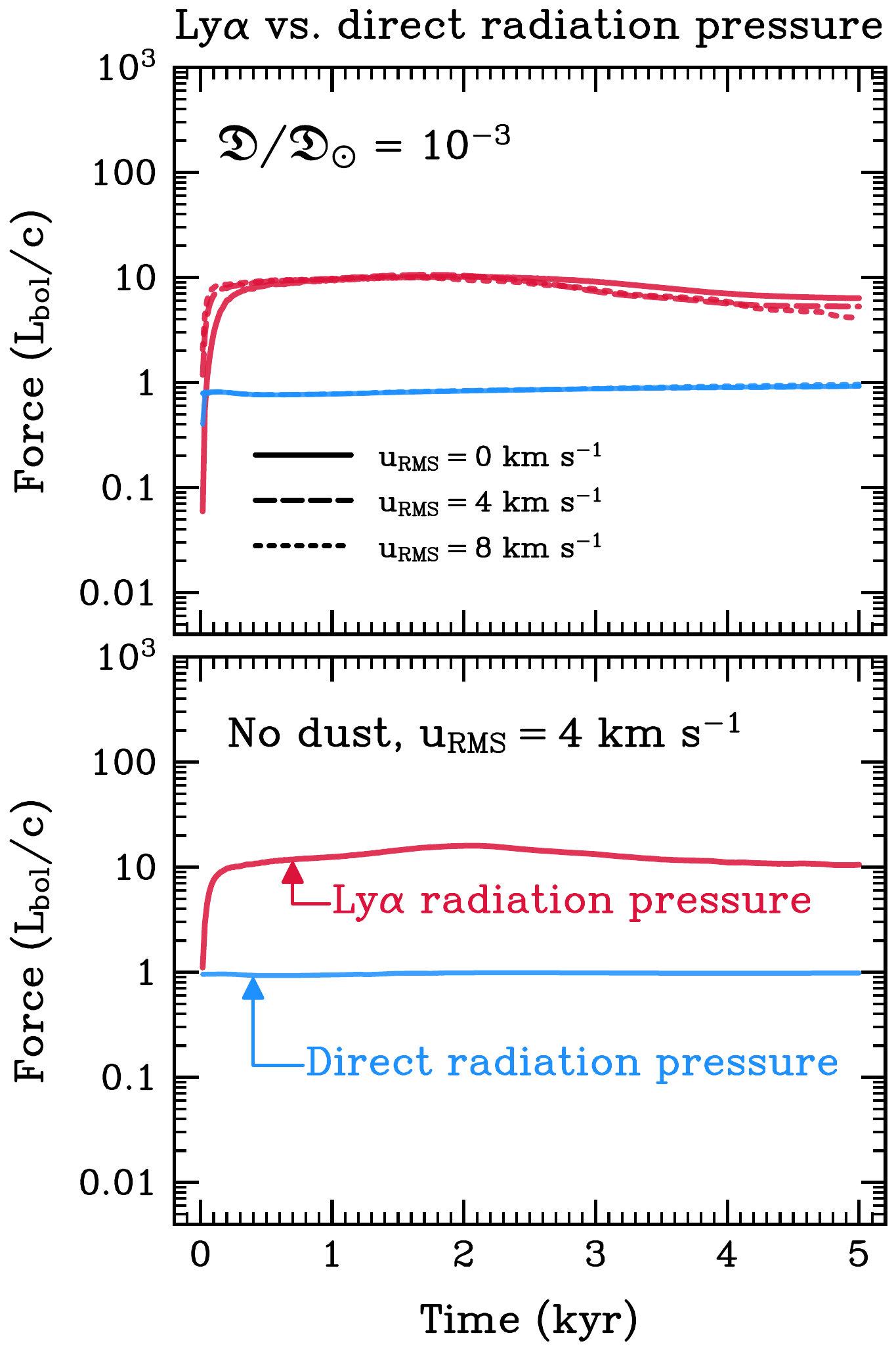}
    \caption{ A comparison between Ly$\alpha$ and direct radiation pressure forces for the isolated star simulations (similar to Fig.~\ref{fig: Feedback comparison 512 sim} for the star cluster simulation \texttt{SCLyaR512}). Radial forces are plotted, and the IR radiation pressure force is not shown since it is negligible ($\sim \textrm{few} \times 10^{-4} \, L_{\rm bol}/c$ for $\mathfrak{D}/\mathfrak{D}_\odot = 10^{-3}$). \textbf{Top panel:} The results for the simulations with initial DtG ratio $\mathfrak{D}/\mathfrak{D}_\odot = 10^{-3}$, and varying levels of initial turbulence. \textbf{Bottom panel:} The results for the dust-free Pop III star simulation (\texttt{StarLyaNoDust}). The physically unrealistic simulation ignoring Ly$\alpha$ destruction, \texttt{StarLyaNodest}, is not shown here. }
    \label{fig: Lya vs direct RP isolated stars}
\end{figure}

\begin{figure}
    \centering
    \includegraphics[width=1.0\columnwidth]{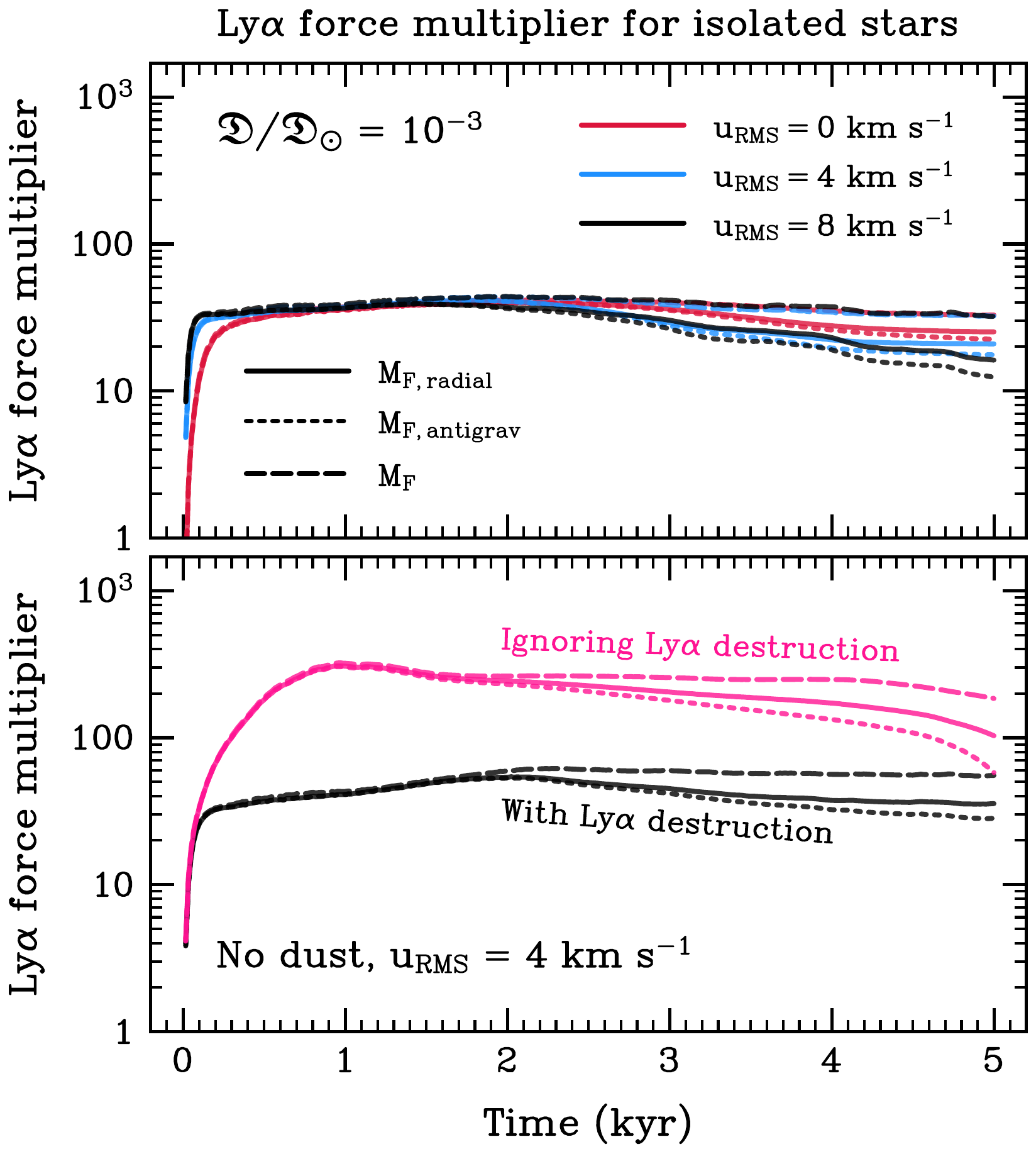}
    \caption{ The evolution of the force multipliers (as defined in Eqs.~\ref{Naive M_F def}--\ref{Antigrav M_F def}), for the isolated star simulations. \textbf{Top panel:} The results for the simulations with initial DtG ratio $\mathfrak{D}/\mathfrak{D}_\odot = 10^{-3}$ are shown, with varying levels of initial turbulence. \textbf{Bottom panel:} The results for the dust-free simulations (\texttt{StarLyaNoDust} and \texttt{StarLyaNoDest}) with fixed initial level of turbulence ($u_{\rm RMS} = 4 \, \rm km \, s^{-1}$), including Ly$\alpha$ destruction (black lines), and ignoring it (pink lines). Ly$\alpha$ destruction is seen to be very important at the high densities of these simulations, reducing the force multiplier from $M_{\rm F,radial} \sim 100-300$ to $M_{\rm F,radial} \sim 35 - 60$. However, despite efficient photon destruction, Ly$\alpha$ feedback remains strong. }
    \label{fig: MF isolated star runs}
\end{figure}

\section{Discussion}
\label{Discussion sec}

\subsection{How robust are our conclusions?}
\label{Discussion: How robust are our conclusions}

We have found that Ly$\alpha$ feedback is an important, often dominant, pre-supernova feedback in dust-poor conditions, most prevalent at Cosmic Dawn. This conclusion is consistent with analytical calculations in more idealized settings \citep[][]{Abe2018, Tomaselli2021, Nebrin2022, Nebrin2024, Smith2025, Stace2025, Manzoni2025}, as well as 1D Ly$\alpha$ MCRT results for dust-free settings \citep[][]{Dijkstra2008, Smith2017}, and 3D MCRT post-processing results \citep[][]{Smith2019, Menon2026}. It is still prudent to assess how robust our conclusions are to any assumptions and modelling choices. We discuss a few caveats here, and their possible impact on our results:
\begin{itemize}[leftmargin=*]
    \item \textbf{Dust modelling and assumptions:} Ly$\alpha$ feedback has long been suspected to be sensitive to dust absorption. Although we find strong and dominant Ly$\alpha$ feedback in the dust-poor regimes of interest here, we still find that dust physics must be treated with care in any realistic modelling of Ly$\alpha$ feedback. To improve realism, \textsc{Lydion} implements both self-consistent dust dynamics, and dust growth and destruction. These factors reduce the overall dust opacity to Ly$\alpha$ in the \textsc{H\,ii} region, which promotes build-up of Ly$\alpha$ pressure. While these processes are expected to be less important at extremely low initial dust-to-gas ratios approaching dust-free conditions, they are likely crucial for an accurate assessment of Ly$\alpha$ feedback in clouds with higher dust abundances relevant for the formation of metal-poor (yet enriched) star clusters and dwarf galaxies. We urge the community to keep this in mind in future studies of Ly$\alpha$ feedback. 

    Greater uncertainty in our modelling is connected to the assumed initial grain size distributions, and the neglect of other processes. On the latter point, \cite{Hoang2019} have argued that larger dust grains can be spun up by radiative torques to the point of breaking, fragmenting into smaller grains \citep[a process known as radiative torque disruption, RATD; for a review, see][]{Hoang2020}. Since smaller grains have higher tensile strengths and are less susceptible to disruption, this could boost the small-grain population, and thereby boost absorption of Ly$\alpha$ in \textsc{H\,ii} regions. However, more recent observations and modelling suggest that RATD is unlikely to be very important in the regimes of interest to us here. \cite{Salgado2016} find evidence of large dust grains ($\gtrsim 0.1 -1 \, \rm \mu m$), and significant Ly$\alpha$ heating of dust, in the Orion Nebula, both contrary to naive expectations from RATD. Similarly, \cite{Silsbee2025} show from the observed population of larger grains in the Solar system that RATD must be more inefficient than early predictions suggested. This has been reinforced in more detailed modelling of RATD by \cite{Hoang2025}, who find that RATD can be inefficient in compact \textsc{H\,ii} regions. In summary, realistic implementations of RATD are unlikely to change conclusions in this paper. 

    Finally, we have assumed an initial grain size distribution from \cite{Weingartner2001} biased towards larger grains ($R_{\rm V} = 5.5$, Fig.~\ref{fig: Dust distribution}). While this can reduce the Ly$\alpha$ dust absorption opacity by a factor of a few compared to their fiducial Milky Way $R_{\rm V} = 3.1$ grain size distribution, it is also degenerate with the assumed initial dust-to-gas ratio. As discussed in Sec.~\ref{Sec: RHD simulations}, for our star cluster simulations, we have assumed $\mathfrak{D}/\mathfrak{D}_\odot =Z/Z_\odot$, which is likely a significant overestimate for the gas metallicity $Z/Z_\odot = 0.01$. A more realistic star cluster simulation setup would likely have $\mathfrak{D}/\mathfrak{D}_\odot \sim 0.1\, Z/Z_\odot$ (Fig.~\ref{Dust-to-gas vs observations}), which would cancel the increased dust absorption from a hypothetical dust population governed by smaller grains. Large grains, however, are the general expectations for the very dense environments studied here ($n_{\rm H} \gtrsim 10^5 - 10^6 \, \rm cm^{-3}$). Dust coagulation, not simulated in the current version of \textsc{Lydion}, will tend to produce large grains in dense clouds \citep[e.g.][]{Ormel2009, Silsbee2020}, similar to our assumed initial dust size distribution. Depletion of small grains is also predicted at high redshifts, further biasing the dust size distribution towards larger grains \citep[][]{Narayanan2026}. Taken together with the low dust abundances in our simulations, we do not expect further improved dust modelling (e.g. dust coagulation with more dust bins) to change the basic conclusions of this paper.  

    \item \textbf{Lyman-$\boldsymbol{\alpha}$ emission:} In \textsc{Lydion} we have, as is standard practice, considered Ly$\alpha$ emission from recombinations, and collisional excitation from the ground state of \textsc{H\,i}. \cite{Raiter2010} have found that in metal-poor and dense gas, the Ly$\alpha$ luminosity can be boosted significantly \citep[for further discussions, see][]{Dijkstra2014}. This is due to collisional mixing of the $2p$ and $2s$ levels \citep[][]{Guzman2017, Nebrin2024}, and higher fractions of \textsc{H\,i} atoms in the $n = 2$ state in hotter metal-poor \textsc{H\,ii} regions. The former effect is already taken into account in \textsc{Lydion}, by using the density-dependent Ly$\alpha$ emission probabilities from \cite{Storey1995} (see Fig.~\ref{fig: Emission probability}). The latter effect could boost photoionizations by photons of lower energy, thereby increasing the rate of recombination-produced Ly$\alpha$ photons.
    
    Judging from fig.~10 in \cite{Raiter2010}, these effects could boost the Ly$\alpha$ luminosity by a factor $\sim 2$ for Pop II stars and star clusters, or more ($\sim 3-4$) for a top-heavy IMF dominated by hotter stars. Since we already take one of the effects into account (the density dependence), and focus on Pop II star clusters and not overly massive stars ($35 \, \rm M_\odot$), it is unlikely that we have \textit{underestimated} the Ly$\alpha$ radiative forces (from this effect alone) by more than a factor $\sim 2$.\footnote{All else being equal, this affects only the luminosity $L_{\rm Ly\alpha}$, not the force multiplier $M_{\rm F}$, and therefore changes the force $M_{\rm F} L_{\rm Ly\alpha} /c$.}

    \item \textbf{Use of the M1 approximation, and how it compares to Monte Carlo RT:} \textsc{Lydion} is based on a two-moment method for Ly$\alpha$, stellar, and IR radiative transfer, using the approximate M1 closure of \cite{Levermore1984}. Although the M1 model is widely used in star and galaxy formation simulations \citep[e.g.][]{Rosdahl2013, Rosdahl2015, Kannan2019, Hopkins2020_Radfeedback, Chan2021, Fukushima2021, Grudic2021, Wikbing2022, Deng2024, Kimura2025, Chan2026}, one may ask whether this induces significant errors in the calculations of Ly$\alpha$ radiation pressure forces when compared to more accurate (but slower) MCRT calculations. We argue that the answer is \textit{no} for the following reasons. First, we have directly verified \textsc{Lydion} predictions against MCRT results, showing excellent agreement. Second, radiative transfer with M1 is only expected to show modest inaccuracies in regions with $\tau \sim 1$. However, the optical depth at line center to Ly$\alpha$ photons is typically very large (e.g. $\tau_0 \sim 10^8 - 10^{11}$ in dense, cold regions). Even in \textsc{H\,ii} regions the optical depth is large ($\tau_0 \sim 10^4 - 10^5$), and the emission uniform enough, to render the Ly$\alpha$ intensity nearly isotropic and well-described by M1. As pointed out by \cite{Kimm2018}, this distinguishes Ly$\alpha$ feedback from IR radiation pressure feedback, which straddles more closely to the regime where M1 can induce modest errors ($\tau_{\rm IR} \sim 1$). Furthermore, in those regimes where M1 becomes somewhat inaccurate, it tends to \textit{underestimate} the radiation pressure forces compared to more accurate calculations with MCRT (see Fig.~\ref{fig: Cone test performance LYDION vs MCRT}), variable-Eddington tensor, and ray-tracing methods \citep[][]{Davis2014, Rosdahl2015, Tsang2015, Zhang2017, Smith2020_Arepo}. Thus, if any noticeable inaccuracies were to survive in our M1 treatment (despite the good agreement with MCRT in testing), it is most likely that we have modestly underestimated the strength of Ly$\alpha$ feedback, particularly when ionization fronts begin to break out and create optically thin sightlines. This would only serve to reinforce our basic conclusions regarding the importance of Ly$\alpha$ feedback.   

    \item \textbf{Neglect of stellar wind feedback:} Although \textsc{Lydion} can simulate stellar wind feedback, we have neglected it for performance reasons in this paper, since even a single cell with fast, shock-heated gas can shrink the allowed CFL time-steps by a factor $\sim (10^8 \, \textrm{K}/10^4 \, \textrm{K})^{1/2} = 100$.\footnote{Local time-stepping, with or without adaptive mesh refinement, could alleviate this problem \citep[][]{Gnedin2018_CFL}.} In local star-forming clouds, stellar wind feedback can play an important role in the overall pre-supernova feedback. Turbulent mixing and conduction is expected to produce momentum-driven winds, that can interact in complicated ways with radiative feedback in \textsc{H\,ii} regions \citep[e.g.][]{Lancaster2021, Geen2023, Lancaster2025_1, Lancaster2025_2, Rodriguez2026_winds}. The corresponding momentum input is comparable to direct radiation pressure ($\sim L_{\rm bol} /c$) for stars of Solar metallicity \citep[e.g.][]{Grudic2018}. At lower stellar metallicities, metal line-driven winds becomes ineffective, with momentum injection dropping as $\Dot{p}_{\rm wind} \propto Z_{\star}^{0.99}$ \citep[][]{Hopkins2023}. Thus, at the metallicities considered in this paper, we expect stellar winds to impart forces of order $\lesssim 0.01 \, L_{\rm bol}/c$. This is far weaker than Ly$\alpha$ radiation pressure in our simulations, $\sim (2 - 16 )\, L_{\rm bol}/c$ (Figs.~\ref{fig: Feedback comparison 512 sim} \& \ref{fig: Lya vs direct RP isolated stars}). The neglect of stellar wind feedback is therefore justified in the metal-poor regime studied in this paper. However, in future studies of Ly$\alpha$ feedback in more metal-enriched clouds, it would be ideal to include stellar wind feedback. This is not only for its direct feedback role, but also because of the associated dust destruction in hot gas \citep[e.g.][]{Kirsanova2023}.

    \item \textbf{Mass of the star cluster, and the role of photoionization feedback:} In our star cluster simulations, we have fixed the mass of the cluster to $10^4 \, \rm M_\odot$, corresponding to a low-mass cluster. The gas mass is also only $\sim 1900 \, \rm M_\odot$ (Fig.~\ref{fig: starforming gas}). These choices are mostly for numerical reasons. In particular, since we are dealing with 2D simulations, and not simulating star formation itself, we cannot model the 3D growth of isolated \textsc{H\,ii} regions that later merge into a single large \textsc{H\,ii} region, nor the increase in feedback from additional star formation. To capture the latter merged phase of the \textsc{H\,ii} region, we are therefore forced to choose initial conditions where the Strömgren sphere is $\gtrsim$ the size of the cluster, which in turn puts some constraints on the assumed gas density, cluster size and mass. Thus, while the density of gas in our setup is reasonable, the cluster mass is lower than that of typical massive globular cluster progenitors, $\sim 10^5 - 10^6 \, \rm M_\odot$ \citep[e.g.][]{Vanzella2023, Adamo2024, Vanzella2026}. This in turn means that we are likely underestimating the relative importance of Ly$\alpha$ pressure compared to photoionization (PI) feedback. It is expected that radiation pressure feedback becomes increasingly dominant compared to PI feedback around massive star clusters and/or on small scales \citep[][]{Krumholz2009_HII}. Observations show this to be the case for direct and IR radiation pressure in compact \textsc{H\,ii} regions \citep[e.g.][]{Barnes2021_RadPressure, Olivier2021}. 
    
    To see this for Ly$\alpha$ feedback, one can compare the radially directed Ly$\alpha$ radiation pressure force $F_{\rm Ly\alpha} = M_{\rm F,radial} L_{\rm Ly\alpha}/c$ to the gas pressure force $F_{\rm gas} \simeq n_{\rm II} k_{\rm B} T_{\rm II} \, 4 \pi R_{\rm II}^2$, where $R_{\rm II}$ is the radius of the \textsc{H\,ii} region, and $n_{\rm II}$ and $T_{\rm II}$ are the total number density and temperature of the ionized gas, respectively. For an order of magnitude estimate, consider pure hydrogen gas, so that $n_{\rm II} = 2 n_{\rm HII}$. In photoionization equilibrium, $(4\pi/3) k_{\rm rec,B} n_{\rm HII}^2 R_{\rm II}^3 = \Dot{Q}_{\rm LyC}$, where $k_{\rm rec,B}$ is the Case B recombination coefficient, and $\Dot{Q}_{\rm LyC} \simeq 5 \times 10^{46} \, (M_{\star}/1 \, \rm M_{\odot}) \, s^{-1}$ is the LyC photon emission rate of a star cluster of mass $M_{\star}$, assuming a Kroupa IMF \citep[see][and references therein]{Nebrin2024}. Eliminating $n_{\rm HII}$ in favor of $R_{\rm II}$, we then find for the ratio between forces:
    \begin{align}
        \dfrac{F_{\rm Ly\alpha}}{F_{\rm gas}} &\simeq~ \dfrac{M_{\rm F,radial}}{2c\sqrt{12\pi} k_{\rm B}T_{\rm II}} \dfrac{k_{\rm rec,B}^{1/2}L_{\rm Ly\alpha}}{(\Dot{Q}_{\rm LyC} R_{\rm II})^{1/2}} \\ &\simeq~ \dfrac{M_{\rm F,radial}}{3c\sqrt{12\pi R_{\rm II}} k_{\rm B}T_{\rm II}} \, k_{\rm rec,B}^{1/2}\Dot{Q}_{\rm LyC}^{1/2} E_{\rm Ly\alpha} \, , \nonumber
    \end{align}
    where, on the second line, we have used $L_{\rm Ly\alpha} \simeq (2/3) \Dot{Q}_{\rm LyC} E_{\rm Ly\alpha}$ for the recombination-driven Ly$\alpha$ luminosity. Numerically we find:\footnote{We have used $k_{\rm rec,B} \simeq 2.56 \times 10^{-13} \, T_{\rm II,4}^{-0.83} \, \rm cm^3 \, s^{-1}$ \citep[][]{Draine2011}, evaluated at $T_{\rm II} \sim 2 \times 10^4 \, \rm K$.}
    \begin{align}
        \dfrac{F_{\rm Ly\alpha}}{F_{\rm gas}} &\simeq~ 1.2 \, \left( \dfrac{M_{\rm F,radial}}{10} \right) \label{Ratio Lya to PI eq} \\ &\times~ \left( \dfrac{M_\star}{10^4 \, \rm M_{\odot}}\right)^{1/2} \left( \dfrac{R_{\rm II}}{0.2 \, \rm pc}\right)^{-1/2}  \, . \nonumber
    \end{align}
    Thus, in the early stages of our star cluster simulation, we expect Ly$\alpha$ pressure to be greater than or comparable to the gas pressure, which is indeed seen in our simulations (Fig.~\ref{Lya vs no Lya figure 512 D0.01 T and PLya}). At later times, when the \textsc{H\,ii} region has grown larger, we expect PI feedback to eventually dominate (also seen in Fig.~\ref{Lya vs no Lya figure 512 D0.01 T and PLya}). However, if we consider a more typical massive and dense star cluster at Cosmic Dawn \citep[e.g.][]{Kimm2016, Ricotti2016, Vanzella2023, Adamo2024, Vanzella2026}, with $M_\star = 10^6 \, \rm M_\odot$ and $R_{\rm II} = 1 \, \rm pc$, we expect that $F_{\rm Ly\alpha}/F_{\rm gas} \simeq 5.4 \, (M_{\rm F,radial}/10)$, and Ly$\alpha$ feedback would clearly dominate over PI, as well as direct and IR radiation pressure, and stellar wind feedback (as discussed earlier). Thus, for the force multipliers observed in our RHD simulations ($M_{\rm F,radial} \sim 10 - 40$ for the simulations with $\mathfrak{D}/\mathfrak{D}_\odot = 10^{-3} - 0.01$), Ly$\alpha$ feedback is predicted to be the dominant pre-supernova feedback process during the formation of metal-poor star clusters at Cosmic Dawn. 

    \item \textbf{1D vs.\ 2D vs.\ 3D:} Our simulations are two-dimensional, which raises the question of how strongly the results depend on dimensionality. At early times, the radiation field, ionization structure, and gas dynamics remain close to spherically symmetric, implying that 1D simulations should be sufficient for broad parameter exploration in this regime. The 2D simulations are nevertheless useful because they capture non-radial motions, gas instabilities, and the timing of asymmetric breakout. We find that the diffusive and nearly isotropic nature of Ly$\alpha$ radiative transfer tends to evenly distribute perturbations across the \HII region boundary, preventing instabilities from growing freely. Once breakout occurs, however, the morphology can become substantially asymmetric, even though the global force multiplier is only reduced below $\sim 10$ when the cloud is already nearly disrupted. 
    
    Fully 3D calculations may introduce additional effects absent in axisymmetry, including porous escape channels, azimuthal fragmentation, pockets of accelerated gas, and fallback or recycling of dense disrupted material. However, recent 3D MCRT post-processing results by \cite{Menon2026} show that strong Ly$\alpha$ radiation pressure survives in more complicated 3D star-forming environments, with force multipliers broadly consistent with those found in our simulations \citep[also see][]{Smith2019}. Further 2D/3D Ly$\alpha$ RHD studies are therefore needed to quantify how multidimensional phenomena affect the late-time morphology and coupling efficiency of Ly$\alpha$ feedback.
\end{itemize}
Overall, we conclude that our result that Ly$\alpha$ feedback is dynamically important in dust-poor gas is likely to be robust. Where there are caveats (e.g. Ly$\alpha$ emission in metal-poor gas, RT methodology, and dust modelling), it is more likely that we have underestimated Ly$\alpha$ feedback, rendering our conclusions relatively conservative.

\subsection{Broader implications and future work}

\begin{table*}
\centering
\caption{A (non-exhaustive) list of pre-supernova feedback implementations in recent (last $\leq 10 \, \rm yrs$) high-resolution galaxy and star formation simulations, that have at least some focus on high redshifts and/or dust-poor conditions. The table lists whether a given simulation implements radiation pressure (RP) from Ly$\alpha$, stellar (direct), and infrared photons. Whether stellar wind feedback is implemented or not is also listed; here, we refer specifically to momentum and/or energy injection by massive-star winds. We further note that, for simulations of Pop III stars, stellar winds and infrared radiation pressure are both expected to be negligible, so that their neglect in Pop III simulations should not be taken as an omission of important physics.  }
\begin{tabular}{c c c c c c c}
\hline
\hline
\noalign{\vskip 2pt}
Simulation & Ly$\alpha$ RP? & Direct RP & Infrared RP? & Photoionization? & Stellar winds? & Pop II/III \\

\noalign{\vskip 3pt}
\hline
\hline
\noalign{\vskip 2pt}

\textsc{Fire-2/3}$^\dagger$  & No  & Yes & Yes & Yes & Yes & Pop II/III  \\

\textsc{Thesan-Zoom}$^\dagger$  & No  & Yes & Yes & Yes & Yes & Pop II/III   \\

\textsc{Megatron}$^\dagger$  & Subgrid  & Yes & Yes & Yes & No & Pop II/III   \\

\textsc{Lyra}$^\dagger$  & No  & No & No & No & No & Pop II/III   \\

\textsc{Griffin}$^\dagger$  & No  & No & No & Subgrid & Yes & Pop II   \\

\textsc{Edge}$^\dagger$  & No  & Yes & Yes & Yes & Yes & Pop II   \\

\textsc{Siege}$^\dagger$ & No & No & No & No & Yes & Pop II \\

\citet{Kimm2016}  & No  & Yes & No & Yes & No & Pop II   \\

\citet{Kimm2017}  & No  & Yes & Yes & Yes & No & Pop II/III  \\

\citet{Kimm2018}  & Subgrid  & Yes & Yes & Yes & No & Pop II  \\

\citet{Ricotti2016}  & No  & No & No & Yes & No & Pop II/III  \\

\citet{Jeon2021}  & No  & No & No & No & No & Pop II/III  \\

\citet{Garcia2023}  & No  & No & No & Yes & No & Pop II/III  \\

\\

\textsc{Starforge}$^\dagger$  & No  & Yes & Yes & Yes & Yes & Pop II/III  \\

\textsc{Popsicle}$^\dagger$  & No  & Yes & No & Yes & No & Pop III  \\

\citet{Han2022}  & Subgrid  & Yes & Yes & Yes & Yes & Pop II  \\

\citet{Menon2024} & No & Yes & Yes & Yes & No & Pop II \\

\citet{Jaura2022}  & No  & Yes & No & Yes & No & Pop III  \\

\citet{Latif2022}  & No  & Yes & No & Yes & No & Pop III  \\

\citet{Sugimura2023}  & No  & No & No & Yes & No & Pop III  \\

\citet{Fukushima2021}  & No  & Yes & No & Yes & No & Pop II  \\

\citet{Fukushima2020}  & No  & Yes & No & Yes & No & Pop II/III  \\

\citet{Park2023}  & No  & Yes & No & Yes & No & Pop III  \\

\citet{Chon2025}  & No  & No & No & Yes & No & Pop II/III  \\

\noalign{\vskip 2pt}
\hline
\hline
\end{tabular}
\vspace{1 pt}\\
\raggedright
{\footnotesize
$^\dagger$: A selection of relevant references for these simulations are: \textsc{Fire-2}/\textsc{Fire-3} \citep[][]{Hopkins2018_FIRE2, Wheeler2019, Hopkins2020_Radfeedback, Hopkins2023, Meziani2026}, \textsc{Thesan-zoom} \citep[][]{Kannan2025_ThesanZoomIntro, Zier2025_PopIII, Shen2026}, \textsc{Megatron} \citep[][]{Katz2026_Megatron, Storck2026}, \textsc{Lyra} \citep[][]{Gutcke2021, Gutcke2022_LyraII, Gutcke2022_LyraIII, Brown2025}, \textsc{Griffin} \citep[][]{Lahen2020_Griffin, Lahen2025_Mergers}, \textsc{Edge} \citep[][]{Agertz2020_EDGE, Andersson2025_EDGE_INFERNO, Rey2025_EDGE}, \textsc{Siege} \citep[][]{SiegeI, SiegeIII, SiegeIV}, \textsc{Starforge} \citep[][]{Grudic2021, Meziani2026}, \textsc{Popsicle} \citep[][]{Sharda2025, Sharda2025_Popsicle2}.  }
\label{Galaxy/star formation simulations literataure}
\end{table*}

Our results entail an urgent need for more comprehensive Ly$\alpha$ RHD simulations and methods to accurately interpret observations of early galaxies and star clusters from \textit{JWST}, and to understand star formation during Cosmic Dawn. If the forces found here persist across a broader range of cloud properties, then Ly$\alpha$ pressure may alter the disruption of natal clouds, the stellar initial mass function, the escape of ionizing radiation, the clustering of early star formation, and the conditions required for efficient formation of bound stellar systems \citep[e.g.][]{McKeeTan2008, Abe2018, Kimm2018, Nebrin2022, Nebrin2024}.


Several lines of research are especially promising. First, broader parameter surveys are needed to determine how Ly$\alpha$ feedback depends on cloud mass, surface density, metallicity, dust abundance, ionizing luminosity, and stellar population properties. Such studies can establish when Ly$\alpha$ feedback is dominant, or when it may be suppressed by dust absorption, photon destruction, or rapid leakage through low-column-density channels. Second, Ly$\alpha$ feedback should be explored around Pop III stars, accreting black holes, and compact metal-poor star clusters, where hard spectra, high gas densities, and low dust opacities may all modify the Ly$\alpha$ luminosity and coupling efficiency. Third, the results of detailed Ly$\alpha$ RHD calculations should be distilled into subgrid models for galaxy formation simulations \citep[][]{Kimm2018, Nebrin2024}, where the process is currently almost always omitted. In Table~\ref{Galaxy/star formation simulations literataure} we provide a summary of pre-supernova feedback implementations in  contemporary high-resolution simulations of galaxy and star formation in dust/metal-poor environments. Among these state-of-the-art simulations, the vast majority neglect Ly$\alpha$ feedback, either citing high computational demands and/or uncertainties \citep{Hopkins2020_Radfeedback, Grudic2021, Jaura2022, Sharda2025, Meziani2026}, or not mentioning the process at all, despite its potential dominance (most studies).\footnote{Early discussions of feedback-free starbursts did not take Ly$\alpha$ feedback into account either \citep[][]{Dekel2023}, despite the focus on the environments where it is expected to matter the most. Ly$\alpha$ feedback is expected to introduce complications for the concept of feedback-free starbursts at high redshifts \citep[][]{Nebrin2022, Nebrin2024, Ferrara2025_FeedbackFreeLya, Manzoni2025}. }

A subset of simulations in \textsc{ramses} includes a subgrid prescription of Ly$\alpha$ feedback developed by \cite{Kimm2018}. In these simulations, Ly$\alpha$ feedback is predicted to be important, reducing both the efficiency of star cluster formation and the overall star formation efficiency \citep[][]{Kimm2018, Katz2026_Megatron}. This represented an important first step toward exploring Ly$\alpha$ feedback in galaxy simulations. However, as discussed in \cite{Nebrin2024}, the adopted normalization differs from the (MCRT-verified) analytical solution by a factor of $\sim 3$, and the model does not account for important destruction mechanisms (e.g. $2p \rightarrow 2s$), or photon leakage along low-column-density channels. A more general and accurate analytical solution for the force multiplier $M_{\rm F}$ was derived in \cite{Nebrin2024}, verified with MCRT. This improved prescription could enable more accurate subgrid treatments of Ly$\alpha$ feedback in simulations, taking into account velocity gradients, dust absorption, turbulence, and Ly$\alpha$ destruction. However, a detailed calibration against realistic simulations with \textsc{Lydion} and MCRT, and implementation in 3D simulations, is beyond the scope of this paper.

The methodology introduced here also provides a path toward more practical general-purpose Ly$\alpha$ RHD. In \textsc{Lydion}, future improvements could include adaptive mesh refinement, GPU acceleration, more detailed dust modelling such as coagulation and grain-size evolution, and protostellar evolution and feedback to follow the earliest stages of compact \textsc{H\,ii} region formation. More broadly, the combination of moment-based spatial transport with Fokker-Planck frequency redistribution may inform efficient Ly$\alpha$ RT algorithms for on-the-fly use in 3D simulations \citep[e.g.][]{Smith2018DDMC, Byrohl2025}. The long-term goal is to move Ly$\alpha$ feedback from an idealized, specialized calculation into the standard toolkit of star and galaxy formation simulations.


    


\section{Summary and conclusion}
\label{Summary sec}

Despite nearly a century of vigorous debate on its potential essential importance, Lyman-$\alpha$ radiation pressure has not been studied self-consistently in any multi-dimensional RHD simulation to date. In this paper we have presented the first 2D Ly$\alpha$ RHD simulations, utilizing a new code, \textsc{Lydion}, which incorporates a novel, and relatively computationally cheap, M1 + Fokker-Planck treatment of Ly$\alpha$ radiative transfer \citep[][]{Levermore1984, Rybicki2006}. We also implement detailed treatments of dust dynamics, dust growth and destruction, photochemistry and cooling, and stellar feedback from photoionization, direct and IR radiation pressure.

In our first application of \textsc{Lydion}, we study feedback from metal-poor star clusters and isolated stars, embedded in dust-poor clouds, relevant for the earliest galaxies, stars, and star clusters at Cosmic Dawn. Our results can be summarized as follows:
\begin{enumerate}
    \item \textbf{Lyman-$\boldsymbol{\alpha}$ pressure strongly boosts pre-supernova feedback in dust-poor environments:} For metal-poor clouds ($Z/Z_\odot = 0.01$), with realistic initial dust-to-gas ratios ($\mathfrak{D}/\mathfrak{D}_\odot = 10^{-3} - 0.01$), we find that Ly$\alpha$ pressure dominates the overall radiation pressure feedback, exerting a force $\sim (2-10) L_{\rm bol} /c$, compared to $\sim (0.6 -0.9) L_{\rm bol} /c$ and $< 0.01 L_{\rm bol} /c$ for direct (stellar) and infrared radiation pressure, respectively (Figs.~\ref{fig: Feedback comparison 512 sim} \& \ref{fig: Lya vs direct RP isolated stars}). In our simulations of low-mass star clusters ($10^4 \, \rm M_\odot$), the Ly$\alpha$ pressure also exceeds the gas pressure in the \textsc{H\,ii} region at early times, while being comparable at later times near the ionization front (Fig.~\ref{Lya vs no Lya figure 512 D0.01 T and PLya}). For more massive star clusters ($10^5 - 10^6 \, \rm M_\odot$), we expect Ly$\alpha$ feedback to become more dominant relative to photoionization feedback (Eq.~\ref{Ratio Lya to PI eq}), in which case it would be the strongest pre-supernova feedback process at play in these environments. Our simulation of an isolated Pop III star reinforces the same conclusion, predicting radial Ly$\alpha$ forces $\sim (10 - 16) L_{\rm bol}/c$ and force multipliers $M_{\rm F,radial} \sim 35 - 60$. We find that the Ly$\alpha$ force is concentrated near the ionization front, where the gas becomes strongly super-Eddington ($f_{\rm Edd} \gtrsim 10-100$).  

    \item \textbf{Lyman-$\boldsymbol{\alpha}$ pressure is likely to be a key regulator of the star formation efficiency at Cosmic Dawn:} By strongly boosting the overall pre-supernova feedback, it is likely that Ly$\alpha$ feedback is one of the key processes that regulate the star formation efficiency in dust-poor, star-forming clouds. In our set of star cluster simulations, we find that including Ly$\alpha$ feedback reverses gas collapse and the accumulation of dense gas, likely reducing the star formation efficiency (Fig.~\ref{fig: starforming gas}). We stress that this does not preclude efficient star formation in general -- the added momentum input merely raises the required gas surface density where star formation becomes efficient \citep[e.g.][]{Abe2018, Grudic2018, Nebrin2022, Nebrin2024, Manzoni2025}. Indeed, earlier analytical models of star cluster formation at Cosmic Dawn that incorporated approximate prescriptions for Ly$\alpha$ feedback predicted the formation of dense, bound star clusters in very dense clouds \citep[][]{Abe2018,Nebrin2022, Nebrin2024}, albeit less frequently than runs ignoring Ly$\alpha$ feedback. Detailed studies of the impact of Ly$\alpha$ feedback on the star formation effiency will have to await simulations incorporating star formation itself, coupled to Ly$\alpha$ RHD.

    \item \textbf{Lyman-$\boldsymbol{\alpha}$ feedback is an essential missing ingredient in current star and galaxy formation simulations:} Despite its importance, and the growing focus on high-redshift galaxy formation in the age of the \textit{JWST}, almost all simulations neglect Ly$\alpha$ feedback (Table~\ref{Galaxy/star formation simulations literataure}), either citing the high computational demands of Ly$\alpha$ RHD, or not mentioning the process at all. The results from \textsc{Lydion} in this paper, together with earlier analytical work \citep[][]{Tomaselli2021, Nebrin2024}, 1D simulations \citep[][]{Smith2017}, and 3D post-processing results \citep[][]{Smith2019, Menon2026}, show that Ly$\alpha$ feedback cannot be ignored in any realistic simulation of star formation in dust-poor environments.

    \item \textbf{Realistic dust physics cannot be ignored:} The sensitivity of Ly$\alpha$ feedback to the abundance and absorption properties of dust renders it intertwined with dust physics \citep[e.g.][]{Henney1998, Menon2026}. Thus, for realistic simulations and benchmarks of Ly$\alpha$ feedback beyond Pop III applications, it is essential to incorporate dust growth and destruction, as well as dust dynamics, as done in \textsc{Lydion}. We find that their effect is to lower dust opacities in the \textsc{H\,ii} region (Fig.~\ref{Dust properties figure 512 Lya simulation}), thereby further facilitating build-up of Ly$\alpha$ pressure.   

    \item \textbf{Realistic atomic physics cannot be ignored, which constrains Ly$\boldsymbol{\alpha}$ RHD methodology:} We find that Ly$\alpha$ destruction, mainly facilitated by $2p \rightarrow 2s$ transitions, cannot be ignored in dense, dust-free clouds. In our simulations of isolated massive Pop III stars in metal/dust-free clouds of initial density $n_{\rm H} = 10^6 \, \rm cm^{-3}$, we find that the radially directed Ly$\alpha$ force multiplier $M_{\rm F,radial} \equiv F_{\rm Ly\alpha,radial} /(L_{\rm Ly\alpha}/c)$ drops from $\sim 100-300$ to $\sim 35-60$, when taking Ly$\alpha$ destruction into account (Fig.~\ref{fig: MF isolated star runs}). This result is driven by processes like $\textrm{H}(2p) + \textrm{H}(1s) \rightarrow \textrm{H}(2s) + \textrm{H}(1s)$, identified in \cite{Nebrin2024}, but neglected in earlier works.

    This result puts constraints on Ly$\alpha$ RHD methodology, since it requires capturing the numerous scatterings in the line core, usually neglected with core-skipping algorithms in traditional MCRT methods for performance reasons \citep[for discussions on the impact of core-skipping on Ly$\alpha$ feedback and Ly$\alpha$ destruction, see][]{Nebrin2024, Lorinc2025}. The M1 + Fokker--Planck methodology of \textsc{Lydion} naturally captures these core scatterings, as do proposed rDDMC methods for Ly$\alpha$ RT \citep[][Kimura et al., in prep.]{Smith2018DDMC}. 
    
    We stress that, despite the importance of $2p \rightarrow 2s$ transitions, Ly$\alpha$ pressure remains $\sim (10 - 16) \times \,$ stronger than direct radiation pressure in our dust-free simulation, while also dominating over the gas pressure in the \textsc{H\,ii} region (Fig.~\ref{Isolated stars, T, PLya/P plot}). We anticipate that Ly$\alpha$ feedback is stronger (weaker) in less (more) dense environments, where $2p \rightarrow 2s$ transitions are less (more) important, but leave a thorough exploration to future work.
\end{enumerate}
Taken together, these results show that Ly$\alpha$ radiation pressure is not a small correction to early stellar feedback. In dense, dust-poor environments, it can be the dominant pre-supernova force, substantially altering the dynamics of the gas surrounding young stars and star clusters. The strength of this coupling depends on dust absorption, Ly$\alpha$ leakage, velocity gradients, photon destruction, and the geometry of the surrounding gas, but none of these effects eliminate the build-up of strong Ly$\alpha$ pressure in the regimes studied here \citep[largely confirming analytic predictions by][]{Nebrin2024}.

The broader implication is that nearly all current star and galaxy formation simulations omit a feedback channel that may be central to the regulation of metal-poor star formation at Cosmic Dawn (Table~\ref{Galaxy/star formation simulations literataure}). This omission is especially relevant in the \textit{JWST} era, where observations are beginning to probe precisely the compact, low-metallicity systems in which Ly$\alpha$ feedback should be most effective. Incorporating Ly$\alpha$ feedback into future simulations---whether through direct Ly$\alpha$ RHD, accelerated transport methods, or calibrated subgrid models---is therefore essential for a complete theory of early star formation, ionizing photon escape, and the formation of the first galaxies and stellar systems.


\section{Acknowledgements}
We thank Maria S. Murga, Kazutaka Kimura, William Henney, Todd A. Thompson, Thiem Hoang, Duncan Bossion, Shyam Menon, Mike Grudi{\'c}, and Evan O'Connor for helpful and interesting comments and discussions during the course of the project. We thank Ilian Iliev and Rahul Kannan for kindly providing simulation data for the D-type test comparison. O.N. and G.M. acknowledge support from Swedish Research Council grant 2020-04691. O.N., A.S., and K.L. acknowledge support from the NASA theory grant JWST-AR-08709.

\textit{Software}: \textsc{Lydion} is written in the \textsc{Julia} programming language \citep{Julia2017}, and we have made use of \textsc{matplotlib} \citep{Hunter2007_matplotlib} for the plots in this paper. 

\section{Data availability}

The \textsc{Lydion} code will be made publicly available upon acceptance for publication. The simulation data output from the RHD simulations in this paper will be shared on reasonable request to the corresponding author.

\section{Author contributions}

O.N. led the project conceptualization, development of the \textsc{Lydion} code and methodology, analysis, interpretation, and manuscript preparation. 
A.S. led the development of the MCRT comparison tests, ran the production RHD simulations, and contributed to the analysis, interpretation, and manuscript preparation. 
G.M. contributed to the hydrodynamical and stellar radiative transfer modelling, interpretation of the results, and manuscript preparation. K.L. contributed to the MCRT comparison testing, analysis, and interpretation. D.M. contributed to testing of the \textsc{Lydion} code and initiated exploratory work on GPU and C$^{++}$ implementations of \textsc{Lydion}. 


%




\newpage

\appendix
\onecolumngrid
\section{Implicit method for frequency diffusion, atomic recoil, and Doppler shifts}
\label{frequency diffusion appendix}
Here we outline the solution method for handling frequency diffusion, atomic recoil, and Doppler shifts, i.e.:
\begin{equation}
\dfrac{1}{\Tilde{c}} \dfrac{\partial J}{\partial t}  = \dfrac{\partial}{\partial \nu} \left\{ \dfrac{1}{2} \Delta \nu_{\rm D}^2 \alpha \mathcal{H} \left( \dfrac{\partial J}{\partial \nu} + \mathfrak{R}J \right) \right\} + \dfrac{1}{c} \dfrac{\partial }{\partial \nu}  [ \nu \, (\boldsymbol{\mathsf{D}} \boldsymbol{:} \boldsymbol{\nabla}\boldsymbol{u})J] \equiv \dfrac{\partial F_\nu}{\partial \nu}  \, . \label{Frequency step PDE appendix}
\end{equation}
We discretize this in frequency using an exponential scheme \citep[e.g.][]{Chang1970, Patankar2018}. The exponential scheme ensures that there is upwinding if the frequency advection terms (Doppler shift and recoil) dominate. In the opposite case where frequency diffusion dominates, the exponential scheme gives a central difference discretization. The exponential scheme can be derived by assuming that the frequency flux $F_\nu$ remains constant between bin centers\footnote{We place the bin centers $\Bar{\nu}_k$ at the geometric centers of the bins, i.e. $\Bar{\nu}_k = (\nu_{k+1/2} + \nu_{k-1/2})/2$.} $\Bar{\nu}_k$ and $\Bar{\nu}_{k+1}$, equal to $(F_\nu)_{k+1/2}$. Within bin $k$, we therefore want to solve the differential equation:
\begin{equation}
    K_k \dfrac{\partial J^{(k)}}{\partial \nu} + C_k J^{(k)} = (F_\nu)_{k+1/2} = \textrm{constant} \, , \label{Exponential scheme ODE bin k}
\end{equation}
where $J^{(k)}$ is $J(\nu)$ within bin $k$. We adopt a piecewise-constant approximation for the frequency dependence of the diffusion ($K$) and advection ($C$) coefficients \citep[consistent with the discrete diffusion Monte Carlo method for Ly$\alpha$ by][]{Smith2018DDMC}:
\begin{equation}
    K_k = \left\langle \dfrac{1}{2} \Delta \nu_{\rm D}^2 \alpha \mathcal{H} \right\rangle_k \, , \quad  C_k =
    \left\langle \dfrac{1}{2} \Delta \nu_{\rm D}^2 \alpha \mathcal{H} \mathfrak{R} +  \dfrac{\nu}{c} (\boldsymbol{\mathsf{D}} \boldsymbol{:} \boldsymbol{\nabla}\boldsymbol{u}) \right\rangle_k \,, \label{Bin averaged coefficients definition}
\end{equation}
where the brackets denote bin-averaging. We solve Eq.~(\ref{Exponential scheme ODE bin k}) between $\Bar{\nu}_k$ and the interface $\nu_{k+1/2}$, under the assumption that $J$ change by $J_{k+1/2} - J_k$ over this interval:
\begin{equation}
    J^{(k)}(\nu) = \dfrac{(F_\nu)_{k+1/2}}{C_k} + \dfrac{(J_{k+1/2} - J_k) \, e^{-C_k \nu / K_k}}{e^{-C_k\nu_{k+1/2}/K_k} - e^{-C_k \Bar{\nu}_k / K_k}} \, .
\end{equation}
After evaluating this at $\nu = \nu_{k+1/2}$, with $J^{(k)}(\nu_{k+1/2}) = J_{k+1/2}$, a little algebra gives us an expression for $J_{k+1/2}$:
\begin{equation}
    J_{k+1/2} = \dfrac{(F_\nu)_{k+1/2}}{C_k} \, \phi^L(\mathcal{P}_k) + [1 - \phi^L(\mathcal{P}_k)] J_k \,, \quad \mathcal{P}_k \equiv \dfrac{C_k (\nu_{k+1/2} - \Bar{\nu}_k)}{K_k} \, , \label{J at k+1/2 from left}
\end{equation}
where $\phi^L(x) = 1 - e^{-x}$. We note that $\mathcal{P}_k$ is (half) the Péclet number for the bin, measuring the relative importance of advection and diffusion terms. Next, we want to determine the flux $(F_\nu)_{k+1/2}$ by matching $J_{k+1/2}$ from the left (Eq.~\ref{J at k+1/2 from left}), and right. For the right side of the interface, $\nu \in [\nu_{k+1/2}, \Bar{\nu}_{k+1}]$, we can solve Eq.~(\ref{Exponential scheme ODE bin k}) with $k \rightarrow k+1$ in an analogous manner, under the constraint that $J$ change by an amount $J_{k+1} - J_{k+1/2}$ over this interval, and that the flux there too is $(F_\nu)_{k+1/2}$. We then find:
\begin{equation}
    J^{(k+1)}(\nu) = \dfrac{(F_\nu)_{k+1/2}}{C_{k+1}} + \dfrac{(J_{k+1} - J_{k+1/2}) \, e^{-C_{k+1} \nu / K_{k+1}}}{e^{-C_{k+1}\Bar{\nu}_{k+1}/K_{k+1}} - e^{-C_{k+1} \nu_{k+1/2} / K_{k+1}}} \, .
\end{equation}
We evaluate this at $\nu_{k+1/2}$ and solve for $J_{k+1/2}$:
\begin{equation}
    J_{k+1/2} = \dfrac{(F_\nu)_{k+1/2}}{C_{k+1}}\, \phi^R(\mathcal{P}_{k+1}) + [1 - \phi^R(\mathcal{P}_{k+1})] J_{k+1} \, , \quad \mathcal{P}_{k+1} \equiv \dfrac{C_{k+1} (\Bar{\nu}_{k+1} - \nu_{k+1/2})}{K_{k+1}} \,, \label{J at k+1/2 from right}
\end{equation}
with $\phi^R(x) = 1 - e^x$. Now we match the reconstructions in Eqs.~(\ref{J at k+1/2 from left}) and (\ref{J at k+1/2 from right}) and solve for the interface flux:
\begin{equation}
    (F_\nu)_{k+1/2} = \dfrac{[1 - \phi^R(\mathcal{P}_{k+1})]J_{k+1} - [1 - \phi^L(\mathcal{P}_{k})]J_{k}}{\phi^L(\mathcal{P}_k)/C_k - \phi^R(\mathcal{P}_{k+1})/C_{k+1}} \, .
\end{equation}
We can also get the flux at $k-1/2$ by simply making the replacement $k \rightarrow k-1$:
\begin{equation}
    (F_\nu)_{k-1/2} = \dfrac{[1 - \phi^R(\mathcal{P}_{k})]J_{k} - [1 - \phi^L(\mathcal{P}_{k-1})]J_{k-1}}{\phi^L(\mathcal{P}_{k-1})/C_{k-1} - \phi^R(\mathcal{P}_{k})/C_{k}} \, .
\end{equation}
We can now discretize Eq.~(\ref{Frequency step PDE appendix}) by bin-averaging both sides (i.e. perform the integral $(1/\Delta \nu_k)\int_{\Delta \nu_k} \textrm{d}\nu$ of both sides, where $\Delta \nu_k = \nu_{k+1/2} - \nu_{k-1/2}$ is the bin width). As mentioned in the main text, we also employ implicit (BDF1) time integration for numerical stability. Taken together, we get the following equation for each cell $(i,j)$:
\begin{equation}
\dfrac{1}{\Tilde{c}} \dfrac{ J_{i,j,k}^{n+1} - J_{i,j,k}^{\rm em/abs}}{\Delta t_{\rm RT}}  = \mathcal{F}^+_{i,j,k} J_{i,j,k+1}^{n+1} -\mathcal{F}_{i,j,k} J_{i,j,k}^{n+1} + \mathcal{F}^-_{i,j,k} J_{i,j,k-1}^{n+1} \, , \label{Discretized freq eq}
\end{equation}
where $J^{\rm em/abs}_{i,j,k}$ is the Ly$\alpha$ intensity just after the emission/absorption update, and the leakage coefficients are (suppressing the spatial indices on the right for clarity):
\begin{align}
    \mathcal{F}_{i,j,k}^{+} &=~ \dfrac{1}{\Delta \nu_k}\dfrac{1 - \phi^R(\mathcal{P}_{k+1})}{[\chi^L(\mathcal{P}_{k})\delta \nu_k/K_{k} - \chi^R(\mathcal{P}_{k+1}) \delta \nu_{k+1}/K_{k+1}]} \, , \label{Leakage coeff +}
    \\ \mathcal{F}_{i,j,k} &=~ \dfrac{1}{\Delta \nu_k} \left[ \dfrac{1-\phi^L(\mathcal{P}_k)}{\chi^L(\mathcal{P}_{k}) \delta \nu_k/K_{k} - \chi^R(\mathcal{P}_{k+1})\delta \nu_{k+1}/K_{k+1}} + \dfrac{1-\phi^R(\mathcal{P}_k)}{\chi^L(\mathcal{P}_{k-1})\delta \nu_{k-1}/K_{k-1} - \chi^R(\mathcal{P}_{k})\delta \nu_k/K_{k}} \right] \,, \\ \mathcal{F}_{i,j,k}^{-} &=~ \dfrac{1}{\Delta \nu_k}\dfrac{1 - \phi^L(\mathcal{P}_{k-1})}{[\chi^L(\mathcal{P}_{k-1})\delta\nu_{k-1}/K_{k-1} - \chi^R(\mathcal{P}_{k})\delta \nu_k/K_{k}]} \, . \label{Leakage coeff -}
\end{align}
We have introduced the convenient functions $\chi^{L/R}(\mathcal{P}) \equiv \phi^{L/R}(\mathcal{P})/\mathcal{P}$, and the intervals $\delta \nu_{k+1} = \Bar{\nu}_{k+1} - \nu_{k+1/2}$, $\delta \nu_{k} = \nu_{k+1/2} - \Bar{\nu}_{k}$,  and $\delta \nu_{k-1} = \nu_{k-1/2} - \Bar{\nu}_{k-1}$. The functions $\chi^{L/R}$ and $\phi^{L/R}$ involve exponentials and divisions, which become expensive to evaluate in simulations, since this has to be done for every cell, bin, and time-step. We therefore create look-up tables for these functions of $\mathcal{P}$, which we have found to greatly speed up the calculation of the leakage coefficients in Eqs.~(\ref{Leakage coeff +})--(\ref{Leakage coeff -}). It is straightforward to show that in the absence of velocity gradients and recoil (i.e. $\mathcal{P} \rightarrow 0$), the leakage coefficients above gives a central difference discretization, consistent with \cite{Smith2018DDMC}. In the opposite limit of large velocity gradients or strong recoil (i.e. $\lvert\mathcal{P}\rvert \gg 1$), we recover an upwind scheme. We solve the tridiagonal equations (Eq.~\ref{Discretized freq eq}) using \texttt{LAPACK.gtsv!}, available through the standard \texttt{LinearAlgebra.jl} package of \textsc{Julia}. 

Finally, we elaborate on the calculation of the coefficients $C_k$ and $K_k$, as given in Eq.~(\ref{Bin averaged coefficients definition}). We make the approximations (suppressing spatial indices):
\begin{equation}
    K_k = \dfrac{1}{2} \Delta \nu_{\rm D}^2 \alpha \left\langle \mathcal{H} \right\rangle_k \, , \quad  C_k =
     \dfrac{1}{2} \Delta \nu_{\rm D}^2 \alpha \left\langle \mathcal{H} \right\rangle_k \mathfrak{R}_k +  \dfrac{\Bar{\nu}_k}{c} (\boldsymbol{\mathsf{D}}_k \boldsymbol{:} \boldsymbol{\nabla}\boldsymbol{u})  \,, \label{Bin averaged coefficients definition}
\end{equation}
where $\mathfrak{R}_k = h/k_{\rm B} T - 2 / \Bar{\nu}_k$ is the recoil factor for bin $k$, and we use the bin-averaged Voigt profile:
\begin{equation}
    \langle \mathcal{H} \rangle_k  \equiv \dfrac{1}{\Delta \nu_k} \int_{\Delta \nu_k} \textrm{d}\nu \, \mathcal{H}(\nu) = \dfrac{1}{\Delta x_{k}} \left[\mathcal{I}(x_{k+1/2} , a_{\rm v}) - \mathcal{I}(x_{k-1/2} , a_{\rm v}) \right] \, , \label{Voigt profile}
\end{equation}
where $\Delta x_k = (\nu_{k+1/2} - \nu_{k-1/2})/\Delta \nu_{\rm D}$,  $x_{k\pm1/2} = (\nu_{k\pm1/2} - \nu_{\rm Ly\alpha})/\Delta \nu_{\rm D}$, $a_{\rm v}$ is the Ly$\alpha$ Voigt parameter, and
\begin{equation}
    \mathcal{I}(x,a_{\rm v}) \equiv \dfrac{\sqrt{\pi}}{2}\,\text{erf}(x) - \dfrac{2a_{\rm v}}{\sqrt{\pi}}\int_0^{x} \text{d}y \, e^{y^2-x^2} +a_{\rm v}^2 \, x \, e^{-x^2} \, . \label{expensive integral} 
\end{equation}
Equation~(\ref{Voigt profile}) is obtained from a second-order expansion of $\mathcal{H}$ in $a_{\rm v}$ \citep[][]{Smith2015, Smith2018DDMC}. Because the Dawson integral and other functions of $x$ in Eq.~(\ref{expensive integral}) are expensive to evaluate for every cell, bin, and time-step, we create look-up tables for them, with grid spacing $\delta x = 0.01$ and linear interpolations between grid points. At large arguments, $\lvert x \rvert > 10^3$, we resort to asymptotic expansion of the Dawson integral, and take relevant limits of the other functions (e.g. $\exp(-x^2) \rightarrow 0$). We evaluate $\boldsymbol{\mathsf{D}}_k \boldsymbol{:} \boldsymbol{\nabla}\boldsymbol{u}$ in cylindrical coordinates with axisymmetry ($\partial_\phi =0$, and further no rotation) as follows:
\begin{equation}
    \boldsymbol{\mathsf{D}}_k  \boldsymbol{:} \boldsymbol{\nabla}\boldsymbol{u} = (\boldsymbol{\mathsf{D}}_k)_{RR} \, \dfrac{\partial u_R}{\partial R} + (\boldsymbol{\mathsf{D}}_k)_{ZZ} \, \dfrac{\partial u_Z}{\partial Z}  + (\boldsymbol{\mathsf{D}}_k)_{\phi \phi} \,\dfrac{u_R}{R} +  (\boldsymbol{\mathsf{D}}_k)_{RZ} \left[ \dfrac{\partial u_R}{\partial Z}  +  \dfrac{\partial u_Z}{\partial R} \right] \, . \label{D : grad u}
\end{equation}
In the RHD simulations, the velocity gradient components in a given cell are estimated using the left (L) and right (R) reconstructed velocities at the cell interfaces:
\begin{equation}
    \left( \dfrac{\partial u_{R/Z}}{\partial R} \right)_{i,j} \simeq \dfrac{(u_{R/Z})_{i+1/2,j}^{\rm L} - (u_{R/Z})_{i-1/2,j}^{\rm R}}{R_{i+1/2} - R_{i-1/2}} \, , \quad \quad  \left( \dfrac{\partial u_{R/Z}}{\partial Z} \right)_{i,j} \simeq \dfrac{(u_{R/Z})_{i,j+1/2}^{\rm L} - (u_{R/Z})_{i,j-1/2}^{\rm R}}{Z_{j+1/2} - Z_{j-1/2}} \, .
\end{equation}
And for $u_R /R$ in Eq.~(\ref{D : grad u}), we use its cell-average $\langle u_R /R\rangle$, estimated to second-order accuracy \citep[for similar geometric source term discretizations, see][]{Mignone2014}:
\begin{equation}
    \Big\langle\dfrac{u_R}{R} \Big\rangle_{i,j} \simeq \dfrac{(u_R)_{i+1/2,j}^{\rm L} + (u_R)_{i-1/2,j}^{\rm R}}{R_{i+1/2} + R_{i-1/2}} \, .
\end{equation}

\section{Hydrodynamics: Methodology and tests}
\label{Hydro Appendix}

In this Appendix, we give a brief overview of the hydrodynamics code employed in \textsc{Lydion}, as well as showing a series of validation tests.

\subsection{Methodology}

We adopt WENO3 to achieve higher-order spatial accuracy for the hydrodynamics in smooth flows. We closely follow \cite{Mignone2014} in implementing WENO3 in both cylindrical and Cartesian coordinates, and so in the interest of space we leave out mathematical details here and refer to \cite{Mignone2014}. A directionally unsplit method is employed, with third-order RK3 time-integration \citep{Mignone2007}. The geometric source term for the momentum equation in cylindrical coordinates, $P/R$, is discretized using a third-order accurate method described in \cite{Mignone2014}.

To better ensure stability (e.g. avoid unphysical densities and pressures), we fall back to lower-order methods near strong shocks, or otherwise detected problematic regions, similar to \cite{Verma2019}. More specifically, we define the dimensionless shock indicator:
\begin{equation}
    \Delta P_{i,j} \equiv \dfrac{\lvert P_{i+1,j} - P_{i-1,j} \rvert + \lvert P_{i,j+1} - P_{i,j-1} \rvert}{\min[P_{i,j}, P_{i-1,j}, P_{i+1,j}, P_{i,j-1}, P_{i,j+1}]} \, . \label{Shock indicator}
\end{equation}
If $1 \leq \Delta P_{i,j} < 4$, we flag the cell $(i,j)$ as having a weak shock, and we revert to using the HLL flux function instead of HLLC, but still employ WENO3. The use of HLL here is to minimize the risk of the carbuncle instability, which can occur for shocks along grid lines when HLLC is employed. If $ \Delta P_{i,j} \geq 4$, we flag the cell $(i,j)$ as having a strong shock, and we revert to using second-order piecewise linear reconstruction (PLM) instead of WENO3 near this shock, with a MinMod slope limiter as well as the HLL instead of the HLLC flux. We have found that WENO3 can still give unstable results near strong density gradients. Because of this, we also revert to PLM + HLL near strong gradients in the gas density and/or dust bin densities, defining $\Delta \rho$ and $\Delta \rho_{\rm d,\beta}$ analogous to Eq.~(\ref{Shock indicator}), with fallback in case $\max(\Delta \rho, \Delta \rho_{\rm d,\beta} ) > 100$.

To approach third-order spatial accuracy, one has to distinguish between cell-averaged and cell-center values of conservative and primitive variables. We convert between the two following \cite{Mignone2014}. Near sharp gradients, this can sometimes lead to unstable or unphysical values. In those cases, we revert to the common second-order approximation that cell-center and cell-averaged variables are equal. More specifically, we use an approach inspired by \cite{Balsara2012}. To illustrate this, let us consider the conversion from cell-averaged to cell-central values of conservative variables $\rho$, $\rho u_R$, $\rho u_Z$, and $E$, in reconstruction along the $R$-axis. If we have a candidate cell-center density $(\rho_{\rm c})_{i,j}$, we correct it as follows:
\begin{equation}
    (\rho_{\rm c})_{i,j} = \tau (\rho_{\rm c})_{i,j} + (1-\tau)\rho_{i,j} \, ,
\end{equation}
where $0 \leq \tau \leq 1$ is a correction factor, and $\rho_{i,j}$ the cell-averaged density. For smooth flows, $\tau$ is typically unity, and there is no deviation from the third-order prediction. If there are strong gradients however, the correction factor should approach $0$. We determine $\tau$ such that $(\rho_{\rm c})_{i,j}$ has values in the range:
\begin{equation}
    0.6 \, \min(\rho_{i,j}, \rho_{i+1,j}, \rho_{i-1,j})< (\rho_{\rm c})_{i,j} < 1.4 \, \max(\rho_{i,j}, \rho_{i+1,j}, \rho_{i-1,j}) \, .
\end{equation}
This allows for smooth extrema, as long as they are within a reasonable range. Corresponding bounds are taken for the other conservative variables (taking into acocunt the sign for momenta), and then we choose $\tau$ to be the \textit{minimum} value computed, and derive the corresponding cell-central values of all conservative variables using this $\tau$. Furthermore, we also revert to $\tau = 1$ if the corresponding pressure or density is still unphysical, or if the cell has been flagged as containing a shock. Finally, in the interface reconstruction, we only accept a candidate WENO3 reconstruction if the interface values are physical, and do not differ too much from cell-average values \citep{Titarev2004}. If a candidate WENO3 interface reconstruction is rejected, we revert to PLM, and if that still fails, we revert to a simple first-order piecewise-constant reconstruction. Besides the fallback methods, we note that our methodology is at most second-order spatially accurate since we do not distinguish between interface-center fluxes and interface-averaged fluxes.

\subsection{Tests of the hydrodynamics code}
Here we present a selection of standard tests of the hydrodynamics module in \textsc{Lydion}. 

\subsubsection{Sedov-Taylor blast wave}

\begin{figure*}
    \centering
    \begin{minipage}{0.49\textwidth}
        \centering
        \includegraphics[width=\textwidth]{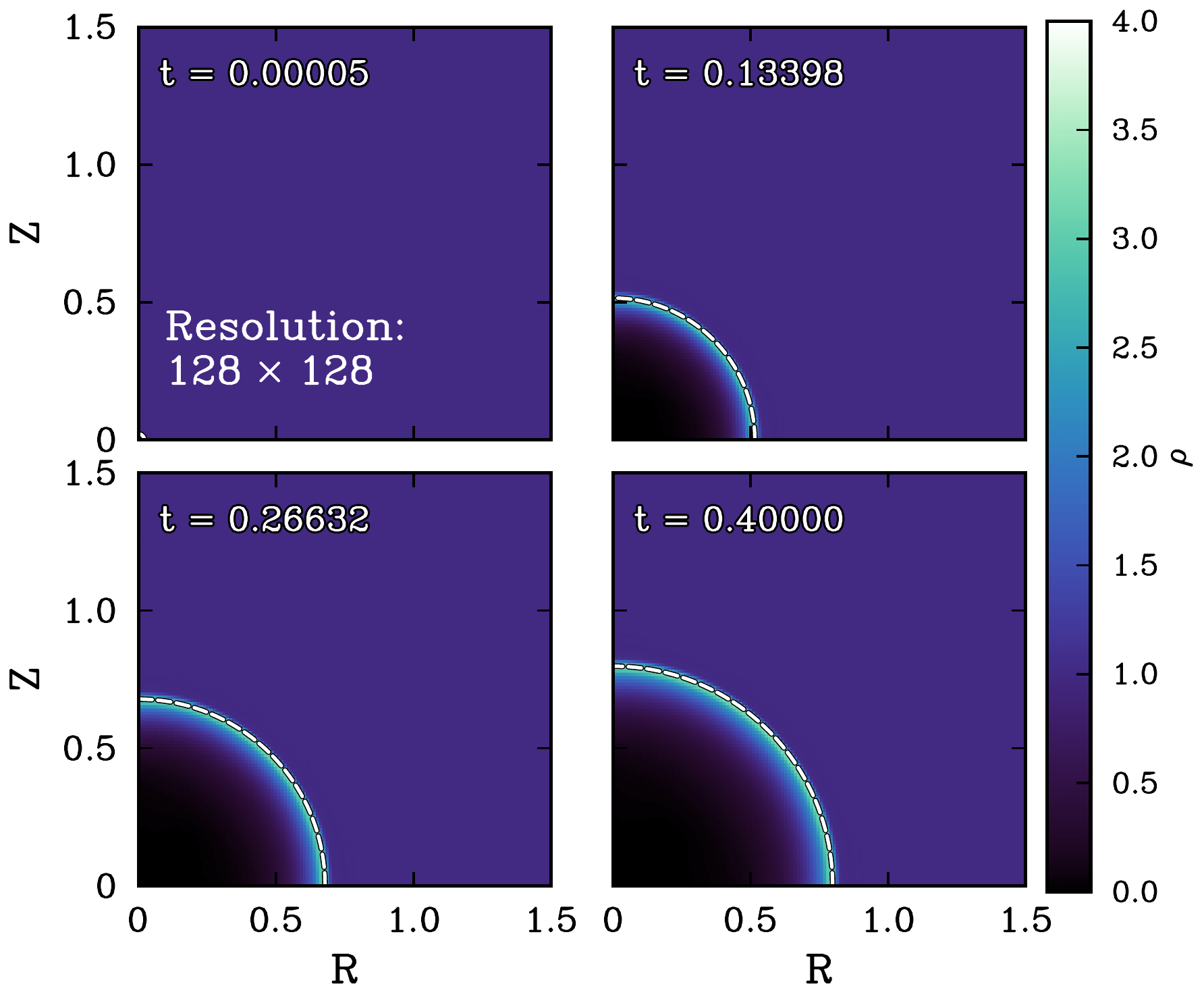}
        \vspace{0.3em}
        {\small (a) $128^2$ resolution}
    \end{minipage}
    \begin{minipage}{0.49\textwidth}
        \centering
        \includegraphics[width=\textwidth]{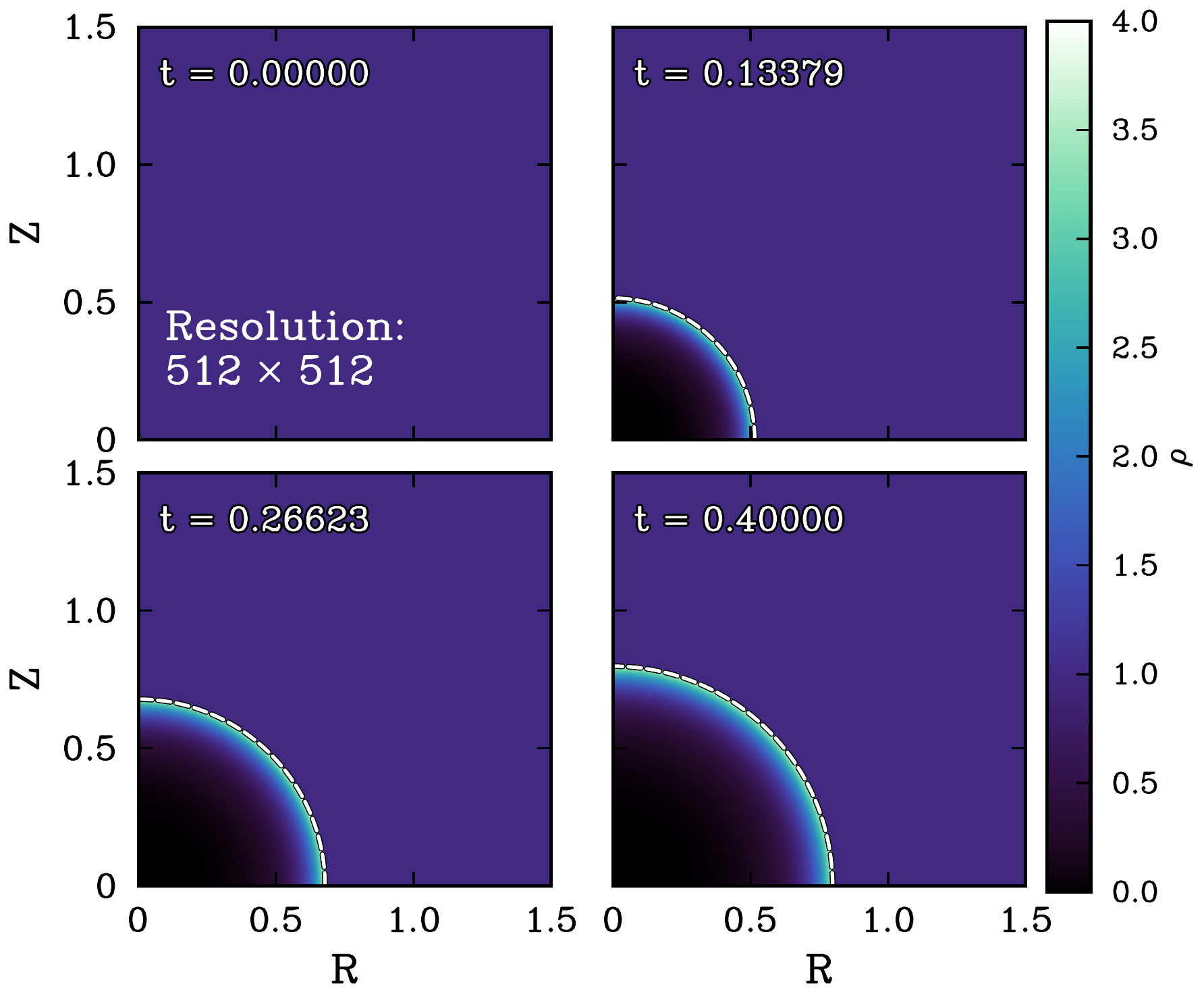}
        \vspace{0.3em}
        {\small (b) $512^2$ resolution}
    \end{minipage}
    \caption{Snapshots from the Sedov–Taylor blast wave test, for two resolutions: $128^2$ (a) and $512^2$ (b). In each panel, the dashed white line shows the expected spherically symmetric blast wave radius from the Sedov-Taylor solution, $R_{\rm ST} = 1.15167\, (E_{\rm blast} t^2/\rho_0)^{1/5}$ \citep[e.g.][]{Ostiker1988}.  }
    \label{SedovTest}
\end{figure*}

\begin{figure*}
\centering
\includegraphics[width=0.8\textwidth]{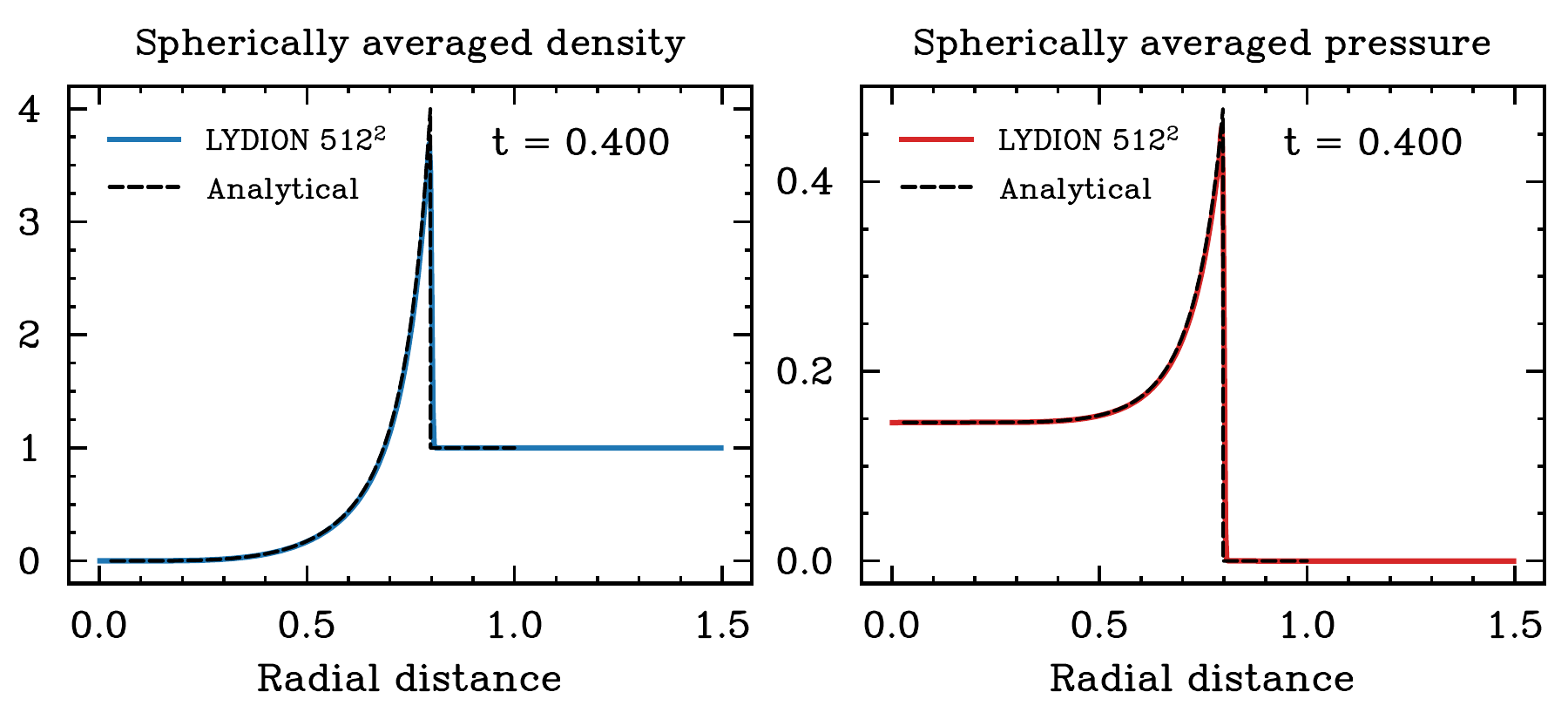}
\caption{ The spherically averaged density (left) and pressure (right) profiles at $t = 0.4$, for the Sedov-Taylor test at $512^2$ resolution (solid lines). Dashed lines show the analytical solution for comparison. }
\label{fig: Sedov Taylor density pressure profiles}
\end{figure*}

To test the ability of \textsc{Lydion} to handle strong shocks and to maintain reasonable symmetry, we set up a spherical Sedov-Taylor blast wave test in cylindrical coordinates $(R,Z)$. A uniform, initially static, medium of ambient density $\rho_0 = 1$ and ambient pressure $P_0 = 10^{-10}$ is set up. The size of the simulation volume is $0 \leq R \leq 1.5$ and $0 \leq Z \leq 1.5$, with symmetry across $Z = 0$ (i.e. only upper half of the blast is simulated). The adiabatic index is fixed to $\gamma = 5/3$, and we impose reflective boundary conditions on all boundaries. Within a central, approximately spherical, region of radius $\sim 5$ cells, we initiate a blast of energy $E_{\rm blast} = 1$ by raising the pressure in this region accordingly. The simulation is evolved up to $t = 0.4$. Snapshots from two simulations with resolution $128^2$ and $512^2$ are shown in Fig.~\ref{SedovTest}. For both runs, we find a spherically symmetric blast wave, with a shock radius in good agreement with the analytical solution. In Fig.~\ref{fig: Sedov Taylor density pressure profiles} we plot the spherically averaged density and pressure profiles at $t = 0.4$ from the $512^2$ simulation, and compare to the analytical predictions from the Sedov-Taylor solution. We find excellent agreement with the analytical solution, showing that \textsc{Lydion} can correctly handle strong shocks. 

\subsubsection{Sod shock tube}

\begin{figure*}
\centering
\includegraphics[width=0.8\textwidth]{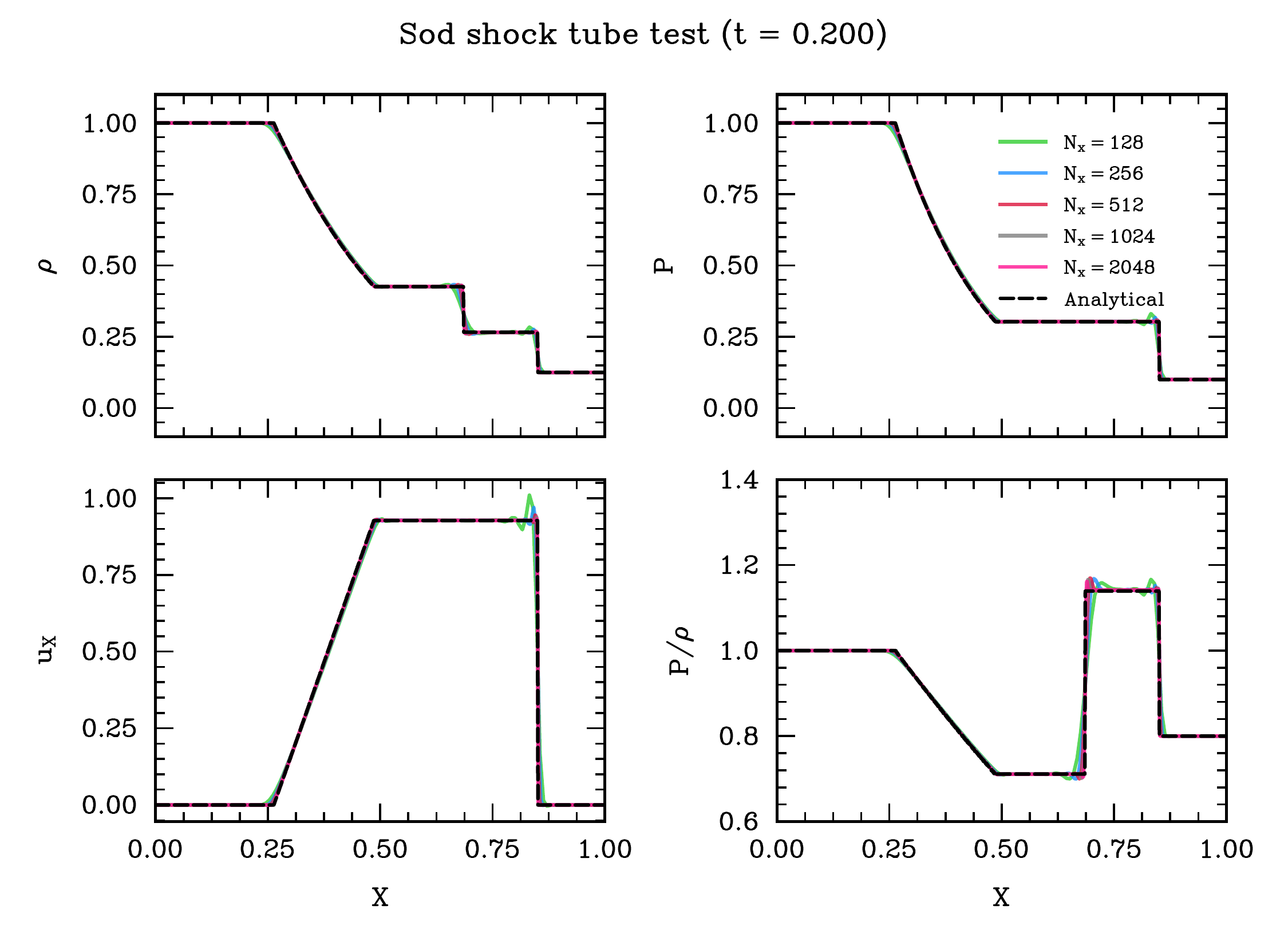}
\caption{ Results for the Sod shock tube test at $t = 0.2$ for different resolutions. The numerical results for $\rho$ (upper left panel), $P$ (upper right panel), $u_X$ (lower left panel), and $P/\rho$ (proxy for temperature, lower right panel), along with the exact analytical solution (dashed lines) are shown. }
\label{Sod test fig}
\end{figure*}

Next we set up a sod shock tube test, to check the ability of the code to capture both shocks and contact discontinuities. A uniform 2D Cartesian grid $(X,Y)$ is set up, with resolution ranging from $128 \times 32$ to $2048 \times 32$, and box size $0 \leq X \leq 1$ and $0 \leq Y \leq 1$. Initial conditions are $\rho(X\leq 0.5) = P(X \leq 0.5) = 1$ and $\rho(X >0.5) = 0.125$, $P(X > 0.5) = 0.1$, with $u_X = u_Y = 0$ everywhere, and an adiabatic index $\gamma = 7/5$. The results at $t = 0.2$ along the $X$-axis at $Y \simeq 0.5$ are plotted in Fig.~\ref{Sod test fig}, along with the exact analytical solution. Overall we find good agreement with the analytical solution, with better convergence at higher resolution.

\subsubsection{Rayleigh-Taylor instability}

\begin{figure*}
\centering
\includegraphics[
    width=0.8\textwidth,
    trim={1.2cm 1.4cm 0.0cm 0cm},
    clip
]{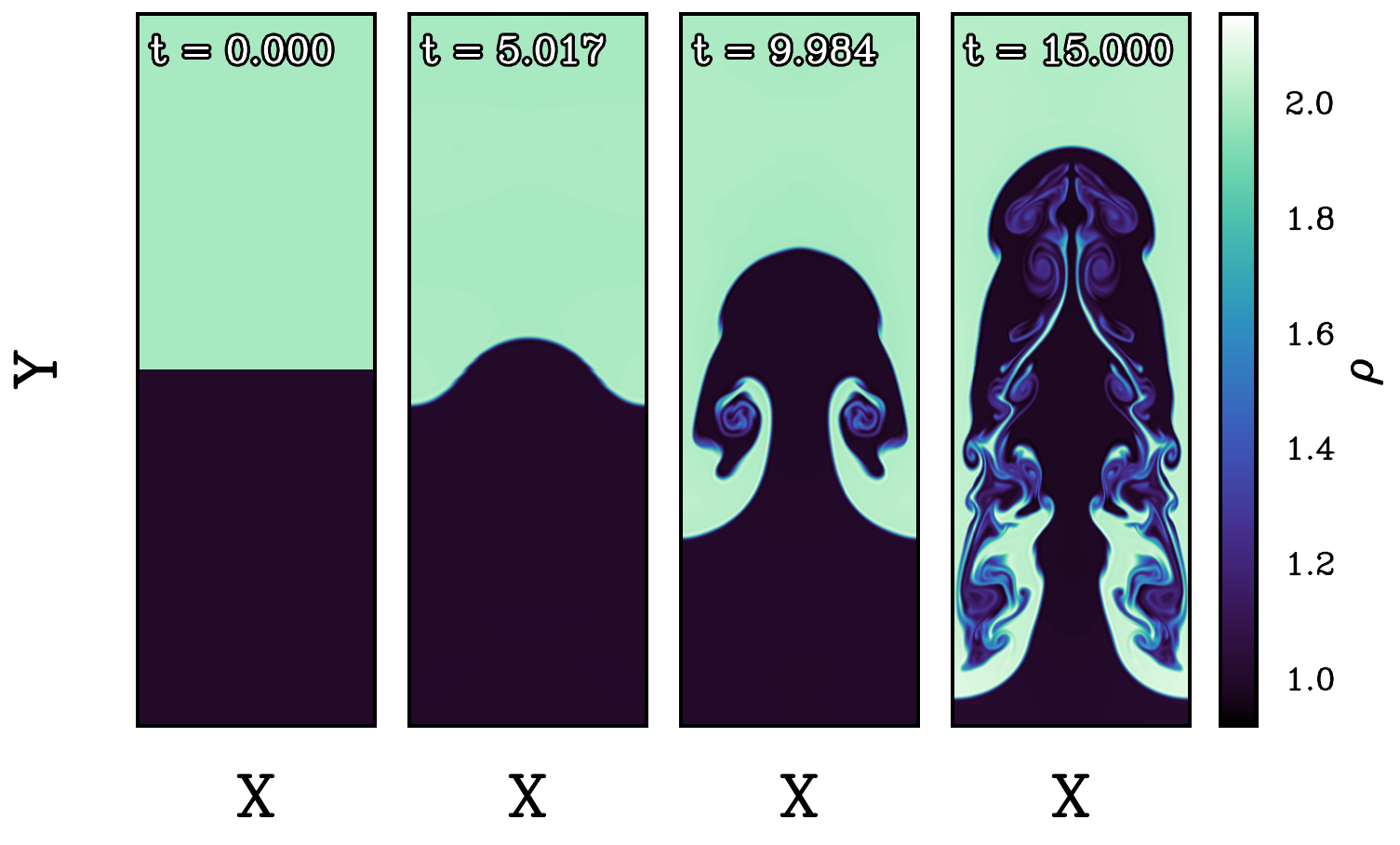}
\caption{Four snapshots of the density $\rho$ from the Rayleigh-Taylor instability test, for a resolution $256 \times 768$.}
\label{RTI test fig}
\end{figure*}

Next, we perform a Rayleigh-Taylor instability (RTI) test in a 2D Cartesian setup, of box length $(X_{\rm max},Y_{\rm max}) = (0.5, 1.5)$.\footnote{The setup for the test follows \url{https://www.astro.princeton.edu/~jstone/Athena/tests/rt/rt.html}. } A uniform, time-independent gravitational field $-\boldsymbol{\nabla}\Phi = - 0.1\, \boldsymbol{\hat{Y}}$ is set up, with periodic and reflective boundary conditions in the $X$ and $Y$ directions, respectively. The adiabatic index is set to $\gamma = 7/5$. The initial gas density is $\rho = 1$ in the bottom half ($Y < 0.5 \, Y_{\rm max}$), and $\rho = 2$ in the top half ($Y > 0.5 \, Y_{\rm max}$). The initial gas pressure is set to $P = 2.5 - 0.1 ( Y - 0.5 \,Y_{\rm max})\rho$, for initial (approximate) hydrostatic equilibrium. The equilibrium is perturbed by an initial velocity in the $Y$-direction:
\begin{equation}
    u_Y(X,Y) = \dfrac{1}{4} \times0.01 \, \big\{ 1 + \cos[4\pi(X - 0.5\, X_{\rm max})]\big\} \big\{ 1 + \cos[3\pi (Y-0.5 \, Y_{\rm max})] \big\} \, .
\end{equation}
The resolution is $256 \times (3 \times 256)$, and the simulation evolved up to $t = 15$. Four snapshots from the test are shown in Fig.~\ref{RTI test fig}. We find that onset of the RTI is correctly captured, as are the qualitative features.\footnote{Consistent with the findings of \cite{Fleischmann2019}, we have found that the test result is subject to symmetry breaking if the initial conditions are not perfectly symmetric, or if other floating point errors are not properly handled. We have therefore carefully set up the test to maintain symmetry by mirroring initial conditions (rather than evaluating cos), and also followed \cite{Fleischmann2019} in the use of bracketing in the HLLC flux to minimize these errors. }

\subsubsection{Kelvin-Helmholtz instability}
\begin{figure*}
\centering
\includegraphics[width=0.7\textwidth,
    trim={1.2cm 1.4cm 0.0cm 0cm},
    clip]{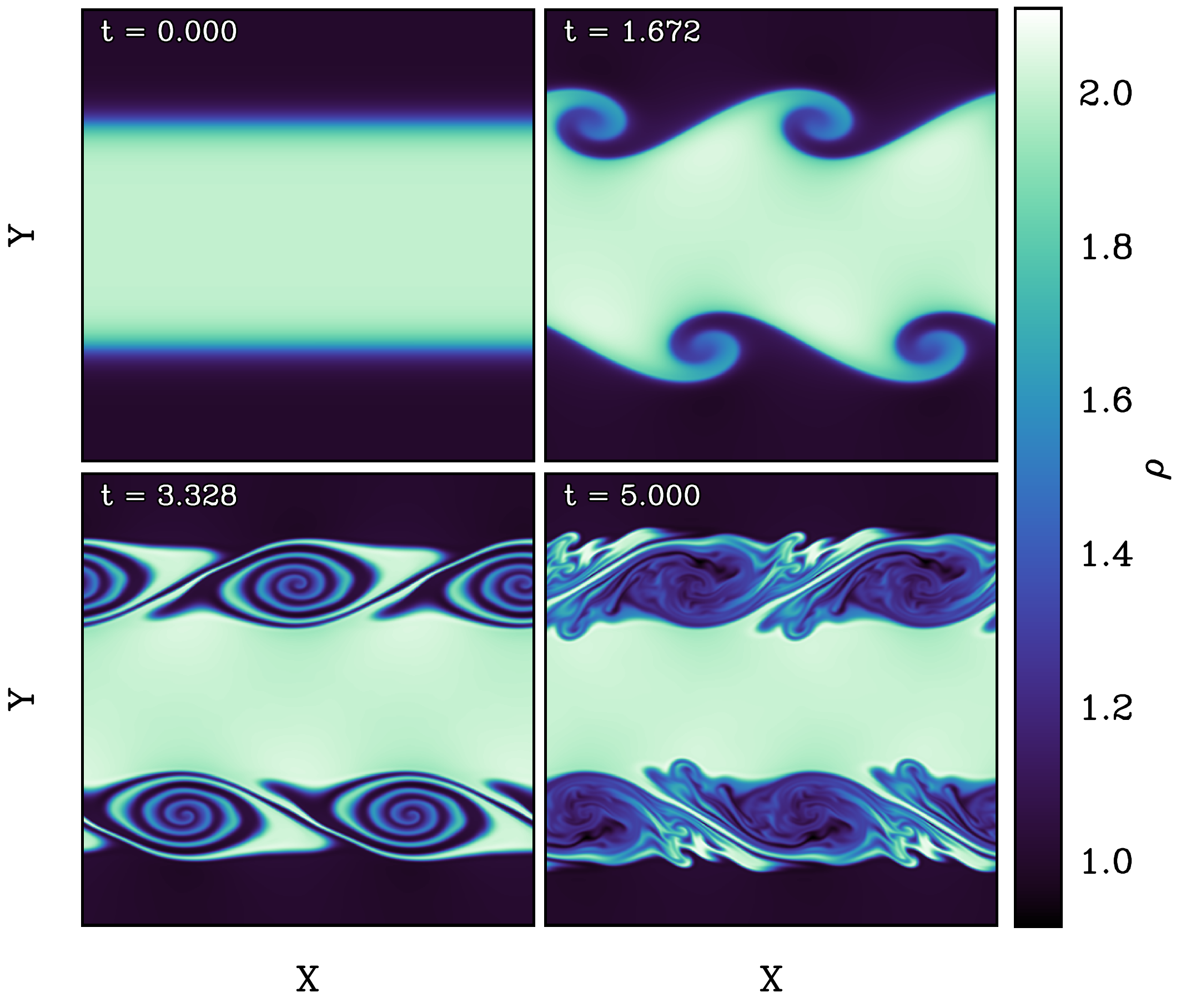}
\caption{Four snapshots of the gas density from the Kelvin-Helmholtz instability test, at a resolution of $512^2$.}
\label{KHI test fig}
\end{figure*}
Finally, a Kelvin-Helmholtz test is set up in 2D Cartesian coordinates, with resolution $512^2$. The initial conditions follow those described in \cite{McNally2012}, and the simulation was evolved up to $t = 5.0$. The adaibatic index is set to $\gamma = 5/3$. Periodic boundary conditions are adopted at the left and right $X$-boundaries, and reflective boundary conditions on the top and bottom $Y$-boundaries. Four snapshots from the test are shown in Fig.~\ref{KHI test fig}, showcasing the expected development of the Kelvin-Helmholtz instability.

\section{Self-gravity}
\label{Poisson equation Appendix}

\subsection{Solving the Poisson equation and computing the acceleration}

We aim to solve the Poisson equation $\nabla^2 \Phi = 4 \pi G \rho_{\rm tot}$ efficiently in 2D. We have found an SOR iterative method \citep{Teyssier2002} to be too slow, and for simplicity we have avoided implementing a faster but more complicated multigrid method. Instead, we adopt a Peaceman-Rachford ADI method \citep[][]{PeacemanRachford1955}, wherein one treats one direction implicitly at a time. For a detailed description of the method applied to the Poisson equation, we refer the reader to \cite{Black1975}, \cite{Norman1986}, and \cite{Stone1992}. In this method, we aim to solve the following related parabolic equation, for each cell $(i,j)$:
\begin{equation}
     \dfrac{\partial \Phi_{i,j}}{\partial \tau } = \langle \nabla^2 \Phi \rangle_{i,j}  - 4 \pi G (\rho_{\rm tot})_{i,j} \, ,\label{Pseudo time Poisson eq}
\end{equation}
up until we have converged to a pseudo-time steady-state where $\partial_{\tau} \Phi \simeq 0$. We discretize Eq.~(\ref{Pseudo time Poisson eq}) using a finite-volume method. At outer (non-reflective) boundaries, we fix the potential using a multipole expansion up to $\ell = 10$ \citep[][]{Black1975, Stone1992}. For a pseudo-time step $n \rightarrow n+1$ of size $\Delta \tau$, we first treat the $R$-direction implicitly (with BDF1 for the implicit pseudo-time integration):
\begin{align}
    \dfrac{ \Phi_{i,j}^{\bullet} - \Phi^n_{i,j}}{\Delta \tau } &=~  \dfrac{1}{\Delta V_{i,j}}\left[ A_{i+1/2,j}\left( \dfrac{\Phi_{i+1,j}^{\bullet} - \Phi_{i,j}^{\bullet}}{\Bar{R}_{i+1} - \Bar{R}_i} \right) - A_{i-1/2,j}\left( \dfrac{\Phi_{i,j}^{\bullet} - \Phi_{i-1,j}^{\bullet}}{\Bar{R}_{i} - \Bar{R}_{i-1}} \right)\right] \label{R-step Poisson} \\
    &+~ \dfrac{1}{\Delta V_{i,j}}\left[ A_{i,j+1/2}\left( \dfrac{\Phi_{i,j+1}^{n} - \Phi_{i,j}^{n}}{\Bar{Z}_{j+1} - \Bar{Z}_j} \right) - A_{i,j-1/2}\left( \dfrac{\Phi_{i,j}^{n} - \Phi_{i,j-1}^{n}}{\Bar{Z}_{j} - \Bar{Z}_{j-1}} \right) \right] \nonumber \\ \nonumber &-~  4 \pi G (\rho_{\rm tot})_{i,j} \, , 
\end{align}
where $(\Bar{R}_i,\Bar{Z}_j)$ cell volume center, $\Delta V_{i,j} = \pi (R_{i+1/2}^2 - R_{i-1/2}^2) (Z_{j+1/2} - Z_{j-1/2})$ is the cell volume, and $A_{i+1/2,j}$ ($A_{i,j+1/2}$) the area of the interface at $R_{i+1/2}$ ($Z_{j+1/2}$). We solve Eq.~(\ref{R-step Poisson}) for $\Phi^\bullet$ using the same tridiagonal solver used in the frequency update for Ly$\alpha$ (Appendix~\ref{frequency diffusion appendix}). Next, to complete the pseudo-time step one treats the $Z$-direction implicitly in an analogous manner:
\begin{align}
    \dfrac{ \Phi_{i,j}^{n+1} - \Phi^\bullet_{i,j}}{\Delta \tau } &=~  \dfrac{1}{\Delta V_{i,j}}\left[ A_{i+1/2,j}\left( \dfrac{\Phi_{i+1,j}^{\bullet} - \Phi_{i,j}^{\bullet}}{\Bar{R}_{i+1} - \Bar{R}_i} \right) - A_{i-1/2,j}\left( \dfrac{\Phi_{i,j}^{\bullet} - \Phi_{i-1,j}^{\bullet}}{\Bar{R}_{i} - \Bar{R}_{i-1}} \right)\right] \label{Z step Poisson} \\
    &+~ \dfrac{1}{\Delta V_{i,j}}\left[ A_{i,j+1/2}\left( \dfrac{\Phi_{i,j+1}^{n+1} - \Phi_{i,j}^{n+1}}{\Bar{Z}_{j+1} - \Bar{Z}_j} \right) - A_{i,j-1/2}\left( \dfrac{\Phi_{i,j}^{n+1} - \Phi_{i,j-1}^{n+1}}{\Bar{Z}_{j} - \Bar{Z}_{j-1}} \right) \right] \nonumber \\ \nonumber &-~  4 \pi G (\rho_{\rm tot})_{i,j} \, . 
\end{align}
We choose the pseudo-time steps following \cite{Black1975} and \cite{Norman1986}, and iterate until $\max_{i,j} (\lvert \nabla^2 \Phi - 4 \pi G \rho_{\rm tot}\rvert/4\pi G \rho_{\rm tot}) < 10^{-6}$. Furthermore, we follow \cite{Stone1992} and only update $\Phi$ in case the density has changed significantly enough to render the current $\Phi$ inaccurate (see their eqs. 86--87, but we adopt a tolerance $10^{-6}$ instead of $10^{-5}$). This procedure can significantly lower the time spent on the gravity solver \citep[also see][]{Grudic2021_gravityperformance}. To compute the gravitational acceleration $\boldsymbol{g} = -\boldsymbol{\nabla}\Phi$, we cell-average $\boldsymbol{g}$ in a manner consistent with the discretization of the Poisson equation (Eqs.~\ref{R-step Poisson}--\ref{Z step Poisson}). Thus, to obtain $(g_{R})_{i,j}$, we make the linear approximation within a cell $(i,j)$:
\begin{align}
    g_R(R,Z) \simeq g_{i-1/2,j} + (g_{i+1/2,j} - g_{i-1/2,j}) \,\dfrac{R-R_{i-1/2}}{\Delta R_i} \, ,
\end{align}
where $\Delta R_i = R_{i+1/2} - R_{i-1/2}$ is the cell width. We then cell-average this profile to yield, after some simplification:
\begin{equation}
    (g_R)_{i,j} = \omega_{i-1/2} g_{i-1/2,j} + \omega_{i+1/2} g_{i+1/2,j} \, , \quad \omega_{i-1/2} = 1 - \dfrac{\Bar{R}_i - R_{i-1/2}}{\Delta R_i} \, , \quad \omega_{i+1/2} = 1 - \omega_{i-1/2} \, .
\end{equation}
One gets an analogous expression for the cell-averaged acceleration in the $Z$-direction, $(g_Z)_{i,j}$. For a uniform Cartesian grid, or far away from the symmetry axis, one finds $\omega_{i-1/2} = \omega_{i+1/2} = 1/2$, consistent with \cite{Mullen2021}. To be consistent with our discretization of the Poisson equation, we further have $g_{i+1/2,j} = - (\Phi_{i+1,j} - \Phi_{i,j})/(\Bar{R}_{i+1} - \Bar{R}_{i})$ and $g_{i-1/2,j} = - (\Phi_{i,j} - \Phi_{i-1,j})/(\Bar{R}_{i} - \Bar{R}_{i-1})$.

\subsection{Near pressure-free collapse test}

\begin{figure*}
    \centering
    \begin{minipage}{0.49\textwidth}
        \centering
        \includegraphics[width=\textwidth]{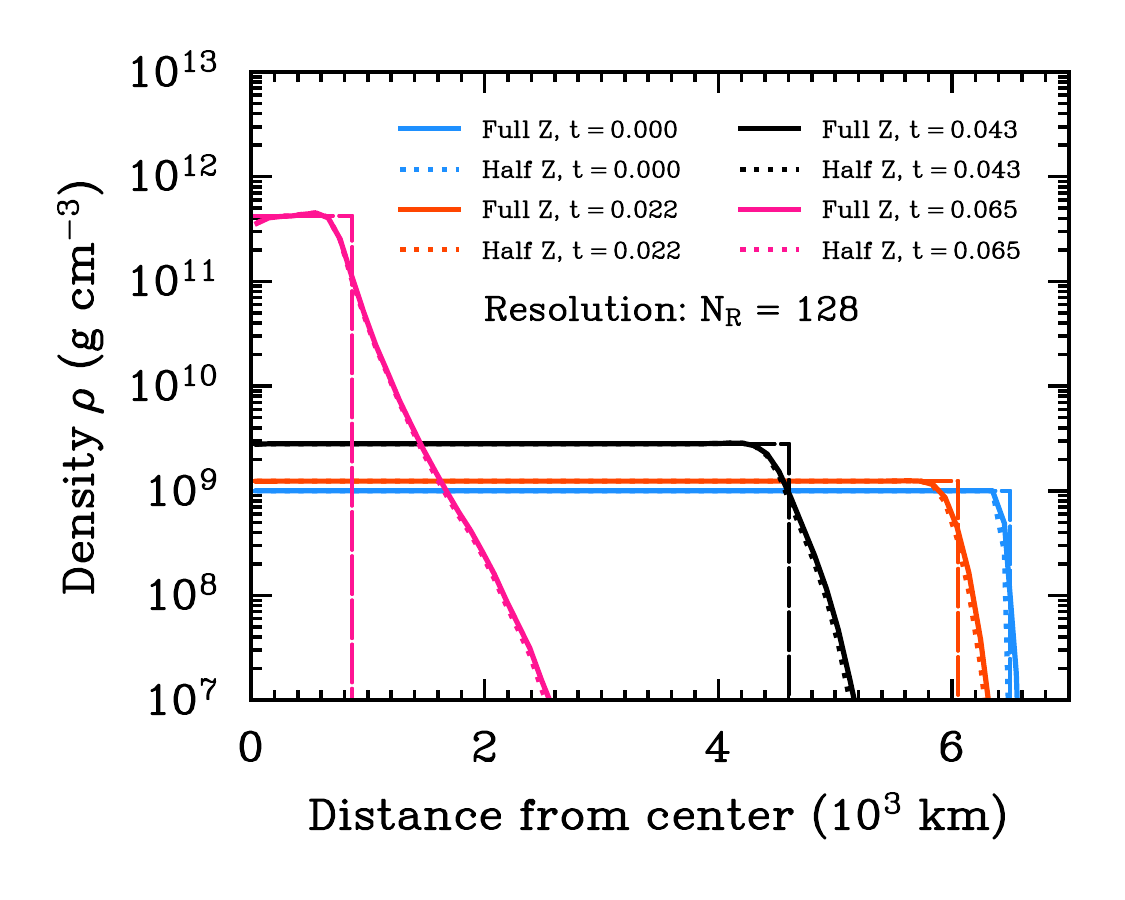}
        \vspace{0.3em}
        {\small (a) $N_{R} =128$ resolution}
    \end{minipage}
    \begin{minipage}{0.49\textwidth}
        \centering
        \includegraphics[width=\textwidth]{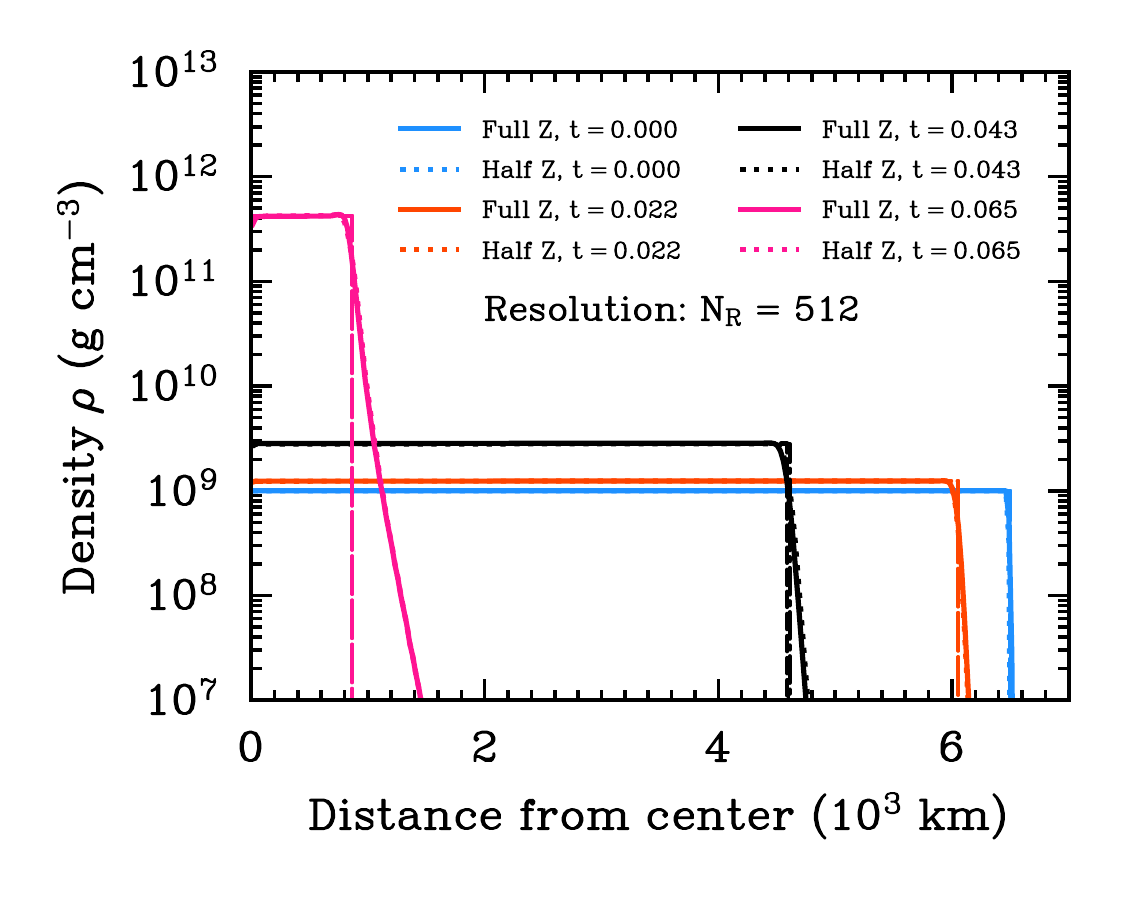}
        \vspace{0.3em}
        {\small (b) $N_{R} = 512$ resolution}
    \end{minipage}
    \caption{Snapshots of the spherically averaged density profile for the nearly pressure-free collapse test, for two resolutions: $N_R = 128$ (a) and $N_Z = 512$ (b). In each panel, the dashed lines show the exact analytical solution for the density profile, for a pressure-free collapse \citep[see eqs. 89--90 in][]{Stone1992}. Solid lines show the numerical results without assumed mirror symmetry ($N_Z =2 N_R$), while dotted lines are from the simulations that only simulate the upper half of the cloud (with $N_Z = N_R$). The physical resolutions (i.e. cell sizes) are identical in each case. We also note that the full grid extends to $1.3 \times 10^4 \, \rm km$.}
    \label{CollapseTest}
\end{figure*}

To test both the Poisson solver and the coupling to hydro, we perform a nearly pressure-free collapse test, following \cite{Skinner2019} \citep[also see e.g.][]{Stone1992, Truelove1998, Almgren2010, Mignone2014}. In particular, we use 2D cylindrical coordinates and a uniform grid, and we set up a uniform spherical cloud of initial radius $R_{\rm cl} = 6500 \, \rm km$, and initial density $\rho_{\rm cl} = 10^9 \, \rm g \, cm^{-3}$. The initial gas density outside is $\rho = 10^{-5} \, \rho_{\rm cl}$, and the grid extends a distance $2 R_{\rm cl}$ away from the cloud center along the $R$ and $Z$ axes. The adiabatic index is $\gamma = 5/3$, and the initial gas pressure is uniform, and taken to be $P_0 = \epsilon (4 \pi G/\gamma) \rho_{\rm cl}^2 R_{\rm cl}^2$, with $\epsilon = 10^{-7}$ to ensure negligible pressure \citep[][]{Skinner2019}. We perform the test both assuming mirror symmetry at $Z = 0$ (i.e. only upper half of the cloud simulated), and for no imposed mirror symmetry (i.e. cloud center placed at $Z = Z_{\rm max}/2$, so the whole cloud is simulated). The simulations are evolved up to $t = 0.065 \,  \textrm{sec} \simeq 0.978 \, t_{\rm ff}$. The results for two resolutions, $N_R = 128$ and $N_R = 512$, are shown in Fig.~\ref{CollapseTest}, with and without assumed mirror symmetry. We find no significant difference whether we assume mirror symmetry or not, indicating that the solver can maintain symmetry throughout the evolution. There is also good overall agreement with the analytical predictions, especially for the peak density. The cloud radius becomes somewhat smeared towards the later stages of the collapse, owing to the non-zero gas pressure and finite resolution. This is also seen in other codes \citep[e.g.][]{Stone1992, Truelove1998, Mignone2014, Skinner2019}, and we find that there is a sharper cloud boundary at higher resolutions. 

\subsection{Hydrostatic equilibrium test}

\begin{figure*}
    \centering
    \begin{minipage}{0.31\textwidth}
        \centering
        \includegraphics[width=\textwidth]{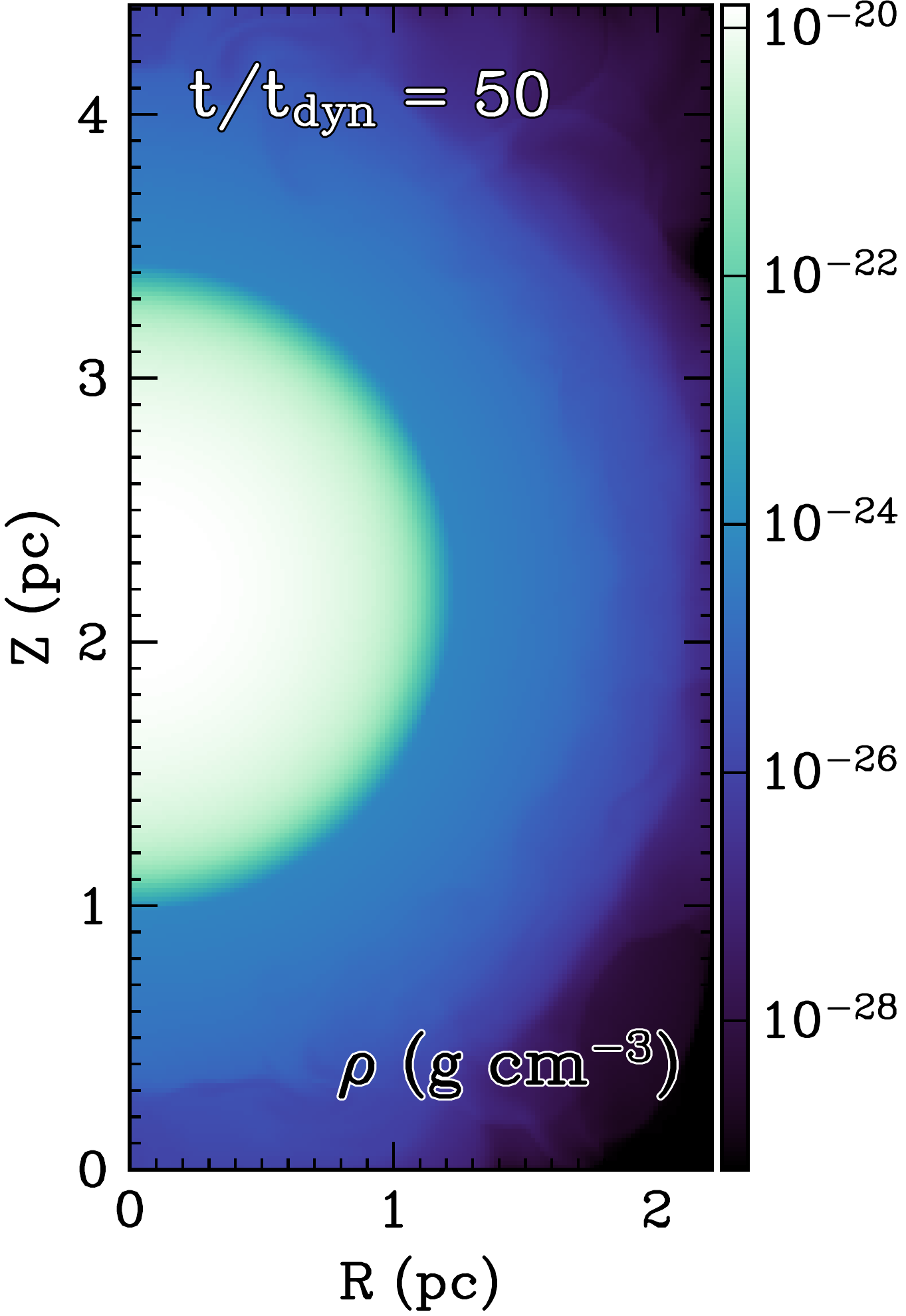}
        \vspace{0.3em}
        {\small (a)}
    \end{minipage}
    \begin{minipage}{0.56\textwidth}
        \centering
        \includegraphics[width=\textwidth]{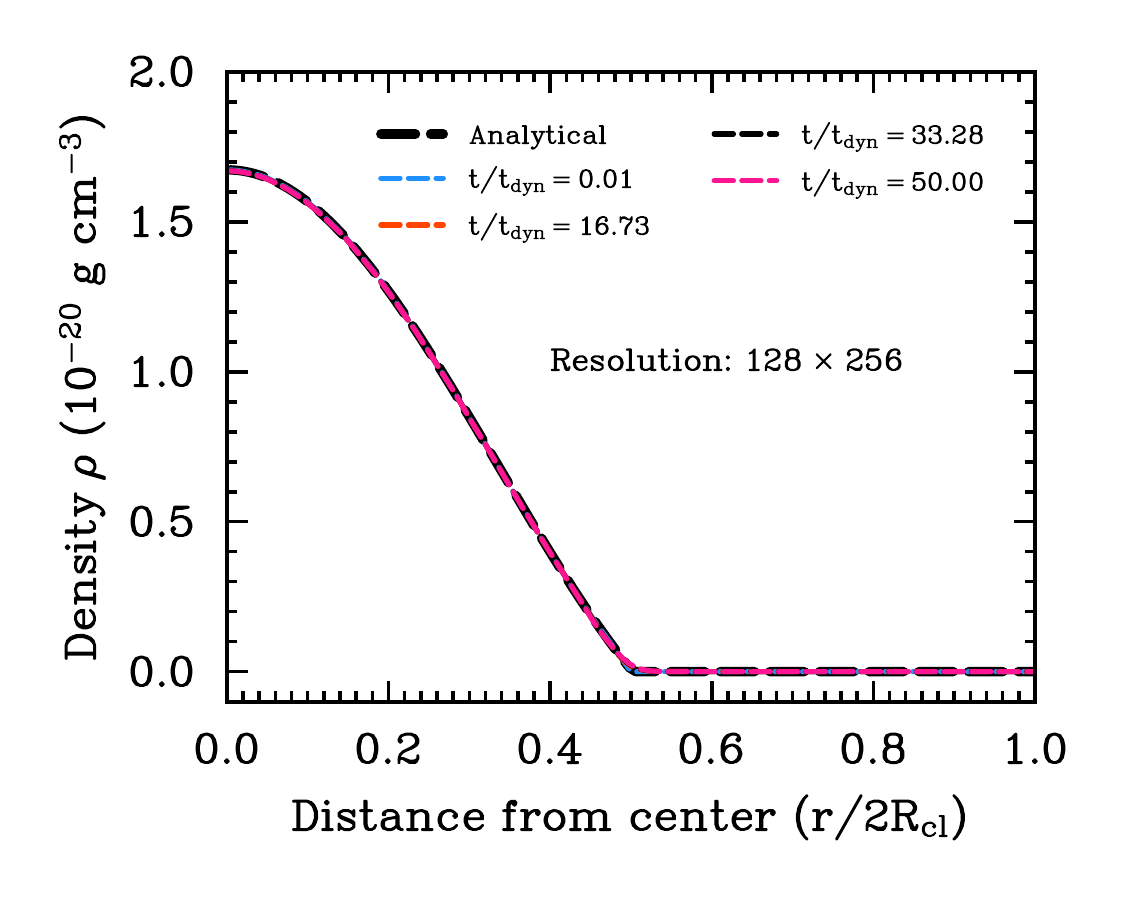}
        \vspace{0.3em}
        {\small (b) }
    \end{minipage}
    \caption{Results for the hydrostatic equilibrium test. Panel (a) shows the gas density at the final time $t = 50 \, t_{\rm dyn}$. Panel (b) shows the evolution of the spherically averaged density profile, compared to the analytical prediction from Eq.~(\ref{polytrope density}). }
    \label{HydrostaticTest}
\end{figure*}

Finally, we test whether hydrostatic equilibrium can be approximately maintained. To do so, we set up a polytropic cloud, with initial pressure $P = K \rho^\Gamma$, polytropic index $\Gamma = \gamma = 2$, and density profile \citep[e.g.][]{Mullen2021}:
\begin{equation}
    \rho(r < R_{\rm cl}) = \rho_{\rm c}\, \dfrac{\sin(\alpha r)}{\alpha r} \,, \quad \alpha = \sqrt{\dfrac{2 \pi G}{P_{\rm c}}} \,\rho_{\rm c} \, . \label{polytrope density}
\end{equation}
The central density is set to $\rho_{\rm c} = 10^4 \, m_{\rm H}$, and the pressure $P_{\rm c} = 10^4 k_{\rm B} \times 100$. The resulting initial cloud has a radius $R_{\rm cl} = \pi/\alpha \simeq 1.1 \, \rm pc$, mass $M_{\rm cl} = (4/\pi) R_{\rm cl}^3 \rho_{\rm c} \simeq 423 \, \rm M_{\odot}$, and dynamical time-scale $t_{\rm dyn} = R_{\rm cl}/v_{\rm esc} \simeq 594 \, \rm kyr$. The density outside the cloud is set to $10^{-5} \, \rho_{\rm c}$. If the implementation of gravity and hydrodynamics is accurate, then the cloud should maintain the equilibrium profile in Eq.~(\ref{polytrope density}) over many dynamical time-scales. We evolve the simulation up to $t = 50 \, t_{\rm dyn}$. The results for a resolution $128 \times 256$ (no imposed mid-plane symmetry) is shown in Fig.~\ref{HydrostaticTest}, and we find that hydrostatic equilibrium (Eq.~\ref{polytrope density}) is very well maintained over the duration of the simulation. We conclude from this test, and the previous one, that the gravity solver in \textsc{Lydion}, and its coupling to hydrodynamics, is accurate enough for our purposes.

\section{Stellar radiative transfer: D-type photoionization test}
\label{Stellar RT appendix}



\begin{figure*}
\centering

\begin{minipage}[t]{0.39\textwidth}
    \centering
    \includegraphics[width=\textwidth]{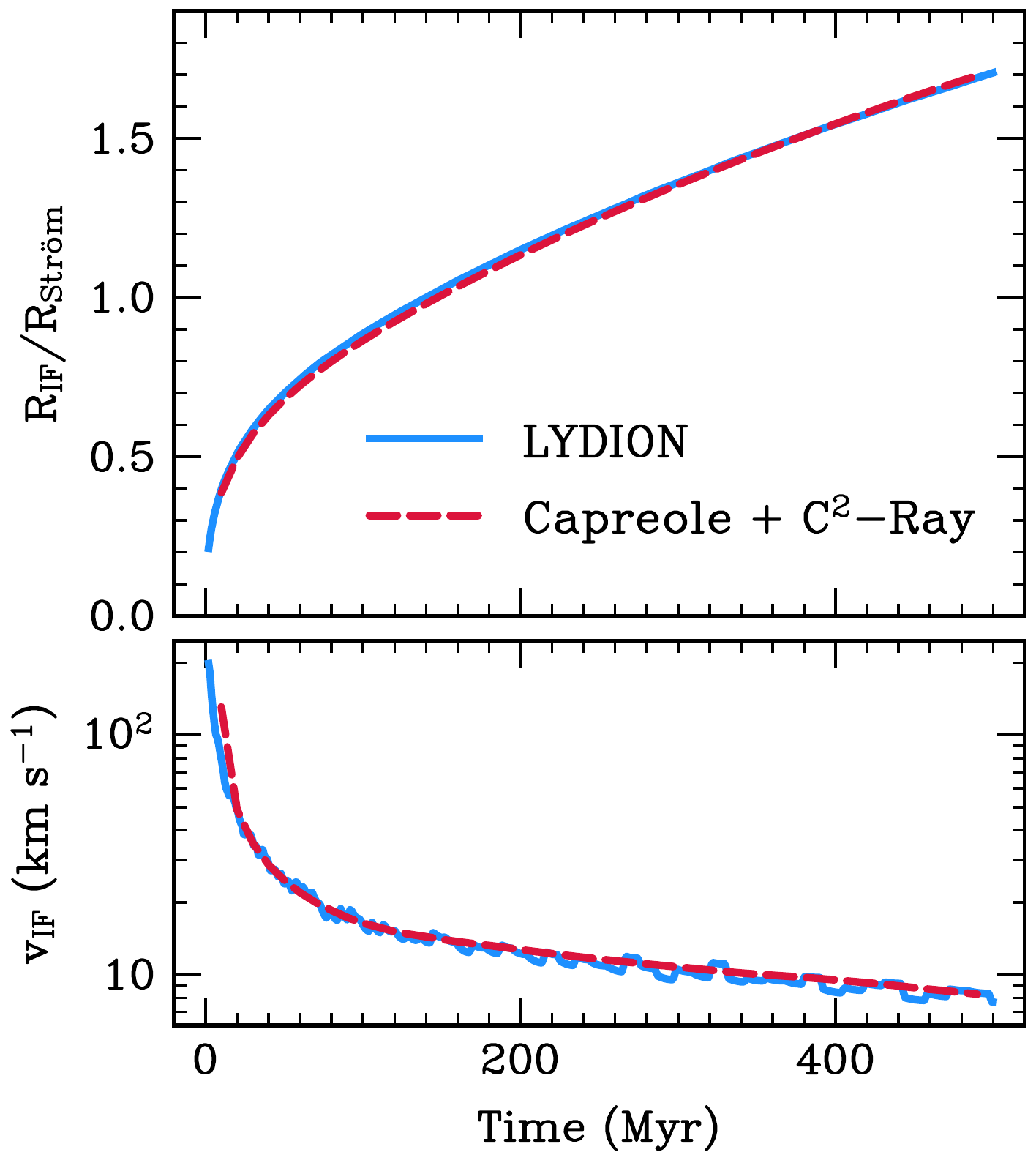}
\end{minipage}
\hfill
\begin{minipage}[t]{0.59\textwidth}
    \centering
    \includegraphics[width=\textwidth]{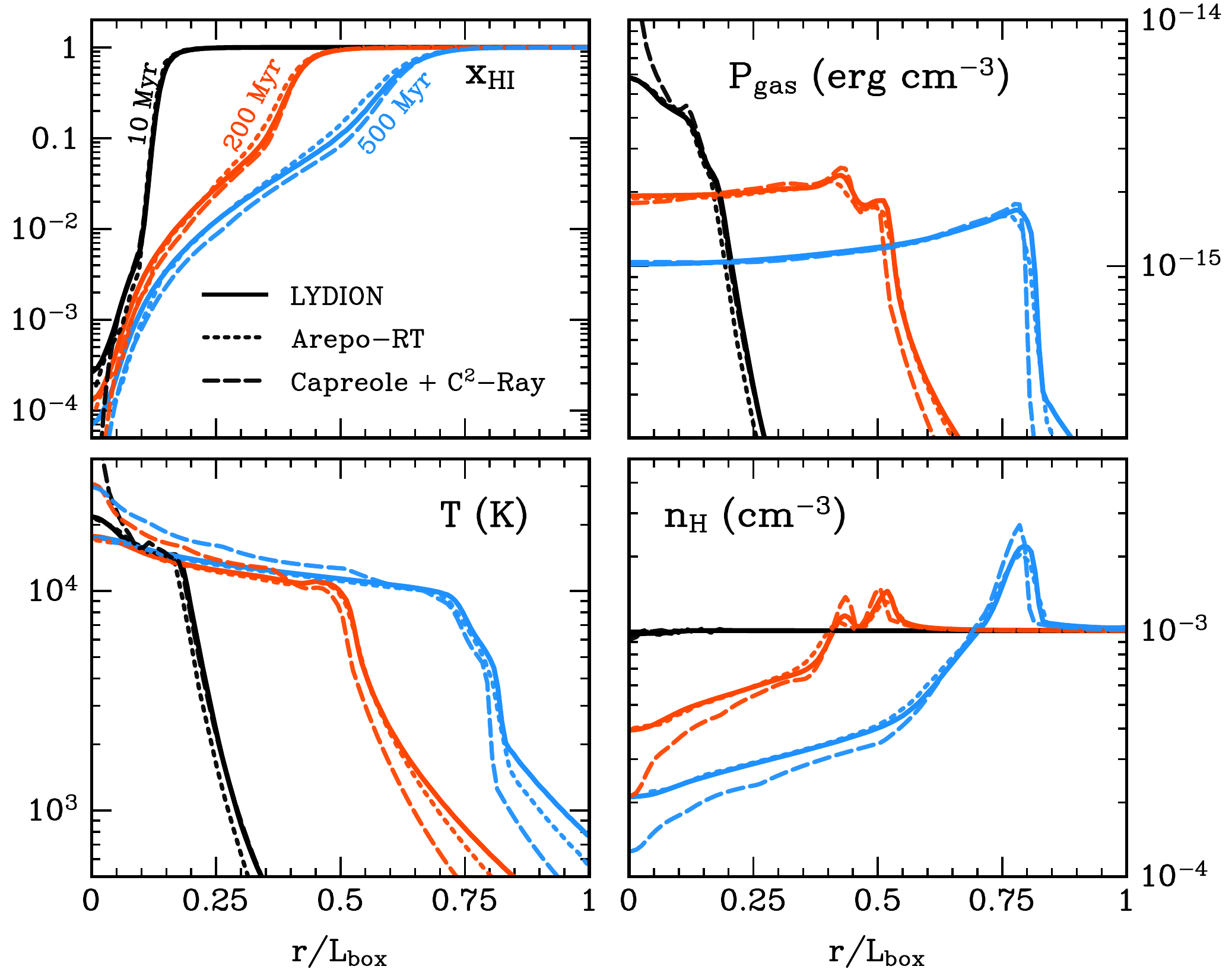}
\end{minipage}

\caption{
D-type photoionization test.
\textbf{Left:} Evolution of the ionization front radius (in units of the Strömgren radius) and velocity compared to the \textsc{Capreole} + C$^2$-ray reference solution.
\textbf{Right:} Predicted \textsc{H\,i} fraction, gas pressure, temperature, and density profiles.
Solid, short-dashed, and long-dashed lines show results from \textsc{Lydion}, \textsc{Arepo-rt}, and \textsc{Capreole} + C$^2$-ray, respectively.
The profile comparison is shown at $t = (10,200,500)\,{\rm Myr}$.
}
\label{fig:D type_test}
\end{figure*}

In this Appendix, we test the ability of \textsc{Lydion} to capture the evolution of \textsc{H\,ii} regions. To do this, we perform test 5 of \cite{Iliev2009}, wherein the R and D-type evolution of an ionization front (IF) in a uniform medium is studied. As in \cite{Iliev2009}, we initialize a $15 \, \rm kpc \times 15 \, \rm kpc$ box of resolution $128^2$, density $n_{\rm H} = 10^{-3} \, \rm cm^{-3}$, and temperature $T = 100 \, \rm K$. The gas is pure hydrogen, i.e. we ignore helium, deuterium, metals, and dust. Radiation pressure is also ignored in this test. We place a source at $(R,Z) = (0,0)$, which emits $5 \times 10^{48} \, \rm s^{-1}$ ionizing photons in the interval $13.6 \, \textrm{eV} \leq h\nu \leq 100 \, \rm eV$, with a $T_{\rm eff} = 10^5 \, \rm K$ black-body spectrum. Following \cite{Rosdahl2013} and \cite{Kannan2019}, who also employ the M1 moment method, we adopt a minimum reduced speed of light of $\Tilde{c}_{\rm min} =0.01 \, c$ for this test, to correctly capture the fast initial R-type IF expansion. The simulations are evolved for $500 \, \rm Myr$. In Fig.~\ref{fig:D type_test} we compare the results from \textsc{Lydion} to those from \textsc{Arepo-rt} \citep[][]{Kannan2019}, as well as the ray-tracing code C$^2$--ray coupled to the \textsc{Capreole} hydrodynamics code \citep[see][]{Mellema2006, Iliev2009}. We plot and compare the evolution of the ionization front radius and velocity (left), and gas profiles at different times (right). We find broad agreement between all codes, with \textsc{Arepo-rt} and \textsc{Lydion} predictions being most similar. This is expected, given the similar M1-based RT method, and identical frequency band division, in both codes.

\section{Infrared radiative transfer}
\label{IR appendix}

\begin{figure*}
\centering
\includegraphics[width=0.9\textwidth]{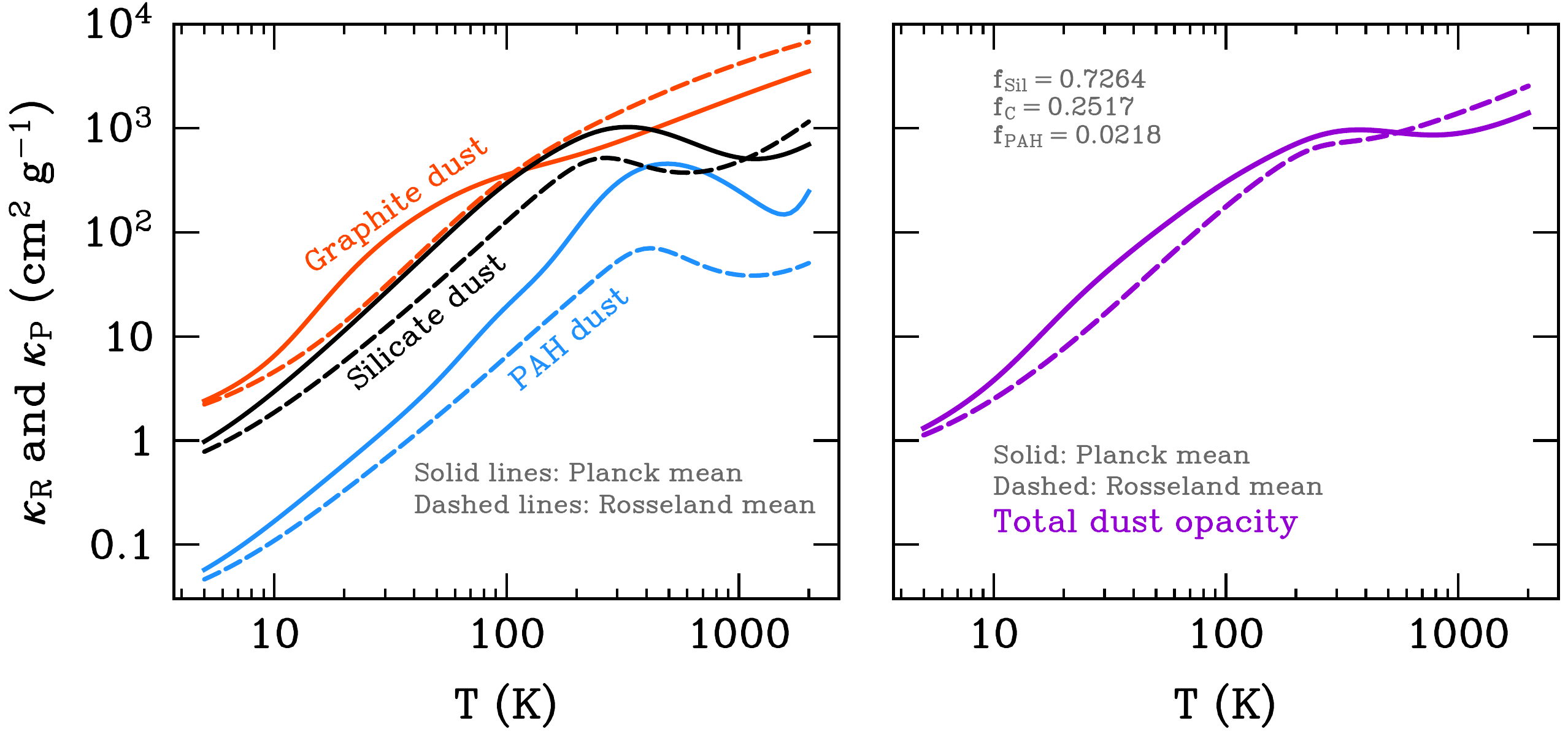}
\caption{Planck and Rosseland-mean dust opacities, as a function of temperature. \textbf{Left panel:} Opacities in a given dust bin (cm$^2$ per gram of dust mass in that bin). \textbf{Right panel:} The total opacities (cm$^2$ per gram of total dust mass), assuming the initial mass distribution for the $R_{\rm V} = 5.5$ dust model of \cite{Weingartner2001}. The opacities scale approximately as $\propto T^2$ for $T \lesssim 200 \, \rm K$, and then reach a plataeu around $10^3 \, \rm cm^2 \, g^{-1}$ \citep[comparable to other dust models, see fig.~17 in][]{HensleyDraine2023}. Note however that in RHD simulations, dust growth and destruction will in general lead to deviations from these plotted total opacities. }
\label{Planck Rosseland opacities fig}
\end{figure*}

In this Appendix we describe and test the implementation of IR transport in \textsc{Lydion}. Let $J_{\rm IR}^{\rm tot} = J_{\rm IR} + J_{\rm CMB}$ be the \textit{total} IR intensity, including the CMB background (which is constant and uniform). The moment equations for the IR band read:\footnote{To obtain the equation for $J_{\rm IR}$, we have noted that $\partial_t J_{\rm CMB} = 0$, and $\boldsymbol{H}_{\rm CMB} =0$. We have further used $T_{\rm CMB}$ for the radiation temperature of the CMB background (in the CMB absorption term), and neglected velocity terms of order $\mathcal{O}(J_{\rm CMB} \boldsymbol{\nabla} \boldsymbol{\cdot}\boldsymbol{u}/c)$.}
\begin{align}
    \dfrac{1}{\Tilde{c}}\dfrac{\partial J_{\rm IR}}{\partial t} + \boldsymbol{\nabla}\boldsymbol{\cdot}\left(\boldsymbol{H}_{\rm IR} + \dfrac{\boldsymbol{u}}{c} J_{\rm IR} \right) &=~ j_{\rm IR} - \sum_{\beta \in (\rm Sil,C,PAH)}\rho_{\rm d,\beta} \left[\kappa_{\rm P,\beta}(T_{\rm rad})J_{\rm IR} + \kappa_{\rm P,\beta}(T_{\rm CMB})\dfrac{\sigma_{\rm SB} T_{\rm CMB}^4}{\pi}\right] \label{J_IR PDE} \\ &-~\boldsymbol{\mathsf{K}}_{\rm IR} \boldsymbol{:}\dfrac{\boldsymbol{\nabla}\boldsymbol{u}}{c} \, , \nonumber  \\ \dfrac{1}{\Tilde{c}}\dfrac{\partial \boldsymbol{H}_{\rm IR}}{\partial t} + \boldsymbol{\nabla}\boldsymbol{\cdot}\left(\boldsymbol{\mathsf{K}}_{\rm IR} + \dfrac{\boldsymbol{u}}{c} \boldsymbol{H}_{\rm IR} \right) &=~  - \left[\sum_{\beta \in (\rm Sil,C,PAH)}\rho_{\rm d,\beta} \kappa_{\rm R,\beta}(T_{\rm rad}) + n_{\rm e} \sigma_{\rm T} \right] \boldsymbol{H}_{\rm IR} \, , \label{H_IR PDE} 
\end{align}
where we are using frequency-averaged quantities, e.g. $J_{\rm IR} \equiv \int_{\rm IR} \textrm{d}\nu \, J_{\rm IR}(\nu)$, and $\sigma_{\rm T} = 6.65 \times 10^{-25} \, \rm cm^{2}$ is the Thomson scattering cross-section. In the above equations we have implicitly assumed that the IR intensity has a black-body shape, such that the $J_{\rm IR}(\nu)$-weighted dust absorption opacity is given by the Planck-mean ($\rho_{\rm d} \kappa_{\rm P}$), evaluated at the radiation temperature $T_{\rm rad}$ (discussed below). Furthermore, in the optically thick regime, the same assumption gives a $\boldsymbol{H}_{\rm IR}(\nu)$-weighted flux opacity equal to the Rosseland-mean ($\rho_{\rm d} \kappa_{\rm R}$). We plot $\kappa_{\rm P}$ and $\kappa_{\rm R}$ in Fig.~\ref{Planck Rosseland opacities fig} for each dust bin/type, computed assuming the \cite{Weingartner2001} $R_{\rm V} = 5.5$ dust model (for consistency with its application for the non-IR dust opacities). 

To proceed, we assume that the only important source of IR emission is emission from each dust bin/type (silicate grains, graphite grains, PAHs). Assuming that the dust temperatures are in equilibrium, the IR emissivity is then (in erg s$^{-1}$ cm$^{-3}$ sr$^{-1}$):
\begin{align}
    j_{\rm IR} &=~ \sum_{\beta\in(\rm Sil,C,PAH)} \rho_{\textrm{d},\beta} \kappa_{\textrm{P},\beta}(T_{\rm d,\beta}) \, \dfrac{\sigma_{\rm SB}T_{\textrm{d},\beta}^4}{\pi} \label{j_IR} \\ &=~  \sum_{\beta\in(\rm Sil,C,PAH)}\rho_{\rm d,\beta} \left[ \sum_{\mathcal{B} \neq \rm IR} \kappa_{\rm \mathcal{B},abs,\beta} J_{\mathcal{B}} + \kappa_{\rm Ly\alpha,abs,\beta}\dfrac{c e_{\rm Ly\alpha}}{4\pi}  + \kappa_{\rm P,\beta}(T_{\rm rad}) J_{\rm IR} + \kappa_{\textrm{P},\beta}(T_{\rm CMB}) \, \dfrac{\sigma_{\rm SB}T_{\textrm{CMB}}^4}{\pi} + \dfrac{\Lambda_{\rm gd,\beta}}{4\pi \rho_{\rm d,\beta}} \right] \, , \nonumber
\end{align}
where $\kappa_{\mathcal{B},\rm abs}$ is the dust absorption cross-section per unit dust mass in non-IR band $\mathcal{B}$ (Table~\ref{Dust optical properties non IR}), $e_{\rm Ly\alpha}$ is the Ly$\alpha$ energy density, and $\Lambda_{\rm gd}$ is the net dust collisional heating rate (described below). This equation can be solved for the dust temperature $T_{\rm d,\beta}$ at any time and for each dust bin/type, which we do numerically using a bisection root-finding method for robustness. For the radiation temperature $T_{\rm rad}$, we use the approach of \cite{Grudic2021}, wherein the energy and photon count from each separate black-body spectrum is matched by a corresponding single black-body of temperature $T_{\rm rad}$. This yields the following approximate update rule for $T_{\rm rad}$ over a subcycling substep $\bullet \rightarrow \bullet\bullet$ of size $\Delta t_{\rm sub}$:
\begin{equation}
    T_{\rm rad}^{\bullet \bullet} \simeq \dfrac{J_{\rm IR}^\bullet + \Tilde{c} \Delta t_{\rm sub} \sum_{\beta \in \rm (Sil,C,PAH)} \, j_{\rm IR,\beta}^{\bullet}}{J_{\rm IR}^{\bullet}/T_{\rm rad}^{\bullet} + \Tilde{c} \Delta t_{\rm sub}\sum_{\beta \in \rm (Sil,C,PAH)} \, j_{\rm IR,\beta}^{\bullet}/T_{\rm d,\beta}^{\bullet}} \, . 
\end{equation}
Since $j_{\rm IR,\beta} \propto \rho_{\rm d,\beta} \kappa_{\rm P,\beta}(T_{\rm d,\beta})\, T_{\rm d,\beta}^4$, this estimate is biased to the hottest grains. Using Eq.~(\ref{j_IR}) in Eq.~(\ref{J_IR PDE}) yields: 
\begin{equation}
    \dfrac{1}{\Tilde{c}}\dfrac{\partial J_{\rm IR}}{\partial t} + \boldsymbol{\nabla}\boldsymbol{\cdot}\left(\boldsymbol{H}_{\rm IR} + \dfrac{\boldsymbol{u}}{c} J_{\rm IR} \right) = \sum_{\beta }\rho_{\rm d,\beta} \left[  \sum_{\mathcal{B} \neq \rm IR}  \kappa_{\rm \mathcal{B},abs,\beta} J_{\mathcal{B}} + \kappa_{\rm Ly\alpha,abs,\beta}\dfrac{c e_{\rm Ly\alpha}}{4\pi} \right]  + \sum_{\beta }\dfrac{\Lambda_{\rm gd,\beta}}{4\pi} - \boldsymbol{\mathsf{K}}_{\rm IR} \boldsymbol{:}\dfrac{\boldsymbol{\nabla}\boldsymbol{u}}{c} \, . 
\end{equation}
We solve this form of the equation for $J_{\rm IR}$, and Eq.~(\ref{H_IR PDE}) for $\boldsymbol{H}_{\rm IR}$. As for the other bands, operator-splitting is used to first isolate and solve the pure advection terms. We use the same asymptotic diffusion fix for IR photons as we do for Ly$\alpha$ (see Eq.~\ref{Corrected GLF flux J} and associated discussion), in order to treat the case where $\tau_{\rm cell} \gtrsim 1$ \citep[for similar approaches for M1 transport applied to the IR, see][]{Rosdahl2015, Kimura2025}. After the advection update, we subcycle the source terms (together with the other bands, and thermochemistry) using a semi-implicit BDF1 method. Thus, over a subcycle update $\bullet \rightarrow \bullet\bullet$ of step-size $\Delta t_{\rm sub}$, and after the candidate update of (non-IR) $J_{\mathcal{B}}$ and the Ly$\alpha$ energy density (from emission and absorption), we solve, in the following order:
\begin{align}
    \dfrac{J_{\rm IR}^{\bullet\bullet} - J_{\rm IR}^{\bullet}}{\Tilde{c}\Delta t_{\rm sub}} &=~ \sum_{\beta}  \rho_{\rm d,\beta} \left( \sum_{\mathcal{B} \neq \rm IR} \kappa_{\rm \mathcal{B},abs,\beta} J_{\mathcal{B}}^{\bullet \bullet} + \kappa_{\rm Ly\alpha,abs,\beta}\dfrac{c e_{\rm Ly\alpha}^{\bullet \bullet}}{4\pi} \right)  + \sum_{\beta}\dfrac{1}{4\pi} \Lambda_{\rm gd,\beta}(T_{\rm d,\beta}^{\bullet},T^{\bullet}) \, ,  \\ \dfrac{\boldsymbol{H}_{\rm IR}^{\bullet\bullet} - \boldsymbol{H}_{\rm IR}^{\bullet}}{\Tilde{c}\Delta t_{\rm sub}} &=~ - \left[\sum_{\beta} \rho_{\rm d,\beta} \kappa_{\rm R,\beta}(T_{\rm rad}^{\bullet\bullet})+n_{\rm e}\sigma_{\rm T} \right] \boldsymbol{H}_{\rm IR}^{\bullet \bullet} \, .
\end{align}
We reject the update, and re-do the whole subcycle step with a smaller step-size, in case $J_{\rm IR}$, $\boldsymbol{H}_{\rm IR}$, or $T_{\rm d,\beta}$ (as obtained from Eq.~\ref{j_IR} with $J^{\bullet\bullet}_{\mathcal{B}}$, $J^{\bullet\bullet}_{\rm IR}$, and $e_{\rm Ly\alpha}^{\bullet\bullet}$) changed by more than $10\%$. The contribution to the collisional heating rate from dust bin $\beta$ is \citep[e.g.][]{Burke1983, Draine2011}:\footnote{A more accurate treatment would involve summing over all major gas species, and taking the grain charge into account \citep[][]{DraineSutin1987}. We have implicitly baked these complications in $\Bar{\alpha}$. Collisional heating is typically a very minor contribution to the dust temperature \citep[][]{Draine2011}, so we have neglected these second-order details.}
\begin{align}
    \Lambda_{\rm gd,\beta}(T_{\rm d,\beta},T) &=~ n_{\rm H} \int \textrm{d}a_{\rm gr} \, \dfrac{\partial n_{\textrm{gr}, \beta}}{\partial a_{\rm gr}} \pi a_{\rm gr}^2 \Bar{\alpha} \left(\dfrac{8 k_{\rm B}T}{\pi m_{\rm H}} \right)^{1/2} 2 k_{\rm B}(T - T_{\rm d,\beta}) \\ &=~  \dfrac{3}{4}n_{\rm H} \,  \dfrac{\rho_{\textrm{d},\beta}\Bar{\alpha}}{\rho_{\textrm{gr},\beta} \Bar{a}_{\textrm{gr},\beta}} \left(\dfrac{8 k_{\rm B}T}{\pi m_{\rm H}} \right)^{1/2} 2 k_{\rm B}(T - T_{\rm d,\beta}) \nonumber \, .
\end{align}
Here $\Bar{\alpha}$ is the accomodation factor, which we take to be $0.5$ \citep[][]{Burke1983, Kannan2020}. On the second line we have defined the area-weighted grain size (of a given bin/type) $\Bar{a}_{\rm gr} \equiv \langle a_{\rm gr}^3 \rangle / \langle a_{\rm gr}^2 \rangle$, where $\langle \Bar{a}_{\rm gr}^n \rangle \equiv [\int \textrm{d}a_{\rm gr} \, a_{\rm gr}^{n} \partial n_{\rm gr}/\partial a_{\rm gr}] / [\int \textrm{d}a_{\rm gr} \, \partial n_{\rm gr}/\partial a_{\rm gr}]$. For the assumed $R_{\rm V} = 5.5$ dust model of \cite{Weingartner2001} and $a_{\rm gr} = 10^{-3} \, \mu \rm m$ size boundary between the graphite and PAH dust bins, we compute $\Bar{a}_{\rm gr,Sil} = 0.138 \,  \rm \mu m$, $\Bar{a}_{\rm gr,C} = 0.0530 \, \rm \mu m$, and $\Bar{a}_{\rm gr,PAH} = 5.62 \times 10^{-4} \, \rm \mu m$. The grain material densities are $\rho_{\rm gr,Sil} = 3.5 \, \rm g \, cm^{-3}$ and $\rho_{\rm gr,C/PAH} = 2.24 \, \rm g \, cm^{-3}$ \citep{Weingartner2001}.

\section{Dust physics}
\label{Dust physics}

\subsection{Dust-gas coupling from drag and radiation pressure}
\label{Dust dynamics appendix}

As described in the main text,  we treat each dust bin as a pressureless fluid that experiences forces from radiation pressure, gravity, and gas drag \citep[for similar modelling, see][]{Paardekooper2006, Ishiki2018, Verrier2025}. Dust advection is operator split and treated the together with, and similar to, the gas dynamics. After gas and dust advection, the relevant operator-split equations for dust and gas read:
\begin{align}
    \dfrac{\partial \rho_{\rm d}}{\partial t} &=~ \dfrac{\rho_{\rm d,\beta}}{t_{\rm gr,\beta,0}}\left(1 - \dfrac{\rho_{\rm d,\beta}}{\rho_{\rm d,\beta} + \rho_{Z,\beta}} \right) - \dfrac{\rho_{\rm d,\beta}}{t_{\rm dest,\beta}} \, , \label{Dust continuity} \\ \dfrac{\partial (\rho_{\rm d} \boldsymbol{u}_{\rm d})_{\beta}}{\partial t} &=~ \boldsymbol{f}_{\rm d,\beta} + \mathcal{K}_{\beta}(\boldsymbol{u}-\boldsymbol{u}_{\rm d,\beta}) \, , \\ \dfrac{\partial (\rho \boldsymbol{u})}{\partial t} &=~ \boldsymbol{f}_{\rm g} + \sum_{\beta \in \rm (Sil,C,PAH)}\mathcal{K}_{\beta}(\boldsymbol{u}_{\rm d,\beta}-\boldsymbol{u}) \, ,
\end{align}
where $\boldsymbol{f}_{\rm d,\beta}$ is the radiation pressure force per unit volume on dust bin $\beta$, $\boldsymbol{f}_{\rm g}$ is the radiation pressure force per unit volume on the gas, $\mathcal{K}_{\beta}$ is the gas drag coefficient for dust bin $\beta$, and the right-hand side of Eq.~(\ref{Dust continuity}) describes dust growth and destruction. These terms are subcycled together with the photo-thermochemistry. Within a given subcycle step, we further operator split dust growth/destruction from  the radiation pressure + drag (RPD), and start with the latter. When considering RPD, we keep the densities constant during the update of dust and gas velocities, so that we have to solve:
\begin{equation}
    \rho_{\textrm{d},\beta} \dfrac{\textrm{d}\boldsymbol{u}_{\rm d,\beta}}{\textrm{d}t}    =  \boldsymbol{f}_{\textrm{d},\beta} - \mathcal{K}_\beta (\boldsymbol{u}_{\textrm{d},\beta} - \boldsymbol{u}) \, ,  \quad \quad \rho \dfrac{\textrm{d}\boldsymbol{u}}{\textrm{d}t}  =  \boldsymbol{f}_{\rm g}  + \sum_{\beta} \mathcal{K}_\beta(\boldsymbol{u}_{\textrm{d},\beta} - \boldsymbol{u}) \, .  \label{du/dt dust and gas Appendix}
\end{equation}
It is convenient to work with relative velocities, because allow us to can control their magnitudes (for numerical reasons), while also reducing the size of matrices to be inverted (from $4\times4$ to $3\times3$ for three dust bins). We define $\boldsymbol{u}_{\textrm{rel},\beta} \equiv \boldsymbol{u}_{\textrm{d},\beta}-\boldsymbol{u}$. We then have (using Eq.~\ref{du/dt dust and gas Appendix} for $\textrm{d}\boldsymbol{u}/\textrm{d}t$):
\begin{equation}
 \dfrac{\textrm{d}\boldsymbol{u}_{\rm rel,\beta}}{\textrm{d}t}= \dfrac{\boldsymbol{f}_{\textrm{d},\beta}}{\rho_{\textrm{d},\beta}} -\dfrac{\boldsymbol{f}_{\rm g}}{\rho} - \sum_{\beta' \neq \beta}\dfrac{\mathcal{K}_{\beta'}}{\rho}\, \boldsymbol{u}_{\textrm{rel},\beta'}   - \mathcal{K}_\beta\left( \dfrac{1}{\rho_{\textrm{d},\beta}} + \dfrac{1}{\rho}  \right) \boldsymbol{u}_{\textrm{rel},\beta} \, .
\end{equation}
For stability reasons, we aim to solve this over a subcycle step $\bullet \rightarrow \bullet\bullet$ of size $\Delta t_{\rm sub}$ using an implicit BDF1 method:
\begin{equation}
 \dfrac{\boldsymbol{u}_{\rm rel,\beta}^{\bullet\bullet} - \boldsymbol{u}_{\rm rel,\beta}^{\bullet}}{\Delta t_{\rm sub}}= \dfrac{\boldsymbol{f}_{\textrm{d},\beta}}{\rho_{\textrm{d},\beta}} -\dfrac{\boldsymbol{f}_{\rm g}}{\rho} - \sum_{\beta' \neq \beta}\dfrac{\mathcal{K}_{\beta'}}{\rho}\, \boldsymbol{u}_{\textrm{rel},\beta'}^{\bullet \bullet}   - \mathcal{K}_\beta\left( \dfrac{1}{\rho_{\textrm{d},\beta}} + \dfrac{1}{\rho}  \right) \boldsymbol{u}_{\textrm{rel},\beta}^{\bullet \bullet} \, .
\end{equation}
This can be rearranged to the following matrix equation for $\boldsymbol{u}^{\bullet\bullet}_{\rm rel,\beta}$ (using the Einstein summation convention):
\begin{align}
 \boldsymbol{\mathsf{M}}_{\beta',\beta} \boldsymbol{u}_{\rm rel, \beta}^{\bullet\bullet} &=~ \boldsymbol{A}_{\beta} \, , \\ \boldsymbol{\mathsf{M}}_{\beta',\beta} &=~ \left[1 + \Delta t_{\rm sub} \mathcal{K}_{\beta'} \left( \dfrac{1}{\rho_{\textrm{d},\beta'}} + \dfrac{1}{\rho}  \right) \right]\delta_{\beta',\beta} + (1-\delta_{\beta',\beta})\Delta t_{\rm sub} \dfrac{\mathcal{K}_\beta}{\rho} \, , \\ \boldsymbol{A}_\beta &=~ \boldsymbol{u}_{\textrm{rel},\beta}^{\bullet} + \left(\dfrac{\boldsymbol{f}_{\textrm{d},\beta}}{\rho_{\textrm{d},\beta}} - \dfrac{\boldsymbol{f}_{\rm g}}{\rho} \right) \Delta t_{\rm sub} \, .
\end{align}
The updated relative velocities are then obtained as $\boldsymbol{u}_{\rm rel,\beta}^{\bullet\bullet} = \boldsymbol{\mathsf{M}}_{\beta',\beta}^{-1} \boldsymbol{A}_{\beta}$. If the updated relative velocity is $\lvert \boldsymbol{u}_{\rm rel,\beta}^{\bullet\bullet} \rvert < 100 \, \rm km \, s^{-1}$, we accept $\boldsymbol{u}_{\rm rel,\beta}^{\bullet \bullet} $. Otherwise, we scale down the components of the relative velocity accordingly. This prevents extremely large relative velocities to affect the CFL time-step, and is physically motivated because sputtering will destroy fast-moving grains anyway (as discussed below). Once we have updated the relative velocities, we obtain the absolute gas and dust velocities from the updated center-of-mass and relative velocities:
\begin{align}
    \boldsymbol{u}_{\rm CoM}^{\bullet\bullet} &=~ \boldsymbol{u}_{\rm CoM}^\bullet + \dfrac{1}{\rho_{\rm tot}}\left(\boldsymbol{f}_{\rm g} + \sum_\beta \boldsymbol{f}_{\rm d,\beta}\right) \Delta t_{\rm sub} \, , \\ \boldsymbol{u}^{\bullet\bullet} &=~ \boldsymbol{u}_{\rm CoM}^{\bullet\bullet} - \dfrac{1}{\rho_{\rm tot}} \sum_\beta \rho_{\rm d,\beta} \boldsymbol{u}_{\rm rel,\beta}^{\bullet\bullet} \, , \\ \boldsymbol{u}_{\rm d,\beta}^{\bullet\bullet} &=~ \boldsymbol{u}_{\rm CoM}^{\bullet\bullet} + \boldsymbol{u}_{\rm rel,\beta}^{\bullet\bullet} - \dfrac{1}{\rho_{\rm tot}} \sum_{\beta'} \rho_{\rm d,\beta'} \boldsymbol{u}_{\rm rel,\beta'}^{\bullet\bullet} \, ,
\end{align}
where $\rho_{\rm tot} = \rho + \sum_\beta \rho_{\rm d,\beta}$ is the total dust + gas density, and $\boldsymbol{u}_{\rm CoM} = [\rho \boldsymbol{u} + \sum_\beta (\rho_{\rm d}\boldsymbol{u})_\beta ]/\rho_{\rm tot}$ the center-of-mass velocity. As discussed below, the drag coefficient in general depends on $\lvert \boldsymbol{u}_{\rm rel} \rvert$, giving rise to a non-linear set of equations. We therefore use Picard iteration, going through the above steps with updated estimates of $\mathcal{K}_\beta$ each iteration. If this converges, we accept the update.\footnote{By iterating, we improve on the modelling of \cite{Ishiki2018} by ensuring that the values of $\mathcal{K}_\beta$ are in fact evaluated in an implicit manner.} If not, we retry the whole subcycle step with a smaller $\Delta t_{\rm sub}$. It is straightforward to show that for strong drag (large $\mathcal{K}_\beta$), the velocities are coupled ($\boldsymbol{u}_{\rm d,\beta}^{\bullet\bullet} \simeq \boldsymbol{u}^{\bullet\bullet}$), with the total (dust + gas) radiation pressure affecting both components. In contrast, in the low-drag regime, the dust is decoupled from the gas, and could be partially evacuated from dense \textsc{H\,ii} regions by radiation pressure on the dust \citep[][]{Draine2011_HII, Akimkin2015, Akimkin2017, Ishiki2018}.

The drag coefficient is $\mathcal{K} \equiv \lvert \langle n_{\rm gr} \boldsymbol{F}_{\rm drag} \rangle \rvert/\lvert \boldsymbol{u}_{\rm rel}\rvert$, where $\boldsymbol{F}_{\rm drag}$ is the drag force on a single grain, and $n_{\rm gr}$ is the grain number density, with $\langle \rangle$ denoting an average over the grain size distribution (for a given bin). The drag force on a single grain of size $a_{\rm gr}$ is \citep{DraineSalpeter1979_dustdrag}:
\begin{equation}
    \lvert \boldsymbol{F}_{\rm drag} \rvert = 2 \pi a_{\rm gr}^2 k_{\rm B}T \sum_{i} n_{i} \left[\mathcal{G}_{0}(s_i) +  Z_{i}^2 \phi^2\ln \lvert \Lambda/Z_i\rvert \, \mathcal{G}_{2}(s_i)\right] \, , \label{Fdrag }
\end{equation}
where the sum is over the gas species $i \in (\rm \textsc{H\,i}, H_2, \textsc{H\,ii}, He\,\textsc{i}, He\,\textsc{ii}, He\,\textsc{iii}, D\,\textsc{ii}, C\,\textsc{ii}, O\,\textsc{ii}, \textit{e}^- )$,\footnote{Other species, e.g. CO and \textsc{O\,i}, have a negligible impact on the drag compared to the included species. } and $Z_i$ is the charge of species $i$ in units of the elementary charge $e$ (so, e.g., the electron has $Z_{\rm e} =-1$). For the other terms, $s_i \equiv \lvert \boldsymbol{u}_{\rm rel}\rvert/\sqrt{2k_{\rm B}T/m_i}$,
\begin{equation}
     \mathcal{G}_{0}(s_i) \simeq \dfrac{8 s_i}{3\sqrt{\pi}} \, \sqrt{1 + 9\pi s_i/64} \, , \quad \mathcal{G}_{2}(s_i) \simeq \dfrac{s_i}{3\sqrt{\pi}/4 + s_i^3} \,, 
\end{equation}
$\phi \equiv Z_{\rm gr} e^2 /a_{\rm gr} k_{\rm B} T$, and $\Lambda \equiv (3/2a_{\rm gr} e \lvert \phi \rvert) \sqrt{k_{\rm B}T/\pi n_{\rm e}}$. Evidently, a large grain charge $Z_{\rm gr}$ can strongly boost the drag (via $\phi^2$ in Eq.~\ref{Fdrag }), and more easily keep the gas and dust velocities well-coupled. For the drag coefficient we get, after averaging over the grain size distribution:\footnote{To obtain this result we have assumed that $\phi^2$ is approximately constant with respect to $a_{\rm gr}$ for a given grain type, which is typically the case \citep[][]{Weingartner2001_Photoelectric, Tielens2005}.}
\begin{equation}
    \mathcal{K}_\beta = 2 \times \dfrac{3 \rho_{\text{d},\beta} k_{\rm B}T}{4 \Bar{a}_{\text{gr},\beta} \rho_{\text{gr},\beta}} \, \sum_i n_i \left[\mathcal{G}'_{0}(s_{i,\beta}) +  Z_{i}^2 \phi_{\beta}^2\ln \lvert \Lambda_\beta/Z_i\rvert \, \mathcal{G}'_{2}(s_{i,\beta})\right] \, ,
\end{equation}
with $\Bar{a}_{\rm gr,\beta}$ given in Appendix~\ref{IR appendix}, and $\mathcal{G}' \equiv \mathcal{G}/\lvert\boldsymbol{u}_{\rm rel}\rvert$. The grain charge $Z_{\text{gr},\beta}$ of dust bin/type $\beta \in (\rm Sil,C,PAH)$ is determined from a balance between photoelectric + ion collisional charging, and electron collisional charging \citep[e.g.][]{Tielens2005, Draine2011}:
\begin{equation}
   \sum_i \pi \Bar{a}_{\text{gr},\beta}^2 n_i \mathcal{S}_i \left(\dfrac{8 k_{\rm B}T}{\pi m_i} \right)^{1/2} \Tilde{\mathcal{J}}(Z_{\rm gr,\beta}, Z_i, \Bar{a}_{\text{gr},\beta},T) + J_{\text{pe},\beta} - \pi \Bar{a}_{\text{gr},\beta}^2 n_{\rm e} \mathcal{S}_{\rm e} \left(\dfrac{8 k_{\rm B}T}{\pi m_{\rm e}} \right)^{1/2} \Tilde{\mathcal{J}}(Z_{\rm gr,\beta}, -1, \Bar{a}_{\text{gr},\beta},T) = 0 \, . \label{Grain charge equation}
\end{equation}
Here $i \in (\rm \textsc{H\,ii}, He\textsc{\,ii}, He\textsc{\,iii}, \textsc{D\,ii}, \textsc{C\,ii}, \textsc{O\,ii})$, and $\Tilde{\mathcal{J}}$ is a dimensionless factor capturing the Coulomb enhancement/suppression of the collisional rates for charged grains, given in \cite{DraineSutin1987}. The sticking coefficients of ions and electrons are taken to be $\mathcal{S}_i = 1$ and $\mathcal{S}_{\rm e} = 0.5 (1 - e^{-\Bar{a}_{\textrm{gr},\beta}/\ell_{\rm e}})$ with $\ell_{\rm e} = 10 \, \textrm{Å}$, respectively \citep{Umebayashi1983, Weingartner2001_Photoelectric}. The photoelectric charging rate $J_{\rm pe}$ (in units of $e$ per second) as a result of UV absorption is
\begin{align}
    J_{\textrm{pe},\beta} &=~ \pi \Bar{a}_{\textrm{gr},\beta}^2 \int \textrm{d}\nu \, \dfrac{4 \pi J(\nu)}{h\nu} Q_{\text{abs},\beta}(\nu) Y_{\textrm{pe},\beta}(h\nu,Z_{\textrm{gr},\beta},\Bar{a}_{\textrm{gr},\beta}) \label{J_pe} \\ &\simeq~ \pi \Bar{a}_{\textrm{gr},\beta}^2 \sum_{\mathcal{B} \in (\rm FUV, LW, EUV1, EUV2, EUV3)}  \dfrac{4 \pi J_{\mathcal{B}}}{E_{\mathcal{B}}} \, \Bar{Q}_{\textrm{abs},\mathcal{B},\beta} \,Y_{\textrm{pe},\beta}(E_{\mathcal{B}},Z_{\textrm{gr},\beta},\Bar{a}_{\textrm{gr},\beta}) \nonumber \\ &+~ \pi \Bar{a}_{\textrm{gr},\beta}^2 \, \dfrac{ c e_{\rm Ly\alpha}}{E_{\rm Ly\alpha}} \, Q_{\textrm{abs},\beta}(E_{\rm Ly\alpha}/h) \, Y_{\textrm{pe},\beta}(E_{\rm Ly\alpha}, Z_{\rm gr,\beta}, \Bar{a}_{\textrm{gr},\beta}) \nonumber \, .
\end{align}
On the second line we have discretized the integral as a sum over bands, and on the third line included the contribution of Ly$\alpha$ photons. For the non-Ly$\alpha$ bands, we use the photon number-weighted $Q_{\rm abs}$ (denoted $\Bar{Q}_{\rm abs}$), assuming a black-body spectrum with $T_{\rm eff} = 4.8 \times 10^4 \, \rm K$. The computed absorption efficiencies for silicate, graphite, and PAH dust, evaluated at $\Bar{a}_{\rm gr,\beta}$, are given in Table~\ref{Qabs for UV bands}. The photoelectric yield is taken to be $Y_{\rm pe} = Y_{\rm a} Y_2$, with $Y_{\rm a}$ being the yield following photon absorption, and $Y_2$ being the probability that the electron actually escape the grain. We compute $Y_{\rm a}(h\nu, a_{\rm gr})$ using the results for silicate and graphite grains by \cite{Kimura2016},\footnote{For PAHs we adopt the graphite results from \cite{Kimura2016}. Recently, \cite{Hrodmarsson2025} also applied the model of \cite{Kimura2016} to photoelectric charging of PAHs, finding good agreement with data.} whereas $Y_2$ is computed using the earlier modelling of \cite{Weingartner2001_Photoelectric}, but with work functions $W$ updated to be consistent with \cite{Kimura2016}. Our modelling of $Y_{\rm pe}$ closely follows the implementation in the \textsc{DustEM} code.\footnote{See the user guide at \url{https://www.ias.u-psud.fr/DUSTEM/dustem_code.php}, and recent related discussion in \cite{Hrodmarsson2025}.} We solve Eq.~(\ref{Grain charge equation}) for $Z_{\rm gr,\beta}$ using a bisection method, and then we further limit the charge to be above the limit for field emission, following \cite{Weingartner2001_Photoelectric}.

\begin{table*}
\centering
\caption{Dust photon number-weighted absorption efficiencies $Q_{\rm abs}$ for all UV bands, relevant for computing the photoelectric charging rate (Eq.~\ref{J_pe}), the PAH photodissociation rate (Eq.~\ref{PAH photodissociation mass loss rate}), and the photoelectric heating rate (Eq.~\ref{Final PEH rate}). The efficiencies have been evaluated at the characteristic grain sizes $a_{\rm gr,\beta} = \Bar{a}_{\rm gr,\beta}$ for each dust bin/type $\beta$. For the non-Ly$\alpha$ bands we adopt black-body photon number-weighted absorption efficiencies: $\Bar{Q}_{\rm abs,\mathcal{B}} \equiv [\int_{\mathcal{B}} \textrm{d}\nu \, B_\nu Q_{\rm abs}(\nu)/\nu] / [\int_{\mathcal{B}} \textrm{d}\nu \, B_\nu/\nu]$, with an assumed effective temperature $T_{\rm eff} = 4.8 \times 10^4 \, \rm K$.}
\begin{tabular}{c c c c c c c}
\hline
\hline
\noalign{\vskip 2pt}
Dust type & $\Bar{Q}_{\rm abs,FUV}$ & $\Bar{Q}_{\rm abs,LW}$ & $\Bar{Q}_{\rm abs,EUV1}$ & $\Bar{Q}_{\rm abs,EUV2}$ & $\Bar{Q}_{\rm abs,EUV3}$ & $Q_{\rm abs}(E_{\rm Ly\alpha}/h)$  \\
\noalign{\vskip 2pt}
\hline
\hline
\noalign{\vskip 2pt}

Silicate dust & $1.18$ & $1.12$ & $0.958$ & $0.933$ & $0.928$ & $1.20$   \\

Graphite dust & $1.33$ & $1.34$ & $1.07$ & $0.784$ & $0.623$ & $1.50$ \\ 

PAHs & $0.0479$ & $0.0989$ & $0.186$ & $0.0494$ & $0.0150$ & $0.0558$

\\
\noalign{\vskip 2pt}
\hline
\hline
\end{tabular}
\vspace{1 pt}\\
\raggedright
\label{Qabs for UV bands}
\end{table*}

\subsection{Dust growth and destruction}
\label{Dust growth destruction appendix}

Because Ly$\alpha$ feedback is sensitive to the presence of dust in and around \textsc{H\,ii} regions, we implement dust growth and destruction in \textsc{Lydion}. We consider dust growth by accretion of gas-phase metals, and dust destruction by sublimation, thermal and non-thermal sputtering, and photodissociation of PAHs. In a given cell, the dust bin density $\rho_{\rm d,\beta}$ evolves according to:
\begin{align}
    \dfrac{\partial \rho_{\rm d,\beta}}{\partial t} &=~ \dfrac{\rho_{\rm d,\beta}}{t_{\rm gr,0}}\left(1 - \dfrac{\rho_{\rm d,\beta}}{\rho_{\rm d,\beta} + \rho_{\rm X}} \right) - \dfrac{\rho_{\rm d,\beta}}{t_{\rm dest}} \, , \quad \quad \beta \in \rm (Sil,C,PAH) \,, \label{Dust destruction/growth ODE} \\ t_{\rm dest}^{-1} &=~ (t_{\rm subl}^{-1} + t_{\rm sput, th}^{-1} + t_{\rm sput,nth}^{-1} + t_{\rm chemsp}^{-1} + t_{\rm pd}^{-1})^{-1} \, ,
\end{align}
where $\rho_{\rm X}$ is the limiting gas-phase metal density for dust bin/type $\beta$, $t_{\rm gr,0}$ is the gas accretion time-scale in the low-dust depletion limit, and $t_{\rm dest}$ is the dust destruction time-scale considering sublimation ($t_{\rm subl}$), thermal sputtering ($t_{\rm sput.th}$), non-thermal sputtering ($t_{\rm sput,nth}$), chemical sputtering ($t_{\rm chemsp}$), and photodissociation ($t_{\rm pd}$). These time-scales are discussed in more detail below. We solve Eq.~(\ref{Dust destruction/growth ODE}) for each dust bin within the overall photo-thermochemistry subcycling, after photon absorption, using a BDF1 update, such that over a substep $\bullet \rightarrow \bullet \bullet$ of step-size $\Delta t_{\rm sub}$:
\begin{equation}
    \dfrac{\rho_{\rm d,\beta}^{\bullet \bullet} - \rho_{\rm d,\beta}^{\bullet}}{\Delta t_{\rm sub}} = \dfrac{\rho_{\rm d,\beta}^{\bullet \bullet}}{t_{\rm gr,0}}\left(1 - \dfrac{\rho_{\rm d,\beta}^{\bullet \bullet}}{\rho_{\rm d,\beta}^{\bullet} + \rho_{\rm X}^{\bullet}} \right) - \dfrac{\rho_{\rm d,\beta}^{\bullet\bullet}}{t_{\rm dest}} \, .
\end{equation}
In this equation we have used the conservation of $\rho_{\rm d} + \rho_{\rm X}$ to express this sum at the old substep. The above equation is a quadratic one for $\rho_{\rm d,\beta}^{\bullet \bullet}$, which can be readily solved:
\begin{equation}
    \rho_{\rm d,\beta}^{\bullet \bullet} = - \dfrac{\mathcal{A}}{2} + \sqrt{\left( \dfrac{\mathcal{A}}{2} \right)^2 + \mathcal{B}} \, , \quad \textrm{where} \quad \mathcal{A} = (\rho_{\rm d,\beta}^{\bullet} + \rho_{Z,\beta}^{\bullet})\left(\dfrac{t_{\rm gr,0}}{\Delta t_{\rm sub}} + \dfrac{t_{\rm gr,0}}{t_{\rm dest}} -1\right) \, , \quad \mathcal{B} = \dfrac{\rho_{\rm d,\beta}^{\bullet}(\rho_{\rm d,\beta}^{\bullet} + \rho_{Z}^{\bullet}) t_{\rm gr,0}}{\Delta t_{\rm sub}} \, .
\end{equation}
This update is accepted if the relative change in $\rho_{\rm d,\beta}$ was $\leq 10\%$ -- if not, the whole subcycle step is redone with a smaller step-size. If the update is accepted, the gas-phase metal density is also updated using conservation, $\rho_{Z}^{\bullet \bullet} = (\rho_{\rm d}^{\bullet } + \rho_Z^{\bullet}) - \rho_{\rm d}^{\bullet \bullet}$. 

The dust growth and destruction time-scales are estimated as follows:
\begin{itemize}[leftmargin=*]
    \item \textit{\textbf{Dust growth by gas accretion}}: We estimate the dust accretion growth time-scale as follows. A single grain of a given type (silicate, graphite, or PAH), and grain size $a_{\rm gr}$, accretes mass at a rate $\Dot{m}_{\rm gr} = 4\pi a_{\rm gr}^2 v_{\rm rel,X} \rho_{\rm X} \mathcal{S}_{\rm X}/f_{\rm X}$, where $v_{\rm rel,X} = \sqrt{k_{\rm B}T/2\pi m_{\rm X}}$, $\mathcal{S}_{\rm X}$ is the effective sticking coefficient, and X is the rate-limiting species, making up a fraction $f_{\rm X}$ of the grain mass \citep[][]{Hirashita2012}. For silicate and graphite grains, \cite{Hirashita2012} considers the rate-limiting species to be silicon and carbon, with $f_{\rm X} = 0.166$ and $f_{\rm X} = 1$, respectively. We treat PAHs similarly to graphite grains, with carbon being the rate-limiting species. The dust growth time-scale is then determined from $1/t_{\rm gr,\beta} \equiv (1/\rho_{\rm d,\beta}) \int \textrm{d}a_{\rm gr} \, (\partial n_{\rm gr,\beta}/\partial a_{\rm gr})  \, \Dot{m}_{\rm gr,\beta}$. This yields $1/t_{\rm gr,\beta} = 3 v_{\rm rel,X} \rho_{\rm X} \mathcal{S}_{\rm X} / f_{\rm X} \Bar{a}_{\rm gr,\beta} \rho_{\rm gr,\beta}$, where $\Bar{a}_{\rm gr} \equiv \langle a_{\rm gr}^3 \rangle / \langle a_{\rm gr}^2 \rangle$  (see Appendix~\ref{IR appendix} for values). The growth time-scale $t_{\rm gr,0,\beta}$ in the low-depletion limit (which is what enters Eq.~\ref{Dust destruction/growth ODE}) is then $t_{\rm gr,0,\beta} = f_{\rm X} \Bar{a}_{\rm gr,\beta} \rho_{\rm gr,\beta} / 3 v_{\rm rel,X} (\rho_{\rm d,\beta} + \rho_{\rm X}) \mathcal{S}_{\rm X} $. For the sticking coefficient $\mathcal{S}_{\rm X}$ we use a prescription based on the computational results from \cite{Bossion2024}. For the sticking onto amorphous carbon grains, \cite{Bossion2024} compute chemisorption sticking coefficients of $\mathcal{S}_{\rm CI} \sim 0.1 - 0.4$ depending on temperature, and $\mathcal{S}_{\rm CO} = 0$. While sticking of CO and CH can happen by physisorption, the CO and CH would be subject to thermal desorption and photodesorption near massive stars, and hence not contribute to net grain growth in regions of interest \citep{Bossion2024}.\footnote{\cite{Bossion2024} report a computed physisorption sticking coeffient dropping from $\mathcal{S}_{\rm CO} \sim 0.8$ to $0$ above $T = T_{\rm d} \sim \textrm{few} \times 100 \, \rm K$. \cite{Stadler2024} on the other hand measure the sticking coefficient experimentally, finding a low value of $\sim 0.05$ ($\sim 0.1$) on dry (ice-covered) carbon nanoparticles, at $10 \, \rm K$. We are not aware of measurements for sticking of CH (as a proxy for CH$_{\rm x}$), but the thermal desorption time-scale for graphite grains can be estimated as $t_{\rm desorb} \sim 10^{-12} \, \exp(E_{\rm D}\, / T_{\rm d}) \, \rm sec$ \citep[with $E_{\rm D} \simeq 645 \, \rm K$, see][]{Hasegawa1992}, which is $< 1 \, \rm yr$ for $T_{\rm d} \gtrsim 15 \, \rm K$. While ice-covered grains will have higher $E_{\rm D}$ \citep[$E_{\rm D} \simeq 2044 \, \rm K$, ][]{Sil2024}, photodesorption in \textsc{H\,ii} regions can quickly remove the ice in the first place \citep{Oberg2009, Hollenbach2009}. } In the absence of detailed data for ions, and to be conservative, we assume that \textsc{C\,ii} have the same sticking coefficient as \textsc{C\,i}. Thus, for the net growth of carbonaceous grains (graphite and PAH dust), we take the effective sticking coefficient (or accretion efficiency) to be:
    \begin{equation}
        \mathcal{S}_{\rm C} = \mathcal{S}_{\rm CI}(T)[x_{\rm CI} + x_{\rm CII} \Bar{\mathcal{J}}(Z_{\rm gr,C/PAH}, 1, \Bar{a}_{\rm gr,C/PAH},T)] \,,
    \end{equation}
    where $x_{\rm CI/CII} \equiv n_{\rm CI/CII}/n_{\rm C}$, and $\Bar{\mathcal{J}}$ accounts for the Coulomb modification of the capture cross-section for ions accreting onto charged grains \citep[see Eq.~\ref{Grain charge equation} and][]{DraineSutin1987}. While \cite{Bossion2024} computed $\mathcal{S}_{\rm CI}$ assuming that the dust and gas temperatures are equal, that is generally not the case in \textsc{Lydion}. However, preliminary results sugges that the sticking coefficient is most sensitive to the gas temperature $T$ (D. Bossion, private communication), and we therefore interpolate the data from \cite{Bossion2024} for $\mathcal{S}_{\rm CI}$ as a function of $T$. We are not aware of any recent data for the sticking coefficient of Si-species on silicate grains \citep[although that may soon change, see][]{Hansson2025}, and Si chemistry is not tracked in the current version of \textsc{Lydion}. We therefore set $\mathcal{S}_{\rm Si} = \mathcal{S}_{\rm C}$ for simplicity. We note however that the growth time-scale, for the grain size distribution assumed here, is mainly set by accretion onto carbonaceous dust, and so we do not expect the overall dust growth rate to be sensitive to the specific value of $\mathcal{S}_{\rm Si}$. 
\end{itemize}
We implement the following dust destruction processes in \textsc{Lydion}:
\begin{itemize}[leftmargin=*]
    \item \textit{\textbf{Dust destruction by sublimation}}: Hot grains can lose atoms by sublimation. We estimate the sublimation time-scale as \citep[][]{Guhathakurta1989, Waxman2000, Hoang2020}:
    \begin{align}
    t_{\rm subl,Sil}(T_{\rm d,Sil}) &=~ \dfrac{1}{3} \times 6.36 \times 10^{3} \left(\dfrac{\Bar{a}_{\rm gr,Sil}}{0.1 \, \rm \mu m} \right) e^{68100\, \textrm{K} \,(1/T_{\rm d,Sil} - 1/1800 \, \rm K )} \, \rm sec \, , \label{sublimation time silicate} \\ t_{\rm subl,C/PAH}(T_{\rm d,C/PAH}) &=~ \dfrac{1}{3} \times 1.36 \left(\dfrac{\Bar{a}_{\rm gr,C/PAH}}{0.1 \, \rm \mu m} \right) e^{81200 \, \textrm{K} \,(1/T_{\rm d,C/PAH} - 1/3000 \, \rm K )} \, \rm sec \, . \label{sublimation time graphite}
    \end{align}
    Here we have used the expressions for $a_{\rm gr}/\lvert \Dot{a}_{\rm gr} \rvert$ from \cite{Hoang2020}, with the extra factors of $1/3$ coming from $m_{\rm gr} / \lvert \Dot{m}_{\rm gr} \rvert = a_{\rm gr} / 3 \lvert \Dot{a}_{\rm gr} \rvert$. The above time-scales become relatively short ($< 1000 \, \rm yrs$) if $T_{\rm d} \gtrsim 1300 \, \rm K$ for silicate dust, and $T_{\rm d} \gtrsim 1500 \, \rm K$ for carbonaceous dust. 

    \item \textit{\textbf{Dust destruction by thermal sputtering}}: Thermal sputtering of larger grains (silicate and graphite) is implemented using the fits of total yields $Y_{\rm tot}$ (in $\mu$m yr$^{-1}$ cm$^3$) from \cite{Choban2026} to the data of \cite{Nozawa2006}. The destruction time-scale (in yrs) for silicate and graphite dust is then estimated as $t_{\rm sput,th} = (\Bar{a}_{\rm gr}/ 1 \, \mu\textrm{m}) / 3 n_{\rm HII} Y_{\rm tot}$. Thermal sputtering of silicate and graphite dust will only be important in gas with $T \gtrsim 10^5 - 10^6 \, \rm K$, usually not encountered in \textsc{H\,ii} regions in \textsc{Lydion} unless stellar wind feedback is simulated. Thermal sputtering of PAHs, on the other hand, can be efficient down to $\sim 10^4 \, \rm K$ \citep[][]{Micelotta2010, CALIMA2026}. The sputtering time-scale for PAHs is estimated using fits of yields from \cite{Micelotta2010} for PAHs of $N_{\rm C} = 100$ carbon atoms. Although $N_{\rm C} = 468 \, (\Bar{a}_{\rm gr,PAH}/10^{-3} \, \rm \mu m)^3 \simeq 83$ in \textsc{Lydion}, the yields of \cite{Micelotta2010} depend only weakly on $N_{\rm C}$ over $N_{\rm C} = 50 - 100$. The PAH sputtering time-scale is then estimated as $t_{\rm sput,th} = N_{\rm C} / (\mathcal{R}_{\rm electrons} + \mathcal{R}_{\rm H} + \mathcal{R}_{\rm He} + \mathcal{R}_{\rm C})$, where $\mathcal{R}_i$ is the sputtering rate (in s$^{-1}$) from species $i$. 

    \item \textit{\textbf{Dust destruction by non-thermal sputtering}}: Non-thermal sputtering occurs when there is a significant relative velocity between gas and grains ($\gtrsim 50 \, \rm km \, s^{-1}$), which can be marginally important in compact \textsc{H\,ii} regions near stars, where drift velocities can be large \citep[][]{Draine2011_HII, Akimkin2015, Akimkin2017}. We implement non-thermal sputtering using the fitted yields $Y_{\rm tot}$ from \cite{Hu2019} for silicate and carbonaceous grains, and estimate the destruction time-scale as $t_{\rm sput,nth} = \Bar{a}_{\rm gr} / 3 n_{\rm H} Y_{\rm tot}$. In absence of calculations for PAHs, we use the yields for carbonaceous grains for PAHs too. Given that thermal sputtering is efficient down to lower temperatures than for graphite grains \citep{Micelotta2010}, this probably underestimates the rate of non-thermal sputtering of PAHs. On the other hand, PAHs are more easily destroyed by photodissociation, and so in practice, our choice here is unlikely to have a big effect on PAH destruction.

    \item \textit{\textbf{Graphite dust destruction by chemical sputtering}}: Several authors have noted that carbonaceous dust in particular can be destroyed by chemical sputtering. Collisions with hydrogen atoms can produce hydrocarbons (e.g. methane, CH$_4$) that leave the grain, chipping away at the carbon mass of the grain. \cite{Draine1979_ChemicalSputtering}, while skeptical of earlier high estimated chemical sputtering yields, still found that graphite grains in dense ($n_{\rm H} \gtrsim 10^5 \, \rm cm^{-3}$) \textsc{H\,ii} regions could be destroyed by this process. \cite{BarNun1980} provide experimental evidence for efficient hydrocarbon formation on graphite, even at low dust temperatures. \cite{Lenzuni1995} surveyed more recent literature and provided updated yields, which were also recently implemented by \cite{Borderies2025}. We implement the carbon yield for graphite grains by \cite{Lenzuni1995}, which is a function of the graphite dust temperature ($T_{\rm d,C}$):
    \begin{equation}
        Y_{\rm chemsp}(T_{\rm d,C}) = 4.9 \times 10^{-3} \, \exp \left[ - \dfrac{\lvert T_{\rm d,C} - 580 \, \textrm{K} \rvert^{1.65}}{(125 \, \textrm{K})^{1.65}} \right] \, .
    \end{equation}
    Thus, in this treatment, chemical sputtering is most important in regions where the graphite dust is heated to $T_{\rm d,C} \sim \textrm{few} \times 100 \, \rm K$. Like \cite{Draine1979_ChemicalSputtering}, we assume that both \textsc{H\,i} and \textsc{H\,ii} can chemically sputter graphite grains, because \textsc{H\,ii} will capture an electron on impact. The graphite grain mass loss is then $\Dot{m}_{\rm gr} = -4 \pi a_{\rm gr}^2 v_{\rm rel,H}(n_{\rm HI} + n_{\rm HII}\Bar{\mathcal{J}})Y_{\rm chemsp} m_{\rm C}$, where $v_{\rm rel,H} = \sqrt{k_{\rm B} T / 2\pi m_{\rm H}}$, $m_{\rm C}$ is the mass of a carbon atom, and $\Bar{\mathcal{J}}$ accounts for Coulomb interactions with charged grains (see earlier discussion). This gives a grain destruction time-scale for graphite dust of $t_{\rm chemsp} = m_{\rm gr} / \lvert \Dot{m}_{\rm gr} \rvert$, where $m_{\rm gr} = 4\pi a_{\rm gr}^3 \rho_{\rm gr,C}/3$ in the graphite dust mass, and this is evaluated at the characteristic graphite grain size, $a_{\rm gr} = \Bar{a}_{\rm gr,C}$. We are not aware of chemical sputtering yields for PAHs, so we ignore it to be conservative, and note that photodissociation of PAHs (described below) will tend to destroy PAHs in \textsc{H\,ii} regions anyway.

    \item \textit{\textbf{PAH destruction by photodissociation}}: When PAHs absorb UV photons, and fail to dissipate the energy via IR emission, it can lead to loss of carbon atoms, primarily via C$_2$H$_2$ (acetylene). We adopt a model of PAH photodissociation based on the works of \cite{Micelotta2010}, \cite{Murga2019, Murga2020}, and \cite{CALIMA2026}, with some minor modifications. Following \cite{Micelotta2010}, let $\varphi_n$ be the probability of PAH fragmentation between the $n$:th and $(n+1)$:th IR photon emission. Then $\varphi_n = p_{n+1} \prod_{i=0}^n (1 - p_i)$, where $p_n$ is the probability of dissociation for the $n$:th event. The probability of the PAH losing an acetylene molecule after $n_{\rm max}$ IR photons emitted is then $P_{\rm C_2H_2} = \sum_{n=0}^{n_{\rm max}} \varphi_n$. To make the problem analytically tractable, \cite{Micelotta2010} considers $p_n$ to be constant, at some representative value $\Bar{p}$. Doing so yields:\footnote{In the analogous derivation by \cite{CALIMA2026}, there are small errors and typos. For $\Bar{p} \ll 1$, we get $P_{\rm C_2 H_2} \simeq (n_{\rm max} +1) \Bar{p}$. In contrast \cite{CALIMA2026} uses $1/(n_{\rm max} - 1)$ in front of $k_{\rm IR}$ in the main text (likely typo), and $1/(n_{\rm max} + 1)$ in front of $k_{\rm IR}$ in their appendix (which is also incorrect). However, in practice the latter has a negligible effect on the estimated $P_{\rm C_2 H_2}$ compared to the treatment here.}
    \begin{equation}
        P_{\rm C_2 H_2} = 1 - (1 - \Bar{p})^{n_{\rm max}+1} \, .
    \end{equation}
    For a PAH of $N_{\rm C}$ carbon atoms, we take $n_{\rm max} = N_{\rm C}/5$ \citep{CALIMA2026}, which is an excellent approximation to the results of \cite{Micelotta2010}. For the PAH dust bin in \textsc{Lydion}, $N_{\rm C} = 468 \, (\Bar{a}_{\rm gr,PAH}/10^{-3} \, \rm \mu m)^3 \simeq 83$ \citep[][]{Weingartner2001_Photoelectric}. Considering the competition between acetylene loss, loss of hydrogen atoms and molecules, photoionization, and IR emission, we further have:
    \begin{equation}
        \Bar{p} = \dfrac{k_{\rm C_2H_2} [1 - Y_{\rm a}(h\nu)]}{k_{\rm C_2H_2} + k_{\rm H } + k_{\rm H_2} + k_{\rm IR}} \, ,
    \end{equation}
    where $Y_{\rm a}$ is the raw photoelectric yield for the PAH (see the end of Appendix~\ref{Dust dynamics appendix} for details). The rate of IR photon emission is set to $k_{\rm IR} = 100 \, \rm s^{-1}$ following \cite{Micelotta2010}. Maria S. Murga (private communication) has verified that this choice for $k_{\rm IR}$ is an acceptable approximation (to within a factor $\sim 2$) for the adopted PAH size and EUV3 photons that we have found to be most consequential for PAH destruction \citep[for lower photon energies, also see][]{Murga2019}. The rest of the rates $k_i$ (in s$^{-1}$) are computed according to \citep[][]{Micelotta2010, CALIMA2026}:
    \begin{equation}
        k_i(T_{\rm av}) = \dfrac{k_{\rm B}T_{\rm av}}{h} \, \exp\left(1 + \dfrac{\Delta S_i}{\mathcal{R}} \right) \, \exp\left(- \dfrac{E_{0,i}}{k_{\rm B}T_{\rm av}} \right)
    \end{equation}
    where $\mathcal{R}$ is the universal gas constant, and we adopt the constants $\Delta S_i$ and $E_{0,i}$ from table 1 in \cite{Murga2020}. In accordance with the constant-$\Bar{p}$ approximation, the temperature $T_{\rm av}$ is a geometric average of the effective PAH temperature $T_{\rm eff}$ after $n = 0$ and $n = n_{\rm max}$ IR photon emissions, following absorption of a photon with energy $h\nu$, where:
    \begin{equation}
        T_{\rm eff}(h\nu, n) \simeq 2000 \, \textrm{K} \, \left[ \dfrac{(h\nu - n \Delta \epsilon_{\rm IR})/\textrm{1 eV}}{N_{\rm C}}\right]^{0.4} \left[1 - \dfrac{0.2 \,E_{0,i}}{(h\nu - n \Delta \epsilon_{\rm IR})} \right] \, ,
    \end{equation}
    with $\Delta \epsilon_{\rm IR} = 0.16 \, \rm eV$ being the average energy of the emitted IR photons \citep[][]{Micelotta2010}. After computing $P_{\rm C_2 H_2}$ for all UV photon bands (FUV and higher energy bands, and including Ly$\alpha$), the rate of mass loss from a PAH of size $a_{\rm gr}$ is:
    \begin{equation}
        \Dot{m}_{\rm PAH} \simeq - m_{\rm C_2 H_2} \, \left[\sum_{\mathcal{B}} \dfrac{4 \pi J_{\mathcal{B}}}{E_{\mathcal{B}}}\, \Bar{Q}_{\rm abs, \mathcal{B}} \pi a_{\rm gr}^2 P_{\rm C_2 H_2, \mathcal{B}} + \dfrac{c e_{\rm Ly\alpha}}{E_{\rm Ly\alpha}} \,Q_{\rm abs}(E_{\rm Ly\alpha} / h) \, \pi a_{\rm gr}^2 P_{\rm C_2 H_2, Ly\alpha} \right] \, , \label{PAH photodissociation mass loss rate}
    \end{equation}
    where $m_{\rm C_2 H_2}$ is the mass of an acetylene molecule, and the absorption efficiencies $Q_{\rm abs}$ are given in Table~\ref{Qabs for UV bands}. The destruction time-scale for a PAH of mass $m_{\rm PAH}$ is therefore $t_{\rm pd} = m_{\rm PAH} / \lvert\Dot{m}_{\rm PAH} \rvert$, where $m_{\rm PAH} = 4 \pi a_{\rm gr}^3 \rho_{\rm gr,C}/3$, and we evaluate this at the characteristic PAH size $a_{\rm gr} = \Bar{a}_{\rm gr,PAH}$.
\end{itemize}

\subsection{Photoelectric heating}
\label{PhotoelectricHeatingAppendix}

In this Appendix we discuss the implementation of photoelectric heating in \textsc{Lydion}. Since \textsc{Lydion} implements dust growth and destruction, and dust dynamics for individual bins, self-consistency demands that the photoelectric heating (PEH) rate cannot be computed using fits that assume a fixed grain size distribution. Thus, we estimate the PEH rate in a self-consistent manner as follows. The PEH rate contribution from grains of size $a_{\rm gr}$, charge $Z_{\rm gr}$, and number density $n_{\rm gr}$ is \citep{Weingartner2001_Photoelectric}:
\begin{align}
    \Gamma_{\rm pe}(a_{\rm gr}) &=~ n_{\rm gr} \pi a_{\rm gr}^2 \int \textrm{d}\nu \, \dfrac{4\pi J(\nu)}{h\nu} \, Q_{\rm abs}(a_{\rm gr},h\nu) Y_{\rm pe}(a_{\rm gr},Z_{\rm gr},h\nu)  \langle E\rangle \, , \\ \langle E\rangle &=~ \int_{E_{\rm min}}^{E_{\rm max}} \textrm{d}E \, f_{E}(E) E \, .
\end{align}
Here $Y_{\rm pe}$ is the photoelectric yield, and $f_{E}(E)$ is the probability distribution for the energy of the escaping electron. Furthermore, the integration limits are:
\begin{equation}
    E_{\rm min} = \begin{cases}
			0, & \text{if $Z_{\rm gr} \geq 0$}\\
            E_{\rm min}(Z_{\rm gr}<0,a_{\rm gr}), & \text{if $Z_{\rm gr} < 0$}
		 \end{cases} \,, \quad E_{\rm max} = h\nu - h\nu_{\rm pet} + E_{\rm min} \, .
\end{equation}
We can make a simplification as follows. \cite{Weingartner2001_Photoelectric} defines $f_E(E) = f_E^0(E)/Y_2$, where $Y_2 = \int_{\max[E_{\rm low},0]}^{E_{\rm high}} \textrm{d}E \, f_E^0(E)$ is the probability that a photoelectron escapes the grain. Furthermore, they take $Y_{\rm pe} = Y_{\rm a} Y_2$, where $Y_{\rm a}$ is the raw yield of electrons per absorbed photon (which may still fall back to the grain). Thus,  
\begin{align}
    \Gamma_{\rm pe}(a_{\rm gr}) &=~ n_{\rm gr} \pi a_{\rm gr}^2 \int \textrm{d}\nu \, \dfrac{4\pi J(\nu)}{h\nu} \, Q_{\rm abs}(a_{\rm gr},h\nu) Y_{\rm a}(a_{\rm gr},h\nu)  \langle \Tilde{E}\rangle \, , \\ \langle \Tilde{E}\rangle &=~ \int_{E_{\rm min}}^{E_{\rm max}} \textrm{d}E \, f_{E}^0(E) E \, .
\end{align}
In \textsc{Lydion} we use a discretized frequency band distribution, and so we make the approximation: 
\begin{equation}
    \Gamma_{\rm pe}(a_{\rm gr}) \simeq n_{\rm gr} \pi a_{\rm gr}^2 \sum_{\mathcal{B}} \, \dfrac{4\pi J_{\mathcal{B}}}{E_{\mathcal{B}}} \, \Bar{Q}_{\rm abs,\mathcal{B}}(a_{\rm gr},E_{\mathcal{B}}) Y_{\rm a}(a_{\rm gr},E_{\mathcal{B}})  \langle \Tilde{E}\rangle_{\mathcal{B}} \, .
\end{equation}
Next, we also average over grain sizes (for a given grain type or bin). We use the result $\langle n_{\rm gr} \pi a_{\rm gr}^2\rangle = (3/4) (\rho_{\rm d}/\rho_{\rm gr})/\Bar{a}_{\rm gr}$, where $\Bar{a}_{\rm gr} \equiv \langle a_{\rm gr}^3\rangle/\langle a_{\rm gr}^2\rangle$. Then the total PEH rate from all dust bins becomes:
\begin{equation}
    \Gamma_{\rm pe} \simeq \dfrac{3}{4}\sum_{\beta \in (\rm Sil, C, PAH)}\dfrac{\rho_{\textrm{d},\beta}}{\rho_{\textrm{gr},\beta} \Bar{a}_{\textrm{gr},\beta}} \sum_{\mathcal{B}} \, \dfrac{4\pi J_{\mathcal{B}}}{E_{\mathcal{B}}} \, \Bar{Q}_{\rm abs,\mathcal{B},\beta} Y_{\rm a}(\Bar{a}_{\textrm{gr},\beta},E_{\mathcal{B}})  \langle \Tilde{E}\rangle_{\mathcal{B}} \, , \label{Final PEH rate}
\end{equation}
We have further made the simplifying approximation that $Q_{\rm abs}Y_{\rm a}\langle \Tilde{E}\rangle$ is approximately constant, or at least mostly contributed by grains around the area-weighted mean grain size. This may be a rough approximation for large bins (silicates and graphite in our case), but since the heating is dominated by PAHs, which have a fairly narrow bin in \textsc{Lydion}, we do not expect this to cause great inaccuracy. To proceed, let us evaluate $\langle \Tilde{E}\rangle_{\mathcal{B}}$. \cite{Weingartner2001} gives the parabolic distribution:
\begin{equation}
    f_{E}^0(E) = \dfrac{6(E-E_{\rm low})(E_{\rm high} - E)}{(E_{\rm high}-E_{\rm low})^3} \, .
\end{equation}
It is easiest to consider positive and negatively charged grains separately. Starting with positive or neutral grains ($Z_{\rm gr} \geq 0$), then $E_{\rm min} = 0$, and $E_{\rm max} = \Delta \equiv h\nu_{\mathcal{B}} -h\nu_{\rm pet}$. So we have:
\begin{align}
    \langle \Tilde{E}\rangle_{\mathcal{B}}(Z_{\rm gr}\geq 0) &=~ \dfrac{6}{(E_{\rm high} - E_{\rm low})^3} \int_0^{\Delta} \textrm{d}E \, E(E-E_{\rm low}) (E_{\rm high} - E) \\ &=~ \dfrac{6}{(E_{\rm high} - E_{\rm low})^3} \left[\dfrac{1}{3}(E_{\rm high} + E_{\rm low})\Delta^3 - \dfrac{1}{4}\Delta^4  - \dfrac{1}{2}E_{\rm low}E_{\rm high}\Delta^2\right] \, .
\end{align}
We can simplify this result further. \cite{Weingartner2001_Photoelectric} gives $E_{\rm high} = \Delta$ for $Z_{\rm gr} \geq 0$. Thus, after a little algebra,
\begin{equation}
    \langle \Tilde{E}\rangle_{\mathcal{B}}(Z_{\rm gr}\geq 0) = \dfrac{\Delta^3(\Delta - 2E_{\rm low})}{2(\Delta - E_{\rm low})^3} \, , \quad E_{\rm low} = - \dfrac{(Z_{\rm gr} +1)e^2}{a_{\rm gr}} \, .
\end{equation}
where we have used $E_{\rm low}$ for $Z_{\rm gr} \geq 0$ given near eq.~(10) in \cite{Weingartner2001}. Next consider negative grains, $Z_{\rm gr} < 0$. In this case, \cite{Weingartner2001_Photoelectric} give $E_{\rm low} = E_{\rm min}$, and $E_{\rm high} = E_{\rm min} + \Delta = E_{\rm max}$, and so:
\begin{align}
    \langle \Tilde{E}\rangle_{\mathcal{B}}(Z_{\rm gr} < 0) &=~ \dfrac{6}{\Delta^3} \int_{E_{\rm min}}^{E_{\rm max}} \textrm{d}E \, E(E-E_{\rm min}) (E_{\rm max} - E) \\ &=~ \dfrac{6}{\Delta^3} \int_{0}^{\Delta} \textrm{d}x \, (x+E_{\rm min})x (\Delta - x) \, .
\end{align}
The integral (ignoring $6/\Delta^3$) evaluates to $(1/12)\Delta^4 + (1/6)E_{\rm min}\Delta^3$, and so we arrive at the simple result:
\begin{equation}
    \langle \Tilde{E}\rangle_{\mathcal{B}}(Z_{\rm gr} < 0) = E_{\rm min} + \dfrac{1}{2}\Delta \,, \quad \textrm{where} \quad E_{\rm min} = - \dfrac{(Z_{\rm gr} + 1)e^2}{a_{\rm gr}} \left[ 1 + \left( \dfrac{27 \, \text{\AA}}{a_{\rm gr}}\right)^{0.75} \right]^{-1} \, ,
\end{equation}
where we have used the result for $E_{\rm min}(Z_{\rm gr}<0)$ from eq.~(7) in \cite{Weingartner2001_Photoelectric}. The above result has a simple interpretation: for the assumed parabolic distribution, the escaping electron has an average energy $E_{\rm min} + \Delta /2$.

\section{Photochemistry \& cooling}
\label{Photochemistry appendix}

\subsection{Molecular Hydrogen Photochemistry \& Cooling}
\label{Mol hydrogen chemistry appendix}

Molecular hydrogen is an important coolant in primordial gas where the first stars form, and can also destroy Ly$\alpha$ photons. Here we briefly outline the modelling of H$_2$ chemistry in \textsc{Lydion}. The H$_2$ fraction $x_{\rm H_2} \equiv n_{\rm H_2}/n_{\rm H}$, with $n_{\rm H} \equiv n_{\rm HI} + n_{\rm HII} + 2n_{\rm H_2}$, evolve according to:
\begin{align}
    \Dot{x}_{\rm H_2} ~&=~ [\mathcal{R}_{\rm dust, H_2} + k_{\rm H^-,H_2}n_{\rm e} + k_{\rm 3B,H_2}n_{\rm HI}(n_{\rm HI} + n_{\rm H_2}/8)]x_{\rm HI} \label{xH2 dot} \\ ~&-~ (k_{\rm diss, HI}n_{\rm HI} + k_{\rm diss, HII}n_{\rm HII} +  k_{\rm diss, H_2}n_{\rm H_2} + k_{\rm diss, e}n_{\rm e} + \Gamma_{\rm LW}^{\rm H_2} + \Gamma_{\rm ion}^{\rm H_2} ) x_{\rm H_2} \nonumber \, .
\end{align}
In this equation, $\mathcal{R}_{\rm dust, H_2}$, $k_{\rm H^-, H_2}$, and $k_{\rm 3B,H_2}$ are the rate coefficients for H$_2$ formation on dust grains, in gas via H$^{-}$, and in 3-body reactions, respectively. Finally, $k_{\text{diss},j}$, $\Gamma_{\rm LW}^{\rm H_2}$, and $\Gamma_{\rm ion}^{\rm H_2}$ are the rate coefficients for collisional dissociation of H$_2$ by species $j$, Lyman-Werner photodissociation of H$_2$, and photoionization of H$_2$, respectively.

The gas-phase formation of H$_2$ primarily takes place via H$^-$, in two steps:
\begin{alignat}{2}
  \mathrm{H}+e^-          ~&\;\to\; &~ \mathrm{H}^- + h\nu &\quad \text{\citep[rate coefficient $k_1$, fit from][]{Galli1998}}\,,\\
  \mathrm{H}+\mathrm{H}^- ~&\;\to\; &~ \mathrm{H}_2 + e^- &\quad \text{\citep[rate coefficient $k_2$, fit from][]{Kreckel2010}}\,.
\end{alignat}
The hydrogen anion H$^-$ can also be destroyed before producing H$_2$, including from photodetachment by photons with energy $h\nu > 0.754 \, \rm eV$ \citep[][]{McLaughlin2017}, and in collisions with \textsc{H\,i}, \textsc{H\,ii}, and $e^-$:
\begin{alignat}{2}
  \mathrm{H}^- + h\nu ~         &\;\to\; &
        ~\mathrm{H} + e^-            &\quad (\text{rate coefficient }\Gamma_{\mathrm{pd}}^{\mathrm{H}^-})\,,\\
  \mathrm{H}^- + \mathrm{H} ~   &\;\to\; &
        ~\mathrm{H} + \mathrm{H} + e^- &\quad \text{\citep[rate coefficient $k_{16}$, fit from][]{Glover2010}}\,,\\
  \mathrm{H}^- + \mathrm{H}^+ ~ &\;\to\; &
        ~\mathrm{H} + \mathrm{H}      &\quad \text{\citep[rate coefficient $k_5$, fit from][]{Glover2010}}\,,\\
  \mathrm{H}^- + \mathrm{H}^+ ~ &\;\to\; &
        ~\mathrm{H}_2^+ + e^-      &\quad \text{\citep[rate coefficient $k_{17}$, fit from][]{Glover2010}}\,,\\
  \mathrm{H}^- + e^- ~ &\;\to\; &
        ~\mathrm{H} + \mathrm{H} + e^- + e^-      &\quad  \text{\citep[rate coefficient $k_{15}$, fit from][]{Glover2010}}\, .
\end{alignat}
Since the dominant rates governing H$^-$ are rapid, we can safely assume that it has the equilibrium abundance \citep[e.g.][]{Tegmark1997, Park2021, Hopkins2023, Nebrin2023_starbursts}:
\begin{equation}
    n_{\rm H^-} = \dfrac{k_1 n_{\rm HI}n_{\rm e}}{(k_2 + k_{16})n_{\rm HI} + (k_5 + k_{17})n_{\rm HII} + k_{15}n_{\rm e} + \Gamma_{\rm pd}^{\rm H^-}} \, .
\end{equation}
The H$_2$ formation rate via the H$^-$ channel is $k_{2}n_{\rm HI}n_{\rm H^-}$ (cm$^{-3}$ s$^{-1}$), and so the corresponding effective rate coefficient in Eq.~(\ref{xH2 dot}) becomes:
\begin{equation}
    k_{\rm H^-,H_2} = \dfrac{k_2 k_1 n_{\rm HI}}{(k_2 + k_{16})n_{\rm HI} + (k_5 + k_{17})n_{\rm HII} + k_{15}n_{\rm e} + \Gamma_{\rm pd}^{\rm H^-}} \, .
\end{equation}
For the formation of H$_2$ in 3-body reactions, we use the fit $k_{\rm 3B,H_2} = 6 \times 10^{-32}\, T^{-1/4} + 2\times 10^{-31}\, T^{-1/2}$, taken from \cite{Forrey2013}. The H$_2$ formation rate coefficient for grains, $\mathcal{R}_{\rm dust, H_2}$ (in s$^{-1}$), is \citep[e.g.][]{Hollenbach1979, Draine2011}:
\begin{align}
    \mathcal{R}_{\rm dust,H_2} &=~ \dfrac{1}{2} \sum_{\beta \in (\textrm{Sil,C,PAH})} \int \textrm{d}a_{\rm gr} \, \dfrac{\partial n_{\textrm{gr},\beta}}{\partial a_{\rm gr}} \, \pi a_{\rm gr}^2 \left( \dfrac{8k_{\rm B}T}{\pi m_{\rm H}} \right)^{1/2} \mathcal{S}_{\textrm{HI},\beta} \epsilon_{\textrm{H}_2,\beta} \\ &=~ \dfrac{3}{8} \sum_{\beta \in (\textrm{Sil,C,PAH})} \dfrac{1}{\Bar{a}_{\textrm{gr},\beta}} \dfrac{\rho_{\textrm{d},\beta}}{\rho_{\textrm{gr},\beta}} \left( \dfrac{8k_{\rm B}T}{\pi m_{\rm H}} \right)^{1/2} \mathcal{S}_{\textrm{HI},\beta} \epsilon_{\textrm{H}_2,\beta} \, \nonumber ,
\end{align}
where the sum is over silicate, graphite, and PAH dust grains, $\Bar{a}_{\rm gr} = \langle a_{\rm gr}^3\rangle / \langle a_{\rm gr}^2\rangle$  is the surface area-weighted grain size, $\rho_{\rm d}$ the dust density, and $\rho_{\rm gr}$ the grain density. Values for these parameters for the assumed dust model in \textsc{Lydion}, and each dust bin/type $\beta$, can be found in Appendix~\ref{IR appendix}. Finally, $\mathcal{S}_{\rm HI}$ and $\epsilon_{\rm H_2}$ are the sticking coefficient and H$_2$ formation efficiency, respectively. In general, these are expected to vary depending on grain type, but involves a wide range of complicated and uncertain physics \citep[e.g.][]{Cazaux2004, Cazaux2010, Grieco2023, Nebrin2023_starbursts}. For simplicity we adopt the model of \cite{Hollenbach1979} for all types of grains \citep[see also e.g.][]{GloverJappsen2007, Hopkins2023}:
\begin{equation}
    \mathcal{S}_{\textrm{HI},\beta} \epsilon_{\textrm{H}_2,\beta} = \dfrac{1}{[1 + 0.4\,(T_{\rm 2} + T_{\textrm{d},\beta,2})^{1/2} + 0.2 \, T_2 + 0.08 \, T_2^2][1 + 10^4 \, \exp(-600/T_{\textrm{d},\beta})]} \, ,
\end{equation}
where $T_2 \equiv T/100 \, \rm K$, and $T_{\textrm{d},\beta}$ is the dust temperature for grain bin/type $\beta$ (see Appendix~\ref{Dust dynamics appendix}). For the collisional dissociation rates, we use fits compiled in \cite{Glover2010}. The Lyman-Werner photodissociation rate of H$_2$ is:
\begin{equation}
    \Gamma_{\rm LW}^{\rm H_2} = f_{\rm sh,H_2}\,\dfrac{4 \pi J_{\rm LW} \sigma_{\rm H_2,LW}} {\mathcal{E}_{\rm LW}} \, ,
\end{equation}
where $\mathcal{E}_{\rm LW} = 12.26 \, \rm eV$ is the average stellar photon energy in the Lyman-Werner band \citep[][]{Kannan2020}, and $\sigma_{\rm H_2,LW} = 2.47 \times 10^{-18} \, \rm cm^2$ is the band-averaged cross-section in the optically thin limit \citep[][]{Baczynski2015}.\footnote{The effective cross-section from \cite{Baczynski2015} is derived from the optically thin Lyman-Werner photodissociation rate in a straightforward manner, but is a factor $\sim 10$ larger than the cross-section given in \cite{Kannan2020}. To obtain the correct Lyman-Werner photodissociation rate in the optically thin limit, we use the figure from \cite{Baczynski2015}.} When the column density of H$_2$ becomes large, there is significant line overlap, and hence further absorption over what we can track explicitly in our band-averaged approach. We account for this using the self-shielding fit $f_{\rm sh,H_2}$ from \cite{Draine1996}, using a local Sobolev-like estimate of the H$_2$ column density (see Eq.~\ref{Sobolev path length}). 

H$_2$ can also be photoionized by photons with $h\nu > 15.2 \, \rm eV$ \citep[e.g.][]{Baczynski2015}, and by cosmic rays \citep[e.g.][]{Gong2017}. The corresponding rate coefficient is:
\begin{equation}
    \Gamma_{\rm ion}^{\rm H_2} = 4\pi \left( \dfrac{J_{\rm EUV1}\sigma_{\rm H_2,EUV1}}{E_{\rm EUV1}} + \dfrac{J_{\rm EUV2}\sigma_{\rm H_2,EUV2}}{E_{\rm EUV2}} + \dfrac{J_{\rm EUV3}\sigma_{\rm H_2,EUV3}}{E_{\rm EUV3}}\right)\, ,
\end{equation}
where the average stellar photon energies are $(E_{\rm EUV1}, E_{\rm EUV2}, E_{\rm EUV3}) = (18.01,29.89,56.85) \, \rm eV$, and the band-averaged photoionization cross-sections are $(\sigma_{\rm H_2,EUV1}, \sigma_{\rm H_2,EUV2}, \sigma_{\rm H_2,EUV3}) = (5.09,2.42,0.32) \times 10^{-18} \, \rm cm^2$ \citep[][]{Kannan2020}. Finally, as noted above, H$^-$ can be photodetached by photons with $h\nu > 0.754 \, \rm eV$. The corresponding rate is then approximately:
\begin{align}
    \Gamma_{\rm pd}^{\rm H^-} ~&=~  4\pi \left(
    \dfrac{J_{\rm opt}\sigma_{\rm H^-,opt}}{E_{\rm opt}} 
    +
    \dfrac{J_{\rm FUV}\sigma_{\rm H^-,FUV}}{E_{\rm FUV}} 
    + \dfrac{J_{\rm LW}\sigma_{\rm H^-,LW}}{E_{\rm LW}} + \sum_{j = 1}^3  \dfrac{J_{\text{EUV}j}\sigma_{\text{H}^-,\text{EUV}j}}{E_{\text{EUV}j}}\right) \\ ~&+~ \dfrac{c e_{\rm Ly\alpha}\sigma_{\rm H^-,Ly\alpha}}{E_{\rm Ly\alpha}} + \Gamma_{\rm pd,CMB}^{\rm H^-}  \, , \nonumber 
\end{align}
where we have expressed the band-averaged Ly$\alpha$ intensity in terms of the energy density, $\int \textrm{d}\nu \, J(\nu) =  c e_{\rm Ly\alpha}/4\pi$. Since the Ly$\alpha$ line is narrow relative to variations in the cross-section, we adopt the value line center,  $\sigma_{\rm H^-,Ly\alpha} = 5.477\times10^{-18} \, \rm cm^2$ \citep[][]{McLaughlin2017}, and also take $E_{\rm Ly\alpha} = h\nu_{\rm Ly\alpha}$. We have also taken into account photodetachment by CMB photons ($\Gamma_{\rm pd,CMB}^{\rm H^-}$) following \cite{Schauer2019} (see their eqs. A1--A2), which can be important at very high redshifts \citep[e.g.][]{Tegmark1997, Hirano2015_earlystructure}.
\begin{figure*}
\centering
\includegraphics[width=0.6\textwidth]{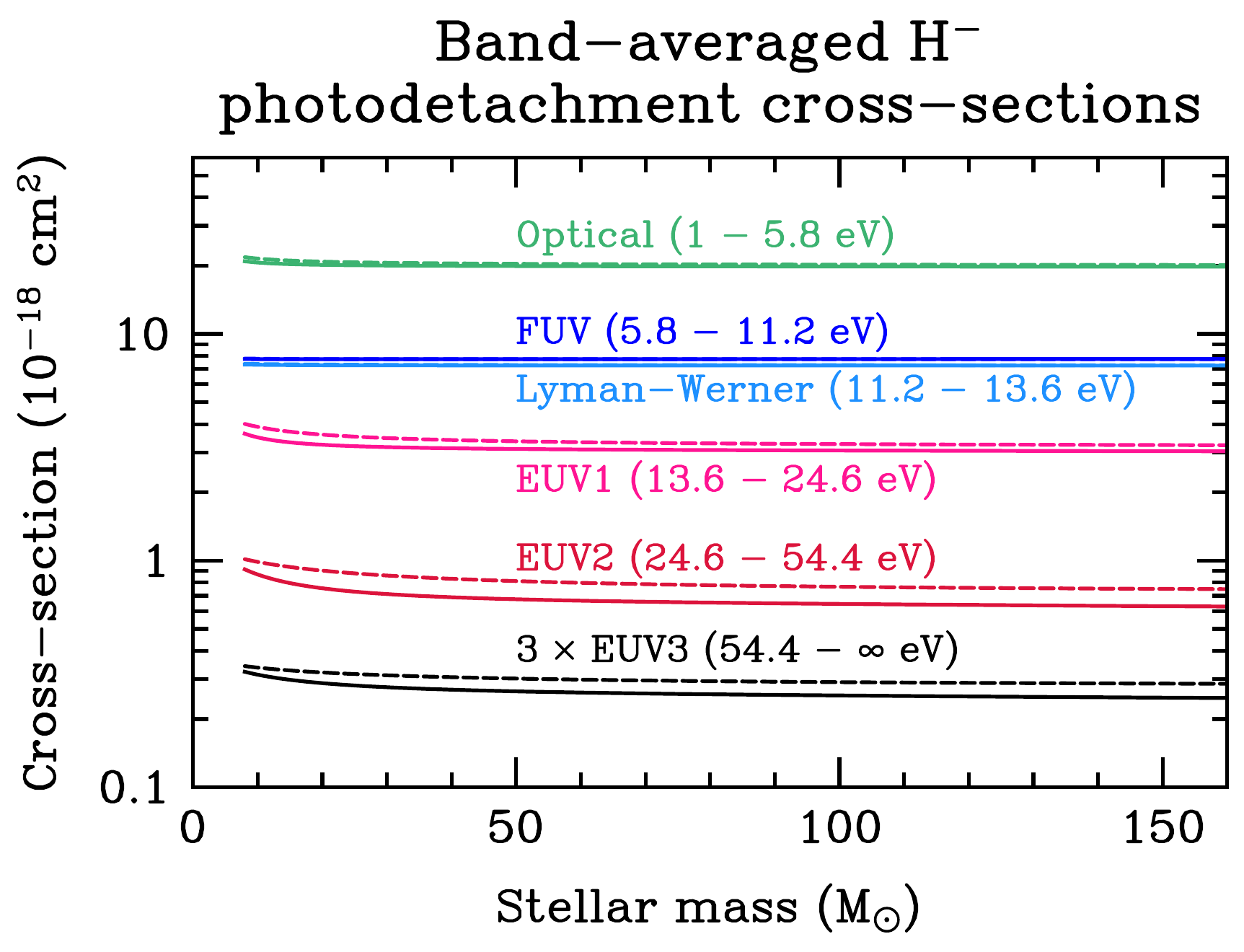}
\caption{The band-averaged photodetachment cross-sections, $\sigma_{\rm H^-,B} \equiv [\int_{\rm B} \textrm{d}\nu \, B_\nu(T_{\rm eff}) \sigma_{\rm H^-}(\nu) / h\nu ] / [\int_{\rm B} \textrm{d}\nu \, B_\nu(T_{\rm eff})  / h\nu]$, assuming black-body spectra for (nearly) zero-metallicity stars (solid lines), and for metal-poor $Z_{\star}/Z_{\odot} = 0.01$ stars (dashed lines). Data for the cross-section is taken from \citet{McLaughlin2017}, and  $T_{\rm eff}$ derived from the fits in \citet{Tanikawa2020}. Note that the cross-section for the Lyman-Werner band is higher than the cross-section at Ly$\alpha$ line center, $\sigma_{\rm H^-,Ly\alpha} = 5.477\times10^{-18} \, \rm cm^2$. This is because of resonances in the cross-section around $\sim 11 \, \rm eV$ \citep[see fig. 1 in][]{McLaughlin2017}.}
\label{Hminus cross sec}
\end{figure*}
H$^-$ photodetachment feedback will be most important in primordial gas, so we evaluate the rest of the band-averaged cross-sections using the cross-section data from \cite{McLaughlin2017}, and assuming black-body spectra with effective temperatures derived from \cite{Tanikawa2020}, for zero-age main sequence stars with $Z_{\star}/Z_{\odot} = 10^{-8}$. For comparison, we also compute the cross-sections for stars with $Z_{\star}/Z_{\odot} = 0.01$. The result is plotted in Fig.~\ref{Hminus cross sec}, and found to be very weakly dependent on the stellar mass and metallicity, especially in the non-ionizing bands which are most important for photodetachment. We therefore adopt the values at $m_{\star} = 30 \, \rm M_{\odot}$ and $Z/Z_{\odot} = 10^{-8}$ for all stars:
\begin{equation}
    (\sigma_{\rm H^-,opt}, \, \sigma_{\rm H^-,FUV}, \, \sigma_{\rm H^-,LW}, \, \sigma_{\rm H^-,EUV1}, \, \sigma_{\rm H^-,EUV2}, \, \sigma_{\rm H^-,EUV3}) = (20, \,7.7,\, 7.3,\,3.2,\,0.71,\,0.092) \times 10^{-18} \, \rm cm^2 \, .
\end{equation}
We model the H$_2$-cooling rate $\Lambda_{\rm H_2}$ (in erg s$^{-1}$ cm$^{-3}$) using the same functional form as in \cite{Galli1998}:
\begin{equation}
    \Lambda_{\rm H_2}(T,n_{\rm H_2},n_{\rm HI}, n_{\rm HeI},n_{\rm HII},n_{\rm e}) = \dfrac{\Lambda_{\textrm{H}_2,n\rightarrow0}}{1 + \Lambda_{\textrm{H}_2,n\rightarrow0}/\Lambda_{\rm H_2,LTE}} \, ,
\end{equation}
where the LTE cooling rate $\Lambda_{\rm H_2,LTE}(T,n_{\rm H_2})$ (important at $n_{\rm H} \gtrsim 10^4 \, \rm cm^{-3}$) is taken from \cite{Hollenbach1979}. For the low-density regime, we have several contributions:
\begin{equation}
    \Lambda_{\textrm{H}_2,n\rightarrow0} = \Lambda_{\rm H_2,HI}(T,n_{\rm H_2},n_{\rm HI}) + \Lambda_{\rm H_2,HeI}(T,n_{\rm H_2},n_{\rm HeI}) + \Lambda_{\rm H_2,HII}(T,n_{\rm H_2},n_{\rm HII}) + \Lambda_{\rm H_2,e}(T,n_{\rm H_2},n_{\rm e}) \, .
\end{equation}
We take $\Lambda_{\rm H_2,HI}$ from \cite{Galli1998}, and the rest of the contributions from \cite{Glover2008}. Although \cite{Glover2008} updates the fit for $\Lambda_{\rm H_2,HI}$ over \cite{Galli1998}, later more accurate calculations have found the fit from \cite{Galli1998} to be slightly more accurate at relevant temperatures \citep[see results and discussion in][]{Coppola2019, Flower2021, Nebrin2023_starbursts}.

\subsection{Deuterium Photochemistry \& Cooling}
\label{Deuterium chemistry appendix}

\begin{table*}
\centering
\caption{Reactions and corresponding rate coefficients for deuterium species. The numbering of rates $k_{50}$--$k_{55}$ follows \cite{McGreer2008}. The units are cm$^{3}$ s$^{-1}$ for collisional rate coefficients ($k$), and $T_{300} \equiv T/300 \, \rm K$.  }
\begin{tabular}{c c c c c}
\hline
\hline
Symbol & Reaction & Rate coefficient & Reference for fit & Original data  \\

\hline
\hline
\vspace{-10 pt}
\\
\vspace{4 pt}

$\Gamma_{\rm ion}^{\rm DI}$  & $\textrm{D} + h\nu \rightarrow \textrm{D}^+ + e^-$ & Same as for \textsc{H\,i} & \cite{Kannan2020} &  \\

$\Gamma_{\rm LW}^{\rm HD}$  & $\textrm{HD} + h\nu \rightarrow \textrm{H}+\textrm{D}$ & See Eq.~\ref{Gamma LW HD} & See text &  \\

$k_{\rm rec}$  & $\textrm{D}^+ + e^- \rightarrow \textrm{D} + h\nu$ & Same as for \textsc{H\,i} & \cite{Rosdahl2013} &  \\

$k_{\rm coll}$  & $\textrm{D} + e^- \rightarrow \textrm{D}^+ + e^-$ & Same as for \textsc{H\,i} & \cite{Rosdahl2013} &  \\


$k_{50}$  & $\textrm{D} + \textrm{H}^+ \rightarrow \textrm{D}^+ + \textrm{H}$ & $2.18 \times 10^{-9} \, T_{300}^{0.355} e^{-43/T}$ & \cite{Faure2024} &  \cite{Savin2002}  \\

$k_{51}$  & $\textrm{D}^+ + \textrm{H} \rightarrow \textrm{D} + \textrm{H}^+$ & $2.18 \times 10^{-9} \, T_{300}^{0.355}$ & \cite{Faure2024} &  \cite{Savin2002}   \\

$k_{52}$  & $\textrm{D}^+ + \textrm{H}_2 \rightarrow \textrm{H}^+ + \textrm{HD}$ & $1.50 \times 10^{-9}$  &  \cite{Faure2024} & \cite{Honvault2013,Honvault2013_err}   \\

$k_{53}$  & $\textrm{H}^+ + \textrm{HD} \rightarrow \textrm{D}^+ + \textrm{H}_2$ & $8.57 \times 10^{-10} \, e^{-405/T}$ &  \cite{Faure2024}  & \cite{Gonzales2022}  \\

$k_{54}$  & $\textrm{D} + \textrm{H}_2 \rightarrow \textrm{H} + \textrm{HD}$ & See reference for fit  & \cite{Glover2008} & \cite{Mielke2003}  \\

$k_{55}$  & $\textrm{H} + \textrm{HD} \rightarrow \textrm{D} + \textrm{H}_2$ & $5.25 \times 10^{-11} \, e^{-4430/T + 173900/T^2}$  &  \cite{GalliPalla2002}$^\dagger$ & \cite{Shavitt1959}   \\

\\

\hline
\hline
\end{tabular}
\vspace{1 pt}\\
\raggedright
$^\dagger$: Following \cite{McGreer2008}, we use $T \rightarrow \max[T,\, 100\,\rm K]$ in this fit to avoid an unphysically large rate coefficient in cold gas.
\label{Deuterium chemistry rates}
\end{table*}

In this Appendix we discuss the treatment of deuterium photochemistry and cooling. We mostly follow \cite{McGreer2008} in terms of the chemical rates and species considered, but also add the effects of Lyman-Werner photodissociation of HD, and photoionization of \textsc{D\,i}. We also use updated rate coefficients for many of the reactions. The reaction rates considered, and the fits adopted, can be found in Table~\ref{Deuterium chemistry rates}. With these rates, the fractional abundances $x_{\rm DI} \equiv n_{\rm DI}/n_{\rm D}$, $x_{\rm DII} \equiv n_{\rm DII}/n_{\rm D}$, $x_{\rm HD} \equiv n_{\rm HD}/n_{\rm D}$ evolve according to:
\begin{align}
    \Dot{x}_{\rm DI} &=~ (k_{\rm rec} n_{\rm e} + k_{51}n_{\rm HI}) x_{\rm DII} + (k_{55}n_{\rm HI} + \Gamma_{\rm LW}^{\rm HD}) x_{\rm HD} - (\Gamma_{\rm ion}^{\rm DI} + k_{\rm coll}n_{\rm e} + k_{50}n_{\rm HII} + k_{54}n_{\rm H_2})x_{\rm DI} \, , \label{xdot DI} \\ \Dot{x}_{\rm DII} &=~ (\Gamma_{\rm ion}^{\rm DI} + k_{\rm coll}n_{\rm e} + k_{50} n_{\rm HII}) x_{\rm DI} + k_{53}n_{\rm HII} x_{\rm HD} - (k_{\rm recB}n_{\rm e} + k_{51} n_{\rm HI} + k_{52}n_{\rm H_2}) x_{\rm DII} \, , \label{xdot DII} \\ \Dot{x}_{\rm HD} &=~ k_{52}n_{\rm H_2} x_{\rm DII} + k_{54} n_{\rm H_2} x_{\rm DI} - (\Gamma_{\rm LW}^{\rm HD}  + k_{53}n_{\rm HII} + k_{55}n_{\rm HI})x_{\rm HD} \, \label{xdot HD}.
\end{align}
This is a set of (approximately) linearly coupled ODE's, which can therefore be solved in a fairly straightforward manner. To proceed, we let $\boldsymbol{x} \equiv (x_{\rm DI}, x_{\rm DII}, x_{\rm HD})^{\rm T}$, and write Eqs.~(\ref{xdot DI})--(\ref{xdot HD}) in matrix form, $\Dot{\boldsymbol{x}} = \mathbf{A}\boldsymbol{x}$, where:
\begin{equation}
    \mathbf{A} = \begin{pmatrix}
    - (\Gamma_{\rm ion}^{\rm DI} + k_{\rm coll}n_{\rm e} + k_{50}n_{\rm HII} + k_{54}n_{\rm H_2}) & k_{\rm rec} n_{\rm e} + k_{51}n_{\rm HI} & k_{55}n_{\rm HI} + \Gamma_{\rm LW}^{\rm HD} \\
    \Gamma_{\rm ion}^{\rm DI} + k_{\rm coll}n_{\rm e} + k_{50} n_{\rm HII} & -(k_{\rm recB}n_{\rm e} + k_{51} n_{\rm HI} + k_{52}n_{\rm H_2}) & k_{53}n_{\rm HII}  \\
    k_{54} n_{\rm H_2} & k_{52}n_{\rm H_2} & - (\Gamma_{\rm LW}^{\rm HD} + k_{53}n_{\rm HII} + k_{55}n_{\rm HI}) 
  \end{pmatrix} \, .
\end{equation}
For stability, we use the BDF1 method to evolve $\boldsymbol{x}$ over a subcycle time-step $\Delta t_{\rm sub}$, so $(\boldsymbol{x}^{n+1} - \boldsymbol{x}^n)/\Delta t_{\rm sub} = \mathbf{A} \boldsymbol{x}^{n+1}$, which yields the solution:
\begin{equation}
    \boldsymbol{x}^{n+1} = (\boldsymbol{1} - \mathbf{A}\Delta t_{\rm sub})^{-1} \boldsymbol{x}^n \, . \label{Deuterium BDF1 update}
\end{equation}
Since $\mathbf{A}$ is only a $3 \times 3$ matrix, the matrix inversion can be performed efficiently. 

To compute the Lyman-Werner photodissociation rate of HD, we assume an effective cross-section $\sigma_{\rm HD,LW} = (1.55/1.39) \times \sigma_{\rm H_2,LW}$ \citep[based on][]{Wolcott2011_HD}, and take into account self-shielding by HD ($f_{\rm sh,HD}$), and shielding of HD by H$_2$ ($f_{\rm sh, H_2, HD}$):\footnote{As noted in the main text, we also take into account shielding (i.e. absorption) by \textsc{H\,i} in the RT equation for the Lyman-Werner intensity $J_{\rm LW}$.}
\begin{equation}
    \Gamma_{\rm LW}^{\rm HD} = f_{\rm sh,HD} f_{\rm sh, H_2, HD} \,\dfrac{4 \pi J_{\rm LW} \sigma_{\rm HD,LW}} {E_{\rm LW}} \, , \label{Gamma LW HD}
\end{equation}
where $E_{\rm LW} = 12.26 \, \rm eV$ is the average stellar photon energy in the Lyman-Werner band \citep[][]{Kannan2020}. Since the HD self-shielding factor $f_{\rm sh,HD}$ is almost identical to that of H$_2$,  $f_{\rm sh,H_2}$ \citep[][]{Wolcott2011_HD, Safranek2012}, we use the same fitting formula for both, taken from \cite{Draine1996}. For $f_{\rm sh,H_2,HD}$ we use the fit from \cite{Wolcott2011_HD} (see their eq. 12). Since the total deuterium abundance is so small, $n_{\rm D}/n_{\rm H} \simeq 2.53 \times 10^{-5}$ \citep[][]{Cooke2018, Planck2018}, we neglect the effect of deuterium in the rate equations for the other tracked species for simplicity (although we do track the small contribution of \textsc{D\,ii} to $n_{\rm e}$). For the HD cooling rate $\Lambda_{\rm HD}$ (in erg s$^{-1}$ cm$^{-3}$), we use the fit:
\begin{equation}
    \Lambda_{\rm HD}(T,n_{\rm HD}, n_{\rm HI}) = \dfrac{\Lambda_{\textrm{HD},n\rightarrow0}}{1 + \Lambda_{\textrm{HD},n\rightarrow0}/\Lambda_{\rm HD,LTE}} \, .
\end{equation}
In this fit, $\Lambda_{\textrm{HD},n\rightarrow0} \propto n_{\rm HD}n_{\rm HI}$ is the low-density cooling rate from HD-\textsc{H\,i} collisions, and $\Lambda_{\rm HD,LTE} \propto n_{\rm HD}$ is the cooling rate in LTE. Our implementation follows \cite{McGreer2008}, and use the fits from \cite{Lipkova2005}.\footnote{As in \cite{McGreer2008}, the LTE cooling rate is obtained by evaluating eq. (4) of \cite{Lipkova2005} at $n_{\rm H} = 10^6 \, \rm cm^{-3}$.}

\subsection{Carbon and Oxygen Photochemistry \& Cooling}
\label{Carbon chemistry appendix}

\begin{table*}
\centering
\caption{Reactions and corresponding rate coefficients for carbon and oxygen species: \textsc{C\,i}, \textsc{C\,ii}, CO, \textsc{O\,i}, \textsc{O\,ii}, and the pseudo-species CH$_{\rm x}$ and OH$_{\rm x}$. The units are cm$^{3}$ s$^{-1}$ for collisional rate coefficients ($k$). The photoionization and photodissociation rates ($\Gamma$) are discussed and computed in the text. }
\setlength{\tabcolsep}{2pt} 
\begin{tabular}{c c c c}
\hline
\hline
Symbol & Reaction & Rate coefficient$^\dagger$ & Reference for fit   \\

\hline
\hline
\vspace{-10 pt}
\\
\vspace{4 pt}

$\Gamma_{\rm ion}^{\rm CI}$  & $\textrm{C} + h\nu \rightarrow \textrm{C}^+ + e^-$ & See Eq.~(\ref{Gamma ion CI eq}) &  See text \\

$\Gamma_{\rm ion}^{\rm OI}$  & $\textrm{O} + h\nu \rightarrow \textrm{O}^+ + e^-$ & See Eq.~(\ref{Gamma ion OI eq}) &  See text \\

$\Gamma_{\rm diss}^{\rm CH_x}$  & $\textrm{CH}_{\rm x} + h\nu \rightarrow \textrm{C} + \textrm{H}$ & See Eq.~(\ref{Gamma CH}) & See text  \\

$\Gamma_{\rm diss}^{\rm OH_x}$  & $\textrm{OH}_{\rm x} + h\nu \rightarrow \textrm{O} + \textrm{H}$ & See Eq.~(\ref{Gamma OH}) & See text   \\

$\Gamma_{\rm diss}^{\rm CO}$  & $\textrm{CO} + h\nu \rightarrow \textrm{C} + \textrm{O}$ & See Eq.~(\ref{Gamma CO}) & See text   \\

$k_{\rm H_3^+,e}$  & $\textrm{H}^+_3 + e^- \rightarrow \text{H and H}_2$ & $1.3 \times 10^{-6} \, T^{-0.52}$ & \cite{Gong2017}     \\

$k_{\rm CI,e}$  & $\textrm{C} + e^- \rightarrow \textrm{C}^+ + e^- + e^-$ & $6.85 \times 10^{-8}(0.193 + u)^{-1} u^{0.25}e^{-u}$ & \cite{GloverJappsen2007}     \\

$k_{\rm OI,e}$  & $\textrm{O} + e^- \rightarrow \textrm{O}^+ + e^- + e^-$ & $3.59 \times 10^{-8}(0.073 + u)^{-1} u^{0.34}e^{-u}$ & \cite{GloverJappsen2007}     \\

$k_{\rm CII,e}$  & $\textrm{C}^+ + e^- \rightarrow \textrm{C} + h\nu$ & $2.995 \times 10^{-9}/[\alpha (1+\alpha)^{1-\gamma} (1+\beta)^{1+\gamma}] + k_{\rm dr}$ & \cite{Gong2017}    \\

$k_{\rm OII,e}$  & $\textrm{O}^+ + e^- \rightarrow \textrm{O} + h\nu$ & See reference for fit & \cite{GloverJappsen2007}    \\


$k_{\rm CI, H_3^+}$  & $\textrm{C} +\textrm{H}^+_3 \rightarrow \textrm{CH}_{\rm x} + \textrm{H}_2$ & See reference for fit & \cite{Gong2017, Gong2018}     \\

$k_{\rm CII,H_2}^{(1)}$  & $\textrm{C}^+ + \textrm{H}_2 \rightarrow \textrm{CH}_{\rm x} + \textrm{H}$ & $2.31 \times 10^{-13}\,T^{-1.3} e^{-23/T}$ & \cite{Gong2017}     \\

$k_{\rm CII,H_2}^{(2)}$  &  $\textrm{C}^+ + \textrm{H}_2 \rightarrow \textrm{C} + \textrm{H} + \textrm{H}$ & $0.99 \times 10^{-13}\,T^{-1.3} e^{-23/T}$ & \cite{Gong2017}     \\

$k_{\rm HII,OI}$  &  $\textrm{H}^+ + \textrm{O} \rightarrow \textrm{O}^+ + \textrm{H}$ & $(1.1\times10^{-11}\,T^{0.517} + 4 \times 10^{-10} \, T^{0.00669} )e^{-227/T}$ & \cite{Gong2017}   \\

$k_{\rm OII,HI}$  &  $\textrm{O}^+ + \textrm{H} \rightarrow \textrm{H}^+ + \textrm{O}$ & $4.99 \times 10^{-11} \, T^{0.405} + 7.5 \times 10^{-10} \, T^{-0.458}$ & \cite{Gong2017}  \\

$k_{\rm OI, H_3^+}$  & $\textrm{O} +\textrm{H}^+_3 \rightarrow \textrm{OH}_{\rm x} + \textrm{H}_2$ & $1.99 \times 10^{-9} \, T^{-0.190} \, r$ & \cite{Gong2017}    \\

$k_{\rm OII,H_2}^{(1)}$  &  $\textrm{O}^+ + \textrm{H}_2 \rightarrow \textrm{OH}_{\rm x} + \textrm{H}$ & $1.6 \times 10^{-9} \, r$ & \cite{Gong2017}     \\

$k_{\rm OII,H_2}^{(2)}$  &  $\textrm{O}^+ + \textrm{H}_2 \rightarrow \textrm{O} + \textrm{H} + \textrm{H}$ & $1.6 \times 10^{-9} \, (1-r)$ & \cite{Gong2017}    \\

$k_{\rm CH_x,OI}$  & $\textrm{CH}_{\rm x} + \textrm{O} \rightarrow \textrm{CO} + \textrm{H}$ & $7.7 \times 10^{-11}$ & \cite{Gong2017}     \\

$k_{\rm CH_x,HI}$  & $\textrm{CH}_{\rm x} + \textrm{H} \rightarrow \textrm{C} + \textrm{H}_2$ & $2.81 \times 10^{-11}\, T^{0.26}$ & \cite{Gong2017}   \\

$k_{\rm OH_x,CI}$  & $\textrm{OH}_{\rm x} + \textrm{C} \rightarrow \textrm{CO} + \textrm{H}$ & $7.95 \times 10^{-10} \, T^{-0.339} e^{0.108/T}$ & \cite{Gong2017}   \\

$k_{\rm OH_x,OI}$  & $\textrm{OH}_{\rm x} + \textrm{O} \rightarrow \textrm{O} + \textrm{O} + \textrm{H}$ & $3.5 \times 10^{-11}$ & \cite{Gong2017}    \\

\\

\hline
\hline
\end{tabular}
\vspace{1 pt}\\
\raggedright
$^\dagger$: In the collisional ionization rates $k_{\rm CI,e}$ and $k_{\rm OI,e}$, we have defined $u \equiv 11.26/T_{\rm eV}$ and $u \equiv 13.6/T_{\rm eV}$, respectively, with $T_{\rm eV}$ being the temperature in units of eV.  In the fit for the recombination rate coefficient $k_{\rm CII,e}$, we have defined: $\alpha \equiv \sqrt{T/6.67 \times 10^{-3}}$, $\beta \equiv \sqrt{T/1.943 \times 10^{6}}$, $\gamma \equiv  
0.7849 + 0.1597 \, \exp(-49550/T)$, and the full expression for the dielectronic recombination rate coefficient $k_{\rm dr}$ can be found in \cite{Gong2017}. Finally, in the two $k_{\rm OII,H_2}$ rates, $r = 6.0 \times 10
^{-10} \, n_{\rm H_2} / (6.0 \times 10
^{-10}\,n_{\rm H_2} + 5.3 \times 10^{-6}T^{-1/2}\,n_{\rm e})$ is a branching ratio. Finally, we note that we use the correct fit from \cite{Gong2018} for $k_{\rm CI,H_3^+}$, which fixes a typo in \cite{Gong2017}. 
\label{Carbon and oxygen chemistry rates}
\end{table*}

In order to properly model cooling by \textsc{C\,i}, \textsc{C\,ii}, CO, and \textsc{O\,i}, we use a simplified model of carbon and oxygen chemistry to track the non-equilibrium abundances of these species. Simplified models (compared to full PDR codes) of CO chemistry of varying complexity and realism exist in the literature \citep[e.g.][]{Nelson1997, Nelson1999, Glover2012_CO, Gong2017, Khatri2024}. In these models, the formation of CO proceeds via the psuedo-species CH$_{\rm x}$ (representing CH, CH$_2$, CH$^+$, CH$_2^+$, CH$_3^+$) and OH$_{\rm x}$ (representing OH, H$_2$O, OH$^+$, H$_2$O$^+$, H$_3$O$^+$). CH$_{\rm x}$ and OH$_{\rm x}$ are mainly formed via H$_3^+$, which in turn is formed by cosmic ray (or photo-) ionization of H$_2$ \citep[e.g.][]{Nelson1999, Draine2011, Gong2017, Khatri2024}. CH$_{\rm x}$ and OH$_{\rm x}$ can also form via \textsc{C\,ii} and \textsc{O\,ii}, respectively \citep[e.g.][]{GloverJappsen2007, Gong2017}. Since we are mainly interested in low-metallicity environments, we will neglect grain-assisted recombinations, and leave such additions to future work. The relevant reactions are given in Table~\ref{Carbon and oxygen chemistry rates}, and largely follow the implementations in \cite{Gong2017} and \cite{Khatri2024}. Using these rates, the corresponding rate equations become:
\begin{align}
    \Dot{n}_{\rm OI} &=~  \underbrace{(k_{\rm OII,e} n_{\rm e} + k_{\rm OII,HI} n_{\rm HI} + k_{\rm OII,H_2}^{(2)} n_{\rm H_2})}_{\mathcal{C}_{\rm OII \rightarrow OI}} n_{\rm OII} + \Gamma_{\rm diss}^{\rm CO} n_{\rm CO} +  \underbrace{(k_{\rm OH_x,OI} n_{\rm OI} + \Gamma_{\rm diss}^{\rm OH_x})}_{\mathcal{C}_{\rm OH_x \rightarrow OI}}n_{\rm OH_x} \label{OI number ODE}\\ &-~ \underbrace{(\Gamma_{\rm ion}^{\rm OI} + k_{\rm OI,e}n_{\rm e} + k_{\rm HII,OI}n_{\rm HII} + k_{\rm CH_x,OI}n_{\rm CH_x} + k_{\rm OI, H_3^+}n_{\rm H_3^+})}_{\mathcal{D}_{\rm OI}}n_{\rm OI} \nonumber \, , \\ 
    \Dot{n}_{\rm OII} &=~ \underbrace{(\Gamma_{\rm ion}^{\rm OI} + k_{\rm OI,e}n_{\rm e} + k_{\rm HII,OI}n_{\rm HII})}_{\mathcal{C}_{\rm OI \rightarrow OII}} n_{\rm OI} - \underbrace{(k_{\rm OII,e} n_{\rm e} + k_{\rm OII,HI}n_{\rm HI} +  k_{\rm OII,H_2}n_{\rm H_2})}_{\mathcal{D}_{\rm OII}} n_{\rm OII} \, , \\
    \Dot{n}_{\rm CI} &=~ \underbrace{(k_{\rm CII,e}n_{\rm e} + k_{\rm CII,H_2}^{(2)}n_{\rm H_2})}_{\mathcal{C}_{\rm CII \rightarrow CI}} n_{\rm CII} + \underbrace{(k_{\rm CH_x}n_{\rm HI} + \Gamma_{\rm diss}^{\rm CH_x} )}_{\mathcal{C}_{\rm CH_x \rightarrow CI}}n_{\rm CH_x} + \Gamma_{\rm diss}^{\rm CO} n_{\rm CO} \\ &-~ \underbrace{(\Gamma_{\rm ion}^{\rm CI} + k_{\rm CI,e} n_{\rm e} + k_{\rm OH_x,CI}n_{\rm OH_x} + k_{\rm CI,H_3^+}n_{\rm H^+_3})}_{\mathcal{D}_{\rm CI}} n_{\rm CI} \nonumber \, , \\
    \Dot{n}_{\rm CII} &=~ \underbrace{(\Gamma_{\rm ion}^{\rm CI} + k_{\rm CI,e} n_{\rm e})}_{\mathcal{C}_{\rm CI \rightarrow CII}}n_{\rm CI} - \underbrace{(k_{\rm CII}n_{\rm e} + k_{\rm CII,H_2}n_{\rm H_2})}_{\mathcal{D}_{\rm CII}} n_{\rm CII} \, .
\end{align}
Finally, for CO we have:
\begin{align}
    \Dot{n}_{\rm CO} &=~ k_{\rm CH_x,OI}n_{\rm CH_x}n_{\rm OI} + k_{\rm OH_x,CI} n_{\rm CI} n_{\rm OH_x} - \Gamma_{\rm diss}^{\rm CO} n_{\rm CO} \, .
\end{align}
We do not write down the rates for $n_{\rm CH_x}$ and $n_{\rm OH_x}$, since they can be obtained from particle conservation:
\begin{equation}
    n_{\rm OH_x} = n_{\rm O} - n_{\rm OI} - n_{\rm OII} - n_{\rm CO} \, , \quad n_{\rm CH_x} = n_{\rm C} - n_{\rm CI} - n_{\rm CII} - n_{\rm CO} \, . \label{CO number ODE}
\end{equation}
Furthermore, we follow \cite{Khatri2024} and assume that H$_3^+$ is in equilibrium,
\begin{equation}
    n_{\rm H^+_3} = \dfrac{\Gamma_{\rm ion}^{\rm H_2} n_{\rm H_2}}{k_{\rm CI,H^+_3}n_{\rm CI} + k_{\rm OI,H^+_3}n_{\rm OI} + k_{\rm H^+_3,e}n_{\rm e} } \, .\label{nH3plus}
\end{equation}
To solve the non-linear coupled Eqs.~(\ref{OI number ODE})--(\ref{CO number ODE}), we use a linearized semi-implicit BDF1 method. In this method, we first make a predictor update $n \rightarrow \bullet$ for the oxygen-carrying species (\textsc{O\,i}, \textsc{O\,ii}, CO), keeping all non-CO carbon abundances and all coefficients fixed at their values at step $n$.\footnote{Typically, the oxygen reactions are faster, so this leads to better convergence.} Defining $x_{\rm OI/OII} \equiv (n_{\rm OI/OII})/n_{\rm O}$ and $x_{\rm CO} \equiv n_{\rm CO}/n_{\rm C}$, and eliminating OH$_{\rm x}$, we get the linearized BDF1 update:
\begin{equation}
    \dfrac{\boldsymbol{x}_{\rm O}^{\bullet} - \boldsymbol{x}_{\rm O}^{n}}{\Delta t_{\rm sub}} = \boldsymbol{A}_{\rm O} + \boldsymbol{\mathsf{B}}_{\rm O} \boldsymbol{x}_{\rm O}^{\bullet} \, ,  \quad \textrm{or:} \quad  \boldsymbol{x}_{\rm O}^{\bullet} = (\boldsymbol{1} - \Delta t_{\rm sub} \boldsymbol{\mathsf{B}}_{\rm O})^{-1} (\boldsymbol{x}_{\rm O}^n + \Delta t_{\rm sub}\boldsymbol{A}_{\rm O}) \, , \label{Oxygen update 1}
\end{equation}
where $\boldsymbol{x}_{\rm O} \equiv (x_{\rm OI}, x_{\rm OII}, x_{\rm CO})^{\rm T}$, and
\begin{align} 
    \boldsymbol{A}_{\rm O} &=~ (\mathcal{C}_{\rm OH_x \rightarrow OI},\, 0, \, \mathcal{C}_{\rm OH_x \rightarrow CO})^{\rm T} \, , \\
    \boldsymbol{\mathsf{B}}_{\rm O} &=~ \begin{pmatrix} -(\mathcal{D}_{\rm OI} + \mathcal{C}_{\rm OH_x \rightarrow OI}) & \mathcal{C}_{\rm OII \rightarrow OI} -\mathcal{C}_{\rm OH_x \rightarrow OI} & (n_{\rm C}/n_{\rm O})(\Gamma_{\rm diss}^{\rm CO} - \mathcal{C}_{\rm OH_x \rightarrow OI}) \\ \mathcal{C}_{\rm OI \rightarrow OII}  & -\mathcal{D}_{\rm OII} & 0 \\ \mathcal{C}_{\rm OI \rightarrow CO} - \mathcal{C}_{\rm OH_x \rightarrow CO} & - \mathcal{C}_{\rm OH_x \rightarrow CO} & -[\Gamma_{\rm diss}^{\rm CO} + (n_{\rm C}/n_{\rm O})\mathcal{C}_{\rm OH_x \rightarrow CO}]  \end{pmatrix} \, , 
\end{align}
with $\mathcal{C}_{\rm OH_x \rightarrow CO} = k_{\rm OH_x,CI} n_{\rm O} x_{\rm CI}^n$, with $x_{\rm CI/CII} \equiv (n_{\rm CI/CII})/n_{\rm C}$ (here and below). Next, we use the predictor values of (\textsc{O\,i}, \textsc{O\,ii}) to feed into an update of the carbon-bearing species (\textsc{C\,i}, \textsc{C\,ii}, CO) (i.e. including a corrector update for CO). The update for $\boldsymbol{x}_{\rm C} \equiv (x_{\rm CI}, x_{\rm CII}, x_{\rm CO})^{\rm T}$ is given by Eq.~(\ref{Oxygen update 1}), but with:
\begin{align} 
    \boldsymbol{A}_{\rm C} &=~ (\mathcal{C}_{\rm CH_x \rightarrow CI},\, 0, \, \mathcal{C}_{\rm CH_x \rightarrow CO})^{\rm T} \, , \\   
    \boldsymbol{\mathsf{B}}_{\rm C} &=~ \begin{pmatrix} -(\mathcal{D}_{\rm CI}+\mathcal{C}_{\rm CH_x \rightarrow CI}) & \mathcal{C}_{\rm CII \rightarrow CI} - \mathcal{C}_{\rm CH_x \rightarrow CI} & \Gamma_{\rm diss}^{\rm CO} - \mathcal{C}_{\rm CH_x \rightarrow CI} \\ \mathcal{C}_{\rm CI \rightarrow CII}  & -\mathcal{D}_{\rm CII} & 0 \\ \mathcal{C}_{\rm CI \rightarrow CO} - \mathcal{C}_{\rm CH_x \rightarrow CO} & - \mathcal{C}_{\rm CH_x \rightarrow CO} & -(\Gamma_{\rm diss}^{\rm CO} + \mathcal{C}_{\rm CH_x \rightarrow CO})  \end{pmatrix} \, , 
\end{align}
with $\mathcal{C}_{\rm CH_x \rightarrow CO} = k_{\rm CH_x,OI} n_{\rm O} x_{\rm OI}^{\bullet}$. Finally, with the corrected update for $x_{\rm CO}$, we renormalize $x_{\rm OI}$ and $x_{\rm OII}$ according to:\footnote{The renormalization follows from the requirement that $1 = f_{\rm renorm}(x_{\rm OI}^{\bullet} + x_{\rm OII}^{\bullet} + x_{\rm OH_x}^{\bullet}) + (n_{\rm C}/n_{\rm O})x_{\rm CO}^{n+1} = x_{\rm OI}^{n+1} + x_{\rm OII}^{n+1} + x_{\rm OH_x}^{n+1} +  (n_{\rm C}/n_{\rm O})x_{\rm CO}^{n+1}$, and eliminating $x_{\rm OH_x}$ by conservation.}
\begin{equation}
    x_{\rm OI}^{n+1} = f_{\rm renorm} x_{\rm OI}^{\bullet} \, , \quad x_{\rm OII}^{n+1} = f_{\rm renorm} x_{\rm OII}^{\bullet} \, , \quad f_{\rm renorm} = \dfrac{1 - (n_{\rm C}/n_{\rm O}) x_{\rm CO}^{n+1}}{1 - (n_{\rm C}/n_{\rm O}) x_{\rm CO}^{\bullet}} \, .
\end{equation}
The overall update $n \rightarrow n+1$ is only expected to be an accurate representation of a full BDF1 update if the renormalization induces modest changes in $x_{\rm OI}$, $x_{\rm OII}$, and $x_{\rm OH_x}$. In practice, this is controlled by the global photo-thermochemistry subcycling: after each chemistry substep, we reject the update and retry with a smaller time-step if selected key abundances change by more than $10\%$. For the C/O network, these quantities include $x_{\rm CO}$, $x_{\rm CII}$, and $x_{\rm OII}$, except at very low metallicities or high temperatures where the C/O chemistry is dynamically unimportant. We do not impose a $10\%$ rule on every individual reaction channel, since the reactions can be extremely stiff (e.g. because of the rapid charge exchange reactions $k_{\rm HII,OI}$ and $k_{\rm OII,HI}$), while the (semi-)implicit update remains stable.

\begin{figure*}
\centering
\includegraphics[width=0.8\textwidth]{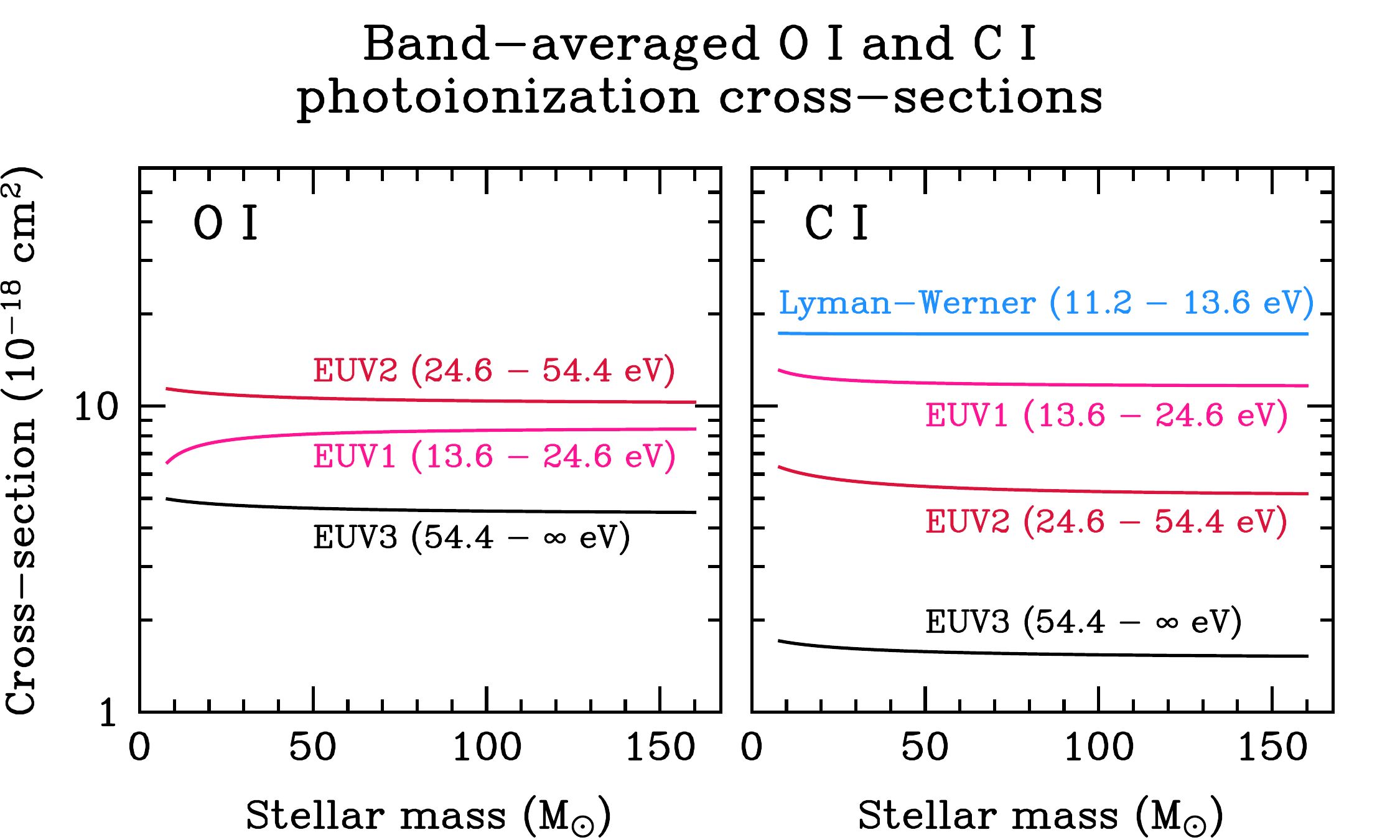}
\caption{The band-averaged \textsc{O\,i} and \textsc{C\,i} photoionization cross-sections for metal-poor $Z_{\star}/Z_{\odot} = 0.01$ stars. Fits for the cross-section were taken from \citet{Verner1996}, and effective temperatures $T_{\rm eff}$ were derived from the fits in \citet{Tanikawa2020}.}
\label{OI and CI cross sec}
\end{figure*}

\begin{figure*}
\centering
\includegraphics[width=1.0\textwidth]{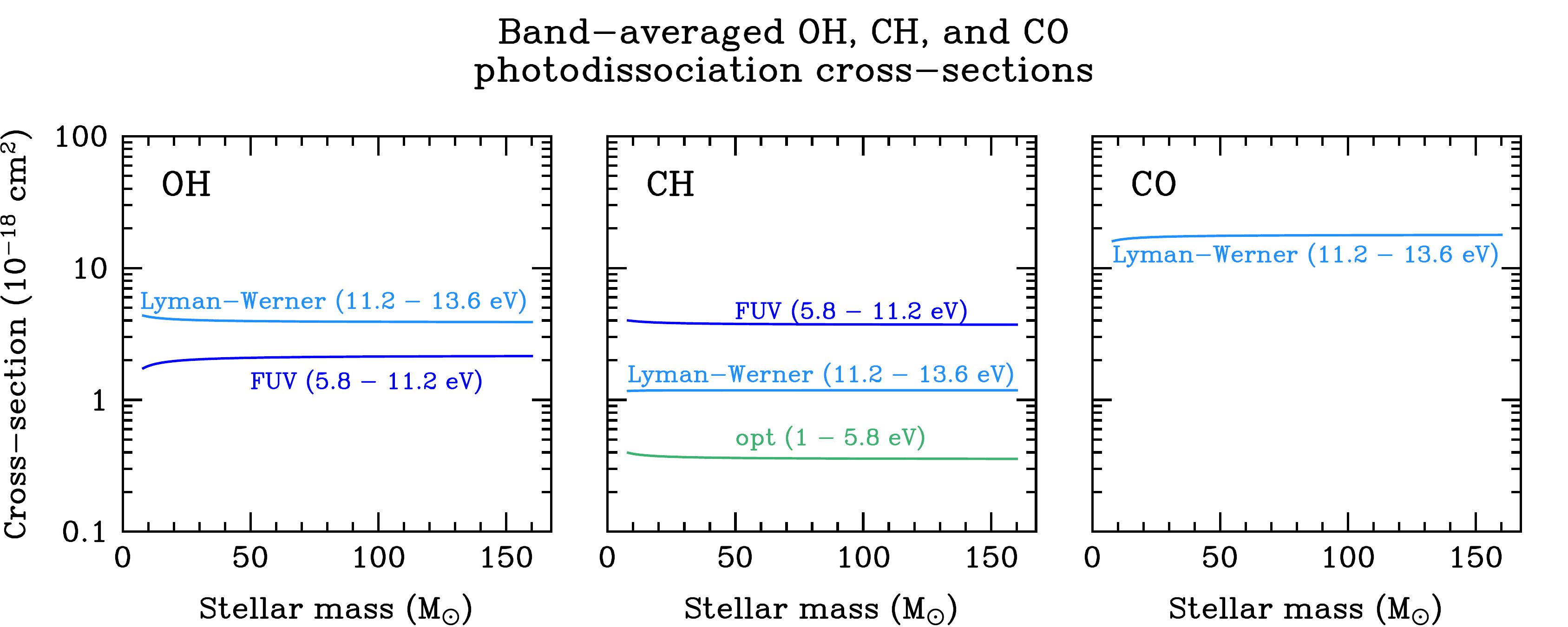}
\caption{The band-averaged OH, CH, and CO photodissociation cross-sections for metal-poor $Z_{\star}/Z_{\odot} = 0.01$ stars. Data for the cross-section were taken from \citet{Heays2017}, and effective temperatures $T_{\rm eff}$ were derived from the fits in \citet{Tanikawa2020}.}
\label{OH, CH, CO cross sec}
\end{figure*}

Next we discuss our implementation of photoionization and photodissociation rates, starting with the former. \textsc{O\,i} and \textsc{C\,i} can be photoionized by photons with $h\nu > 13.62 \, \rm eV$ and $h\nu > 11.26 \, \rm eV$, respectively \citep[][]{Draine2011}. The rates for these processes are therefore:
\begin{align}
    \Gamma_{\rm ion}^{\rm OI} ~&=~ 4\pi \left( \dfrac{J_{\rm EUV1} \sigma_{\rm OI,EUV1}}{E_{\rm EUV1}} + \dfrac{J_{\rm EUV2} \sigma_{\rm OI,EUV2}}{E_{\rm EUV2}} + \dfrac{J_{\rm EUV3} \sigma_{\rm OI,EUV3}}{E_{\rm EUV3}} \right) \, , \label{Gamma ion OI eq}\\ \Gamma_{\rm ion}^{\rm CI} ~&=~ 4\pi \left( \dfrac{J_{\rm LW} \sigma_{\rm CI,LW}}{E_{\rm LW}} + \dfrac{J_{\rm EUV1} \sigma_{\rm CI,EUV1}}{E_{\rm EUV1}} + \dfrac{J_{\rm EUV2} \sigma_{\rm CI,EUV2}}{E_{\rm EUV2}} + \dfrac{J_{\rm EUV3} \sigma_{\rm CI,EUV3}}{E_{\rm EUV3}} \right) \label{Gamma ion CI eq} \, .
\end{align}
We estimate the band-averaged cross-sections using the fits to the energy-dependent cross-sections from \cite{Verner1996}. We further assume black-body spectra with effective temperatures derived from \cite{Tanikawa2020} for zero-age main sequence stars with $Z_{\star}/Z_{\odot} = 0.01$. The result is plotted in Fig.~\ref{OI and CI cross sec}, and is weakly dependent on stellar mass. We therefore adopt the values at $m_{\star} = 30 \, \rm M_{\odot}$:
\begin{align}
    (\sigma_{\rm OI,EUV1}, \, \sigma_{\rm OI,EUV2}, \, \sigma_{\rm OI,EUV3}) ~&=~ (7.86,\,10.8, \, 4.73) \times 10^{-18} \, \rm cm^2 \, , \\ (\sigma_{\rm CI,LW}, \, \sigma_{\rm CI,EUV1}, \, \sigma_{\rm CI,EUV2}, \, \sigma_{\rm CI,EUV3}) ~&=~ ( 17.2, \, 12.1,\,5.68,\,1.61) \times 10^{-18} \, \rm cm^2 \, .
\end{align}
We also compute the photodissociation rates of OH$_{\rm x}$ (represented by OH here), CH$_{\rm x}$ (represented by CH), and CO, from stellar photons and Ly$\alpha$:
\begin{align}
    \Gamma_{\rm diss}^{\rm OH_{\rm x}} ~&=~ 4\pi\left(\dfrac{J_{\rm FUV} \sigma_{\rm OH,FUV}}{E_{\rm FUV}} + \dfrac{J_{\rm LW} \sigma_{\rm OH,LW}}{E_{\rm LW}} \right) +  \dfrac{ c e_{\rm Ly\alpha}\sigma_{\rm OH,Ly\alpha}}{E_{\rm Ly\alpha}} \, , \label{Gamma OH} \\ \Gamma_{\rm diss}^{\rm CH_{\rm x}} ~&=~ 4\pi\left(\dfrac{J_{\rm opt} \sigma_{\rm CH,opt}}{E_{\rm opt}} + \dfrac{J_{\rm FUV} \sigma_{\rm CH,FUV}}{E_{\rm FUV}} + \dfrac{J_{\rm LW} \sigma_{\rm CH,LW}}{E_{\rm LW}} \right) +  \dfrac{ c e_{\rm Ly\alpha}\sigma_{\rm CH,Ly\alpha}}{E_{\rm Ly\alpha}} \, , \label{Gamma CH} \\ \Gamma_{\rm diss}^{\rm CO} ~&=~  \dfrac{4 \pi J_{\rm LW} \sigma_{\rm CO,LW}}{E_{\rm LW}}  \, . \label{Gamma CO}
\end{align}
In the above equations, we have taken into account all bands for which the cross-section is non-zero and non-negligible. We have computed the band-averaged cross-sections in Eqs.~(\ref{Gamma OH})--(\ref{Gamma OH}) using data compiled by \cite{Heays2017}, and assuming black-body spectra for zero-age main sequence stars of metallicity $Z_{\star}/Z_{\odot} = 0.01$ \citep[][]{Tanikawa2020}. The result is shown in Fig.~\ref{OH, CH, CO cross sec}, and seen to be nearly independent of the stellar mass. We therefore adopt the values for $m_{\star} = 30 \, \rm M_{\odot}$ for the stellar RT bands:
\begin{align}
    ( \sigma_{\rm OH,FUV}, \, \sigma_{\rm OH,LW}) ~&=~ ( 2.03,\, 4.02 ) \times 10^{-18} \, \rm cm^2 \, , \\ (\sigma_{\rm CH,opt},\, \sigma_{\rm CH,FUV}, \, \sigma_{\rm CH,LW}) ~&=~ (0.368,\, 3.81, \, 1.18) \times 10^{-18} \, \rm cm^2 \, , \\ \sigma_{\rm CO,LW} ~&=~ 17.3 \times 10^{-18} \, \rm cm^2 \, .
\end{align}
For absorption of Ly$\alpha$, we take $(\sigma_{\rm OH,Ly\alpha},\, \sigma_{\rm CH, Ly\alpha}) = (4.07 \,, 0.05) \times 10^{-18} \, \rm cm^2$ \citep[see table 1 in][]{Heays2017}. We note that the Ly$\alpha$ cross-sections are only accurate to within a factor of $\sim 2$ and $\sim 10$ for OH and CH, respectively \citep[][]{Heays2017}. We model self-shielding of OH$_{\rm x}$, CH$_{\rm x}$, and CO in a highly simplified manner, using the above cross-sections to model absorption by these molecules. This approach is less accurate than interpolations of detailed line-by-line calculations \citep[e.g.][]{Visser2009}, but likely more accurate than the model of \cite{Nelson1997} which neglects self-shielding altogether.

For low-temperature gas, we implement cooling from \textsc{O\,i}, \textsc{C\,i}, \textsc{C\,ii}, and CO, mainly following \cite{Kim2023} and \cite{Deng2024}. For \textsc{O\,i} ($63.18\,\rm \mu m$, $145.5\,\rm \mu m$) and \textsc{C\,i} ($609.13 \, \rm \mu m$, $370.41 \, \rm \mu m$) line cooling, we model the respective atoms as three-level systems, and include collisional rates for $e^-$, \textsc{H\,i}, H$_2$, \textsc{H\,ii}, and He\textsc{\,i}, as compiled in \cite{Draine2011}. Similarly, \textsc{C\,ii} ($157.75 \, \rm \mu m$) cooling is implemented, modelling the system as a two-level atom, taking into account collisions with $e^-$, \textsc{H\,i}, H$_2$, and He\textsc{\,i} \citep[also using rates compiled in][]{Draine2011}. For CO, we model the cooling rate $\Lambda_{\rm CO}$ (in erg s$^{-1}$ cm$^{-3}$) as:
\begin{align}
    \Lambda_{\rm CO}(T,n_{\rm CO},n_{\rm H_2}) = \dfrac{2.16 \times 10^{-27} \, T^{3/2}  n_{\rm H_2} n_{\rm CO} }{1 + (n_{\rm H}/n_{\rm crit,CO})(1 + N_{\rm CO}/N_{\rm CO,crit})} \, ,
\end{align}
where $n_{\rm crit,CO} = 1.9 \times 10^{4} \, T^{1/4} \, \textrm{cm}^{-3}$ and $N_{\rm CO,crit} = 6.66 \times 10^{14} \, T \, \textrm{cm}^{-2}$. The functional form for $\Lambda_{\rm CO}$ is the same as in \cite{Hollenbach1979}, but corrected (by a factor $\sim 2$) in the low-density regime to agree with the fit in \cite{Whitworth2018} (who have the same $T^{3/2}$ scaling).


\bibliography{bibliography}{}
\bibliographystyle{aasjournal}



\end{document}